\documentclass[pof,reprint,showkeys,nofootinbib]{revtex4-1}

\draft % marks overfull lines with a black rule on the right
\usepackage{graphicx}% Include figure files
\usepackage{dcolumn}% Align table columns on decimal point
\usepackage{bm}% bold math
\usepackage[utf8]{inputenc}
\usepackage[T1]{fontenc}
\usepackage{mathptmx}

\usepackage{graphicx}
\usepackage{epstopdf, epsfig}

\usepackage{color,soul}
\usepackage{amssymb}
\usepackage{amsmath}
\usepackage{bm}
\usepackage{multirow}
\usepackage{xcolor}
\usepackage{tikz}
\usepackage{subfig}
\usepackage{subfloat}
\usepackage{multirow}
\usepackage{rotating}
\usepackage{booktabs}
\usepackage{relsize}
\usepackage{nomencl}
\usepackage{etoolbox}
\usepackage{float}
 \usepackage{url}
 \usepackage{hyperref}
 \usepackage{algorithm2e}
\usepackage[export]{adjustbox}
\usepackage{upgreek}
\newcommand{\buf}{\textbf{u}^\mathrm{f}}
\newcommand{\bbf}{\textbf{b}^\mathrm{f}}
\newcommand{\stf}{\bm{\sigma}^{\mathrm{f}}}
\newcommand{\bxf}{\textbf{x}^\mathrm{f}}
\newcommand{\df}{\Omega^\mathrm{f}(t)}
\newcommand{\vf}{\mu^\mathrm{f}}

\begin{document}

% Use the \preprint command to place your local institutional report number 
% on the title page in preprint mode.
% Multiple \preprint commands are allowed.
%\preprint{}

\title{Assessment of unsteady flow predictions using hybrid deep learning based reduced order models} %Title of paper

% repeat the \author .. \affiliation  etc. as needed
% \email, \thanks, \homepage, \altaffiliation all apply to the current author.
% Explanatory text should go in the []'s, 
% actual e-mail address or url should go in the {}'s for \email and \homepage.
% Please use the appropriate macro for the type of information

% \affiliation command applies to all authors since the last \affiliation command. 
% The \affiliation command should follow the other information.

\author{Sandeep Reddy Bukka}
\email{sandeeprb@tcoms.sg}
%\homepage{Your web page}
%\thanks{}
%\altaffiliation{}
\affiliation{Technology Center for Offshore and Marine, Singapore (TCOMS)}

\author{Rachit Gupta}
\email{rachit.gupta@ubc.ca}
\affiliation{University of British Columbia, Vancouver}

\author{Allan Ross Magee}
\email{allan$\_$magee@tcoms.sg}
\affiliation{Technology Center for Offshore and Marine, Singapore (TCOMS)}

\author{Rajeev Kumar Jaiman}
\email{rjaiman@mech.ubc.ca}
\affiliation{University of British Columbia, Vancouver}
% Collaboration name, if desired (requires use of superscriptaddress option in %\documentclass). 
% \noaffiliation is required (may also be used with the \author command).
%\collaboration{}
%\noaffiliation

\date{\today}

\begin{abstract}
In this paper, we present two deep learning-based hybrid data-driven reduced order models for the prediction of unsteady fluid flows. %
These hybrid models rely on recurrent neural networks (RNNs) to evolve low-dimensional states of unsteady flow. The first model projects the high-fidelity time series data from a finite element Navier-Stokes solver to a low-dimensional subspace via proper orthogonal decomposition (POD). The time-dependent coefficients in the POD subspace are propagated by the recurrent net (closed-loop encoder-decoder updates) and mapped to a high-dimensional state via the mean flow field and POD basis vectors. This model is referred as POD-RNN. The second model, referred to as convolution recurrent autoencoder network (CRAN), employs convolutional neural networks (CNN) (instead of POD), as layers of linear kernels with nonlinear activations, to extract low-dimensional features from flow field snapshots. The flattened features are advanced using a recurrent (closed-loop manner) net and up-sampled (transpose convoluted) gradually to high-dimensional snapshots. Two benchmark problems of the flow past a cylinder and flow past a side-by-side cylinder are selected as the test problems to assess the efficacy of these models. For the problem of flow past a single cylinder, the performance of both the models is satisfactory, with CRAN being a bit overkill. However, it completely outperforms the POD-RNN model for a more complicated problem of flow past side-by-side cylinders. Owing to the scalability of CRAN, we  briefly introduce an observer-corrector method for the calculation of integrated pressure force coefficients on the fluid-solid boundary on a reference grid. This reference grid, typically a structured and uniform grid, is used to interpolate scattered high-dimensional field data as snapshot images. These input images are convenient in training CRAN. This motivates us to further explore the application of CRAN models for the prediction of fluid flows. 

\keywords{POD; RNN; CNN; Deep learning; ROM; Autoencoders; Prediction}
\end{abstract}
\maketitle 
%%%%%%%%%%%%%%%%%%%%%%%%%%%%%%%%%%%%%%%%%%%%%%%%%%%%%%%%%%%%%%%%%%%%%%%%%%%%%%%%%%%%%%%%%%%
\section{Introduction}
%%%%%%%%%%%%%%%%%%%%%%%%%%%%%%%%%%%%%%%%%%%%%%%%%%%%%%%%%%%%%%%%%%%%%%%%%%%%%%%%%%%%%%%%%%%%
The fluid flows described by the Navier-Stokes partial differential equations are highly nonlinear and multiscale in nature. While commercial and open-source solvers do a great job in accurate prediction of fluid flows via full-order model (FOM) based on the partial differential equations, the computational costs associated with parametric sweep and dealing with complex problems can lead to unrealistic time scales of computation. 
This has in turn lead to development of several low cost data-driven reduced order models (ROMs), which attempt to replace the high-dimensional (i.e., high-fidelity) solution from the FOM to a low-dimensional (i.e., low-fidelity) solution manifold. The purpose of these low cost models is to accommodate for realistic time scales in computations involving design optimization and real time control. In this work, we are interested in the development of hybrid data-driven ROMs model that can learn the dynamical system well enough to efficiently predict the time series of flow fields using the full-order data.

A large-class of ROMs are projection-based: they seek low-dimensional (coarse grained) representations of an underlying high-dimensional system.  The basic assumption behind these models is that, lower-order representation of higher-dimensional flow field can exist, and it's representation can be identified efficiently within reasonable accuracy. 
% Cite Annual Review of Fluid Mechanics  -- POD is more than 70-year-old technique. Let's not make a big deal out of it.  We should be very brief.
Proper orthogonal decomposition (POD) is one such widely used projection-based techniques for reduced order modeling of fluid flows \cite{berkooz1993proper}. The reason being, the lower-order representations obtained by POD are mathematically optimal for a given dataset.
% Obtaining a low order representation is the first step in building ROM. 
%Galerkin projection (GP) is the most common technique which uses the low order representations to model the flow at future time instants. 
A more extensive work has been reported in the projection-based approach for building ROMs \cite{burkardt2006pod}.  The idea, behind projection based model reduction, is that the Navier-Stokes equations are projected onto the modes obtained primarily by POD and time integration is performed on ODEs obtained by projection, instead of the original PDEs arising from the conservation laws. Although the projection-based approach is computationally efficient, they do not account for local spatial variations in the flow and are known to become unstable under certain conditions even for canonical cases. They are also intrusive in nature which requires extensive modifications to the current state-of-the-art CFD codes.
Projection-based methods have severe computational difficulties when dealing with problems involving dominant flow structures and patterns that propagate over time such as in transport and wave-type phenomena, or convection-dominated  flows \cite{fresca2020comprehensive}.
% Other ROM techniques
Several researchers have also explored non-intrusive approaches based on data-driven techniques. Eigensystem realization algorithm (ERA) is one such method which is purely data-driven and is primarily used for the stability analysis of dynamical systems \cite{yao2017model,reddy2018data,bukka2020stability}. 

% ML in CFD through ROM and data
The ubiquity of machine learning (ML) techniques motivates several researchers in fluid mechanics to implement and adapt in several applications \cite{duraisamy2015new,wang2016physics,tharindu2017}. Deep learning is a subset of ML which refers to the use of highly multilayered neural networks to understand a complex dataset, with the intention of predicting certain features in the dataset. The tremendous success of deep learning \cite{lecun2015deep} in various fields such as imaging, finance, robotics, etc. has prompted its potential application to fluid mechanics problems \cite{wang2018model,mohan2018deep}. Recently, convolutional neural networks (CNNs) \cite{murata2019nonlinear,tharindu2018} were used to develop a novel nonlinear modal decomposition method, which performed superior to the traditional POD. Another aspect is to directly learn the governing equations from higher-order data. This can be of great advantage where the knowledge of the physical laws is hidden. There has been several instances in the literature where ML is applied to learn the hidden governing equations of fluid flow directly from field data  \cite{brunton2016discovering,rudy2019data,raissi2018hidden,champion2019data,choudhury2018developing,long2018pde,pan2018data}.
A relatively new alternative to the traditional neural networks, where the derivative of the hidden state is represented as parameters via neural network, instead of specifying a discrete sequence of hidden layers, is presented in \cite{chen2018neural}. These new families of deep neural network were quickly adapted in the research community as a whole and also specific to fluid mechanics in \cite{maulik2019time}.
%The traditional feedforward neural networks are used in conjunction with convolutional neural networks for the problem of complete flow field prediction for a given aerofoil geometry \cite{sekar2019fast}.

% Different Neural Architectures in FM as such 
 There exist wide variants of deep architectures in the literature, that have been successful in addressing respective flow problems. Deep nets are also a popular choice for real-time fluid simulations in interactive computer graphics. \cite{jeong2015data} describes one such initial attempts. Here, regression forests are employed to approximate the fluid flow equations with the Lagrangian description to predict the position and velocity of the particles. Neural networks are employed in the liquid splash problem by learning the regression of splash formation in \cite{um2018liquid}. A complete data-driven approach based on LSTM networks is proposed in \cite{wiewel2019latent} to infer the temporal evolution of pressure fields coming from the Navier-Stokes solver. \cite{xie2018tempogan} proposed a generative adversial network (GAN) for the problem of reconstructing super resolution smoke flows. Another application of GAN is presented in \cite{kim2019deep}, where the authors developed a novel generative model to synthesize fluid flows from a set of reduced parameters. By the introduction of a novel deformation-aware loss function, the complex behavior of liquids over a parameter range is mapped onto a reduced representation based on space-time deformations in \cite{bonev2017pre}. The accuracy of the deep learning model for the inference of RANS solution is investigated via state-of-the-art U-net architecture with a large number of trained network architectures \cite{thuerey2019}. Bhatnagar \emph{et al} \cite{bhatnagar2019prediction} used CNNs to predict the velocity and pressure field in unseen flow conditions and geometries, for a given pixelated shape of the object. Physics-informed neural networks are proposed in \cite{raissi2019deep} by considering the residual of a PDE as the loss function in the training algorithm. An attempt to develop hybrid models combining the capabilities of POD with deep learning models is also described in \cite{wang2018model}. Carlberg \emph{et al} \cite{carlberg2019recovering} used autoencoders for reconstructing the missing CFD data by using the data saved at few time instances of high-fidelity simulations.

% Addressing Turbulence
% Remove
%Informing turbulence with field data has also received a wider attention. A survey on addressing uncertainties in turbulence (RANS) and improving those in prediction using existing data is reviewed in detail in \cite{duraisamy2019turbulence}. Recently, an attempt to construct a high-resolution flow field from relatively under-resolved turbulent flow field data is described in \cite{fukami2019super}. Maulik \emph{et al} \cite{maulik2019subgrid} used neural networks to propose a data-driven turbulence closure framework which is used for the sub-grid modeling of kraichnan turbulence. Radial basis feed forward neural nets are used in modeling the turbulence of subsonic flows around NACA0012 airfoils \cite{zhu2019machine}. The perturbations in eigenvectors of RANS model are represented via ML based methods in \cite{wu2019representation}. Recurrent Neural Networks (RNNs) \cite{bieker2019deep} are used in devising a novel modal predictive control framework which exploits low-rank features inherent in a fluid flow problem and predict the quantities relevant to the control of flow.

% Note on Koopman theory
In addition to applications specific to fluid dynamics, a more general idea of ROM in nonlinear data-driven dynamics can be understood using ERA, DMD (Dynamic Mode Decomposition) and ML algorithms. In that regard, Koopman theory has gained a lot of traction. It is mainly used for linear control \cite{korda2018linear,peitz2019koopman} and modal decomposition \cite{liu2018decomposition,mezic2013analysis}. An excellent analogy between DMD with Koopman theory is presented in \cite{korda2018}. Marko \emph{et al} \cite{marko2012} surveyed the applications of Koopman theory for nonlinear dynamical systems. Building on the mathematical framework of Koopman theory, several other works that deploy deep learning-based models for nonlinear dynamical systems are \cite{pan2019physics,pan2018long,yeo2019deep,otto2019linearly,lusch2018deep}. 

% Conclusion starts 
Traditional model reduction in fluid mechanics is hence not very different from ML. The motive in either is to build low-dimensional models. The latter has grown from the computer science community with a focus on creating low-dimensional models from black-box data streams. Owing to their non-intrusive nature, they have advantages in adaptability and scalability. Sensors in smart autonomous systems (SASs) act as black-box data streams and provide huge volumes of data. Therefore the marriage of ML with traditional model reduction techniques such as POD seems like a natural evolutionary process. The advantages of both ML and POD, such as scalability and physics-based modeling, can be harnessed via such hybrid ROMs. The role of POD in such a hybrid model is to calculate low-dimensional projections/features, which can be evolved in time via RNNs. It is common knowledge now that CNNs are very good at extracting features. In the recent work of Miyanawala \emph{et al.} \cite{miyanawala2018low}, the authors showcased the advantages of CNN (as a data-driven ROM) for feature extraction in unsteady wake flow patterns. In the quest for more accurate ROM, the concept of autoencoders which has attained great success in image tracking and scene labeling is explored in a series of papers  \cite{lee2018model,erichson2019physics}. A more complex variant of the autoencoder is the convolutional recurrent autoencoder network (CRAN) which is obtained by the integration of CNN with RNN. Recently such models are applied for benchmark fluid flow problem of lid-driven cavity \cite{gonzalez2018deep}. However, the applicability of hybrid ROMs such as POD-RNN and CRAN for the flow past bluff bodies with massively separated flows is yet to be tested and there is no literature providing such analysis to the best of author's knowledge.

The target of the reduced-order models is to process the data coming from sensors and make relevant predictions as required by a particular physical system. To be relevant and be useful for the current automation needs of the industry,
further developments of ROMs must consider the strength of large volumes of data combined with the latest state-of-the-art ML algorithms. The current work is focused on that goal for the development of efficient and reliable ROMs for unsteady fluid flow and fluid-structure interaction.

There appears to be a major gap in the literature in terms of the large applicability of hybrid data-driven ROMs such as POD-RNN and CRAN for fluid-structure interaction problems. The novelty of work lies in testing the ability of hybrid data-driven reduced-order models such as POD-RNN and convolutional recurrent autoencoder networks (CRANs) for the nonlinear unsteady problems of the flow past bluff bodies. We present comprehensive procedures for network architectures and training of POD-RNN and CRAN. For the CRAN, the LSTM is trained to construct the reduced dynamics through the autoencoder process.   Both DL-based ROMs are purely data-driven and rely on a set of FOM snapshots of the Navier-Stokes equations, instead of making an attempt to replace the high-fidelity FOM via deep neural networks.

The article is organized as follows. Section \ref{HDM} deals with the formulation of a high dimensional model that is used to generate the data followed by a mathematical depiction of forward vs inverse problem approach in section \ref{FvI}. The methodology of model reduction via POD-RNN is presented in section \ref{pod-rnn} along with the idea of recurrent neural networks. The mathematical equivalence of autoencoders and POD is discussed in section \ref{AE-POD-me}. Next, the convolutional recurrent autoencoder models are outlined in section \ref{CRAN} with the unsupervised training strategy. The complete applications of these hybrid deep learning models for flow past a cylinder and flow past side-by-side cylinders are discussed in section \ref{APP1STAT} and \ref{APP2SBS} respectively. Finally, concluding remarks are provided in section \ref{conclusion}.  
%%%%%%%%%%%%%%%%%%%%%%%%%%%%%%%%%%%%%%%%%%%%%%%%%%%%%%%%%%%%%%%%%%%%%%%%%%%%%%%%%%%%%%%%%%%%%
\section{High-dimensional  model}\label{HDM}
%%%%%%%%%%%%%%%%%%%%%%%%%%%%%%%%%%%%%%%%%%%%%%%%%%%%%%%%%%%%%%%%%%%%%%%%%%%%%%%%%%%%%%%%%%%%%

A numerical scheme implementing Petrov-Galerkin finite-element and semi-discrete time stepping are adopted in the present work \cite{jaiman2016partitioned,jaiman2016stable}. Let $\Omega^{f}(t) \subset \mathbb{R}^{d}$ be an Eulerian fluid domain at time $t$, where $d$ is the space dimension. The motion of the imcompressible viscous fluid in $\Omega^{f}(t)$ is governed by the incompressible Navier-Stokes equations
% equations with Dirichlet and neumann bc
\begin{align}
\rho^\mathrm{f}\left(\frac{\partial \buf}{\partial t}+\buf\cdot\bm{\nabla}\buf\right)& = \bm{\nabla}\cdot\stf+\bbf \quad \text{on} \quad \Omega^{f}(t), \label{eq1}\\ 
\bm{\nabla}\cdot\buf& = 0 \quad \text{on}\quad \df, \label{eq2}
\end{align}
with the boundary conditions 
\begin{align}
\buf&= \buf_{D} \quad \forall \bxf \in \Gamma^{f}_{D}(t), \label{eq3}\\ 
\stf\cdot\textbf{n}^{\mathrm{f}} &= \textbf{h}^{\mathrm{f}} \quad \forall \bxf \in \Gamma^{f}_{N}(t), \label{eq4} \\
\buf&= \buf_{0} \quad \text{on} \quad \Omega^{f}(0), \label{eq5}
\end{align}
where $\buf = \buf(\textbf{x},t)$ represents the fluid velocity. $\buf_{D}$ and $\textbf{h}^{\mathrm{f}}$ are the Dirichlet and Neumann boundary condition on $\Gamma^{f}_{D}(t)$ and $\Gamma^{f}_{N}(t)$ respectively. Note that $\textbf{n}^{\mathrm{f}}$ is the unit normal on $\Gamma^{f}_{N}(t)$ and $\buf_{0}$ is the initial condition. In Eq. (\ref{eq1}) the partial time derivative is with respect to the Eulerian referential coordinate system. 
Here $\bbf$ represents the body force per unit mass and $\stf$ is the Cauchy stress tensor for a Newtonian fluid which is defined as
\begin{equation}
\stf = -\textit{p}\textbf{I}+\vf\left(\bm{\nabla}\buf+(\bm{\nabla}\buf)^{T}\right),
\end{equation}
where \textit{p}, $\vf$ and \textbf{I} are the hydrodynamic pressure, the dynamic viscosity of the fluid, and the identity tensor, respectively. A rigid-body structure submerged in the fluid experiences unsteady fluid forces and consequently may undergo flow-induced vibrations. In this study, we focus on static bodies and hence do not model the solid movement. The fluid-solid boundary ($\Gamma^{fs}(t)$) is modeled as a no-slip Dirichlet boundary condition Eq. (\ref{eq3})  with $\buf_{D} = 0$. The fluid force along the fluid-solid boundary is computed by integrating the surface traction, from the Cauchy stress tensor, over the first boundary layer elements on the fluid-solid surface. At a time instant, the drag and lift force coefficients $\textbf{C}_{\mathrm{D}}$ and $\textbf{C}_{\mathrm{L}}$ are given by  
\begin{equation}
\begin{aligned}
    \textbf{C}_{\mathrm{D}} = \frac{1}{\frac{1}{2}\rho_{f}U_{\infty}^{2}D} {\int_{\Gamma^{fs}} (\bm{\sigma^{f}}.\mathrm{\mathbf{n}}).\mathrm{\mathbf{n}}_{\mathrm{x}} \mathrm{d}\Gamma },\\
    \textbf{C}_{\mathrm{L}} = \frac{1}{\frac{1}{2}\rho_{f}U_{\infty}^{2}D} {\int_{\Gamma^{fs}} (\bm{\sigma^{f}}.\mathrm{\mathbf{n}}).\mathrm{\mathbf{n}}_{\mathrm{y}} \mathrm{d}\Gamma },
\end{aligned}
\label{force_eqns}
\end{equation}
where $\rho_{f}$, $U_{\infty}$ and $D$ are the fluid density, reference velocity and reference length respectively. For the unit normal $\mathbf{n}$ of the surface, $\mathbf{n}_{\mathrm{x}}$ and $\mathbf{n}_{\mathrm{y}}$ are the x and y Cartesian components, respectively.  

The weak variational form of Eq.~(\ref{eq1}) is discretized in space using $\mathbb{P}_{2}/\mathbb{P}_{1}$ isoparametric finite elements for the fluid velocity and pressure. The second-order backward-differencing scheme is used for the time discretization of the Navier-Stokes system \cite{liu2014stable}. A partitioned staggered scheme is considered for the full-order simulations of fluid-structure interaction \cite{jaiman2011transient}. The above coupled variational formulation completes the presentation of full-order model for high-fidelity simulation. The employed in-house FSI solver has been extensively validated  in \cite{liu2016interaction,mysa2016origin}.
%%%%%%%%%%%%%%%%%%%%%%%%%
\section{Forward vs Inverse problem}\label{FvI}
In the previous section, the numerical methodology for the full order model is outlined. The nonlinear partial differential equations are derived from the principles of conservation and the numerical discretizations are carried out via finite element procedure.  The above approach can also be termed as  \emph{forward problem} for the solution of a physical system.
Given the model differential equations with appropriate initial/boundary conditions, the forward problem by numerical PDE schemes can provide the physical predictions.  
 
The forward problem for a general nonlinear dynamics of a system can be written in an abstract form as
\begin{equation}
\begin{split}
     \frac{d\bf{S}}{dt} &= \bf{F}(\bf{S},\bf{U}),\\
       \bf{Y} &= \bf{G}(\bf{S},\bf{U}),
\end{split}
  \label{eq321}
\end{equation}
where $\bf{S}$ is the state of the system which can be velocity and pressure for the fluid flow problem, $\bf{F}$ contains the information on the evolution of the state, $\bf{U}$ is the input variable to the system, $\bf{Y}$ represents the observable quantities of interest such as forces and $\bf{G}$ is a nonlinear function which maps the state and input of the system to the quantities of the interest. The same above equation results in a nonlinear partial differential equation for the system of fluid flow as described in the previous section. The goal of the forward problem is to compute the function $\bf{F}$ and $\bf{G}$ via principles of conservation and numerical discretizations. The discrete form of the above equation can be written as
\begin{equation}
\begin{split}
    \bf{S}_{n+1} &= \bf{F}(\bf{S}_n,\bf{U}_n),\\
    \bf{Y}_{n} &= \bf{G}(\bf{S}_{n},\bf{U}_n).
\end{split}
    \label{322}
\end{equation}
Another alternative to the above approach is to build the function $\bf{F}$ and $\bf{G}$ via projection-based model reduction, system identification and machine learning methods. This approach is termed as \emph{inverse problem}. 
The inverse problems begin with the available data and aim at estimating
the parameters in the model.
The inference in this inverse problem is purely based on the data and sometimes we might completely ignore the principles of conservation in such methods. The inverse problem aims to construct close approximations $\boldsymbol{\mathcal{F}}$,
 $\boldsymbol{\mathcal{G}}$ to the functions $\bf{F}$, $\bf{G}$ for a given dataset $\boldsymbol{\mathcal{U}} = \{\bf{U}_1, \dots \bf{U}_n\}$, $\boldsymbol{\mathcal{S}} = \{\bf{S}_1,\dots \bf{S}_n\}$ and $\boldsymbol{\mathcal{Y}} = \{\bf{Y}_1,\dots \bf{Y}_n\}$. The general philosophy of the inverse problem is to obtain the functions $\boldsymbol{\mathcal{F}}$, $\boldsymbol{\mathcal{G}}$ by minimizing the loss between the true data and predicted data from the model:
\begin{equation}
\begin{split}
     \|\boldsymbol{\mathcal{F}}(\boldsymbol{\mathcal{S}},\boldsymbol{\mathcal{U}})-\bf{F}(\boldsymbol{\mathcal{S}},\boldsymbol{\mathcal{U}})\| \longrightarrow 0 ,\\
      \|\boldsymbol{\mathcal{G}}(\boldsymbol{\mathcal{S}},\boldsymbol{\mathcal{U}})-\bf{G}(\boldsymbol{\mathcal{S}},\boldsymbol{\mathcal{U}})\| \longrightarrow 0,
\end{split}
\end{equation}
where the data coming from $\bf{F}$, $\bf{G}$ is considered as ground truth data. In practice, the ground truth data is obtained via forward problem/numerical simulations, experiments and field measurements. The ground truth data are obtained from numerical simulations in the current work. The functions  $\boldsymbol{\mathcal{F}}$ and $\boldsymbol{\mathcal{G}}$ represent the approximate (surrogate) reduced models which generate the predicted data. In this paper, we investigate the application of hybrid inverse problems such as POD-RNN and CRAN on canonical problems of flow past a plain cylinder and the flow past side-by-side cylinders. In the upcoming sections, more mathematical details on the methodologies of the inverse problem are presented.
%%%%%%%%%%%%%%%%%%%%%%%%%%%%%%%%%%%%%%%%%%%%%%%%%%%%%%%%%%%%%%%%%%%%%%%%%%%%%%%%%%%%%%%%
%%%%%%%%%%%%%%%%%%%%%%%%%%%%%%%%%%%%%%%%%%%%%%%%%%%%%%%%%%%%%%%%%%%%%%%%%%%%%%%%%%%%%%%%

%%%%%%%%%%%%%%%%%%%%%%%%%%%%%%%%%%%%%%%%%%%%%%%%%%%%%%%%%%%%%%%%%%%%%%%%%%%%%%%%%%%%%%%%%
\section{Model reduction via POD-RNN} \label{pod-rnn}
%%%%%%%%%%%%%%%%%%%%%%%%%%%%%%%%%%%%%%%%%%%%%%%%%%%%%%%%%%%%%%%%%%%%%%%%%%%%%%%%%%%%%%%%%
In the current section, we present a data-driven approach to construct a reduced-order model (ROM) for the unsteady flow field prediction and fluid-structure interaction \footnote{The majority of the content in this section has been presented in our previous work in \cite{reddy2019data,bukka2019data}, It is outlined here again for the sake of completeness}
The application of machine learning in reduced-order modeling of the dynamics of fluid flow has garnered great interest among the research community. They are likewise utilized as a way to show the transient development of the low-dimensional projections obtained from subspace estimation of the full state. As of late, AI procedures are utilized to build start to finish models that learn both the low dimensional projections and their transient advancement. The current work was propelled and along these lines has significant likenesses to past work as far as learning the low-dimensional portrayals and furthermore their transient advancement.

The main idea in these relatively new approaches and traditional projection-based model reduction in general is the following dual-step process:
\begin{itemize}
    \item The calculation of a low-dimensional subspace $\boldsymbol{\Phi}$ embedded in 
    $\mathbb{R}^{N}$ on which majority of the data is represented. This yields, in some sense, an ideal low-dimensional representations $\textbf{A} = f(\textbf{S})$ of the data $\textbf{S}$ in terms of the intrinsic coordinates on $\boldsymbol{\Phi}$, and
    \item The identification of a transient model which effectively evolves the low dimensional representation $\textbf{A}$ on the manifold $\boldsymbol{\Phi}$.

\end{itemize}

\begin{figure}
    \centering
    \includegraphics[width = 0.45\textwidth]{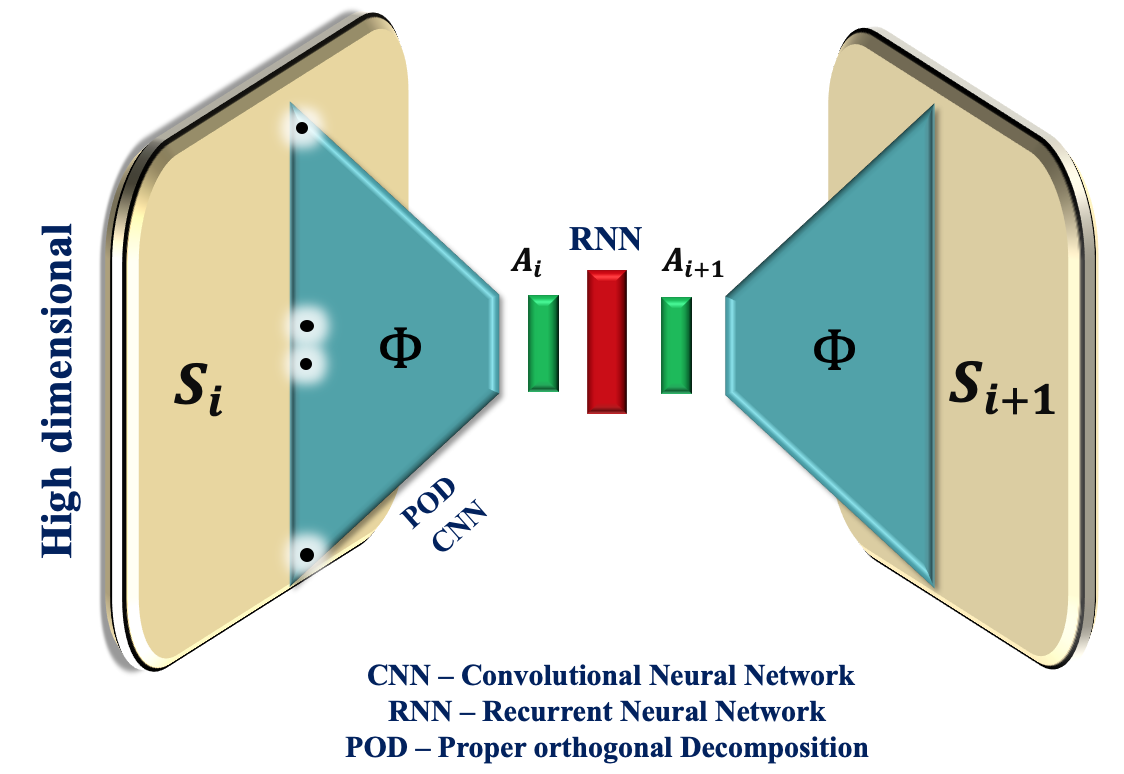}
    \caption{Data-driven reduced order model for prediction of flow field}
    \label{ddframework}
\end{figure}

Fig. \ref{ddframework} depicts the overall framework of the data-driven reduced-order model based on the above process.
%Sandeep: Not sure it's true! Please back it up
The customary methodology of model reduction by means of projection-based technique comes up short for situations where fundamental equations don't exist.
The methodology of demonstrating the transient advancement of the low dimensional representations of the state variable straightforwardly has an extraordinary preferred position for the above situations. A straightforward yet resourceful model for such a methodology is to show the advancement of the low-dimensional representations, acquired for instance through POD, utilizing a RNN
\begin{equation}
    \textbf{A}^{n+1} = \boldsymbol{\mathcal{H}}(\textbf{A}^{n}),
    \label{eq615}
\end{equation}
where the representation vector $\textbf{A}\in \mathbb{R}^{N_{A}}$, is of much smaller size than
the state from which it is learned. This technique has previously been
investigated with regards to low dimensional modeling where $\textbf{A}$ is obtained through POD 
\cite{kani2017dr,wang2018model}, and in the more general case where Eq. (\ref{eq615}) may model the dynamic conduct of unpredictable \cite{ogunmolu2016nonlinear} or turbulent systems \cite{yeo2017model}. In the current section, we first explore the POD-RNN approach of model reduction.
This proposed approach relies on two components:
\begin{itemize}
    \item A projection of the high-dimensional data from the Navier-Stokes equations to a low-dimensional subspace using the proper orthogonal decomposition (POD) and 
    \item Integration of the low-dimensional model with the recurrent neural networks for temporal evolution.
\end{itemize}

For the hybrid ROM formulation, we consider long short-term memory network which is a special variant of recurrent neural networks. The mathematical structure of recurrent neural networks embodies a nonlinear state-space form of the underlying dynamical behavior. This particular attribute of an RNN makes it suitable for nonlinear unsteady flow problems. In the proposed hybrid RNN method, the spatial and temporal features of the unsteady flow system are captured separately. Time-invariant modes obtained by low-order projection embodies the spatial features of the flow field, while the temporal behavior of the corresponding modal coefficients is learned via recurrent neural networks. 
%With regard to a practical marine/offshore engineering demonstration, we have applied and examined the reliability of the proposed data-driven framework for the predictions of vortex-induced vibrations of a flexible offshore riser at high Reynolds number. 

The purpose of the reduced-order model is to predict the unknown future flow fields efficiently without solving the governing differential equations. At first, proper orthogonal decomposition is used for estimating the low-dimensional representation of the flow field. Time-invariant spatial modes obtained from the POD capture the dominant flow patterns. The temporal evolution of these modes is contained in the modal coefficients. Recurrent neural networks are then employed to learn the temporal behavior of the dominant modes obtained from POD. The coupling of traditional POD and the state-of-the-art RNN makes this approach very effective. 

%%%%%%%%%%%%%%%%%%%%%%
\subsection{Proposed architecture} \label{pod-rnn-architecture}
%%%%%%%%%%%%%%%%%%%%%%
The essential and initial component of the POD-RNN model requires a high-fidelity series of snapshots of the flow field which can be obtained by full-order simulations, experiments or field measurements.
Snapshot POD is used to extract the most significant modes and their corresponding POD coefficients. 

\begin{figure}
    \centering
    \includegraphics[width = 0.45\textwidth]{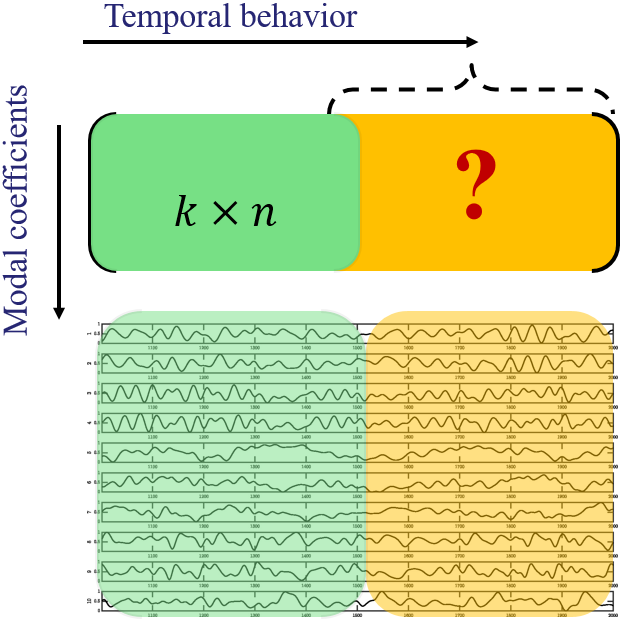}
    \caption{Illustration of prediction of modal coefficients}
    \label{modalpred}
\end{figure}

Let $\boldsymbol{\mathcal{S}} = \left\lbrace\textbf{S}_{1}\;\textbf{S}_{2}\dots\;\textbf{S}_{n}\;\right\rbrace \in \mathbb{R}^{m\times n}$ be the flow field data set where $ \textbf{S}_{i} \in \mathbb{R}^{m}$ is the flow field snapshot at time $t_{i}$ and $n$ is the number of snapshots. $m$ is the number of data points, for example, number of probes in an experiment or field measurements or the number of mesh nodes in a numerical simulation. The target is to predict the future flow fields: $\textbf{S}_{n+1}, \textbf{S}_{n+2}, \dots $ using the data set $\boldsymbol{\mathcal{S}}$. The proposed POD-RNN technique can be divided into two main steps, which are as follows:

\begin{itemize}
\item \textit{Generate the proper orthogonal decomposition (POD) basis for the data set $\boldsymbol{\mathcal{S}}$}

Using the $n$ snapshots given by $\boldsymbol{\mathcal{S}}$, the mean field $(\bar{\textbf{S}} \in \mathbb{R}^{m})$, the POD modes $(\boldsymbol{\Phi} \in \mathbb{R}^{m\times k})$ and the time dependent POD coefficient can be calculated such that

\begin{equation}
\textbf{S}_{i} \approx \bar{\textbf{S}} + \boldsymbol{\Phi} \textbf{A}_i, \quad i = 1,2,\dots,n
\end{equation}
\begin{equation}
     \tilde{\textbf{S}}_{i} \approx  \boldsymbol{\Phi} \textbf{A}_i, \quad i = 1,2,\dots,n
     \label{podrnn1}
\end{equation}

where $ \tilde{\textbf{S}}_{i}$ is the fluctuation matrix which is obtained by removing the mean $\bar{\textbf{S}}$ from instantaneous values $\textbf{S}_{i}$. 
The POD modes are extracted using the eigenvalues $\Lambda_{k\times k} = \mathrm{diag}[\lambda_{1},\lambda_{2},\dots,\lambda_{k}]$ and the eigenvectors $\mathbf{\mathcal{W}} = [w_{1},w_{2},\dots,w_{k}]$ of the covariance matrix $\tilde{\textbf{S}}^{T}\tilde{\textbf{S}} \in \mathbb{R}^{k\times k}$ given by

\begin{equation}
\tilde{\textbf{S}}^{T}\tilde{\textbf{S}}w_{j} = \lambda_{j}w_{j}.
\end{equation}
Here, the POD modes $\boldsymbol{\Phi} = [v_{1},v_{2},\dots,v_{k}]$ are related to $\bm{\Lambda}$ and $\bm{\mathcal{W}}$ by
\begin{equation}
\boldsymbol{\Phi}  = \tilde{\textbf{S}}\bm{\mathcal{W}}\bm{\Lambda}^{-1/2},
\end{equation}
where $\mathbf{A}_{i} = [a_{i1} a_{i2} \dots a_{ik}] \in \mathbb{R}^k$ are the time coefficients of the first $k$ significant modes for the time $t_{i}$. The temporal coefficients of the linear combination are determined by the $\textit{L}^{2}$ inner product between the fluctuation matrix and the POD modes as follows
\begin{equation}
\mathbf{A}_{i}  = \left\langle \tilde{\mathbf{S}}_{i},\boldsymbol{\Phi} \right\rangle.
\end{equation}
The order $k$ can be estimated via the mode energy distribution. The order of the flow field is reduced from $m$ to $k$ via POD. The next step is to predict the matrix $A_{n+1}$. Fig. \ref{POD picture} illustrates the mode extraction from a set of snapshots resulting in modes and their corresponding coefficients with temporal behaviour. 

\item \textit{Train the recurrent neural network}

\begin{figure}
    \centering
    \includegraphics[width = 0.45\textwidth]{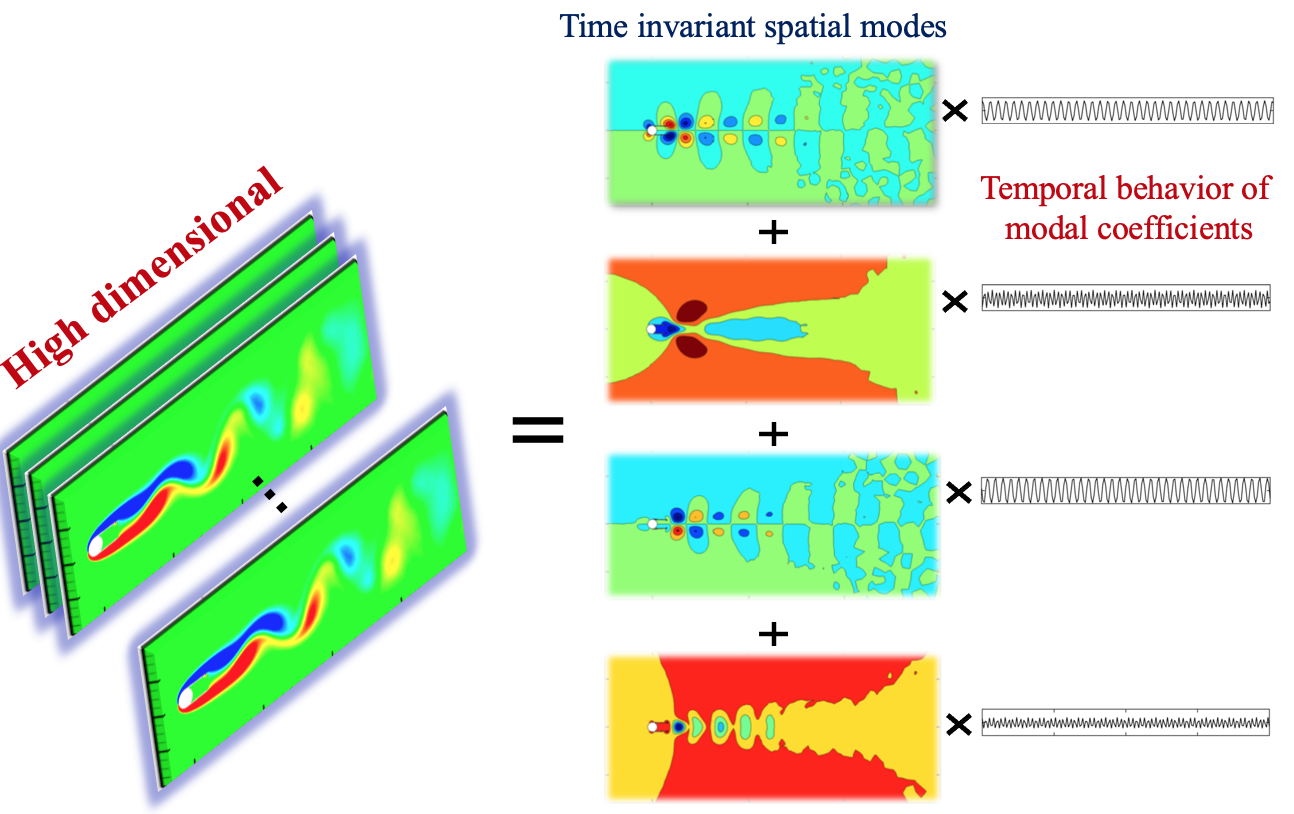}
    \caption{Illustration of POD: Mode extraction of high dimensional data}
    \label{POD picture}
\end{figure}{}
Recurrent neural networks (RNNs) are employed to learn the temporal behaviour of the POD modes. They are generally used in sequence-to-sequence prediction problems. The architecture of RNNs makes it feasible to predict multiple time steps in future at a time. More details on recurrent neural networks are outlined in the upcoming discussions in this section. The purpose of an RNN is to learn a nonlinear function $\boldsymbol{\mathcal{H}}$,  which maps the evolution of modal coefficients, i.e.
\begin{equation}
[\mathbf{A}_{n+s},\mathbf{A}_{n+s-1},\dots,\mathbf{A}_{n+1}] = \boldsymbol{\mathcal{H}}(\mathbf{A}_n,\mathbf{A}_{n-1},\dots,\mathbf{A}_{n-p}),
\label{podrnn2}
\end{equation} 
where $s$ ans $p$ are the output and input sizes of recurrent neural network. Fig. \ref{modalpred} illustrates the prediction of modal coefficients. 
\end{itemize}

Combining Eqs. (\ref{podrnn1}) and (\ref{podrnn2}), we obtain a relationship of a hybrid POD-RNN model in an abstract form
\begin{equation}
     \tilde{\textbf{S}}_{i+1} = \boldsymbol{\Phi} \boldsymbol{\mathcal{H}}(\mathbf{A}_i). 
\end{equation}
The time series of POD modal coefficients are divided into the training and testing sets. The network coefficients are obtained by following the standard learning procedure starting from the training and testing. 

The problem of predicting the evolution of flow-field is now translated to the problem of tracking the temporal evolution of POD modal coefficients. This can be categorized under sequence modeling problems in machine learning. Persistence and retention of the information are required by neural networks in dealing with sequence prediction problems and therefore differ from other canonical learning problems in machine learning.  Traditional neural networks do not have a mechanism of persistence and retention of information. Recurrent neural networks are specifically designed to address the issue of retention of information while training and making physical predictions. The internal state of the model is preserved and propagated through the addition of a new dimension in recurrent neural networks.  Although recurrent neural networks have been successful, several stability issues have been reported. Vanishing gradient is the most common among them. They also cannot learn long term dependencies in the data.

To address the issue of long-term dependency in the data, long short-term memory  (LSTM) networks are utilized for the present work. The LSTM networks are designed with the default characteristic of retaining the information for longer periods and be able to use them for prediction depending on its relevance and context. For example, they can be shown one observation at a time from a sequence and can learn what observations it has seen previously are relevant, and how they can be used to make a prediction \cite{mohan2018deep}.
The fundamental engineering of the LSTM NN is currently illustrated. The LSTM networks are not quite the same as other profound learning structures like convolutional neural organizations (CNN), in that the ordinary LSTM cell contains three gates to control the flow of information. The LSTM manages the progression of information through these gates by specifically including data, eliminating  or letting it through to the following cell.

\begin{figure}
    \centering
    \includegraphics[width = 0.5\textwidth]{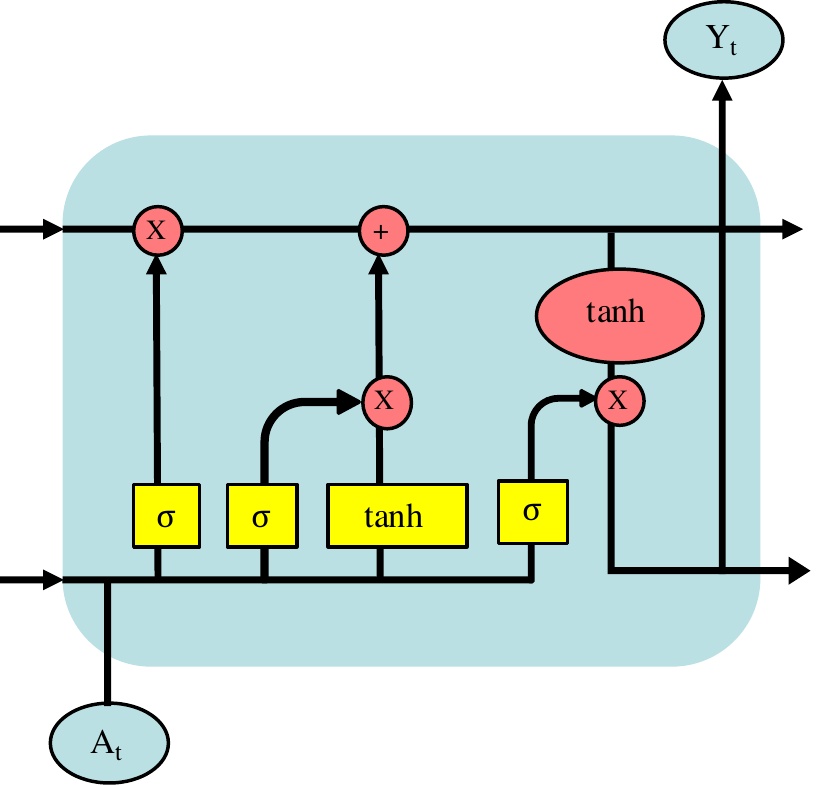}
    \caption{Cell structure of long short-term memory network (LSTM) }
    \label{figlstm}
\end{figure}

The input gate is represented by $\mathbf{i}$, output gate by $\bf{o}$ and forget gate by $\bf{f}$. The cell state is
represented as $\mathbf{X}$ and the cell output is given by $\mathbf{Y}$, while the cell input is denoted as $\mathbf{A}$. The equations to compute its gates and states are as
follows:
\begin{equation}
\begin{split}
    \bf{f}_{t} &= \sigma(\bf{W}_{f}.[\bf{Y}_{t-1},\bf{A}_{t}]+\bf{b}_{f}) \\
\bf{i}_{t} &= \sigma(\bf{W}_{i}.[\bf{Y}_{t-1},\bf{A}_{t}]+\bf{b}_{i}) \\
\tilde{\bf{X}}_{t} &=  \text{tanh}(\bf{W}_{c}.[\bf{Y}_{t-1},\bf{A}_{t}]+\bf{b}_{c})\\
\bf{X}_{t} &= \bf{f}_{t}*\bf{X}_{t-1}+\bf{i}_{i}*\tilde{\bf{X}}_{t} \\
\bf{o}_{t} &= \sigma(\bf{W}_{o}.[\bf{Y}_{t-1},\bf{A}_{t}]+\bf{b}_{o}) \\
\bf{Y}_{t} &= \bf{o}_{t}*\text{tanh}(\bf{X}_{t}),
\end{split}
\label{lstmeq}
\end{equation}
where $\bf{W}$ are the weights for each of the gates and $\tilde{\bf{X}}$ is the updated cell state. These states are proliferated ahead through the model and parameters are refreshed by  backpropagation through time. The forget gate assumes a critical part in dampening the over-fitting by 
not holding all data from the past time steps. Fig. \ref{figlstm} portrays the schematic of a LSTM cell. 
This plan of gates and particular data control is likewise the key motivation behind why LSTMs don't experience the ill effects of the  vanishing gradient issue which tormented customary RNNs. Thus, LSTMs are an amazing asset to 
model non-stattionary datasets. Further insights regarding the LSTM can be found in \cite{olah2015understanding}.

The nonlinear state-space form of the LSTM cell for the current problem can be written as
\begin{equation}
\begin{aligned}
\bf{X}(k+1) = \bf{F}(\bf{X}(k),\bf{A}(k)) ,\\
\bf{A}(k+1) = \bf{G}(X(k+1)),
\end{aligned}
\label{eq18}
\end{equation}
where $\bf{X}$ is the cell state and $\bf{A}$ is the matrix of modal coefficients, whose time history has to be predicted. $\bf{F}$ and $\bf{G}$ represent the nonlinear functions of LSTM cell as given in Eq.~(\ref{lstmeq}) in a mathematically compact form. The Eq. (\ref{eq18}) shares similar structure with Eq. (\ref{322}), which is obtained as the discrete form of the nonlinear dynamical system. The only difference between them is the output of the model, which in the current case is the future of the variable $\bf{A}$ that should be predicted.

\begin{figure} 
    \centering
    \includegraphics[width = 0.45\textwidth]{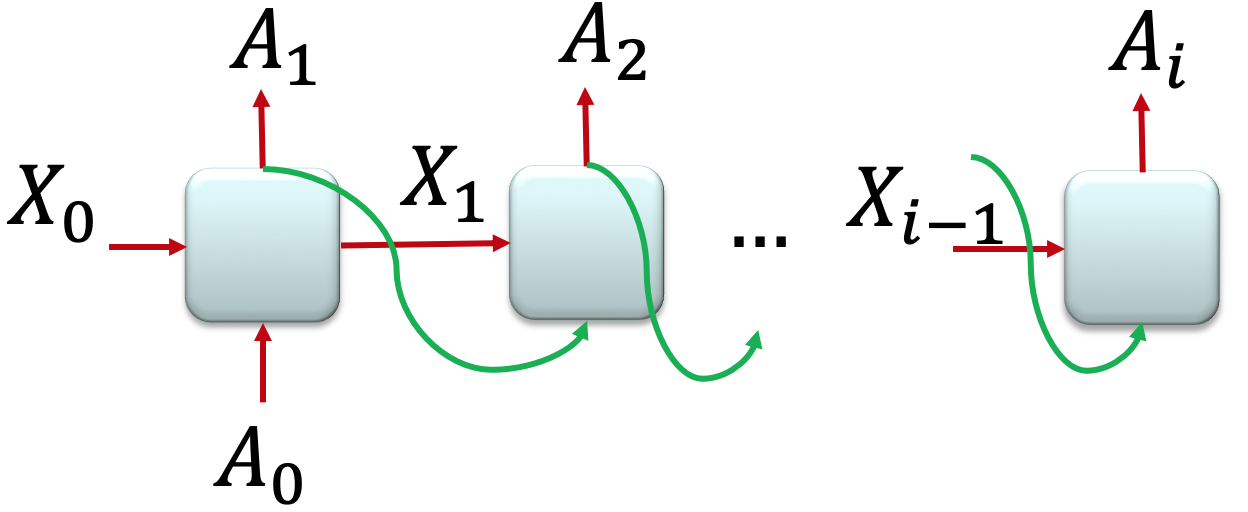}
    \caption{Schematic of closed-loop recurrent neural network}
    \label{closedloop}
\end{figure}

In the current work, we explore two variants of recurrent neural networks
\begin{itemize}
    \item Closed-loop recurrent neural network
    \item Encoder-decoder recurrent neural network
\end{itemize}{}
A schematic of a closed-loop recurrent neural network is depicted in Fig. \ref{closedloop}. It takes one single input and predicts the output and this value is taken as input to the second prediction. In this manner, this architecture forecasts an entire sequence of desirable length with one single input. This approach has a great advantage since it does not require the knowledge of the true data for online prediction. However, it is limited to simple problems such as flow past a plain cylinder. It is ineffective for a more complex problem of the flow past side-by-side cylinders. 

In a quest for more efficient recurrent neural networks which can model the dynamics of a complex system such as flow side-by-side cylinder, the encoder-decoder networks are found to be suitable. The encoder-decoder architectures with long short-term memory networks are widely adopted for both neural machine translation (NMT) and sequence-to-sequence (seq2seq) prediction problems. 
The primary advantage of the encoder-decoder architecture is the ability to handle input and output sequences of text with variable lengths. A single end-to-end model can also be trained directly on the source and target sentences. 
 
\begin{figure} 
\centering
\includegraphics[width =0.45\textwidth]{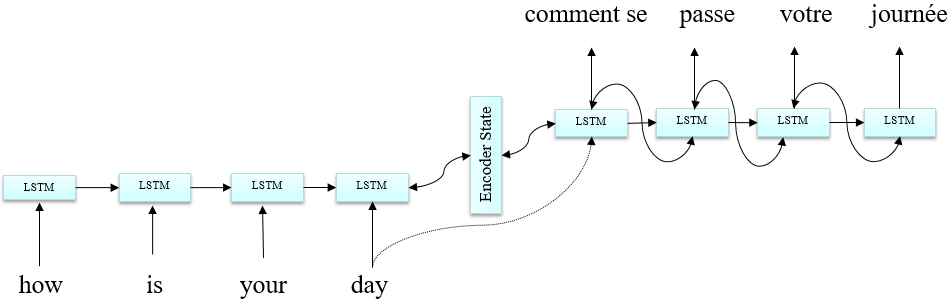}
\caption{Encoder-decoder network for neural machine translation task}
\label{encdncnmt}
\end{figure}

A schematic of the encoder-decoder network for a machine translation example is given in Fig. \ref{encdncnmt}. It takes an input sequence of variable length and encodes the whole information through a series of LSTM cells. There is no direct output in the encoding process. The output is generated via a decoder network through a series of LSTM cells. The input and output time series can have multiple features. The time history of few dominant modes can be used together as a multi-featured data to train the network. In the decoding process, the state and output generated by the previous cell are taken as input for the LSTM cell. 

\begin{figure} 
\centering
\includegraphics[width = 0.45\textwidth]{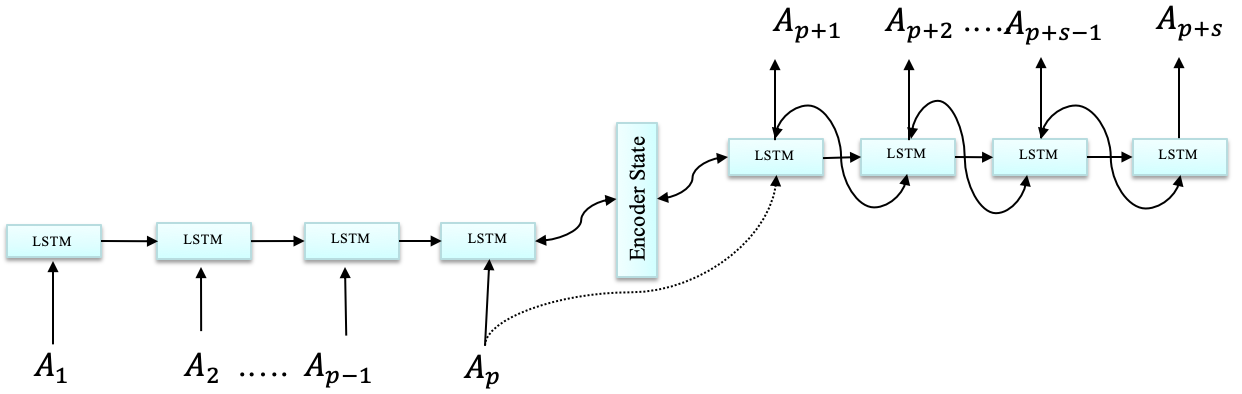}
\caption{Encoder-decoder network for time series prediction}
\label{encdnc}
\end{figure}

In the present work, we use this architecture for time series prediction of the POD modal coefficients (in side-by-side cylinders) as shown in Fig.~\ref{encdnc} where the input sequence will be the known time history of the modal coefficients. The unknown time history of modal coefficients in the future is generated as the output sequence. The output generated by the decoder network can be used as input in a moving window fashion for the next encoding procedure. In this way, a self-sufficient learning model can be generated which can predict the time history of the modal coefficients for a given known time sequence and thereby the prediction of the entire flow field can be achieved. The process of encoding and decoding can be represented mathematically in a compact form using  Eqs.~(\ref{eq18}~-~\ref{eq20}).
  
The encoding process can be represented in the following equation as  
\begin{equation}
\left.\begin{aligned}
\bf{X}(1) &= \bf{F}(\bf{X}(0),\bf{A}(1)) \\
\bf{X}(2) &= \bf{F}(\bf{X}(1),\bf{A}(2)) \\
\vdots \\
\bf{X}(p) &= \bf{F}(\bf{X}(p-1),\bf{A}(p)) \\
\bf{X}_{enc} &= \bf{X}(p)\\
\end{aligned}\right\}, 
\label{eq19}
\end{equation}
where $p$ is the length of the input time history. $\bf{X}_{enc}$ is the encoded state which contains the information pertaining to the input sequence. 
The decoding process can be represented as 
\begin{equation}
\left.\begin{aligned}
\bf{X}(p+1) &= \bf{F}(\bf{X}_{enc},\bf{A}(p)) \\
\bf{A}(p+1) &= \bf{G}(\bf{X}(p+1))\\
\bf{X}(p+2) &= \bf{F}(\bf{X}(p+1),\bf{A}(p+1)) \\
\bf{A}(p+2) &= \bf{G}(\bf{X}(p+2))\\
\vdots \\
\bf{X}(p+s) &= \bf{F}(\bf{X}(p+s-1),\bf{A}(p+s-1)) \\
\bf{A}(p+s) &= \bf{G}(\bf{X}(p+s))\\
\end{aligned}\right\}, 
\label{eq20}
\end{equation}
where $s$ is the length of the output sequence to be predicted. 

%%%%%%%%%%%%%%%%%%%%%%%%%%%%%%%%%%%%%%%%%%%%%%%%%%%%%%%%%%%%%%%%%%%%%%%%%%%%%%%%%%%%%%
\section{Autoencoders and POD} \label{AE-POD-me}
%%%%%%%%%%%%%%%%%%%%%%%%%%%%%%%%%%%%%%%%%%%%%%%%%%%%%%%%%%%%%%%%%%%%%%%%%%%%%%%%%%%%%%
In the section, we attempt to establish a mathematical equivalence between the idea of autoencoders and proper orthogonal decomposition. This particular similarity between the above two methods motivated us to choose convolutional recurrent autoencoder networks as the next natural avenue to build our data-driven reduced order model. 

The manifold hypothesis states that any set of real-world high-dimensional data is spanned by an optimal subspace represented by $\boldsymbol{\Phi}$ which is embedded in $\mathbb{R}^{N}$, where $N$ is large. The above hypothesis forms the core of the data-driven sciences in terms of dimensionality reduction where an approximate low-dimensional representation is constructed to describe the dynamics of high-dimensional data. This particular aspect has contributed to the widespread application of data-driven sciences throughout multiple areas. POD is one of the major dimensionality reduction techniques in this field due to its simplicity and trivial computational efforts. Be that as it may, it has the significant downside of developing just an optimal linear subspace. This is very critical since information tested from perplexing, true frameworks are as a general rule emphatically nonlinear. A wide assortment of methodologies for more precise demonstrating of $\boldsymbol{\Phi}$  have been created throughout the long term, most including utilizing an interwoven of neighborhood subspace acquired through linearizations or higher-order approximations of the state-space.

An under-complete autoencoder can be termed as a  nonlinear generalization of POD. It consists of a solitary multi-layer encoder network.
\begin{equation}
    \textbf{A} = \mathcal{E}(\textbf{S};\boldsymbol{\theta}_{\mathcal{E}});
    \label{eq551}
\end{equation}
where $\hat{\textbf{S}} \in \mathbb{R}^{N}$
is the system state $\textbf{A}\in \mathbb{R}^{N_{A}}$
is the low dimensional embedding, and 
$N_{A} < N$. A decoder network is then used to reproduce $\textbf{S}$ by
\begin{equation}
    \hat{\textbf{S}} = \mathcal{D}(\textbf{A};\boldsymbol{\theta}_{\mathcal{D}}).
\end{equation}
The variables of the above network are obtained by a training procedure which lowers the expected reconstruction error over all training examples.
\begin{equation}
    \boldsymbol{\theta}^{*}_{\mathcal{E}},\boldsymbol{\theta}^{*}_{\mathcal{D}} = \arg \min_{\boldsymbol{\theta}_{\mathcal{E}},\boldsymbol{\theta}_{\mathcal{D}}} \left[ \boldsymbol{\mathcal{L}}(\hat{\textbf{S}},\textbf{S})\right],
    \label{eq552}
\end{equation}
where
$\boldsymbol{\mathcal{L}}(\hat{\textbf{S}},\textbf{S})$
is some measure of discrepancy between
$\textbf{S}$
and its reconstruction 
$\hat{\textbf{S}}$.
The autoencoder is prevented form learning the identity function by restricting the 
$N_{A} < N$.
 
The parameters of 
$\mathcal{E},\mathcal{D}$
and 
$\boldsymbol{\mathcal{L}}(\hat{\textbf{S}},\textbf{S})$
significantly depend on the problem at hand. Indeed, if one chooses a uni-dimensional encoder and decoder of the form
\begin{equation}
    \begin{aligned}
    \textbf{A} = \textbf{W}_{\mathcal{E}}\textbf{S},\\
    \hat{\textbf{S}} = \textbf{W}_{\mathcal{D}}\textbf{A},
\end{aligned}
\end{equation}
where
$\textbf{W}_{\mathcal{E}} \in \mathbb{R}^{N_{A}\times N}$
and
$\textbf{W}_{\mathcal{D}} \in \mathbb{R}^{N\times N_{A}}$,
then with a squared reconstruction error
\begin{equation}
    \begin{aligned}
        \boldsymbol{\mathcal{L}}(\hat{\textbf{S}},\textbf{S}) &= \|\textbf{S}-\hat{\textbf{S}}\|_{2}^{2}\\
        &= \|\textbf{S}-\textbf{W}\textbf{W}^{T}\textbf{S}\|_{2}^{2},
    \end{aligned}
\end{equation}
the autoencoder will obtain the same embedding as the one represented by the first 
$N_{A}$
POD modes if 
$\textbf{W} = \textbf{W}_{\mathcal{D}} = \textbf{W}_{\mathcal{E}}^{T}$. 
However, without additional constraints on 
$\textbf{W}$
i.e.,
$\textbf{W}^{T}\textbf{W} = \textbf{I}_{N_{A}\times N_{A}}$
the columns of 
$\textbf{W}$
do not conform with the rules of orthonormal basis. 

%%%%%%%%%%%%%%%%%%%%%%%%%%%%%%%%%%%%%%%%%%%%%%%%%%%%%%%%%%%%%%%%%%%%%%%%%%%%%%%%%%%%%%%%%%%%
\section{Model reduction via CRAN}\label{CRAN}
%%%%%%%%%%%%%%%%%%%%%%%%%%%%%%%%%%%%%%%%%%%%%%%%%%%%%%%%%%%%%%%%%%%%%%%%%%%%%%%%%%%%%%%%%%%%

In the previous section, we have introduced an overall framework for data-driven reduced-order modeling \footnote{The majority of the content in this section has been presented in our previous work in \cite{bukka2020deep,bukka2019data}, It is outlined here again for the sake of completeness}. While the POD based model reduction is optimal and can generate physically interpretable modes, it is not good enough for practical problems. In such cases, the number of prominent modes will increase and their computation can pose serious challenges. As it will be shown in the application sections, 27 modes are required to capture the dominant energy of the side-by-side cylinders whereas only 4 modes were required for flow past a plain cylinder. Similarly, it is shown in \cite{miyanawala2019decomposition}, that 123 modes were required for flow past a square cylinder at $Re = 22000$. As a natural consequence, the question of completely bypassing POD in building ROM arises and other alternatives for obtaining low-dimensional features such as CNNs will be explored in this section. In Section \ref{AE-POD-me}, an introduction to the concept of autoencoders from ML community and it's mathematical equivalence with POD in a linear setting is presented. This serves as a motivation to explore an advanced data-driven method based on an overall framework of autoencoders. 

An alternative approach, in addition, to the POD-RNN, and also inspired from the overall framework, is a complete data-driven one, where both the low-dimensional embedding of
the state variable and its transient progression are learned via ML techniques. This
method has been investigated in \cite{otto2019linearly} in which an autoencoder is used to learn a low-dimensional embedding of the high-dimensional state,
\begin{equation}
    \textbf{A} = \mathcal{E}(\textbf{S}),
     \label{eq62}
\end{equation}
where $\textbf{S} \in \mathbb{R}^{N}$ high-dimensional state of the system, $\textbf{A}\in \mathbb{R}^{N_{A}},N_{A}<N$, $\mathcal{E}$ is nonlinear enocoder network and a linear recurrent model $\textbf{K}$ is used to evolve the low-dimensional features,

\begin{equation}
    \textbf{A}^{n+1} = \textbf{K}\textbf{A}^{n},
    \label{eq63}
\end{equation}
where $\textbf{K}\in \mathbb{R}^{N_{A}\times N_{A}}$. This approach was first introduced in the context of
learning a dictionary of functions used in extended dynamic mode decomposition to approximate the Koopman operator of a nonlinear system \cite{li2017extended}.

%Based on the framework in \cite{kani2017dr,otto2019linearly,wang2018model}, \cite{gonzalez2018deep} introduced a deep convolutional autoencoder architecture which provides certain advantages in identifying low-dimensional representations of the input data. Fig. \ref{figcnnrnn} depicts the schematic of the model reduction via Convolutional Recurrent Autoencoder Network (CRAN). Benchmark problems of lid-driven cavity and viscous burgers equation are explored via this approach in \cite{gonzalez2018deep}. In the current work, we extend the capabilities of the convolutional recurrent autoencoder model by applying it on more complex problems of fluid-structure interaction. \cite{gonzalez2018deep} contains more details on the model reduction via convolutional recurrent networks and they are also presented here for completeness of the present work.

\begin{figure}
    \centering
    \includegraphics[width = 0.45\textwidth]{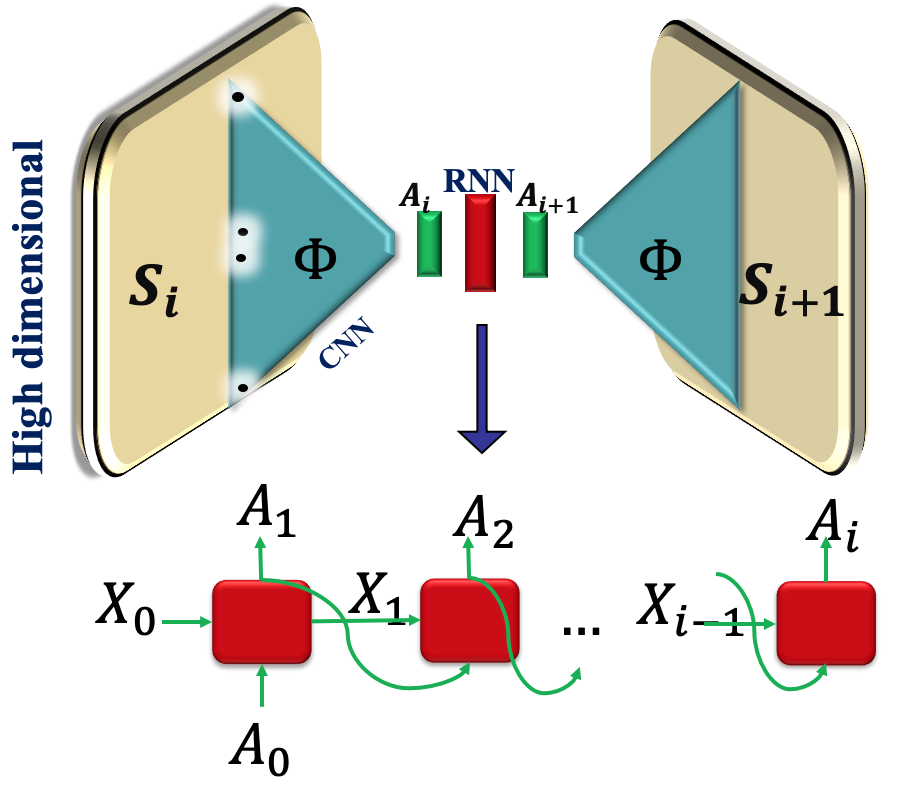}
    \caption{Schematic of convolutional recurrent autoencoder network (CRAN)}
    \label{figcnnrnn}
\end{figure}

Fully-connected autoencoders are used in many applications for dimensionality reduction. However, the high-dimensionality of DNS-level input data can act as a major bottleneck for its application. The high-dimensional input data should be flattened into a 1D array when we consider autoencoders. This might result in the elimination of local spatial relations between the data. In other words, the dominant flow features, often found in FSI simulations might be ignored or might not be captured efficiently. Therefore, the direct application of fully-connected autoencoders on DNS-level input data is generally prohibited.
The current model seeks to extract dominant flow features present in much fluid flow and fluid-structure interaction data through the use of CNNs. Specifically, as opposed to applying a  fully-connected autoencoder implicitly on the  high-dimensional numerical or physical data, it is rather applied on a low dimensional vector obtained from the procedure of feature extraction of a deep CNN acting directly on the high-dimensional data.

\begin{figure}
    \centering
    \includegraphics[width = 0.45\textwidth]{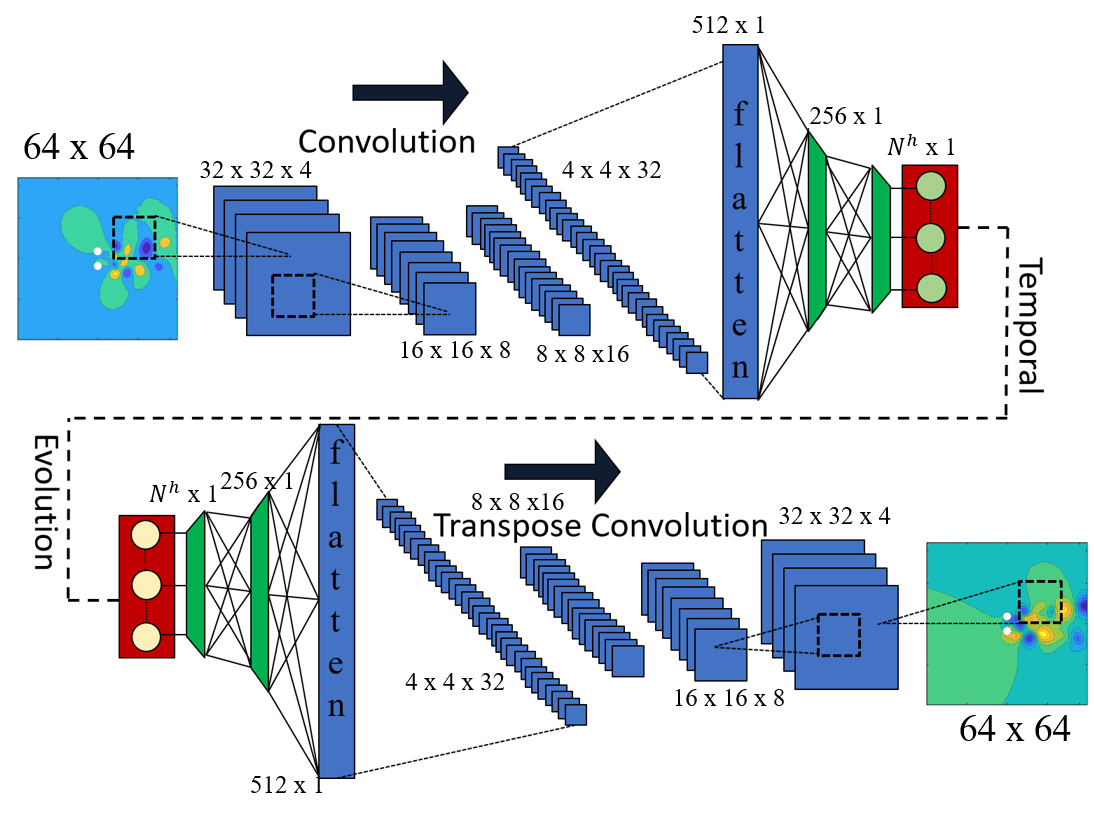}
    \caption{Schematic of 12 layer convolutional recurrent autoencoder model}
    \label{figcnnrnn1}
\end{figure}

\subsection{Proposed architecture}
A 12 layer convolutional autoencoder model is constructed as shown in Fig. \ref{figcnnrnn1}. 
The first 4 layers together form the convolution aspect of the model. The two dimensional input $S \in \mathbb{R}^{N_{x}\times N_{y}}$, with $N_{x} = N_{y} = 64$ is first passed through this convolutional block. All the layers in the convolution block employ a filter bank $\textbf{K}^{f}\in \mathbb{R}^{4 \times 4}$ with the   number of filters in the first layer $f$ increasing from 4 in the first layer to 32 in the fourth layer. A stride of 2 and a padding of 1 is maintained across all layers. 

% let's check the formula 
The process of transpose convolution is carried out in the last four layers of the network. The transpose convolution is not exactly the reverse of regular convolutions but it upsamples the low dimensional representations across four layers gradually to reconstruct the high dimensional state. The output size of the transpose convolutional layer is exactly the inverse of regular convolutional layer and is given by $N_o = (N_i-1)*s + f -2*P$, where $s$ is the stride, $f$ is the filter size, $P$ is the padding, $N_i$ is the input size and $N_o$ is the output size. Table \ref{tabcnnae} outlines the architecture of the convolutional encoder and decoder subgraphs. In this work, we will consider the sigmoid activation function $\sigma(s) = \frac{1}{1+\exp^{-s}}$ for each layer of the CRAN model.

\begin{table}
    \centering
    \begin{tabular}{|c|c|c|c|c|}
    \hline
    Layer & filter size & filters & stride & padding \\ \hline
    1 & $4\times 4$ & 4 & 2 & 1 \\
    2 & $4\times 4$ & 8 & 2 & 1 \\
    3 & $4\times 4$ & 16 & 2 & 1 \\
    4 & $4\times 4$ & 32 & 2 & 1\\
    9 & $4\times 4$ & 16 & 2 & 1\\
    10 & $4\times 4$ & 8 & 2 & 1\\
    11 & $4\times 4$ & 4 & 2 & 1\\
    12 & $4\times 4$ & 1 & 2 & 1\\
    \hline
    \end{tabular}
    \caption{Filter sizes and strides of the layers in convolutional autoencoder network. Layers (1 to 4) correspond to encoder part and layers (9 to 12) correspond to decoder part}
    \label{tabcnnae}
\end{table}
% Sandeep: Let's discuss the below equations and notations

A 4 layer fully connected regular autoencoder with a plain vanilla feed-forward neural network is also used in the architecture. 
The first two layers perform encoding by taking the vectorized form of 32 feature maps from the final layer of convolution block and returns a one-dimensional vector $\textbf{A} \in \mathbb{R}^{N_{A}}$. The last two layers decode the evolved feature vector into a higher dimensional vector which will be reshaped as feature maps and passed into the transpose convolution block. Fig.~\ref{figcnnrnn1} details the whole architecture with the changes in the size of the data as it flows throughout the entire model.  

The key advantage in preferring convolutional autoencoder is its ability in scaling to problems with larger dimensions.
The local coherent spatial structures of the problem at hand are utilized for reducing the model in a nonlinear fashion.

\subsection{Evolution of features}
Now we proceed to model the
evolution of low-dimensional features $\textbf{A}$ in a computationally efficient manner. In terms of an analysis viewpoint, it is beneficial to identify the linear dynamics of $\textbf{A}$. However, we consider a general case of learning feature dynamics in a nonlinear setting.

Consider a set of observations $\{\textbf{s}^{n}\}_{n=0}^{m}$,
$\textbf{s}^{n}\in \mathbb{R}^{N}$ obtained from a CFD simulation or 
through experimental sampling.
An optimal low-dimensional representation $\textbf{A}^{n}\in \mathbb{R}^{N_{A}}$, where 
$N_{A}\ll N$ is obtained for each observation. These low-dimensional representations are obtained via convolutional autoencoder described in the previous section. The model for temporal evolution of these features is constructed in a complete data-driven setting.

LSTM networks are employed to model the evolution of $\textbf{A}$. The details on the LSTM networks are outlined in section \ref{pod-rnn}. 
In a typical machine translation task the size of the hidden states and number of layers is large when compared to the size of the feature vectors. In the current work, a single layer LSTM network is found to be sufficient for evolving the feature vectors $\textbf{A}^{n}$. Note that the development 
of $\textbf{A}$ doesn't need data from the full state $\textbf{S}$ and thereby dodges a  
exorbitant reconstruction 

Instantiating with a pre-computed low-dimensional representation $\textbf{A}^{0}$, one obtains a prediction for the subsequent steps by applying the following iteratively 
\begin{equation}
    \hat{\textbf{A}}^{n+1} = \boldsymbol{\mathcal{H}}(\hat{\textbf{A}}^{n}), \quad n = 1,2,3,\dots
    \label{eq66}
\end{equation}
where $\hat{\textbf{A}}^{1} = \boldsymbol{\mathcal{H}}(\textbf{A}^{0})$ and $\boldsymbol{\mathcal{H}}$ speaks to the activity of Eq. (\ref{lstmeq})  and its subcomponents. A graphical portrayal of this model is outlined 
in Fig. \ref{figcnnrnn}.
%%%%%%%%%%%%
\subsection{Unsupervised training strategy}\label{unsupervised-t-s}
The unsupervised training approach is the key component of the current model. The weights of both 
the convolutional autoencoder network and LSTM recurrent neural network are adjusted in a joint fashion. The main challenge, here, is to prevent overfitting of the CNN and RNN portion of the model. The construction of the training dataset as well as the training and evaluation schematics are presented in this section.

Consider a dataset $\{\textbf{s}^{1},\dots,\textbf{s}^{m}\}$, where 
$\textbf{s}\in \mathbb{R}^{N_{x}\times N_{y}}$ is a 2D snapshot of
some dynamical system (e.g., a pressure field defined on a 2D grid). This dataset is broken up into a set of ${N}_{s}$
finite time training sequences
$\{\textbf{S}^{1},\dots,\textbf{S}^{N_{s}}\}$
where each training sequence
$\textbf{S}^{i}\in \\
\mathbb{R}^{N_{x}\times N_{y}\times N_{t}}$
consists of
$N_{t}$ snapshots.

The fluctuations around the temporal mean are considered as followed generally in the data science community for improving the training.
\begin{equation}
    \textbf{s}^{'n} = \textbf{s}^{n}-\Bar{\textbf{s}},
    \label{eq67}
\end{equation}
where $\Bar{\textbf{s}} = \frac{1}{m}\sum_{n=1}^{m}\textbf{s}^{n}$ is the temporal mean over the complete dataset and $\textbf{s}^{'}$ are the perturbations around this average. Next the scaling of the above data is conducted as below 
\begin{equation}
    \textbf{s}^{'n}_{s} = \frac{\textbf{s}^{'n}-\textbf{s}_{min}^{'}}{\textbf{s}_{max}^{'}-\textbf{s}_{min}^{'}},
    \label{eq68}
\end{equation}
where each $\textbf{s}^{'n}_{s} \in [0,1]^{N_{x}\times N_{y}}$.
The scaling is carried out to ensure that the data lies in the interval of $(0,1)$. This requirement stems from the fact that a sigmoid activation function is used in all the layers of convolutional autoenocoder. The training dataset with the above modifications has the form
\begin{equation}
    \mathcal{S} = \{\textbf{S}^{'1}_{s},\dots,\textbf{S}^{'N_{s}}_{s}\}\in [0,1]^{N_{x}\times N_{y}\times N_{t}\times N_{s}},
    \label{eq69}
\end{equation}
where each training sample $\textbf{S}^{'i}_{s} = [\textbf{s}^{'1}_{s,i},\dots,\textbf{s}_{s,i}^{' N_{t}}]$ is a matrix consisting of the feature-scaled fluctuations.

\begin{figure}
    \centering
    \includegraphics[width = 0.45\textwidth]{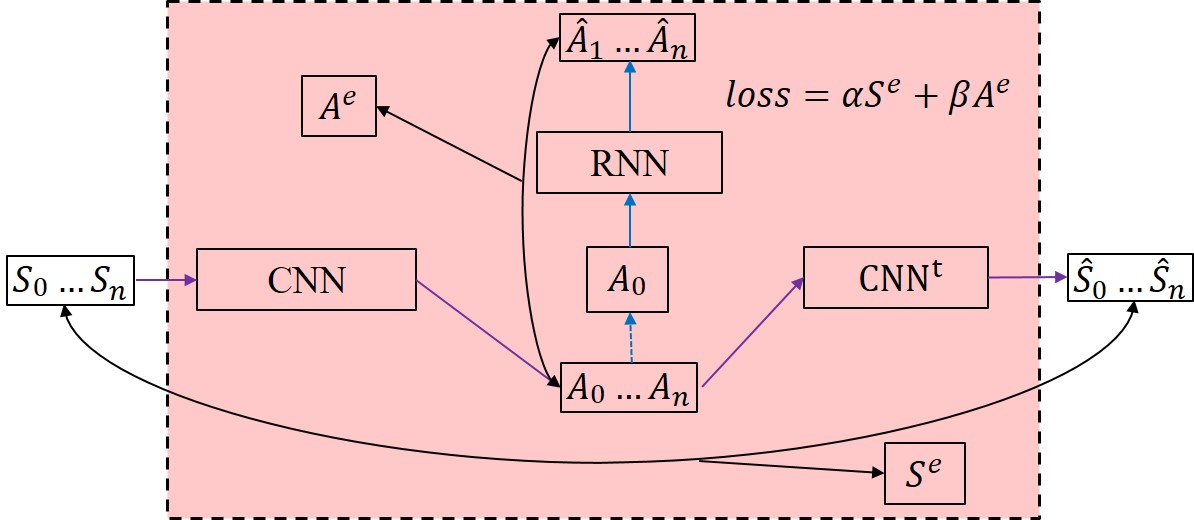}
    \caption{Schematic of training process of convolution recurrent autoencoder network (CRAN)}
    \label{training}
\end{figure}

The process of training entails two stages primarily. The convolutional autoencoder takes an  $N_{b}$- sized batch of the training data $\mathcal{S}^{b}\subset \mathcal{S}$, where $\mathcal{S}^{b}\in [0,1]^{N_{x}\times N_{y}\times N_{t}\times N_{b}}$ 
and outputs the current $N_{b}$ sized batch of low-dimensional representations of the training sequence
\begin{equation}
    \mathcal{A}^{b} = \{\textbf{A}^{1},\dots,\textbf{A}^{N_{b}}\}\in \mathbb{R}^{N_{A}\times N_{t}\times N_{b}},
    \label{eq610}
\end{equation}
where $\textbf{A}^{i} = [\textbf{A}^{1}_{i},\dots,\textbf{A}_{i}^{N_{t}}] \in \mathbb{R}^{N_{A}\times N_{t}}$
and a reconstruction
$\hat{\mathcal{S}}^{b}$ of the original input training batch in the first stage. This is represented as purple arrow in Fig. \ref{training}. 
 The second stage is indicated by the blue arrows in the same Fig. \ref{training}, where the first feature vector of each sequence is used to initialize and iteratively update Eq. (\ref{eq66}) to get a reconstruction  $\hat{\mathcal{A}}^{b}$ of the low-dimensional representations of the training batch $\mathcal{A}^{b}$.

{}

%%%%%%%%%
A loss function is constructed that equally weights the error in the full-state reconstruction and the evolution of the low-dimensional representations. The target of the training is to find the model parameters $\theta$  such that for any sequence 
$\textbf{S}_{s}^{'} = [\textbf{S}_{s}^{'1},\dots,\textbf{S}_{s}^{'N_{t}}]$
and its corresponding low-dimensional representation
$\textbf{A} = [\textbf{A}^{1},\dots,\textbf{A}^{N_{t}}]$, minimizes the following error between the
truth and the predictions:

\begin{equation}
\begin{aligned}
    \mathcal{J}(\theta) &= [\mathcal{L}(\hat{\textbf{S}}_{s}^{'},\textbf{S}_{s}^{'},\hat{\textbf{A}},\textbf{A})] \\
    &=  \left[\frac{\alpha}{N_{t}}\operatornamewithlimits{\sum}_{n=1}^{N_{t}}\frac{\|\textbf{S}_{s}^{'n}-\hat{\textbf{S}}_{s}^{'n}\|_{2}^{2}}{\|\textbf{S}_{s}^{'n}\|_{2}^{2}}+\frac{\beta}{N_{t}}\frac{\|\textbf{A}^{n}-\hat{\textbf{A}}^{n}\|_{2}^{2}}{\|\textbf{A}^{n}\|_{2}^{2}}\right],
\end{aligned}
\label{loss}
\end{equation}
where
$\alpha = \beta = 0.5$. The values of $\alpha$ and $\beta$ are chosen in order to give equal weightage to the errors in full state reconstruction and feature prediction respectively. The predictions are denoted by $\hat{\textbf{S}}_{s}^{'}$. In practice, the error is approximated by averaging
$\mathcal{L}(\hat{\textbf{S}}_{s}^{'},\textbf{S}_{s}^{'},\hat{\textbf{A}},\textbf{A})$ over all samples in a training batch during each backward pass. 
The convolutional autoencoder performs a regular forward pass and also constructs the low-dimensional representation simultaneously at every training step. These low-dimensional representations are used to train the RNN.

The ADAM optimizer \cite{kingma2014adam} is used for updating the weights during the training process. It is an updated version of the canonical SGD method which uses the estimates of first and second moments of the gradients for adaptive  computation of  learning rates. Fig. \ref{training} depicts the overall schematic of the training process for easy comprehension. Tensorflow is used for building this model and the main source code is taken from \cite{gonzalez2018deep}.

\begin{figure}
    \centering
    \includegraphics[width = 0.45\textwidth]{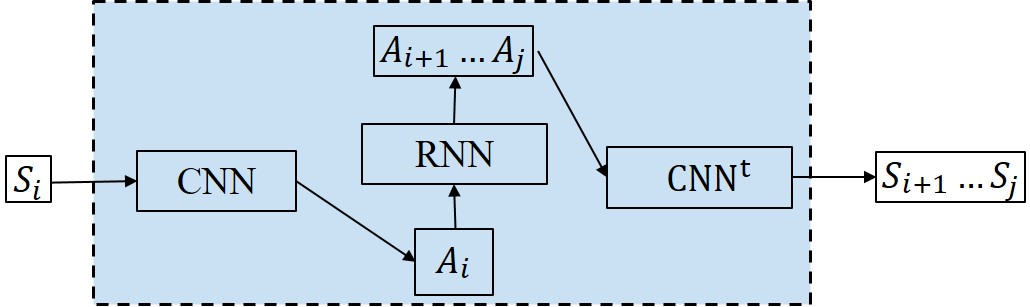}
    \caption{Illustration of prediction process of convolution recurrent autoencoder network (CRAN)}
    \label{prediction}
\end{figure}{}

The online prediction is straight forward and is depicted in Fig. \ref{prediction} for easy comprehension. Firstly. 
 a low-dimensional representation 
$\textbf{A}^{0}\in \mathbb{R}^{N_{A}}$ is constructed using the encoder network for a given initial condition  $\textbf{S}^{0}\in [0,1]^{N_{x}\times N_{y}}$ and a set of trained parameters $\theta^{*}$. The Eq. (\ref{eq66}) is applied iteratively for $N_{t}$ steps with  $\textbf{A}^{0}\in \mathbb{R}^{N_{A}}$ as the initial solution to get predictions of low dimensional representations. Finally, the full-dimensional state is reconstructed  $\hat{\textbf{S}}^{n}$ from the low-dimensional representation $\hat{\textbf{A}}^{n}$ at every time step
or at any specific instance. 

%%%%%%%%%%%%%%%%%
\subsection{Force calculation on the interface} \label{Section:ForceCalc}
\begin{figure}
\includegraphics[width = 0.5\textwidth, left]{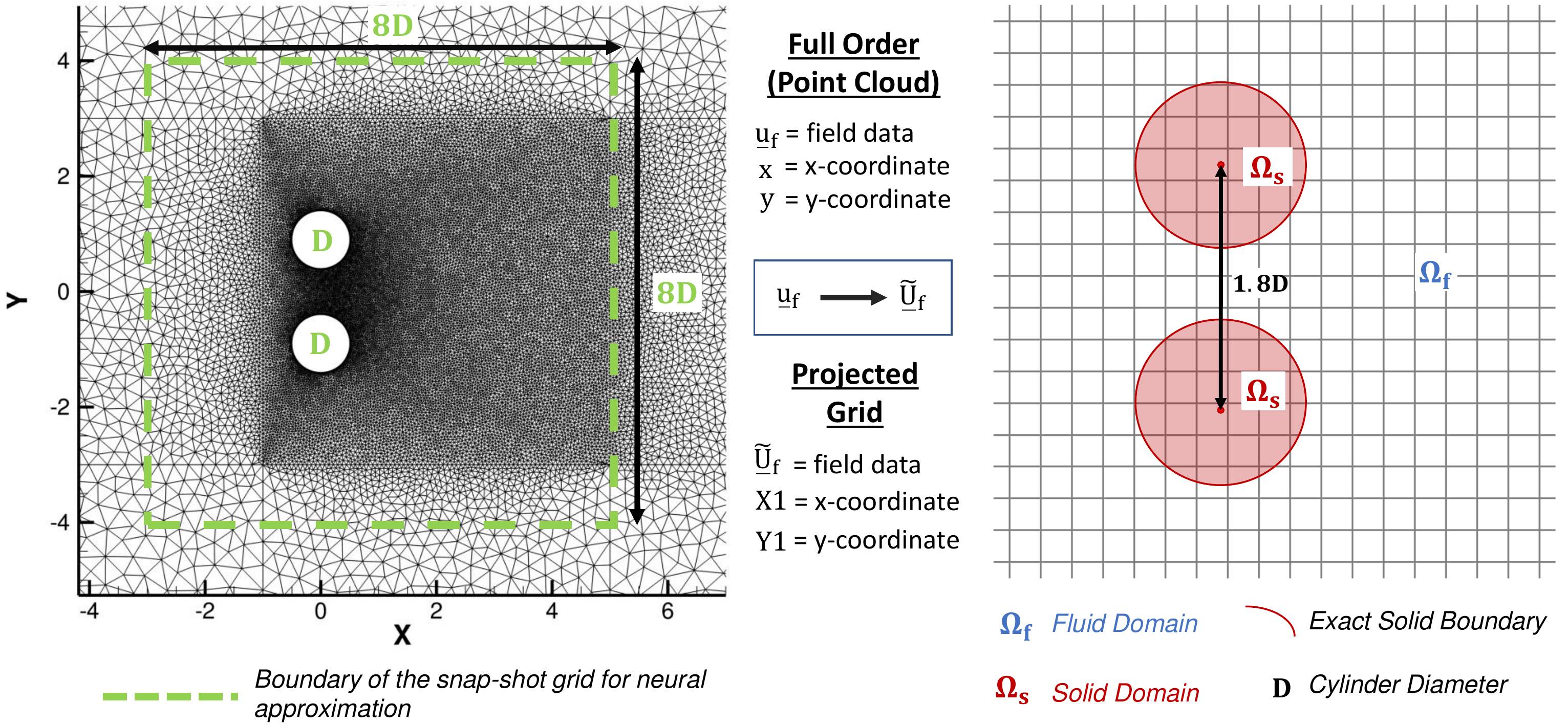}
\caption{Representative unstructured and reference grids for the side-by-side case. Every nodal point, in both grids, has an associated field value along with a 2-d spatial coordinate}
\label{figunstrct-strct}
\end{figure}

%%%%%%%%%%%%%%%%
Since the full-order data from the fluid solver is highly unstructured, these high-dimensional nodal data are first projected as snapshot images onto a uniform (reference) grid. This is an important data processing step, before proceeding with training or testing, as it brings spatial uniformity to the CNN input in the CRAN model. For the present problems, we select a uniform $64\times 64$ as a fixed reference grid for both cases. The interpolation of the 2-d scattered data onto this grid is implemented using the SciPy's $``scipy.interpolate.griddata"$ function \cite{SciPy}. The reference grid dimension is selected to sufficiently enclose the exact fluid-solid interface (FSI) present in the unstructured grid. Fig. \ref{figunstrct-strct} depicts both the unstructured grid (used in the full-order simulations) and the reference (used as an input to the CRAN model) for a side-by-side case. Note that the presence of the rigid body is ensured by masking the entire spatial region covered by the cylinder with a mandatory function which zeroes out the predictions made by the model, including the points inside the cylinder.       

\begin{figure}
\includegraphics[width = 0.5\textwidth, left]{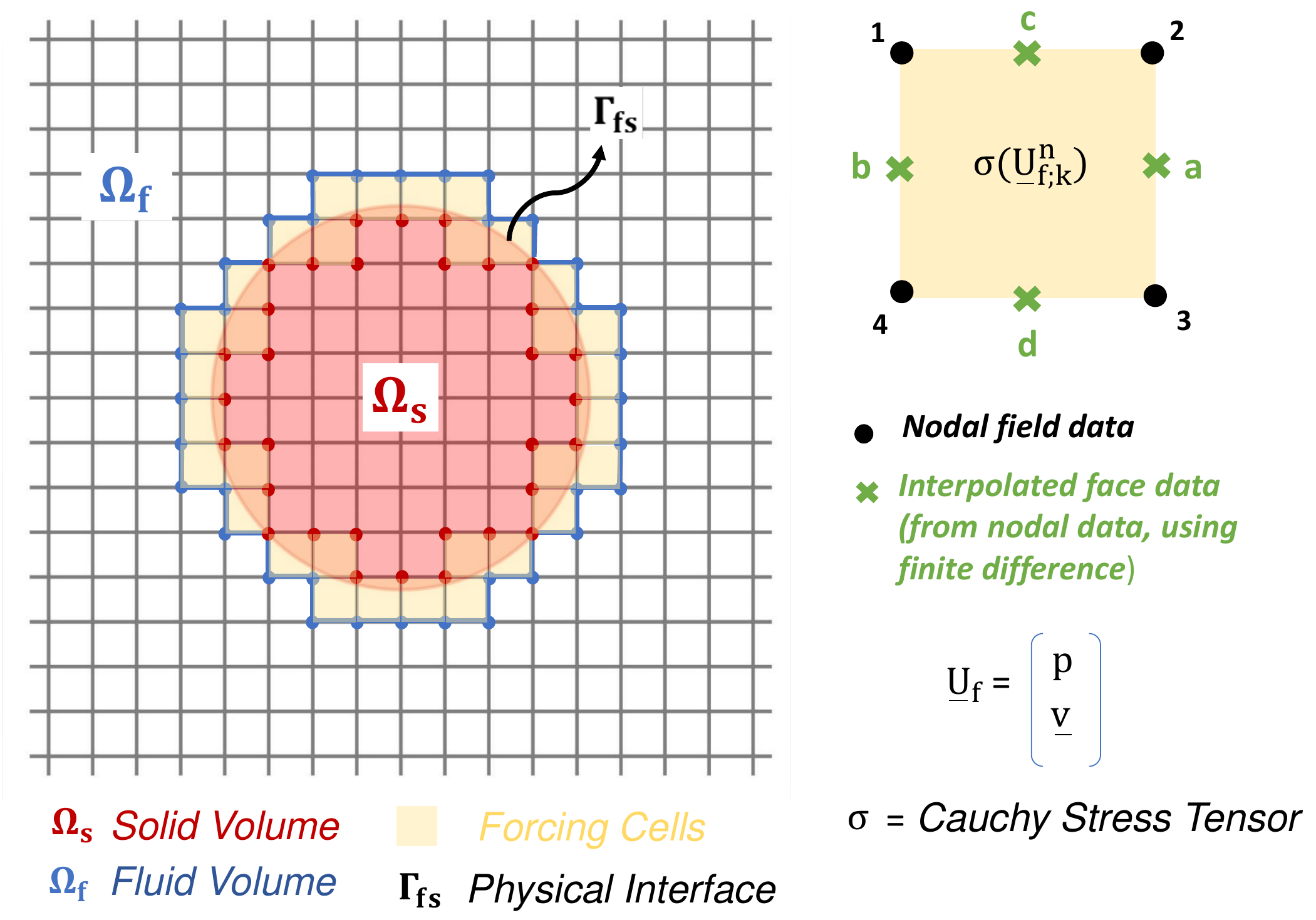}
\caption{Illustration of interpolation process for the reference (uniform) grid of DL-ROM procedure. Three types of cells, solid cells (red), interface cells that contain the exact physical boundary (yellow) and  fluid cells (white) are shown }
\label{figinterface-cells}
\end{figure}

Next, we briefly introduce the idea to calculate integrated pressure force on the exact solid boundary from a fairly coarse grid (reference grid). This is simply done via a two-step process. The first step is to identify all the interface cells for a given solid boundary (see Fig. \ref{figinterface-cells}) and sum the force experienced by these cells. We call the summed force as the total discrete force ($\textbf{F}_{\mathrm{B}}$). The process of calculating the total discrete force is convenient as one can loop over all the cells in the reference grid and mark all those cells that have any one of the cell nodal values as zero. Then the individual cell force is summed for all these marked cells. This will include the contribution from the cells in the interior of the solid domain (with zero force) and the interface cells (non-zero force), giving the discrete force. The second step is a reconstruction mapping ($\psi$) of this discrete force over the exact solid boundary. The reconstruction mapping takes care of the missing physical data that is lost in the interpolation process from a high-resolution grid (full-order) to the reference grid (neural network). This mapping is observed from the training data and corrected in the prediction. We call this entire process as an observer-corrector methodology to calculate the integrated force.    

\begin{figure}
\includegraphics[width = 0.5\textwidth, left]{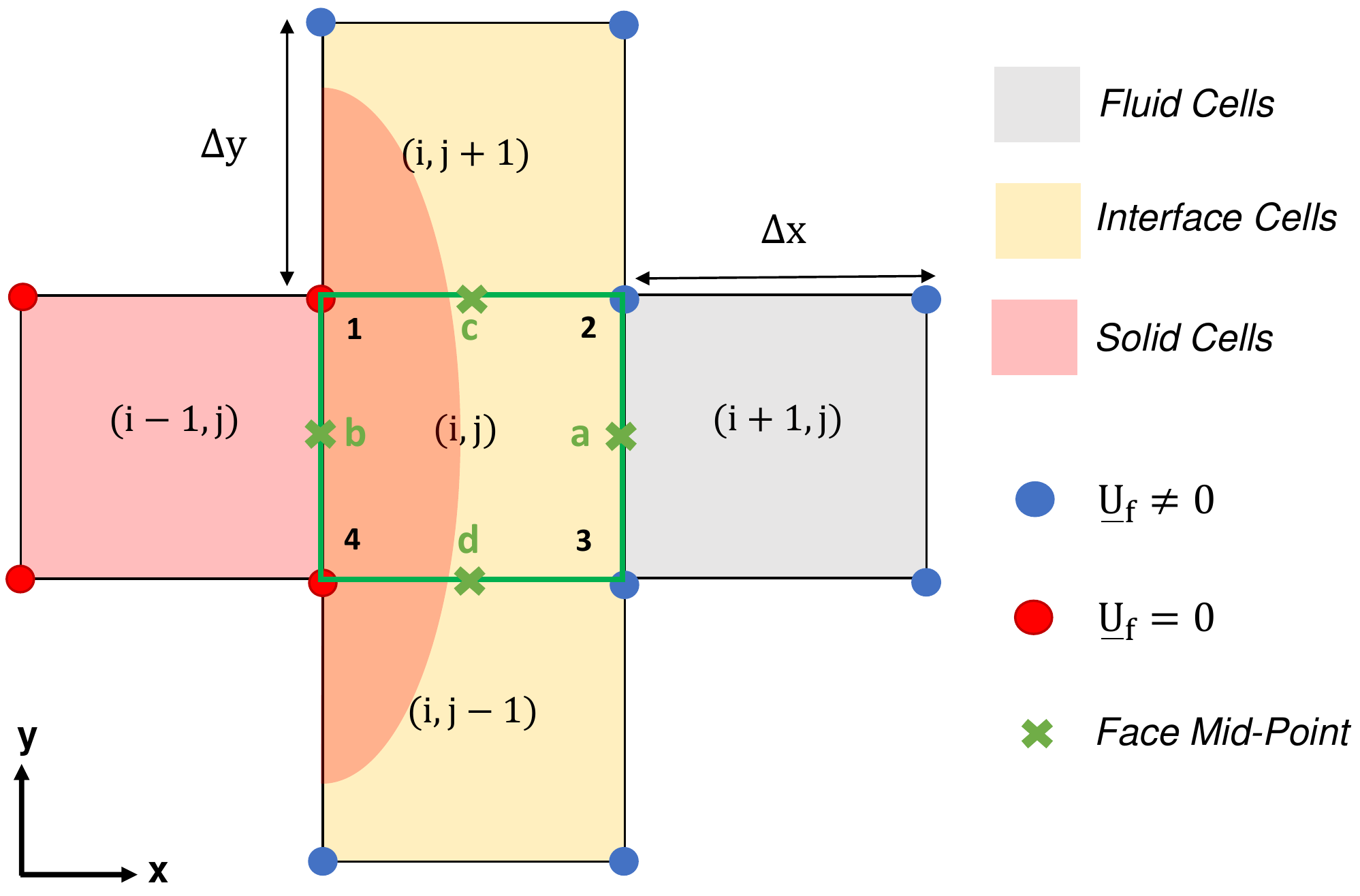}
\caption{A five cell stencil for linear interpolation of gradients using finite difference }
\label{figfd-cell-stencil}
\end{figure}

Let $\textbf{f}_{(\mathrm{i},\mathrm{j})}^{n}$ be the force in any interface cell $(\mathrm{i},\mathrm{j})$ at a time step $n$. The discrete force in this cell is calculated using linear interpolation via Finite Difference. The local face numbering ($\mathrm{a}-\mathrm{d}-\mathrm{b}-\mathrm{c}$) and node numbering ($2-3-4-1$) is shown in Fig. \ref{figinterface-cells}. The pressure and velocity values are approximated at the mid-point of these faces linearly, by simply taking the average of it's end nodal values. This can be further clarified from Fig. \ref{figfd-cell-stencil} and following equation: 
\begin{equation}
\begin{aligned}
     p_{\mathrm{a};(\mathrm{i},\mathrm{j})}^{n} = \frac{p_{2;(\mathrm{i},\mathrm{j})}^{n} + p_{3;(\mathrm{i},\mathrm{j})}^{n}}{2}   ,\quad
     p_{\mathrm{b};(\mathrm{i},\mathrm{j})}^{n} = \frac{p_{4;(\mathrm{i},\mathrm{j})}^{n} + p_{1;(\mathrm{i},\mathrm{j})}^{n}}{2}   ,\\
     p_{\mathrm{c};(\mathrm{i},\mathrm{j})}^{n} = \frac{p_{1;(\mathrm{i},\mathrm{j})}^{n} + p_{2;(\mathrm{i},\mathrm{j})}^{n}}{2}   ,\quad
     p_{\mathrm{d};(\mathrm{i},\mathrm{j})}^{n} = \frac{p_{3;(\mathrm{i},\mathrm{j})}^{n} + p_{4;(\mathrm{i},\mathrm{j})}^{n}}{2}   ,\\
\end{aligned}
\label{faceApprox}
\end{equation}
where $p_{*;(\mathrm{i},\mathrm{j})}^{n}$ represents the pressure at a specific point in the cell at a time-step $n$. This process, similarly, can be repeated for individual velocity components as well.

The discrete force $\textbf{f}_{\mathrm{k}}^{n} $ at a time-step $n$ in a cell $\mathrm{k}$ (or $(\mathrm{i},\mathrm{j})$ from Fig. \ref{figfd-cell-stencil}) for a Newtonian fluid can be written as

\begin{equation}
\begin{aligned}
     \textbf{f}_{\mathrm{k}}^{n} =  (\bm{\upsigma}_{\mathrm{a};\mathrm{k}}^{n} - \bm{\upsigma}_{\mathrm{b};\mathrm{k}}^{n}).\mathrm{\mathbf{n}}_{\mathrm{x}} \mathrm{\Delta} \mathrm{y} + 
                  (\bm{\upsigma}_{\mathrm{c};\mathrm{k}}^{n} - \bm{\upsigma}_{\mathrm{d};\mathrm{k}}^{n}).\mathrm{\mathbf{n}}_{\mathrm{y}} \mathrm{\Delta} \mathrm{x}.
\end{aligned}
\label{faceSigma}
\end{equation}

The Cauchy Stress Tensor $\bm{\upsigma}_{*;\mathrm{k}}^{n} = -p_{*;\mathrm{k}}^{n}\textbf{I} + 2\mathrm{\mu} \textbf{E}_{*;\mathrm{k}}^{n}$ for any point inside the cell $\mathrm{k}$ at instance $n$. Here $-p_{*;\mathrm{k}}^{n}$, $\textbf{I}$, $\mathrm{\mu}$ and $\textbf{E}_{*;\mathrm{k}}^{n}$ denote the fluid pressure, identity tensor, fluid dynamic viscosity and fluid strain rate tensor respectively. $\mathrm{\mathbf{n}}_{\mathrm{x}}$ and $\mathrm{\mathbf{n}}_{\mathrm{y}}$ are unit vectors in the x and y direction, while $\mathrm{\Delta} \mathrm{x} = \mathrm{\Delta} \mathrm{y}$ is the cell size. For the present problems, we only consider the pressure component in the Cauchy Stress Tensor. The viscous component involves the calculation of gradients and can be done using linear interpolation in the five-cell stencil shown in the Fig. \ref{figfd-cell-stencil}. This part has been left for future work for further exploration. With above considerations and using Eqs. (\ref{faceApprox}) and (\ref{faceSigma}), the expression for discrete force is  

\begin{align*}
    \textbf{f}_{\mathrm{k}}^{n} &= 
    \begin{bmatrix}
    p_{\mathrm{a};\mathrm{k}}^{n} - p_{\mathrm{b};\mathrm{k}} & 0 \\ 
    0 & p_{\mathrm{a};\mathrm{k}}^{n} - p_{\mathrm{b};\mathrm{k}} \\
  \end{bmatrix} . 
  \begin{bmatrix}
  1 \\
  0 
  \end{bmatrix} \mathrm{\Delta} \mathrm{y}  
  \end{align*}
  \begin{equation}
  \begin{aligned}
  \quad \: \,+ \,
  \begin{bmatrix}
    p_{\mathrm{c};\mathrm{k}}^{n} - p_{\mathrm{d};\mathrm{k}} & 0 \\ 
    0 & p_{\mathrm{c};\mathrm{k}}^{n} - p_{\mathrm{d};\mathrm{k}} \\
  \end{bmatrix} . 
  \begin{bmatrix}
  0 \\
  1  
  \end{bmatrix} \mathrm{\Delta} \mathrm{x}.
\end{aligned}
\label{presForce}
\end{equation}
\\
Next, a few notes on the reconstruction mapping $\psi$. The coarsening effect (projection) of the full-order data onto the reference grid brings a spatial uniformity in the input state. However, this coarsening can also lead to a loss of accurate forces on the physical FSI, even in the training data. This can be accounted to the fact that the boundary layer is not properly resolved on the reference grid. Hence, the function of  $\psi$ is to observe the loss of data in the training full-order and discrete force. After the prediction of the field on the reference grid by CRAN, this calculated $\psi$ reconstructs to get the full-order prediction force. This will be further clarified in the example problems of a single cylinder and side-by-side cylinders in subsequent sections.  

In summary, the overall process that is followed to predict the integrated value of the force at a time-step $n$ for the fluid-solid interface, 
\begin{itemize}
\item \textit{Calculate the discrete force} $\textbf{f}_{\mathrm{k}}^{n}$ \textit{in each interface cell $k$ using Eq. (\ref{faceSigma})}.

\item \textit{Let $N_{F}$ be the total such cells. Obtain the total discrete force} $\textbf{F}_{\mathrm{B}}^{n}$ \textit{as}
  \begin{equation}
  \begin{aligned}
  \textbf{F}_{\mathrm{B}}^{n} = \sum_{\mathrm{k}=1}^{N_{F}} \textbf{f}_{\mathrm{k}}^{n}.
  \end{aligned}
  \label{totalDiscForce}
  \end{equation}
  
\item \textit{Correct} $\textbf{F}_{\mathrm{B}}^{n}$ \textit{by $\psi$ observed from training data forces to predict the full-order force} $\textbf{F}_{\mathrm{\Gamma}_{\mathrm{fs}}}^{n}$ \textit{as}
\begin{equation}
  \begin{aligned}
  \textbf{F}_{\mathrm{\Gamma}_{\mathrm{fs}}}^{n} \approx \psi \textbf{F}_{\mathrm{B}}^{n}.  
  \end{aligned}
  \label{psitotalDiscForce}
  \end{equation}
\end{itemize}

\newpage 

%%%%%%%%%%%%%%%%%%%%%%%%%%%%%%%%%%%%%%%%%%%%%%%%%%%%%%%%%%%%%%%%%%%%%%%%%%%%%%%%%%%%%%%%%%%%%%%%%%%%%%%%%%%
\section{Application \texorpdfstring{\MakeUppercase{\romannumeral 1}}: Flow past a cylinder}\label{APP1STAT}
%%%%%%%%%%%%%%%%%%%%%%%%%%%%%%%%%%%%%%%%%%%%%%%%%%%%%%%%%%%%%%%%%%%%%%%%%%%%%%%%%%%%%%%%%%%%%%%%%%%%%%%%%%%
% Prob definition and objective

\begin{figure} 
\centering
\subfloat[]{\includegraphics[width =0.4\textwidth]{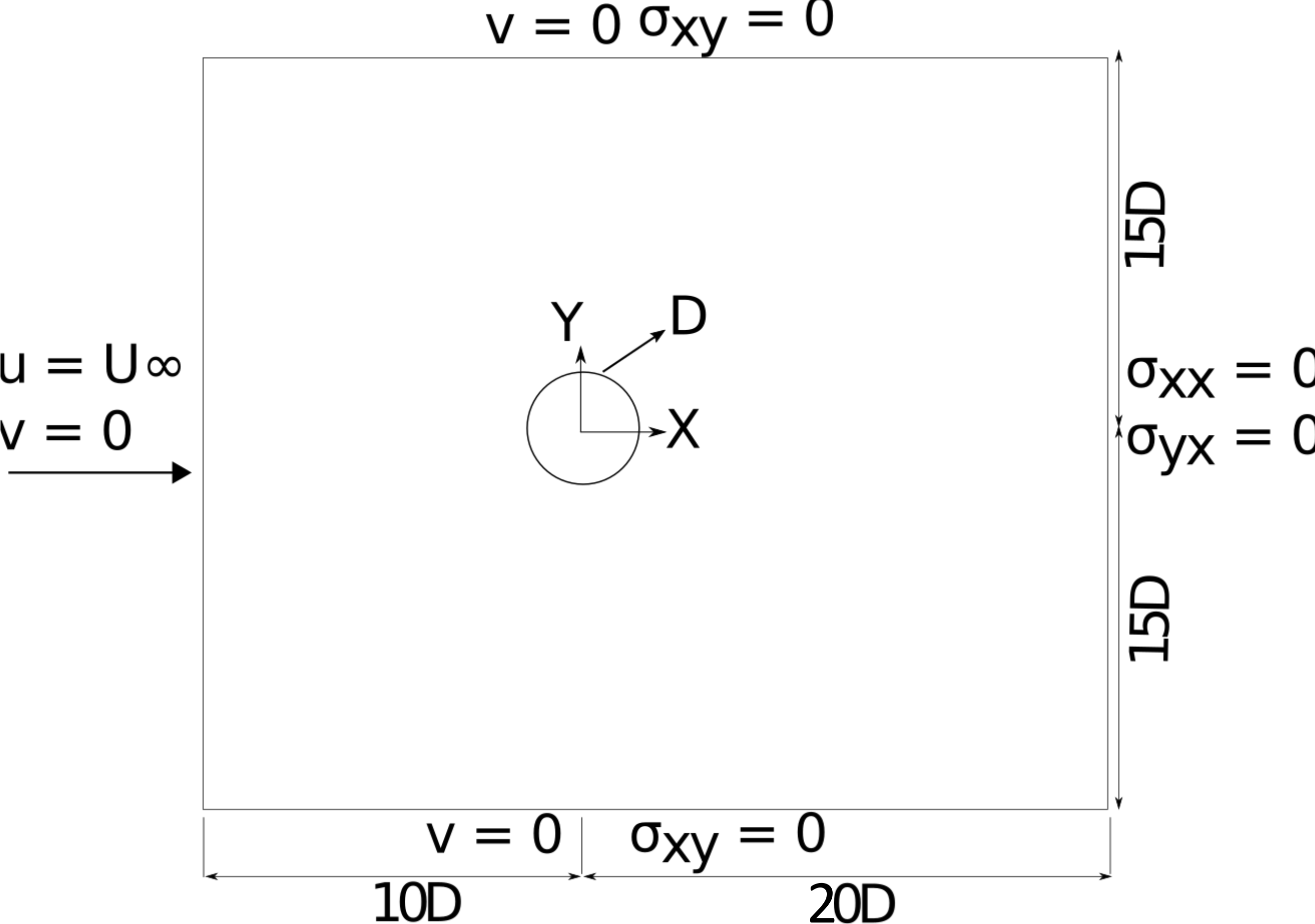}}\\
\subfloat[]{\includegraphics[width = 0.24\textwidth]{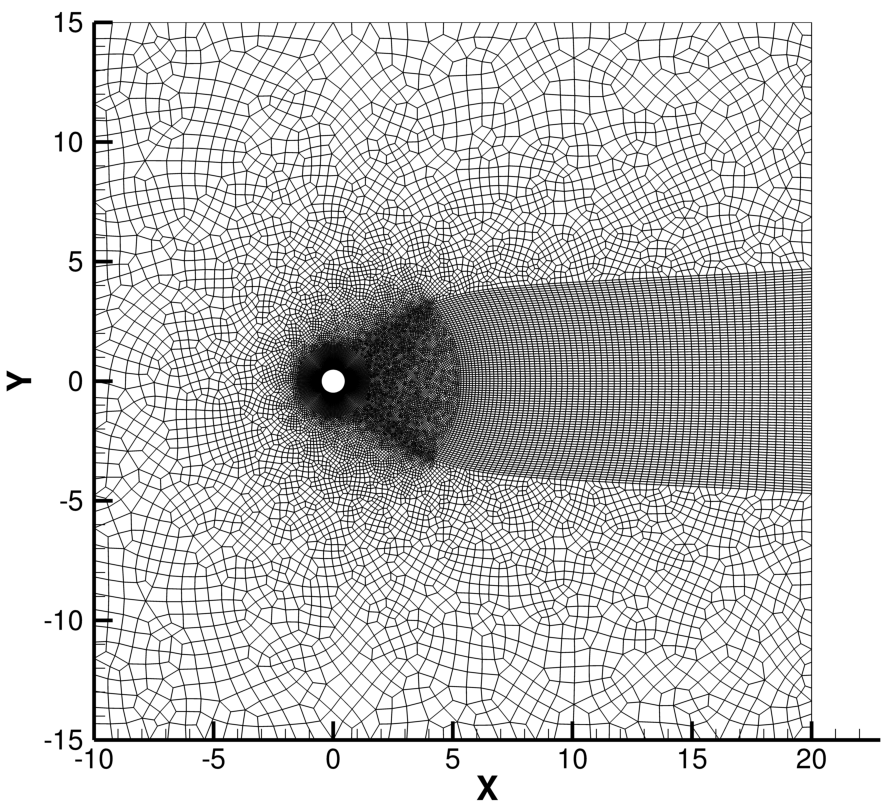}}
\subfloat[]{\includegraphics[width = 0.243\textwidth]{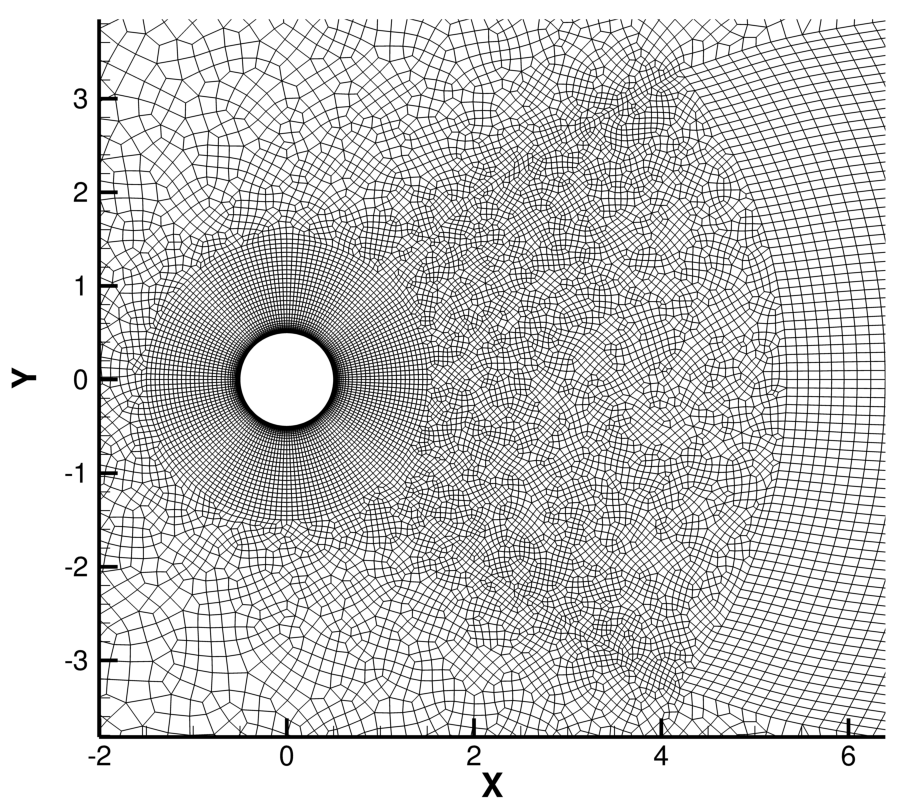}}
\caption{The flow past a cylinder: (a) schematic of the problem set-up, (b) full-domain computational mesh view and (c) close-up computational mesh view}
\label{setup_single}
\end{figure}

As a first demonstration, we apply both POD-RNN and CRAN models for a simplified problem of flow past a cylinder. This includes the prediction of the flow field and the integrated force coefficient on the cylinder for each case.
\\
\textit{Problem objective}: The objective is to sufficiently learn the phenomenon of vortex shedding for flow past a stationary cylinder on a given time-set (training) and extrapolate the behavior based on learned parameters (prediction).
\\
\textit{Full-order data}: The schematic of the problem set-up is the same as depicted in Fig.~\ref{setup_single} (a) where all the information about the domain boundaries and the boundary conditions are clearly outlined. The final mesh which is obtained after following standard rules of mesh convergence is presented in Fig.~\ref{setup_single} (b),(c) which contains a total of $25916$ quadrilateral elements with $26114$ nodes. The numerical simulation is carried out via finite element Navier-Stokes solver to generate the train and test data. The Reynolds number of the problem is set at $Re = 100$. The full-order simulation is carried out for a total of $1125\;tU_{\infty}/D$  with a time-step of $0.25\;tU_{\infty}/D$. A total of $4500$ snapshots of the simulation are collected at every $0.25\;tU_{\infty}/D$ for the pressure field ($P$) and the x-velocity ($U$). Of those 4500 snapshots, 3000 (from 501 to 3500 steps) are used for training and 1000 (from 3501 to 4500 steps) are kept as testing. Thus, the total steps used in the analysis are $N=4000$. 

In this application section, we assess the complete procedure for prediction of flow field and pressure force coefficient for a flow past stationary cylinder using POD-RNN (section \ref{podrnn_stat}) and CRAN (section \ref{cran_stat}). Finally, the performance of both the models is discussed in section \ref{disc_stat}.   

%%%%%%%%%%%%%%%%%%%%%%%%%%%%
\subsection{POD-RNN model}\label{podrnn_stat}
%%%%%%%%%%%%%%%%%%%%%%%%%%%%
The POD-RNN model is applied on the flow past a plain cylinder as follows: 
\begin{enumerate}

% Step 1
\item The POD algorithm discussed in section \ref{pod-rnn-architecture} is applied on all the time snapshots $\boldsymbol{\mathcal{S}} = \left\lbrace\textbf{S}_{1}\;\textbf{S}_{2}\dots\;\textbf{S}_{N}\;\right\rbrace \in \mathbb{R}^{m\times N}$  (here, $m=26114$ and $N=4000$) to get the reduced order dynamics on the given data. In doing so, get the mean field $\bar{\textbf{S}} \in \mathbb{R}^{m}$, the fluctuation matrix $\tilde{\textbf{S}} \in \mathbb{R}^{m\times N}$ and the $N$ spatially invariant POD modes $\boldsymbol{\Phi} \in \mathbb{R}^{m\times N}$. Note that $\boldsymbol{\mathcal{S}}$ can be either pressure field $P$ or x-velocity $U$ snapshots.  

% Step 2
\item Next, extract the eigen values $\Lambda_{N\times N}$ of the covariance matrix $\tilde{\textbf{S}}^{T}\tilde{\textbf{S}} \in \mathbb{R}^{N\times N}$. The absolute value of these eigen values are a measure of energy of the respective POD modes. Cumulative energy and percentage of total modal energies are plotted in Fig.~\ref{ce_te_stat_cyl} for both pressure and x-velocity fields. It is evident from  Fig.~\ref{ce_te_stat_cyl} that most of the system energy is concentrated in the first four to five POD modes (in fact, first two modes contribute $>95\%$ of the total energy) for both pressure and x-velocity data. 
\begin{figure}
\centering
\subfloat[]{\includegraphics[width = 0.245\textwidth]{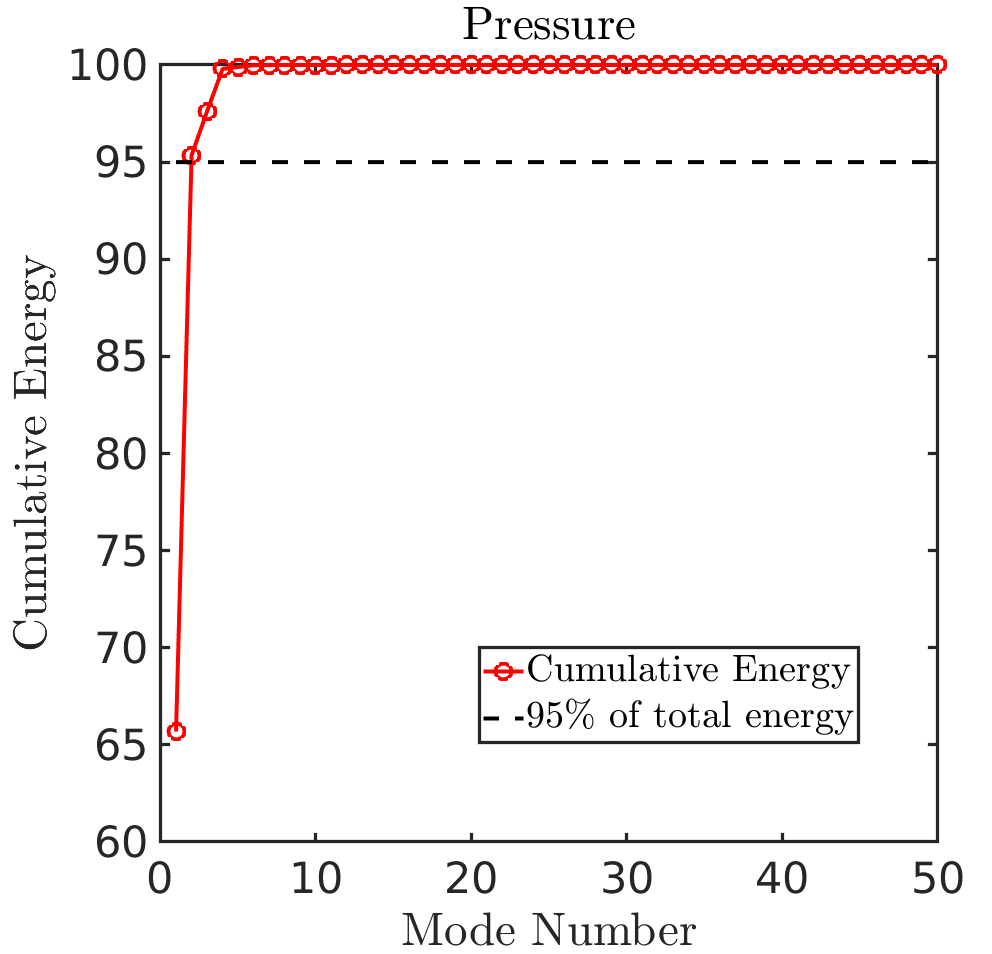}}
\subfloat[]{\includegraphics[width = 0.245\textwidth]{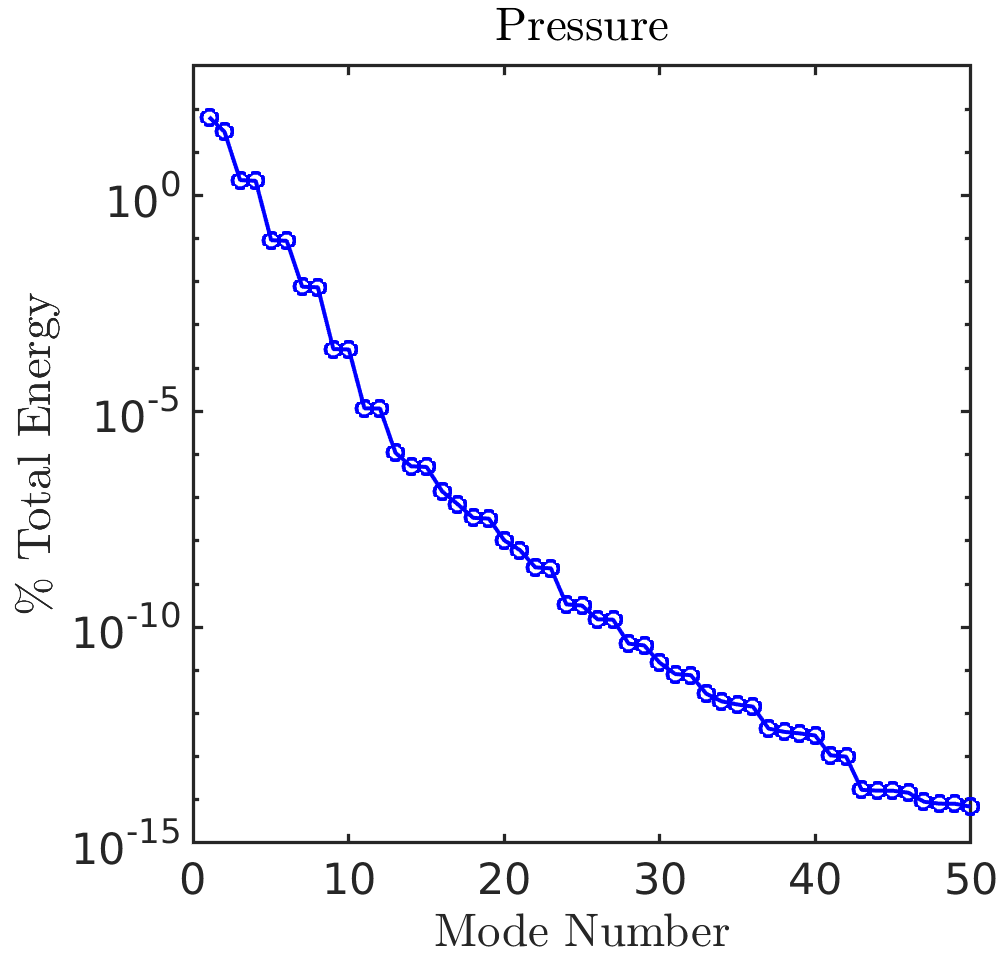}} \\
\subfloat[]{\includegraphics[width = 0.245\textwidth]{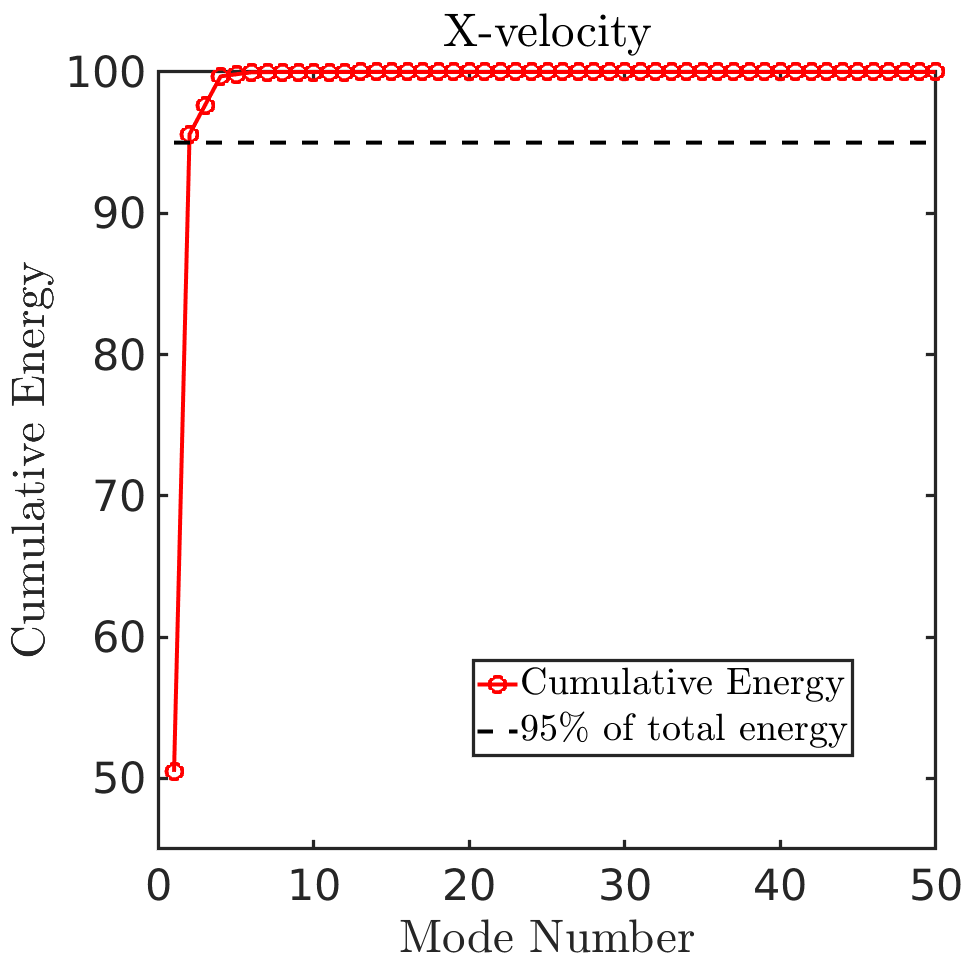}}
\subfloat[]{\includegraphics[width = 0.245\textwidth]{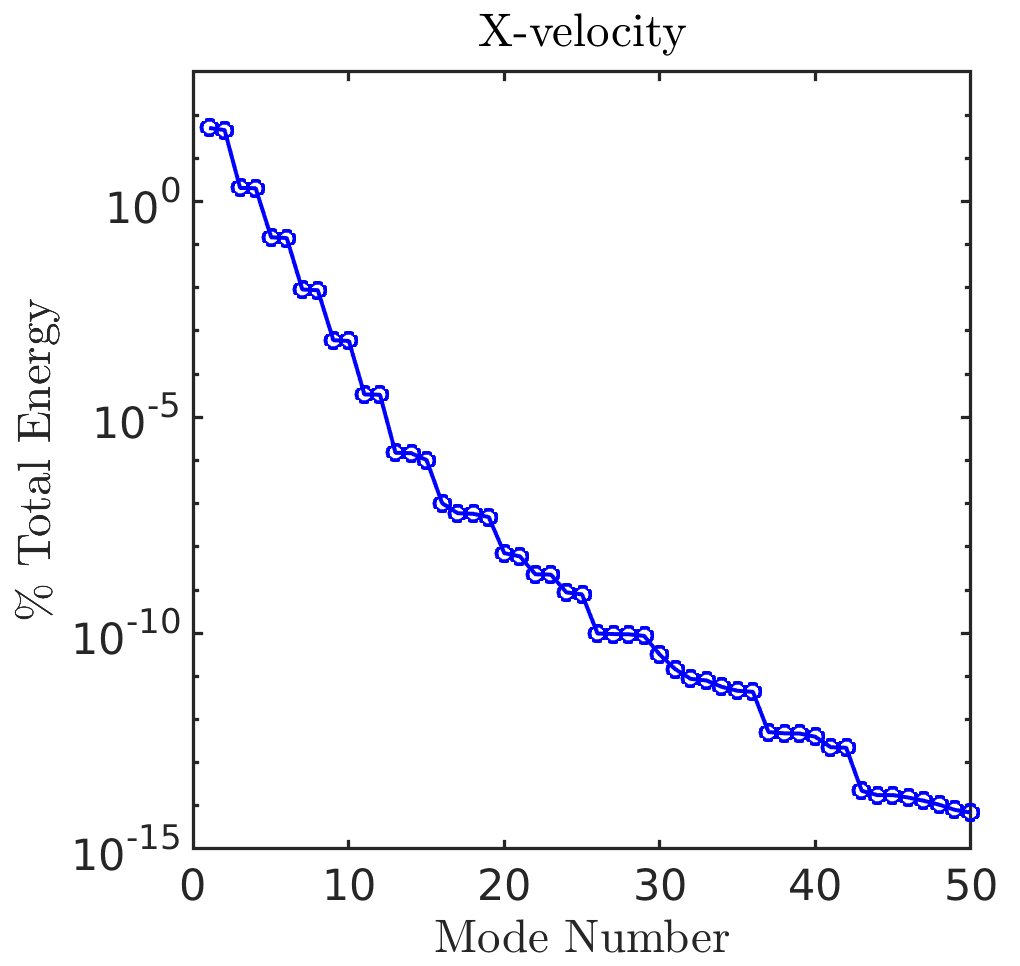}}
\caption{The flow past a cylinder: Cumulative and percentage of modal energies. (a)-(b) for pressure field $P$, (c)-(d) for x-velocity field $U$}
\label{ce_te_stat_cyl}
\end{figure}

% Step 3
\item In order to construct the reduced order system dynamics, select the first few energetic POD modes (here, $k=5$) in the analysis. The $N$ POD modes are now simply approximated with first $k$ modes ($k \ll N$), reducing the POD modes to $\boldsymbol{\Phi} \in \mathbb{R}^{m\times k}$. The first four (most energetic) POD modes for pressure and x-velocity snapshots are depicted in Fig.~\ref{modes_stat_cyl}.

% Step 4
\item The temporal variations of these modes are obtained by $\mathbf{A} = \boldsymbol{\Phi}^{T} \tilde{\textbf{S}}$. Here, $\mathbf{A} \in \mathbb{R}^{k\times N}$, $\boldsymbol{\Phi} \in \mathbb{R}^{m\times k}$ and $\tilde{\textbf{S}} \in \mathbb{R}^{m\times N}$. For instance, the time history of these five modes for pressure and x-velocity field from $501-1500$ ($125-375\;tU_{\infty}/D$) time-steps is depicted in Fig.~\ref{podrnnstatth}. These $N$ temporal modes are divided into training ($n_{tr}=3000$) and testing part ($n_{ts}=1000$ steps). The temporal coefficients from $501-3500$ time-steps are considered for training and $3501-4500$ are kept for testing. 
\begin{figure}
\centering
\subfloat[]{\includegraphics[width =0.48\textwidth]{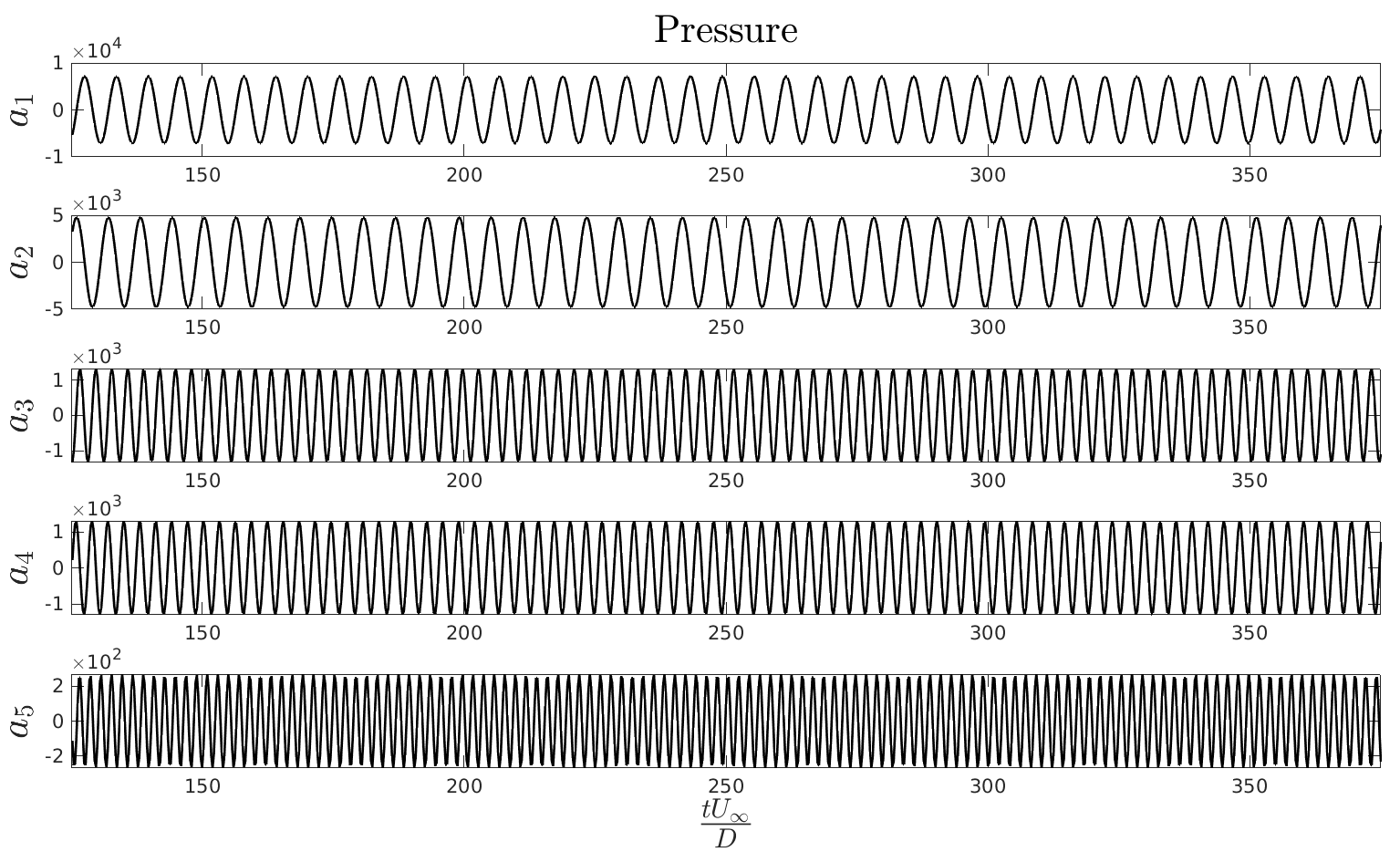}}\\
\subfloat[]{\includegraphics[width =0.48\textwidth]{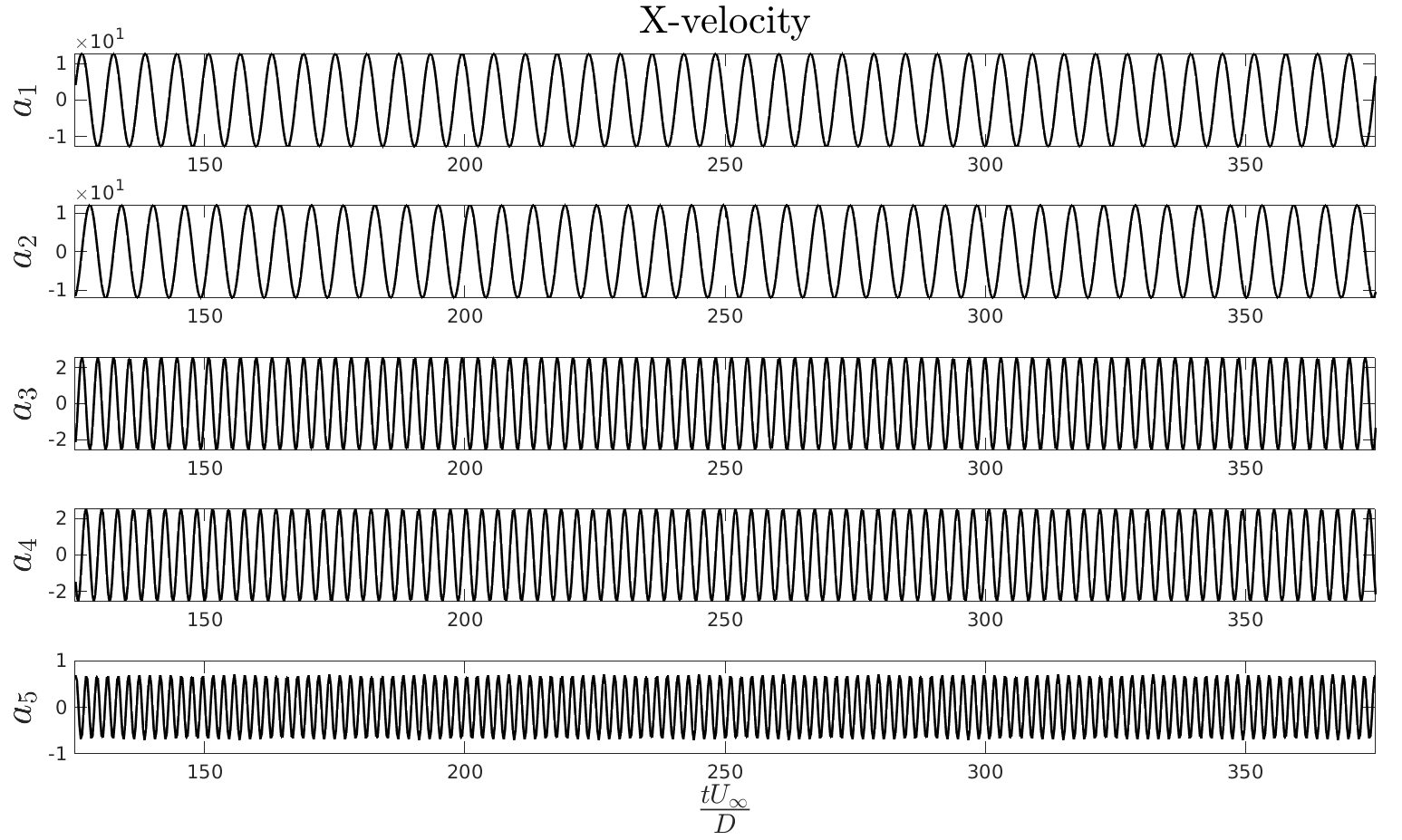}}
\caption{The flow past a cylinder: Time history of modal coefficients (shown from $125$ till $375\;tU_{\infty}/D$) for (a) pressure field $P$ and (b) x-velocity field $U$}
\label{podrnnstatth}
\end{figure}

% Step 5
\item The temporal behavior of the modal coefficients (for both pressure and x-velocity) is learned via a simple closed-loop recurrent neural network (see Fig.~\ref{closedloop}). Note that the modes are normalised between $-1$ to $1$ before proceeding with training. The details of the network  is outlined in the Table \ref{podrnn_na_stat_cyl}. These network hyperparameters are chosen conservatively to give the least error in testing. The input to the closed-loop recurrent neural network is a vector of dimension $k$ (here, $k=5$) so that all $k$ modes are predicted in a single step by utilizing the multi-featured setting of the recurrent neural network. Another advantage of this input dimension is that it can also learn any dependencies that exits in between the modes. As we already know, the modes are not completely independent of each other.  
\begin{table} [H]
\centering
\begin{tabular}{|c|c|}
\hline
Parameter & Value \\ \hline
Input dimension & 5 \\
Output dimension & 5 \\
Hidden dimension & 512 \\
RNN cell & LSTM \\
Number of time-steps & 3500 \\
Number of batches & 1 \\
Initial learning rate & 0.005\\
Learning rate drop period & 125 epochs \\
Learning rate drop factor & 0.2 \\
Iterations & 3000 epochs \\
Optimizer & ADAM \\ \hline
\end{tabular}
\caption{The flow past a cylinder: Network and parameter details for the closed-loop recurrent neural network}
\label{podrnn_na_stat_cyl}
\end{table}

% Step 6
\item The results of the $k=5$ modal predictions from the closed-loop recurrent neural network are plotted in Fig.~\ref{podrnn_th_predict_stat} for (a) pressure and (b) x-velocity. Note that only $k$ modal coefficients at time-step 3500 is used to make prediction for next 1000 steps (from 3501 till 4500). The predicted and true data are denoted by red and black lines respectively in the plot. The prediction agrees reasonably well with the true data and the same is also evident from Fig.~\ref{podrnn_th_predict_stat}. The closed-loop recurrent neural network effectively captured the temporal behavior of the modal coefficients with different frequencies. In addition, the amplitudes and phases of all the modes are completely different. In spite of these drastic differences between the POD modes, one single closed-loop recurrent network is suffice to capture this behavioral pattern. The root mean square error
$RMSE = \sqrt{\frac{\operatornamewithlimits{\sum}_{n=1}^{i=T}(\textbf{A}_{n}-\hat{\textbf{A}}_{n})^2}{T}}$ is tabulated in Table \ref{podrnn_stat_rmse}. Here, $\textbf{A}_{n}$ and $\hat{\textbf{A}}_{n}$ are the true and predicted (normalised) modal coefficients respectively. $T$ are the number of test time-steps (here, $T=1000$). 

\begin{figure}
\centering
\subfloat[]{\includegraphics[width = 0.235\textwidth]{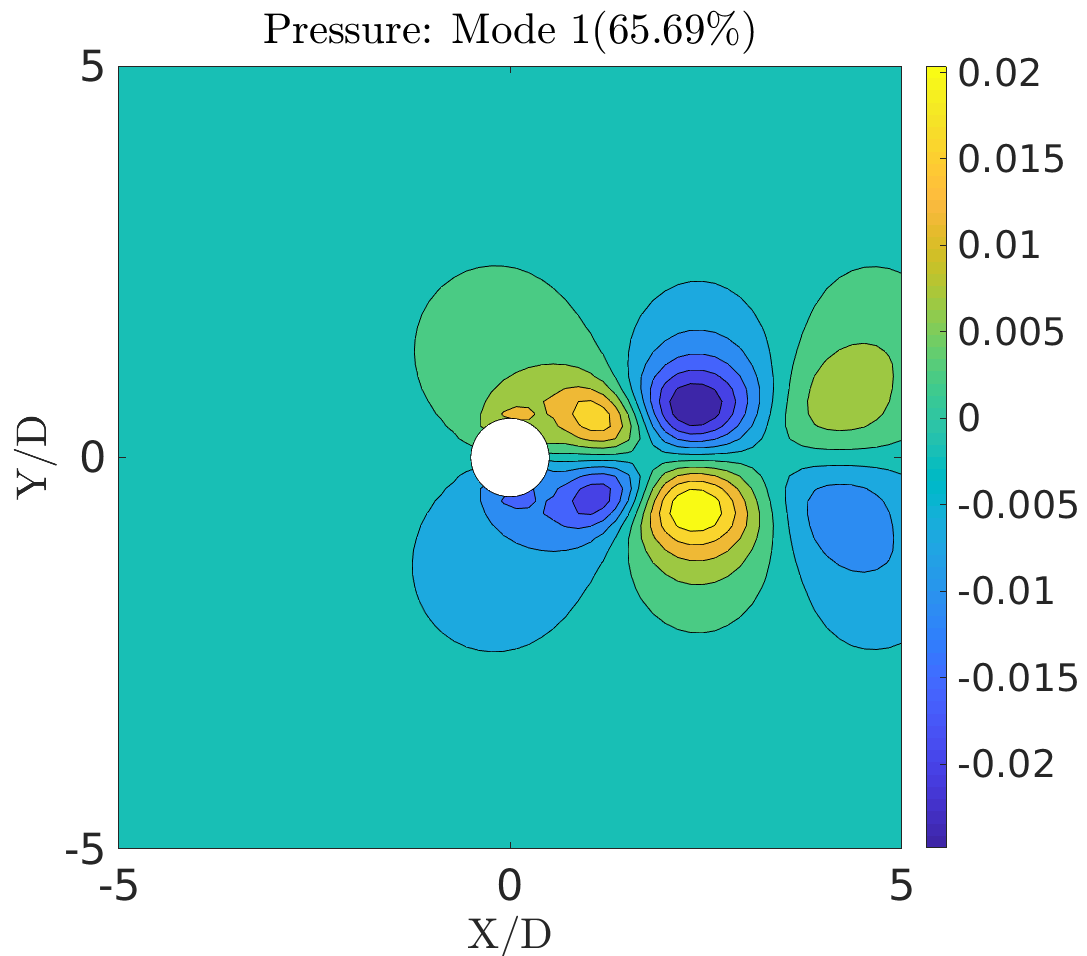}}
\subfloat[]{\includegraphics[width = 0.235\textwidth]{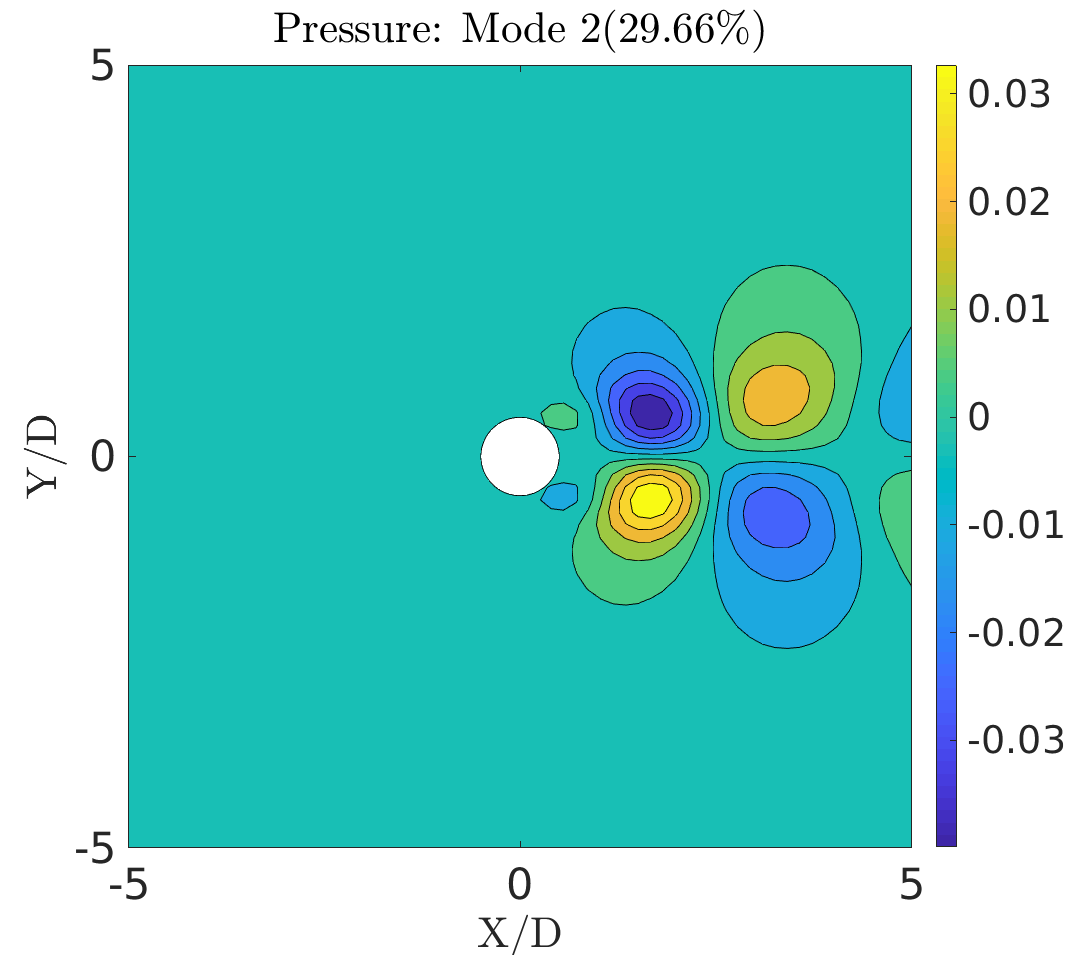}}\\
\subfloat[]{\includegraphics[width = 0.235\textwidth]{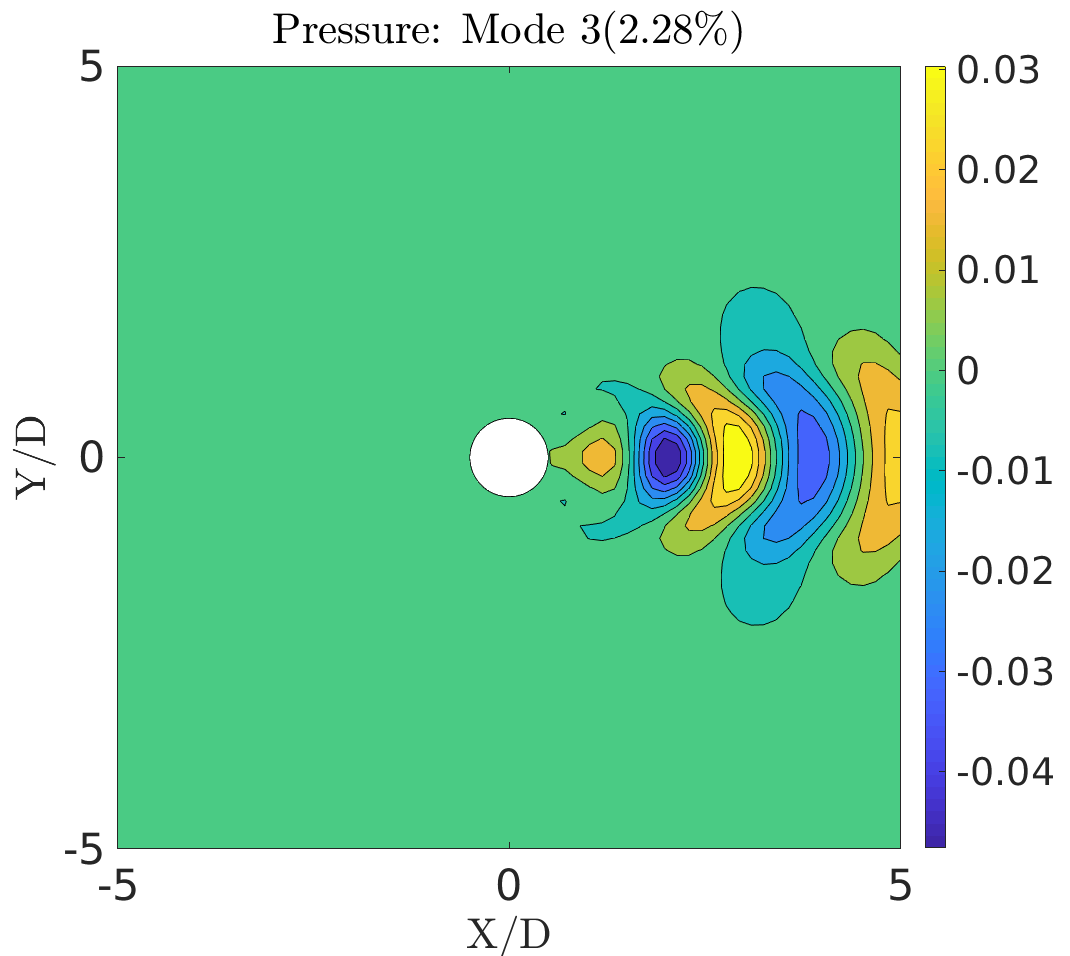}}
\subfloat[]{\includegraphics[width = 0.235\textwidth]{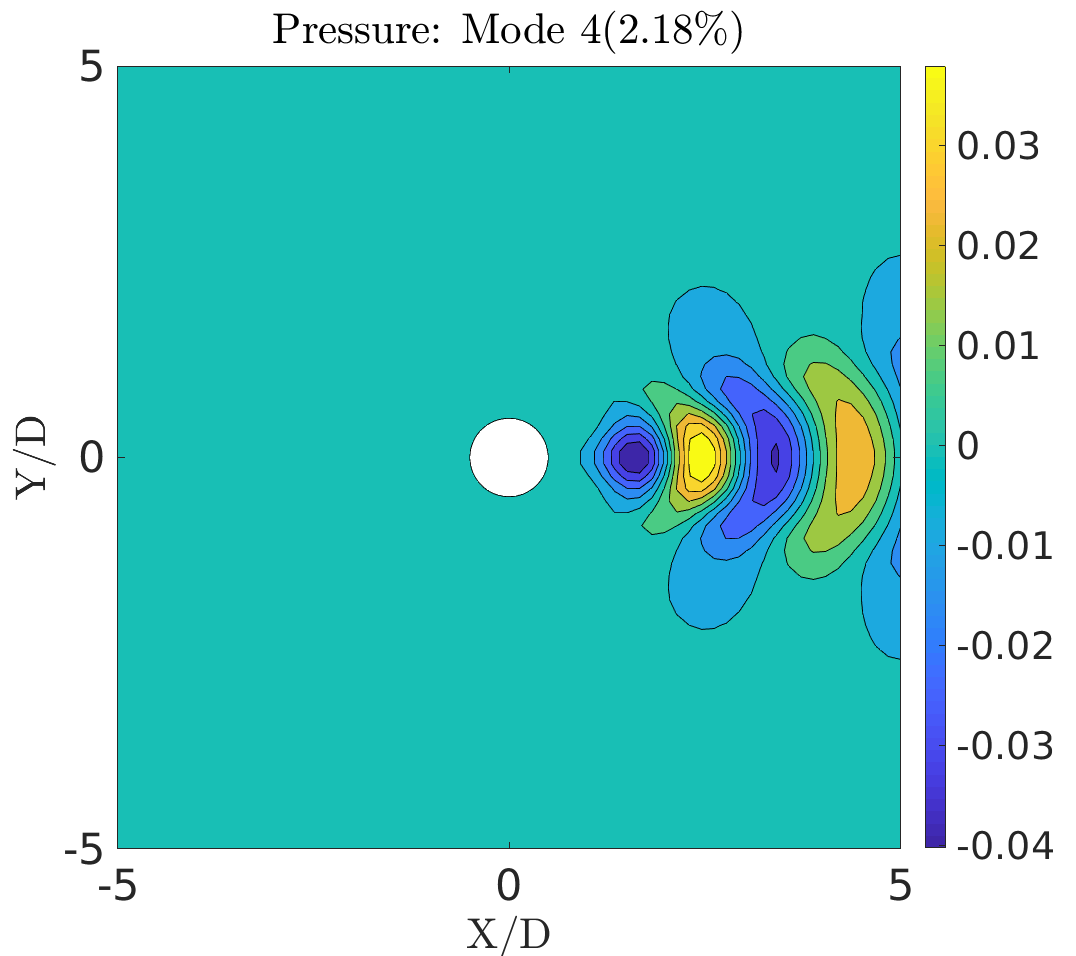}}\\
\subfloat[]{\includegraphics[width = 0.235\textwidth]{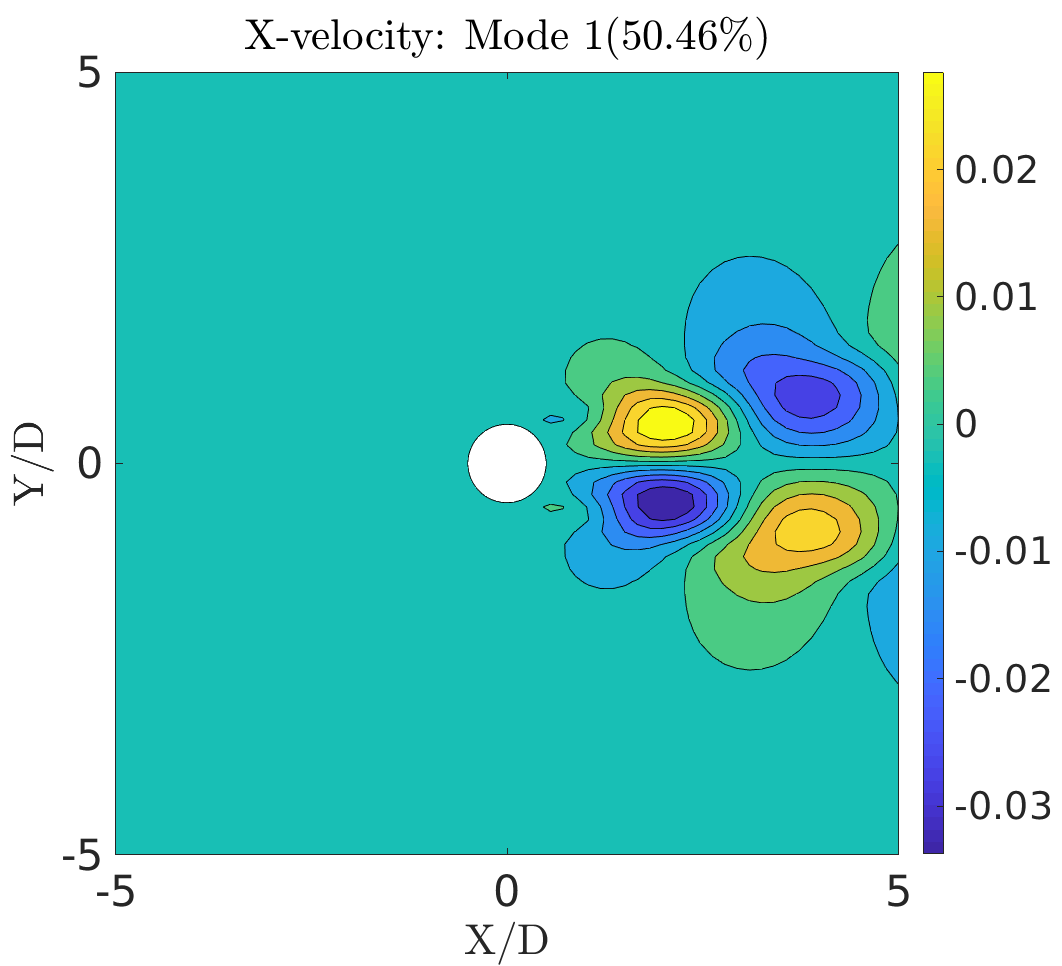}}
\subfloat[]{\includegraphics[width = 0.235\textwidth]{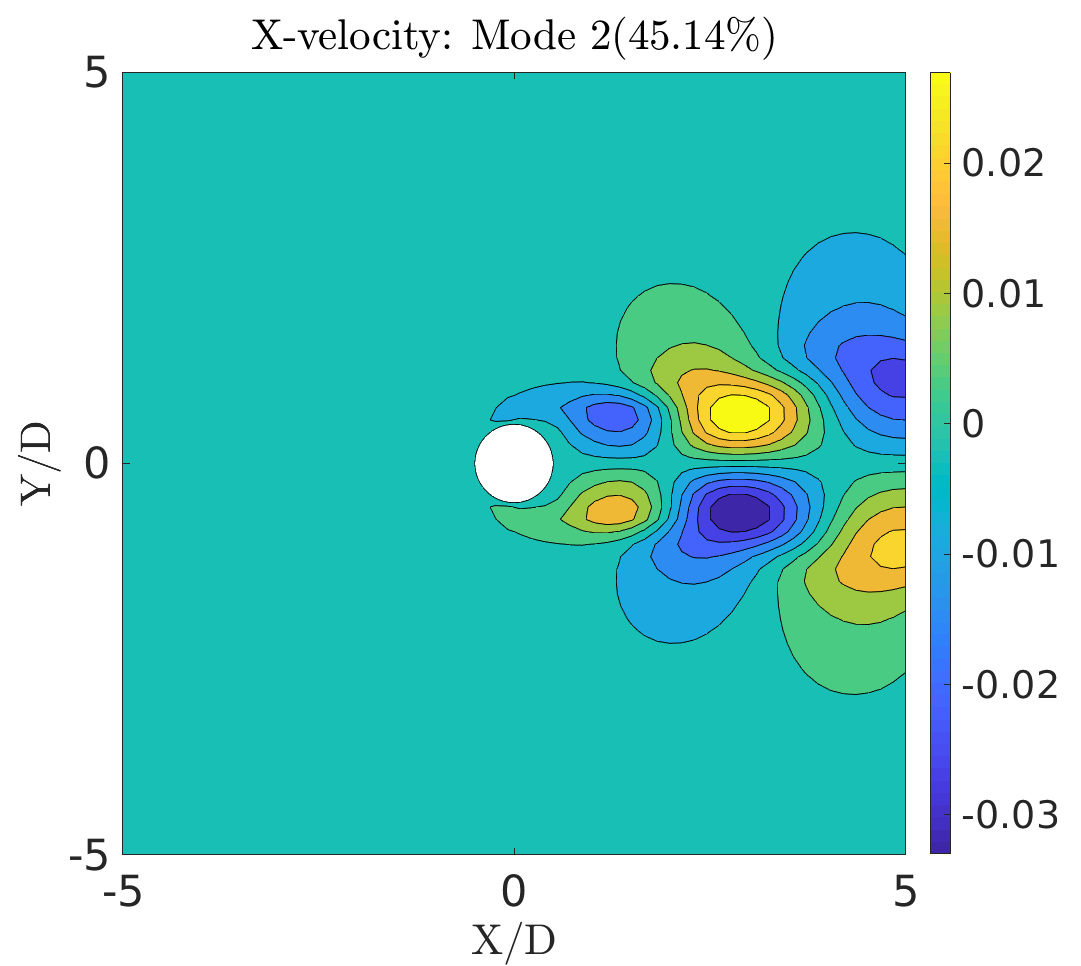}}\\
\subfloat[]{\includegraphics[width = 0.235\textwidth]{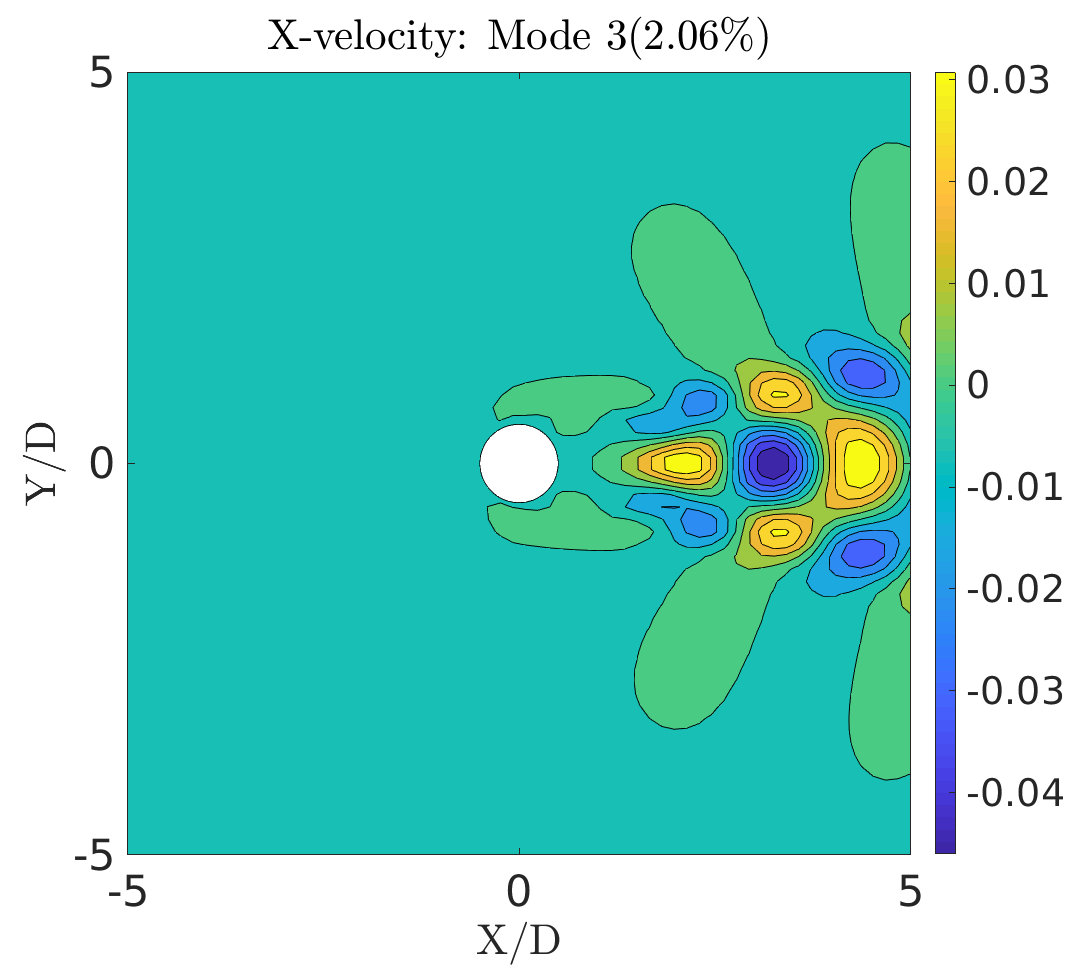}}
\subfloat[]{\includegraphics[width = 0.235\textwidth]{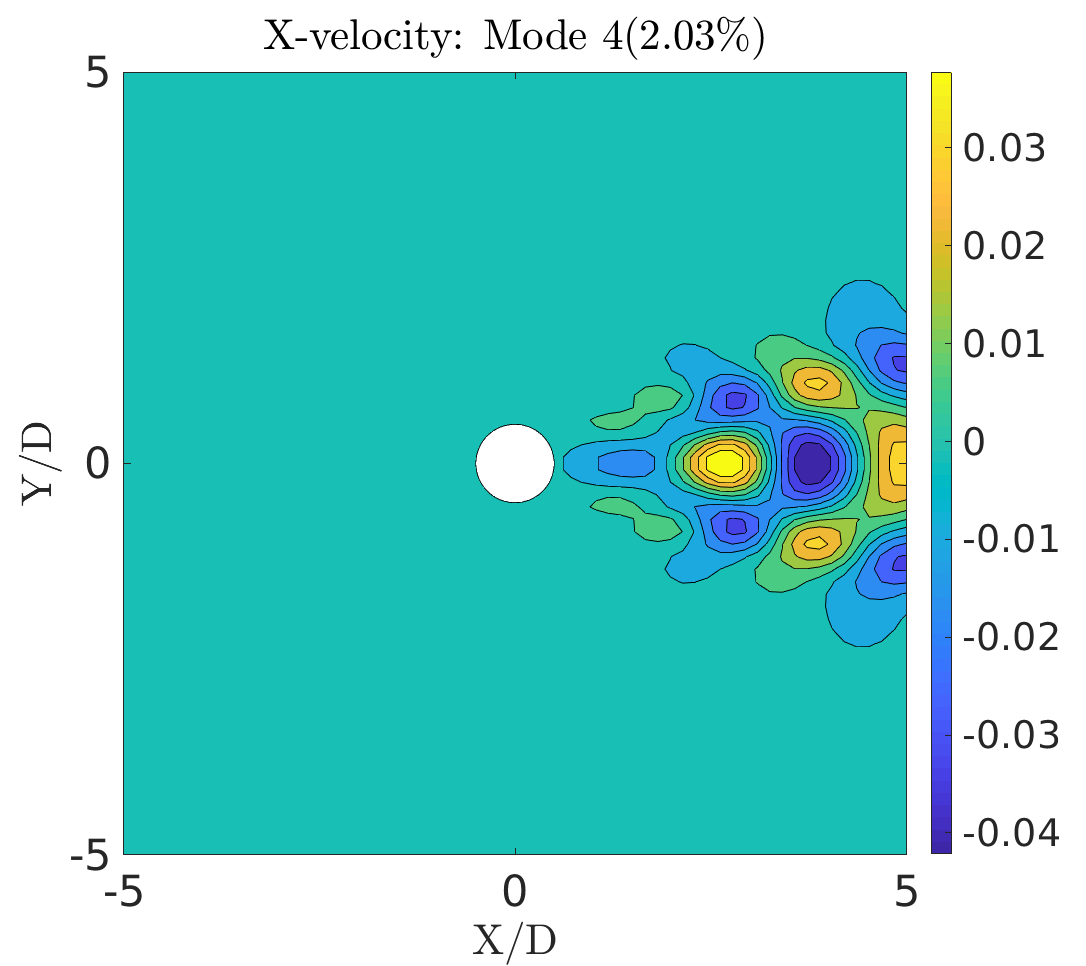}}
\caption{The flow past a cylinder: First four most energetic time-invariant spatial modes obtained from POD along with $\% $ of total energy. (a)-(b)-(c)-(d) for pressure field $P$ and (e)-(f)-(g)-(h) for x-velocity field $U$}
\label{modes_stat_cyl}
\end{figure}

\begin{figure}[H]
\centering
\subfloat[]{\includegraphics[width = 0.48\textwidth]{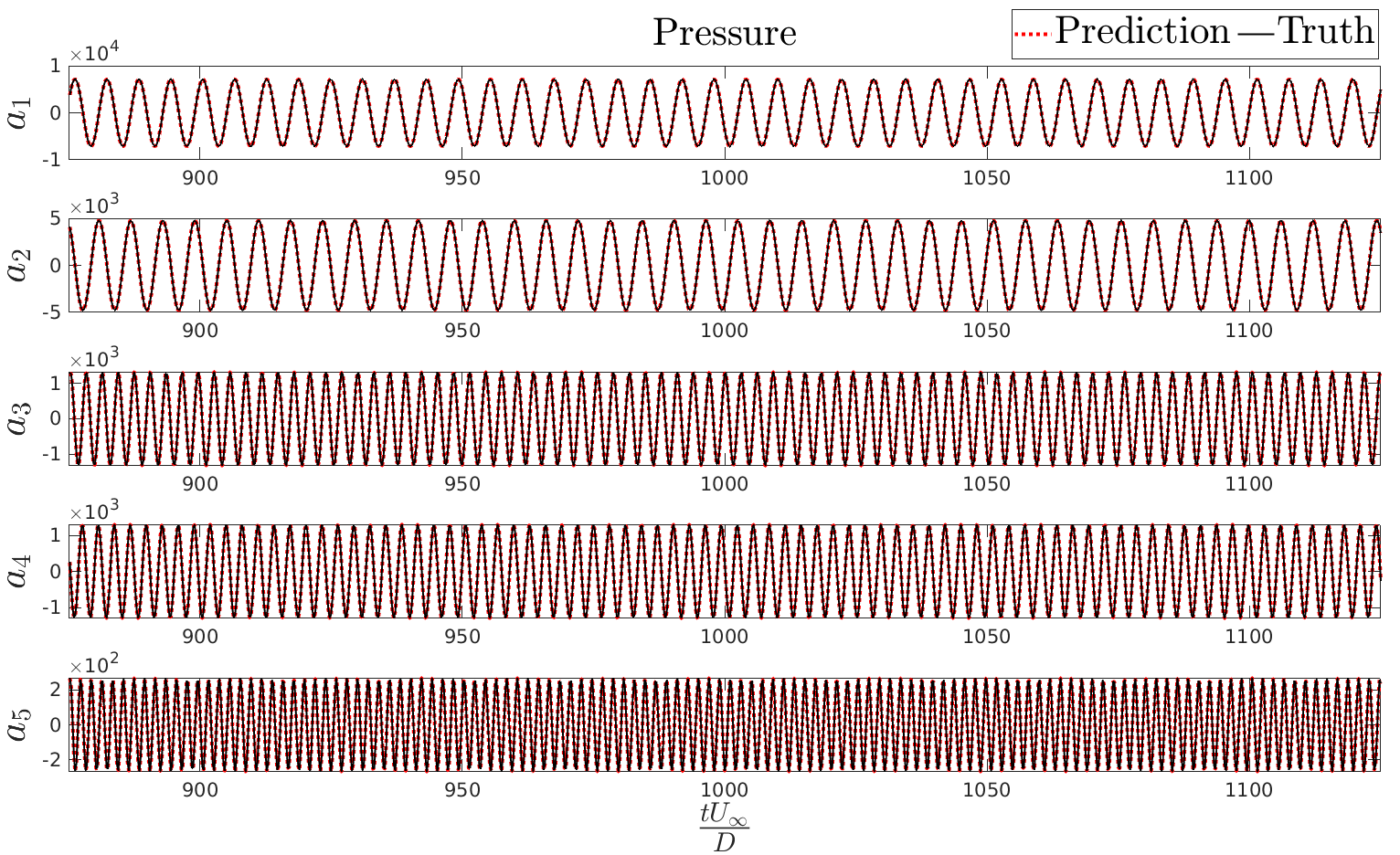}}\\
\subfloat[]{\includegraphics[width = 0.48\textwidth]{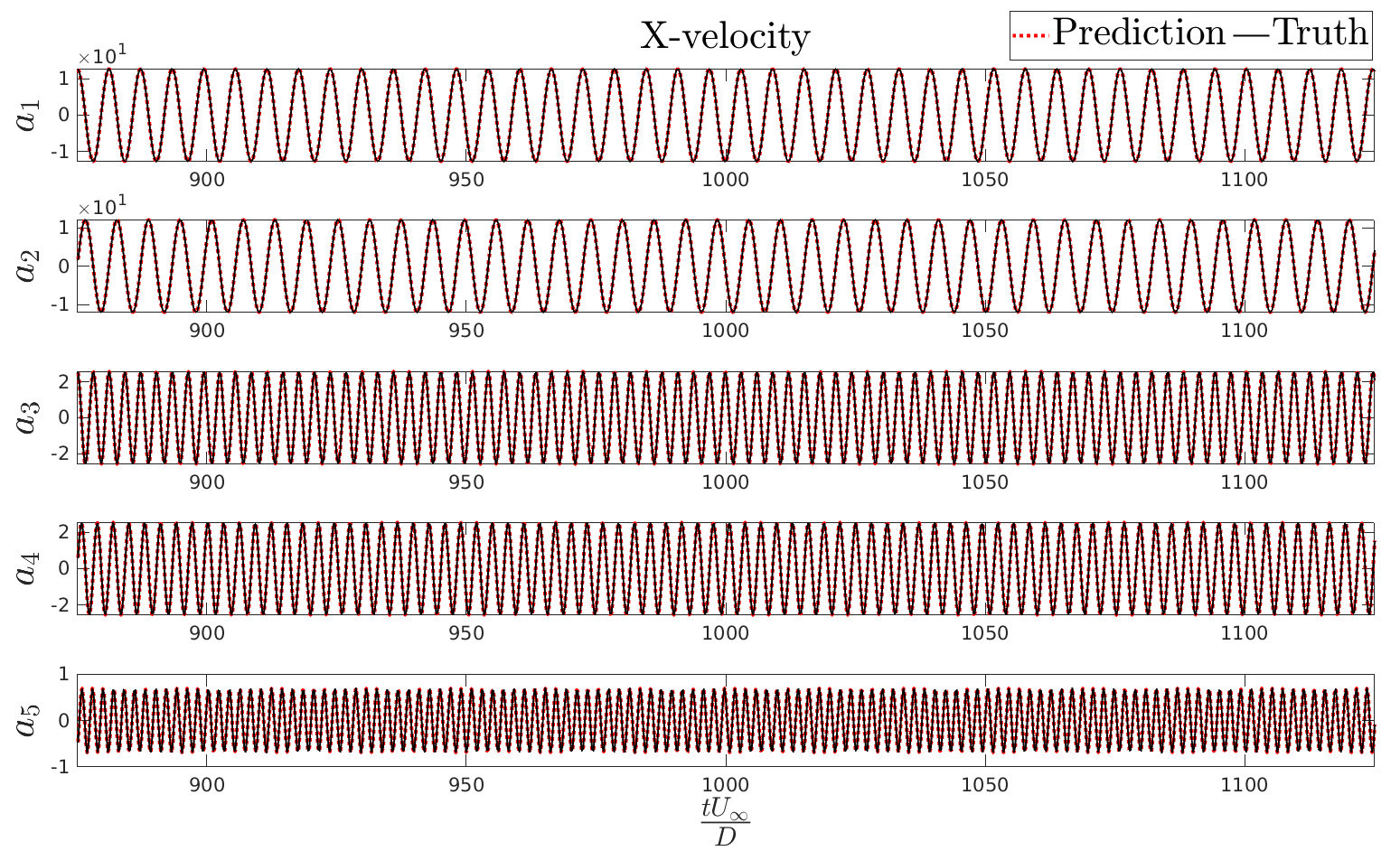}}
\caption{The flow past a cylinder: Temporal evolution (prediction) of modal coefficients (shown from $875$ till $1125\;tU_{\infty}/D$) for (a) pressure field $P$ and (b) x-velocity field $U$}
\label{podrnn_th_predict_stat}
\end{figure}

\begin{table}[H]
\centering
\subfloat[]{
\begin{tabular}{|c|c|}
\hline
 mode & RMSE \\ \hline
 1 &  $5.9849$ x $10^{-4}$ \\  \hline
 2 &  $6.6131$ x $10^{-4}$ \\ \hline
 3 &  $3.4554$ x $10^{-4}$ \\ \hline
 4 &  $3.6082$ x $10^{-4}$ \\ \hline
 5 &  $7.1799$ x $10^{-4}$ \\ \hline
\end{tabular}} 
\hspace{0.9cm}
\subfloat[]{
\begin{tabular}{|c|c|}
\hline
 mode & RMSE \\ \hline
 1 &  $6.4677$ x $10^{-4}$ \\  \hline
 2 &  $6.7578$ x $10^{-4}$ \\ \hline
 3 &  $1.1911$ x $10^{-3}$ \\ \hline
 4 &  $1.1762$ x $10^{-3}$ \\ \hline
 5 &  $1.7746$ x $10^{-3}$ \\ \hline
\end{tabular}}
\caption{The flow past a cylinder: RMSE for the predicted against true (normalised) modal coefficients for (a) pressure (b) x-velocity}
\label{podrnn_stat_rmse}
\end{table}
\end{enumerate}

\begin{figure}[H]
\centering
\subfloat[]
{\includegraphics[width = 0.243\textwidth]{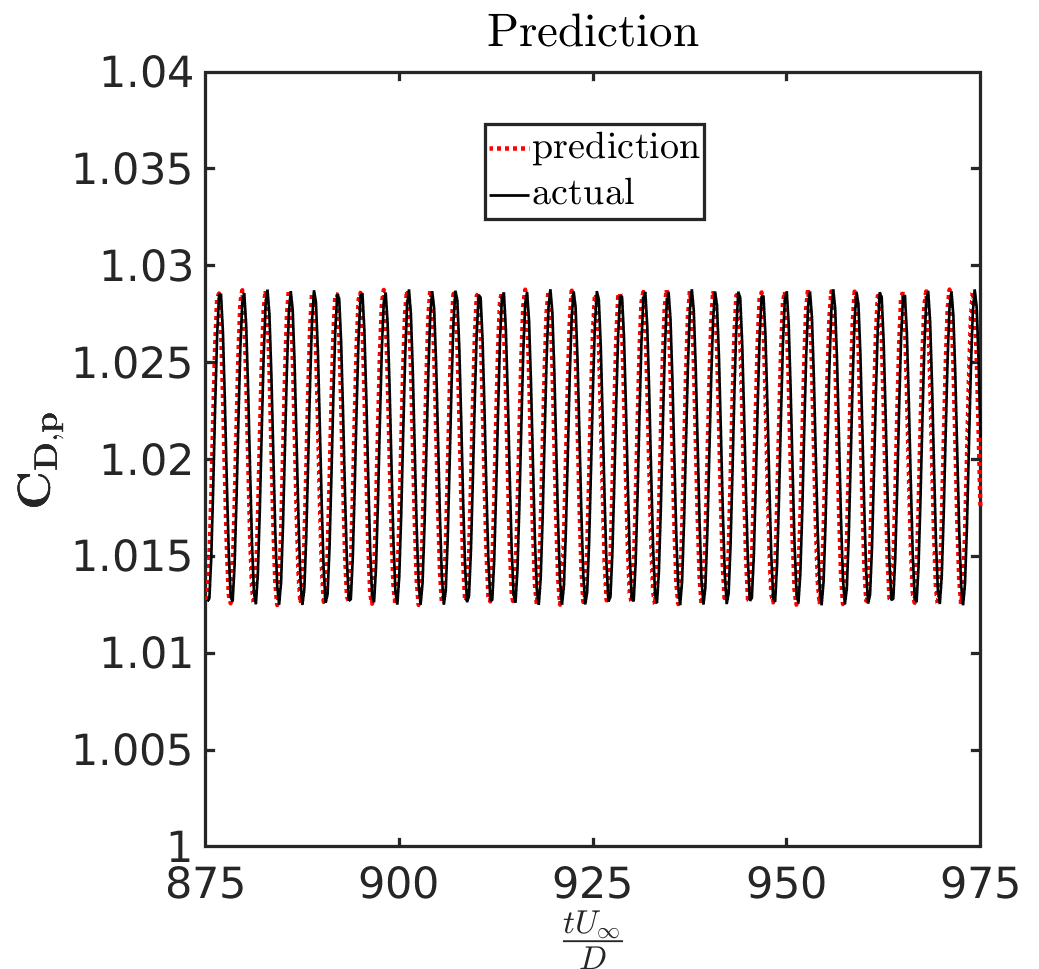}}
\subfloat[]
{\includegraphics[width = 0.233\textwidth]{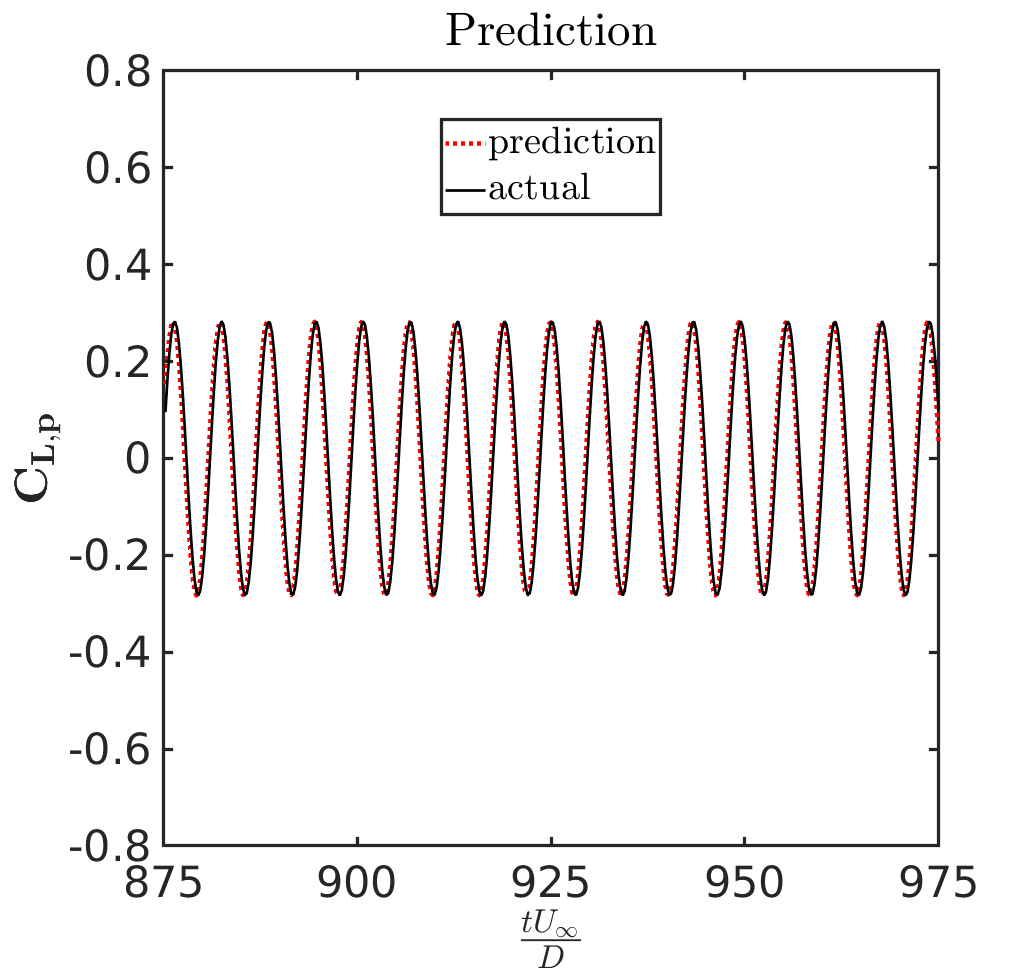}}
\caption{The flow past a cylinder: Predicted and actual (POD-RNN model) (a) drag and (b) lift force coefficients due to the pressure field ($P$) on the cylinder (shown from $875$ till $975\;tU_{\infty}/D$)}
\label{pred-forces-pod-rnn-stat}
\end{figure}

\begin{figure*}
\centering
\subfloat[]{
\includegraphics[width = 0.32\textwidth]{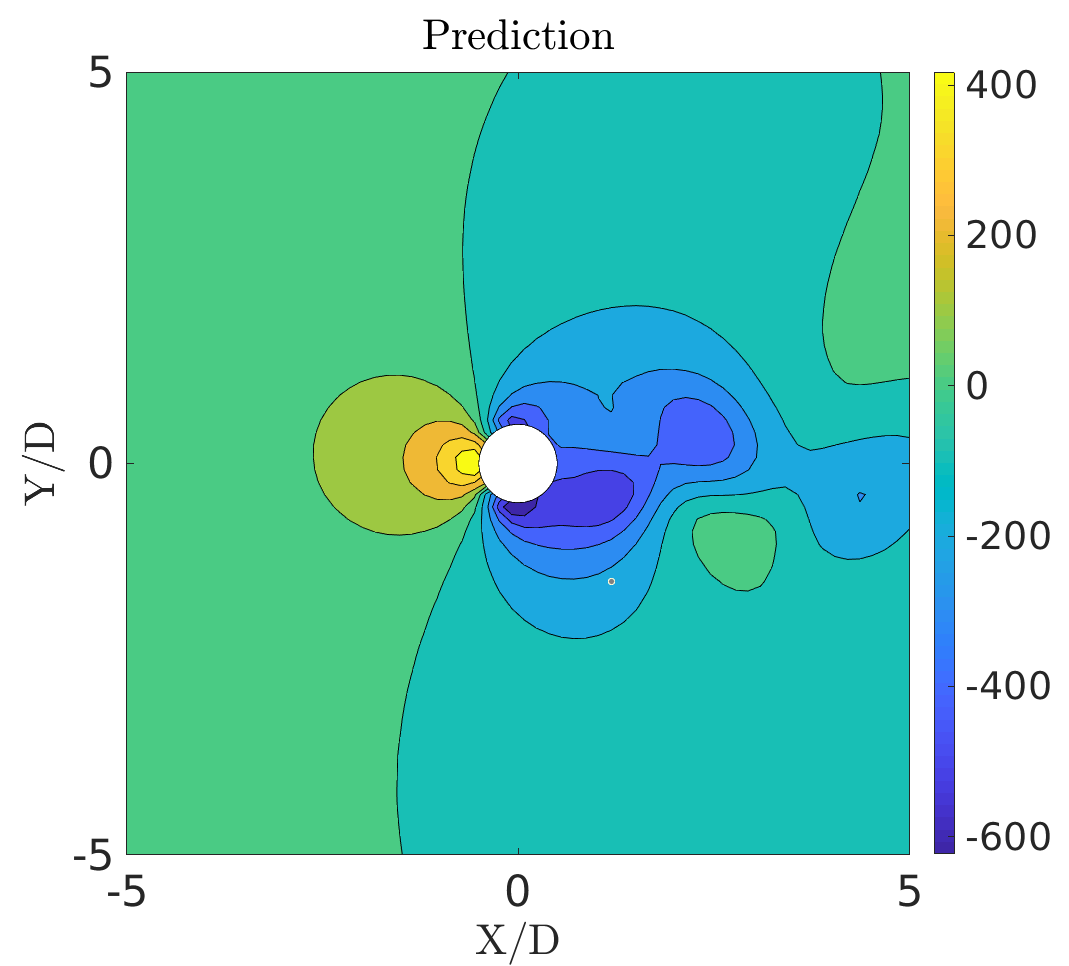}
\hspace{0.01\textwidth}
\includegraphics[width = 0.32\textwidth]{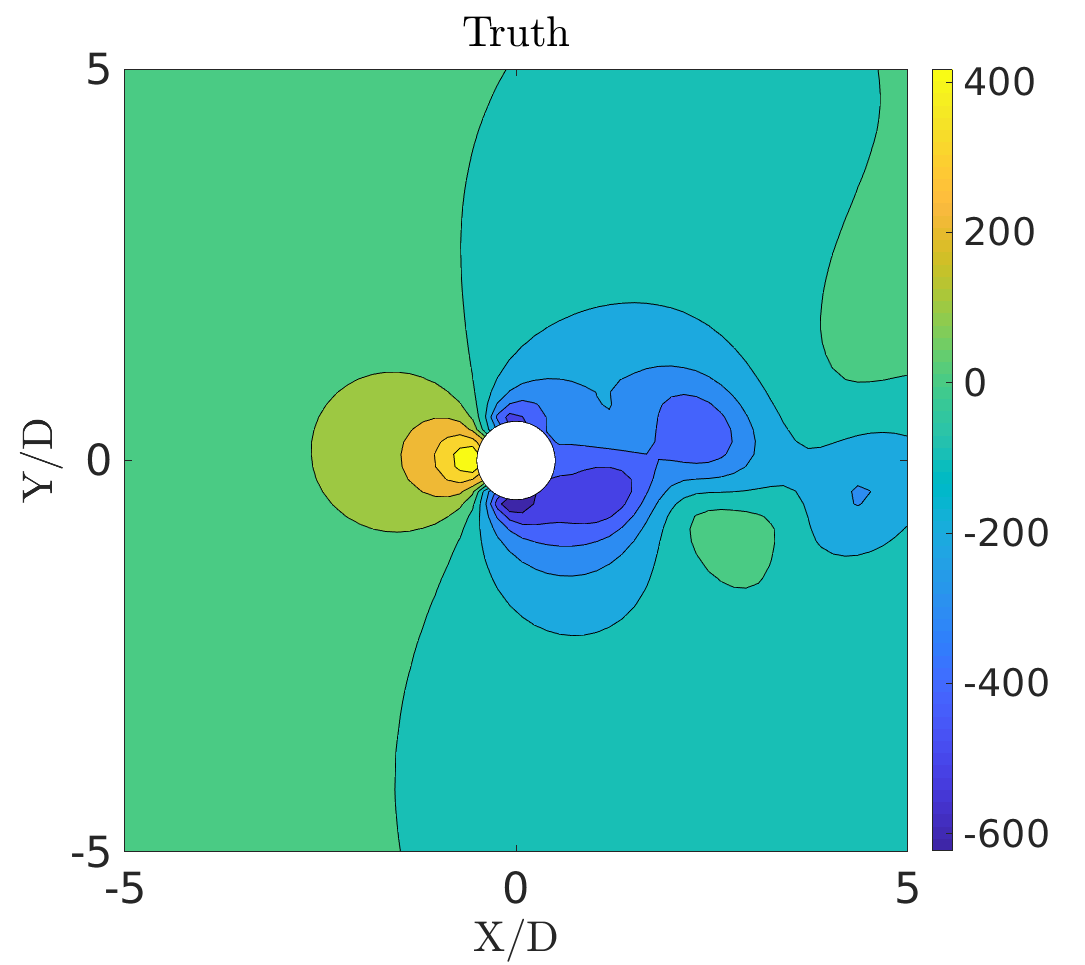}
\hspace{0.01\textwidth}
\includegraphics[width = 0.32\textwidth]{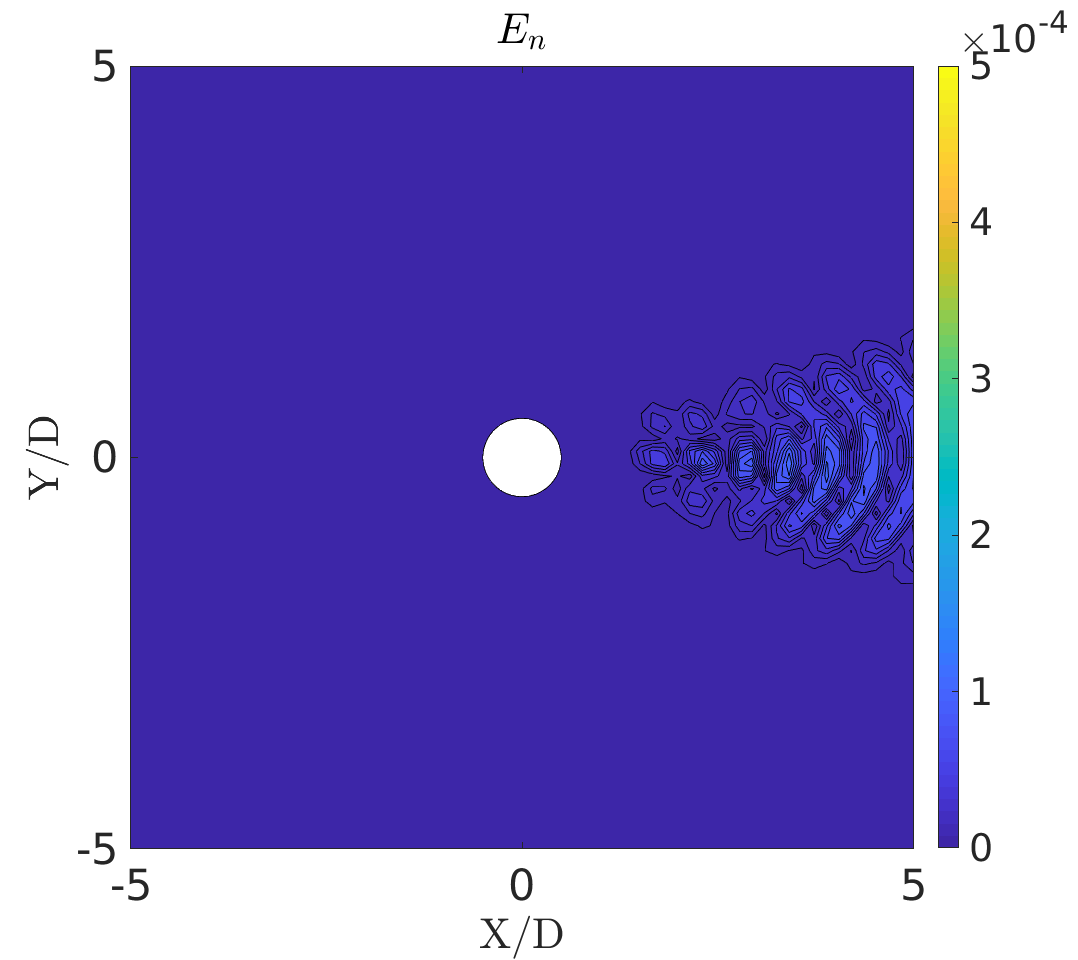}}
\\
\vspace{0.05\textwidth}
\subfloat[]{
\includegraphics[width = 0.32\textwidth]{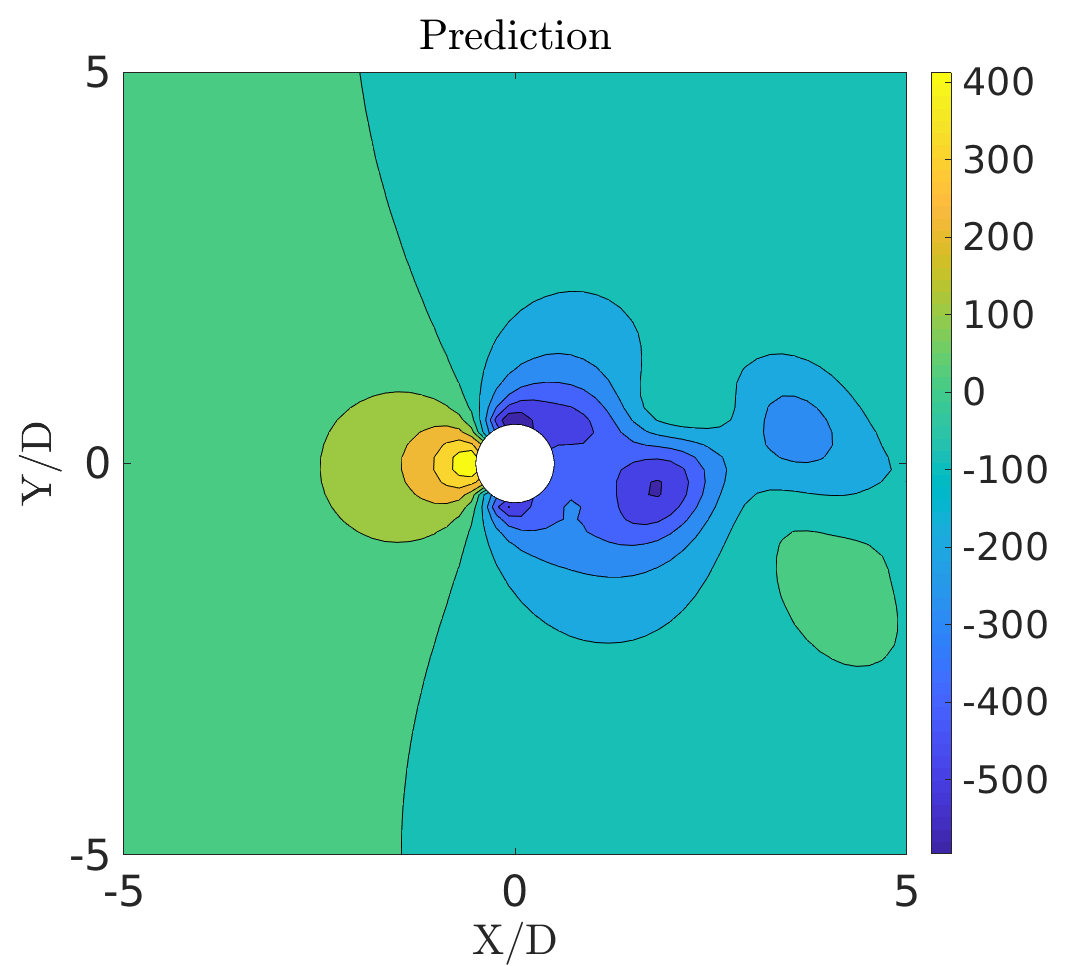}
\hspace{0.01\textwidth}
\includegraphics[width = 0.32\textwidth]{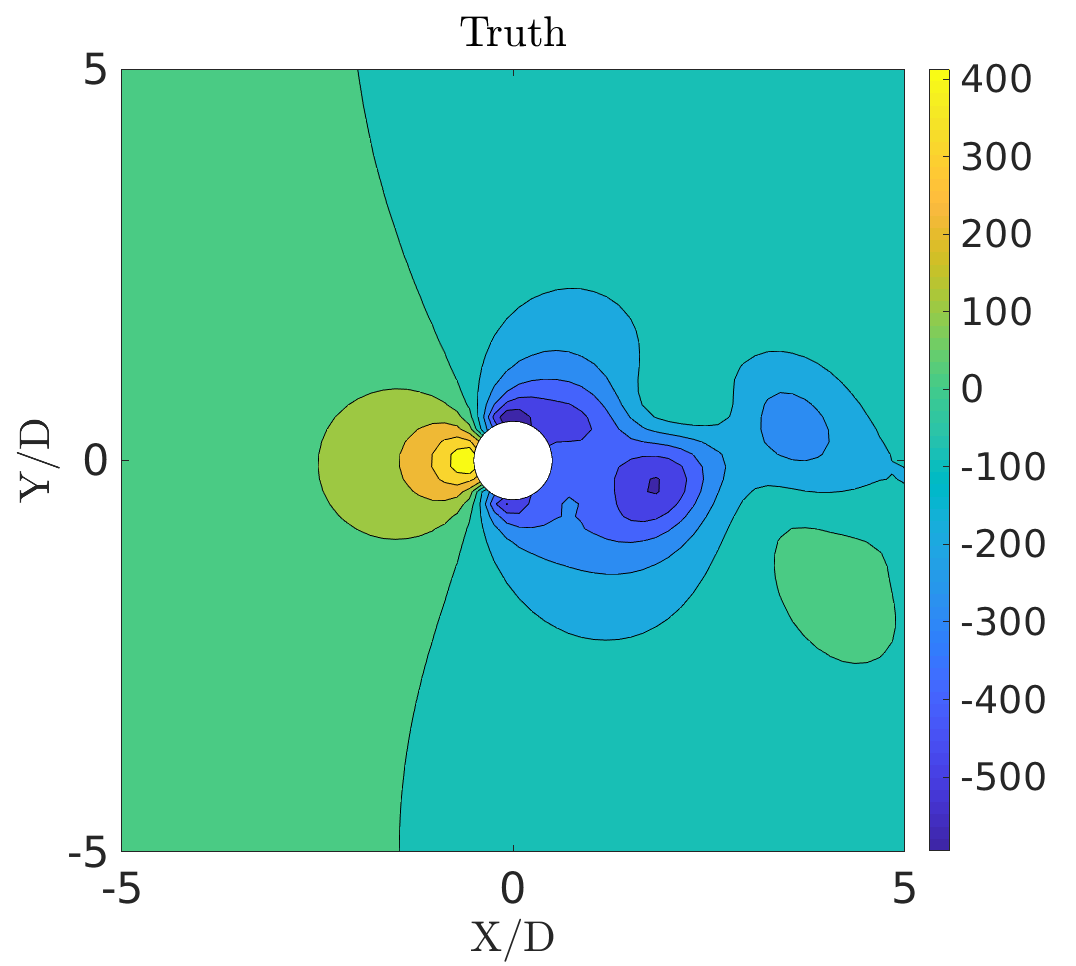}
\hspace{0.01\textwidth}
\includegraphics[width = 0.32\textwidth]{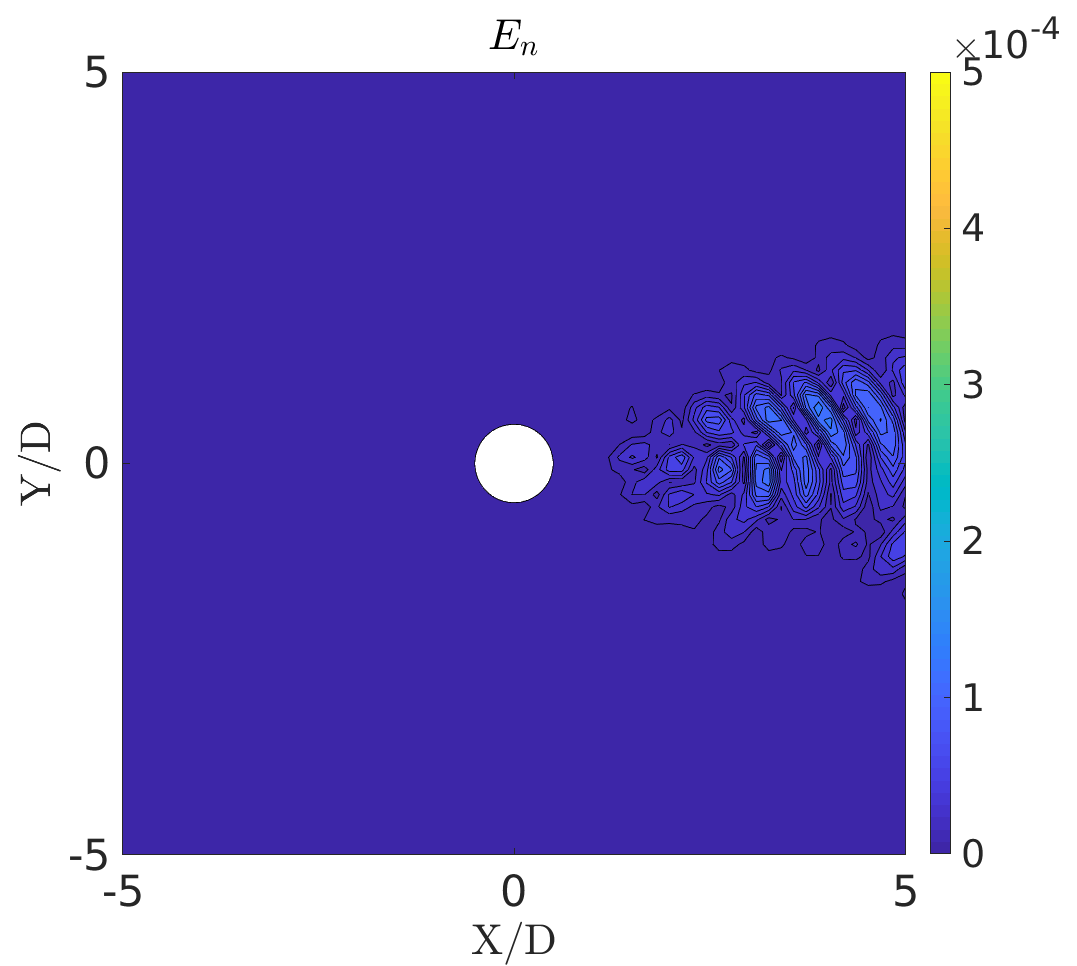}}
\\
\vspace{0.05\textwidth}
\subfloat[]{
\includegraphics[width = 0.32\textwidth]{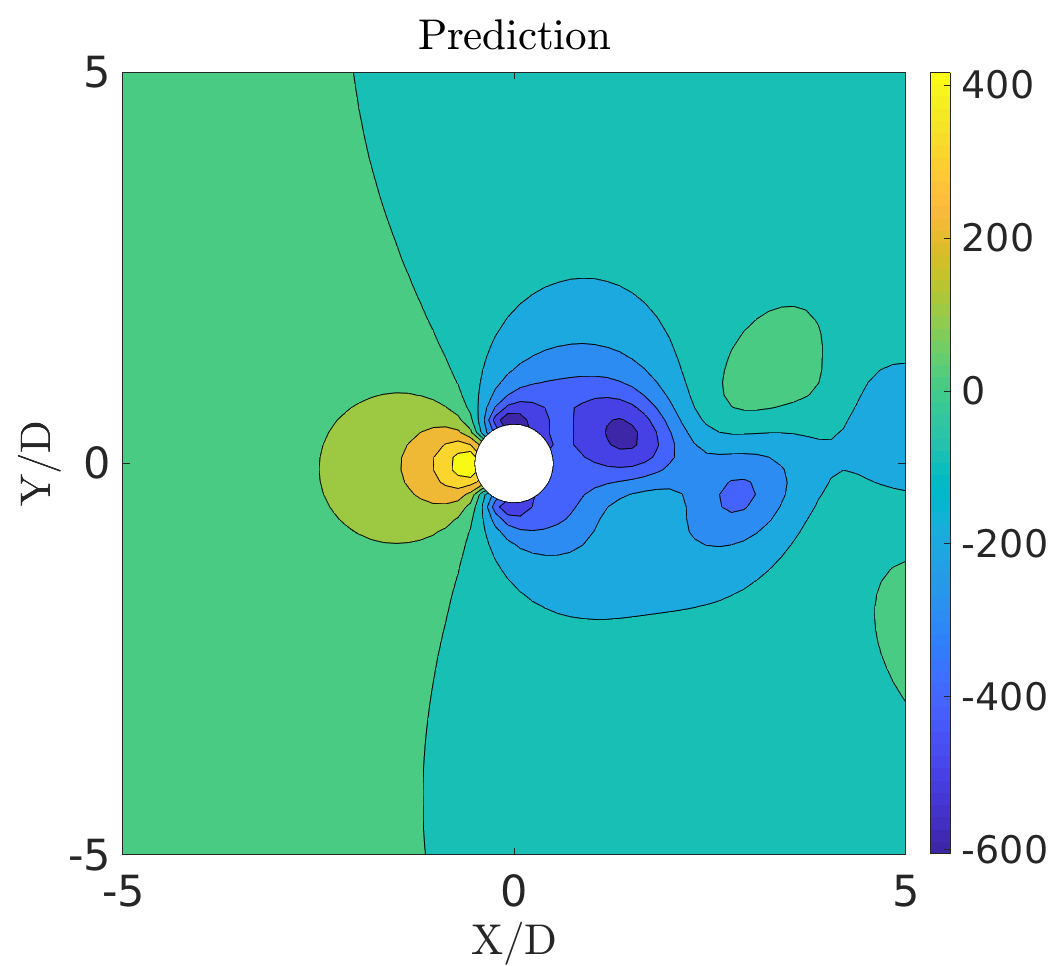}
\hspace{0.01\textwidth}
\includegraphics[width = 0.32\textwidth]{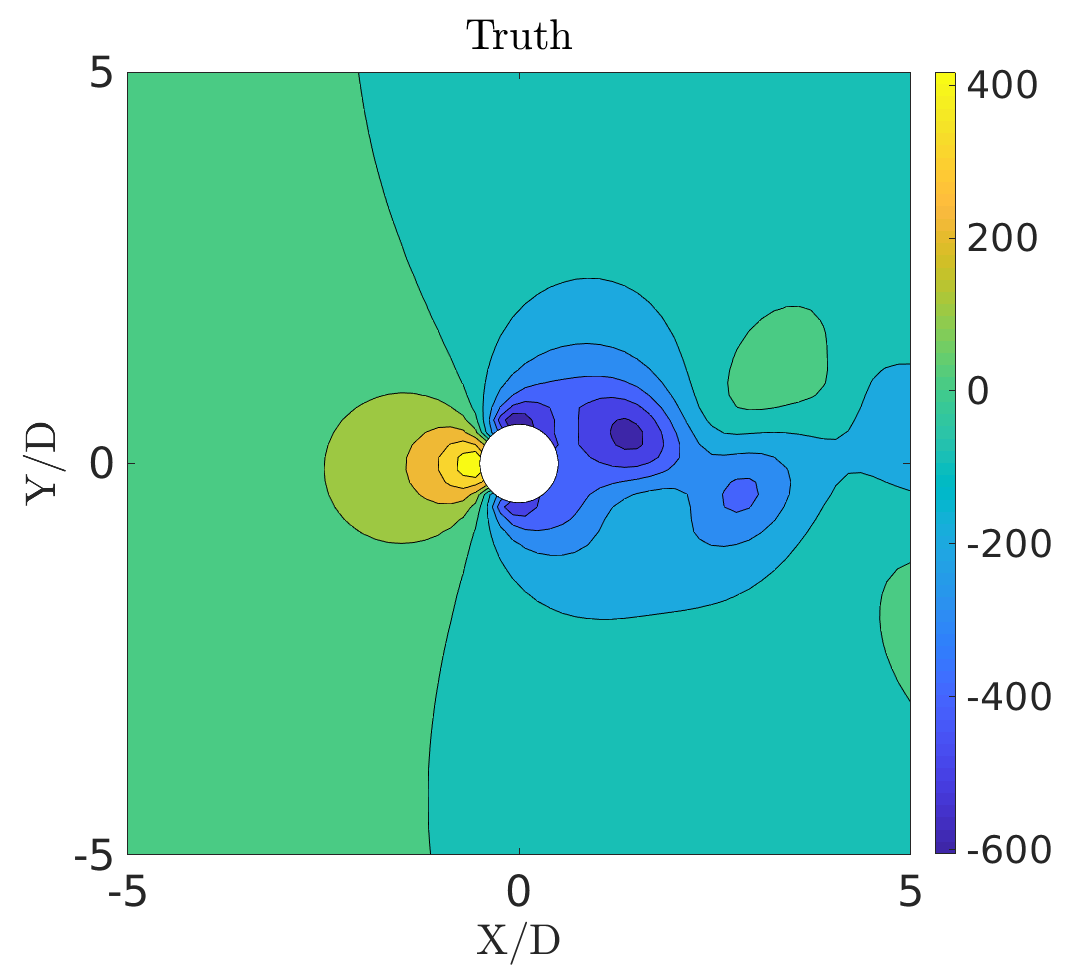}
\hspace{0.01\textwidth}
\includegraphics[width = 0.32\textwidth]{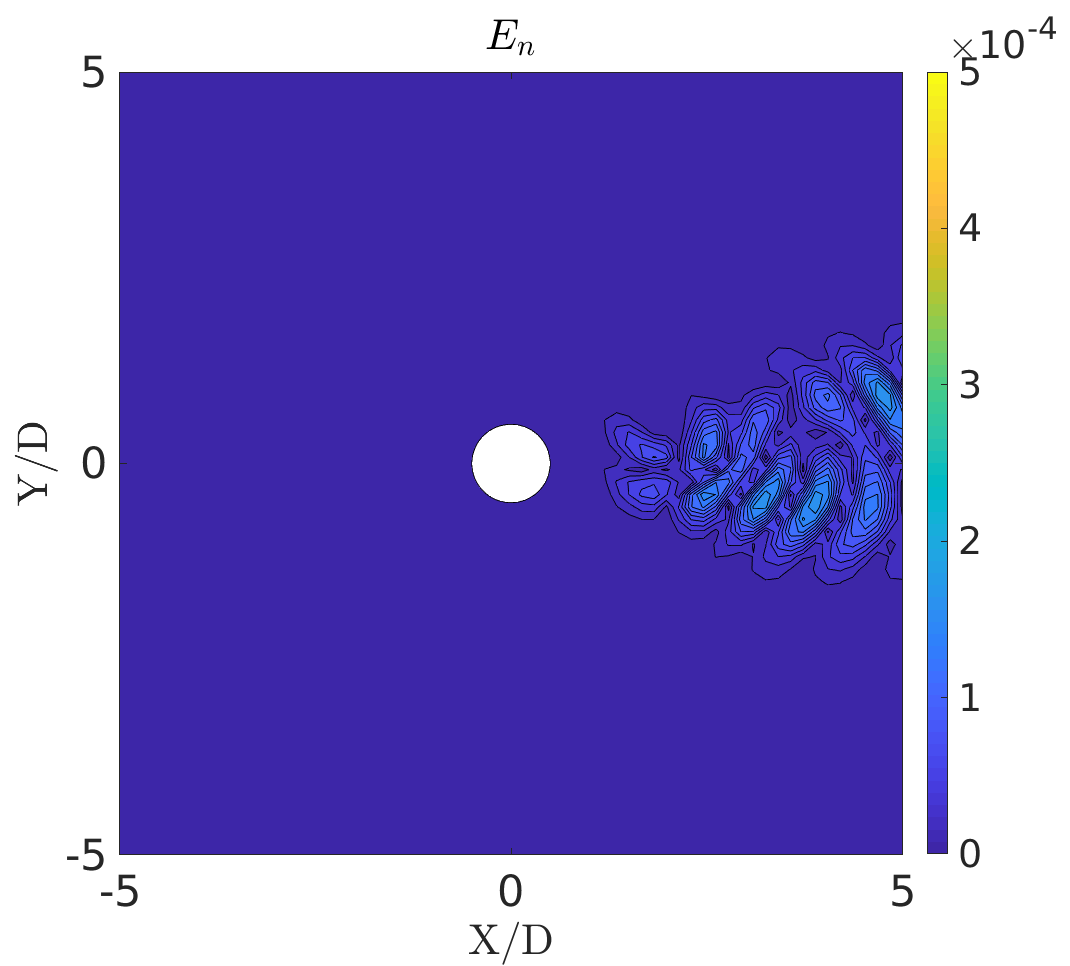}}
\caption{The flow past a cylinder: Comparison of predicted and true fields (POD-RNN model) along with normalized reconstruction error $E_{n}$ at $tU_{\infty}/D =$ (a)  925, (b) 1000, (c) 1075 for pressure field ($P$)}
\label{fig_flow_comp_p_podrnn_stat}
\end{figure*}

\begin{figure*}
\centering
\subfloat[]{
\includegraphics[width = 0.32\textwidth]{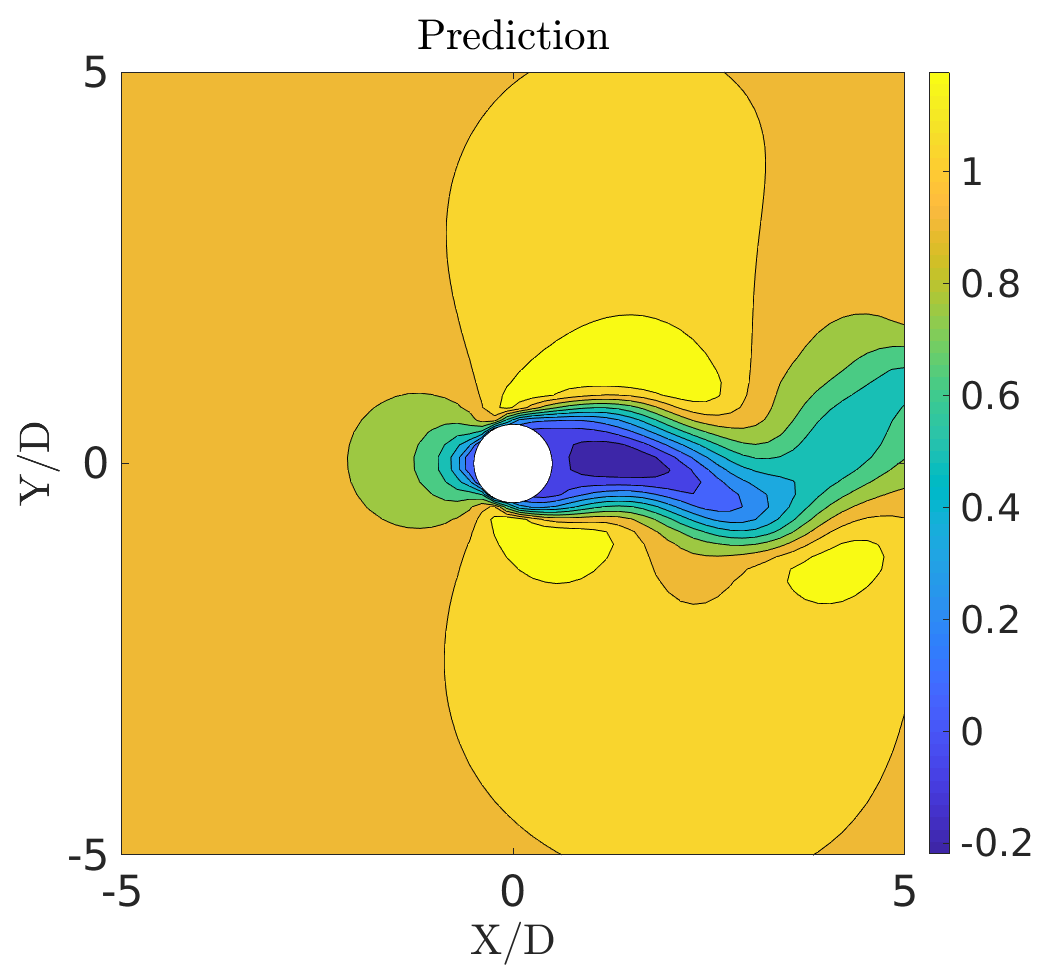}
\hspace{0.01\textwidth}
\includegraphics[width = 0.32\textwidth]{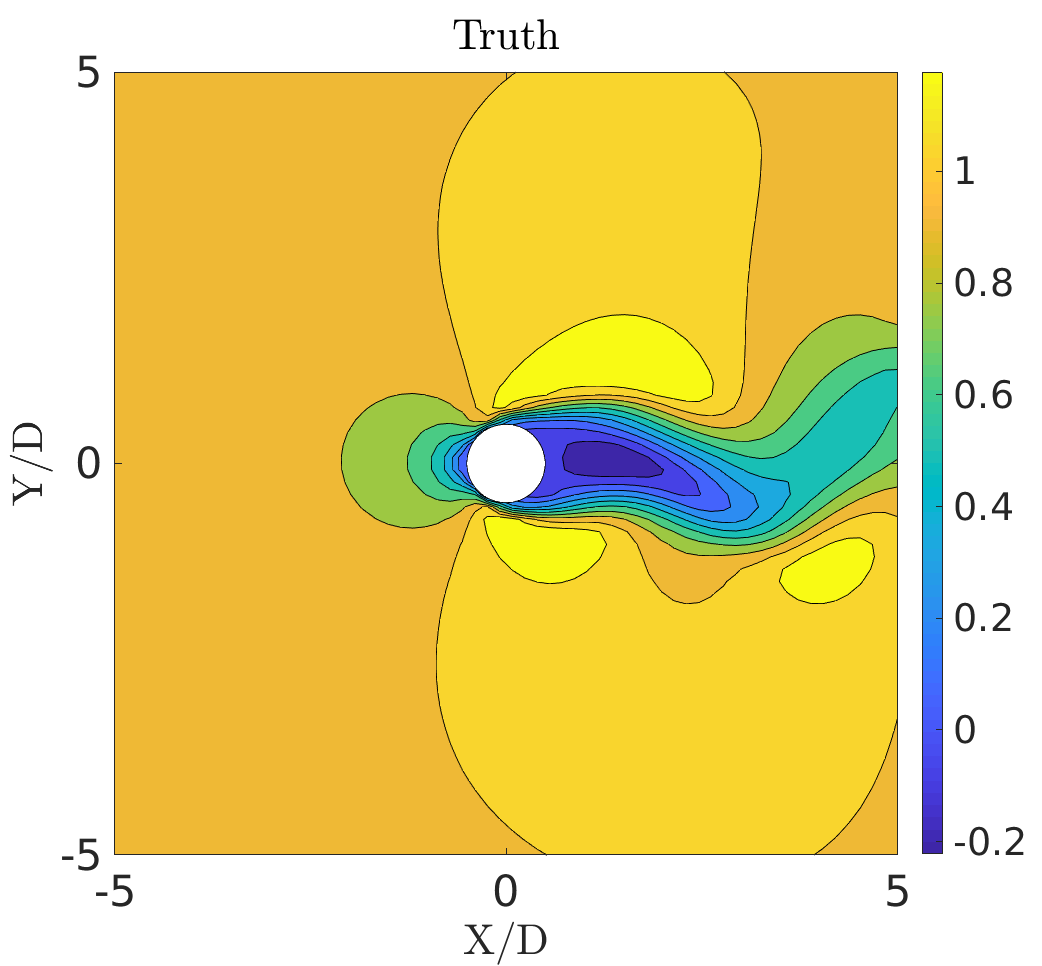}
\hspace{0.01\textwidth}
\includegraphics[width = 0.32\textwidth]{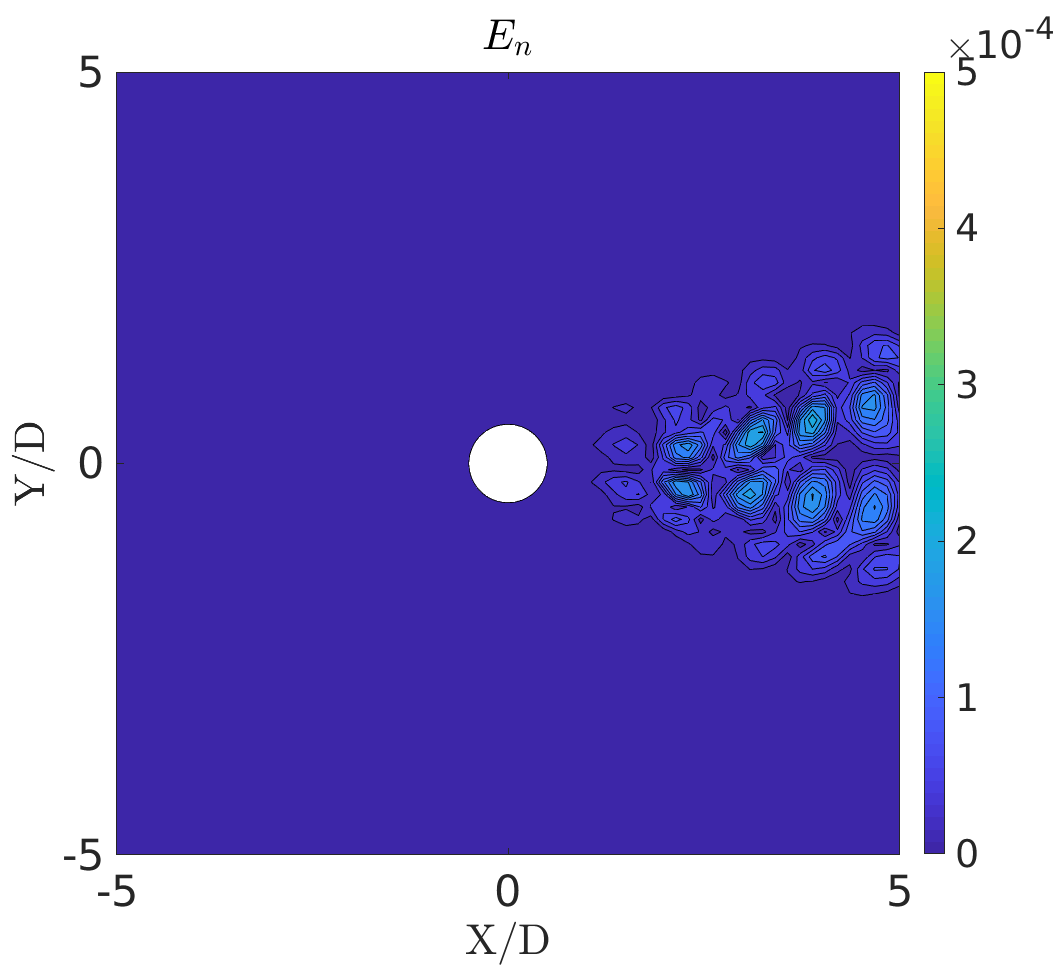}}
\\
\vspace{0.05\textwidth}
\subfloat[]{
\includegraphics[width = 0.32\textwidth]{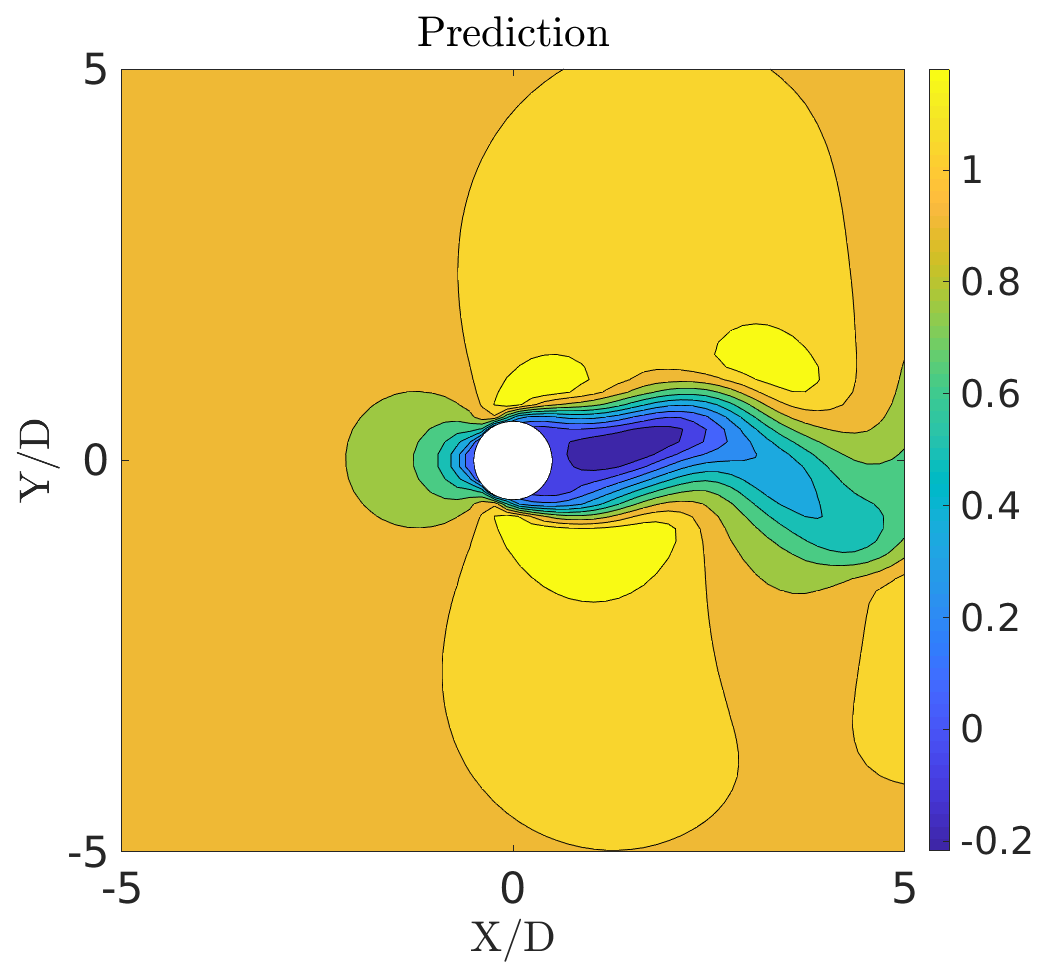}
\hspace{0.01\textwidth}
\includegraphics[width = 0.32\textwidth]{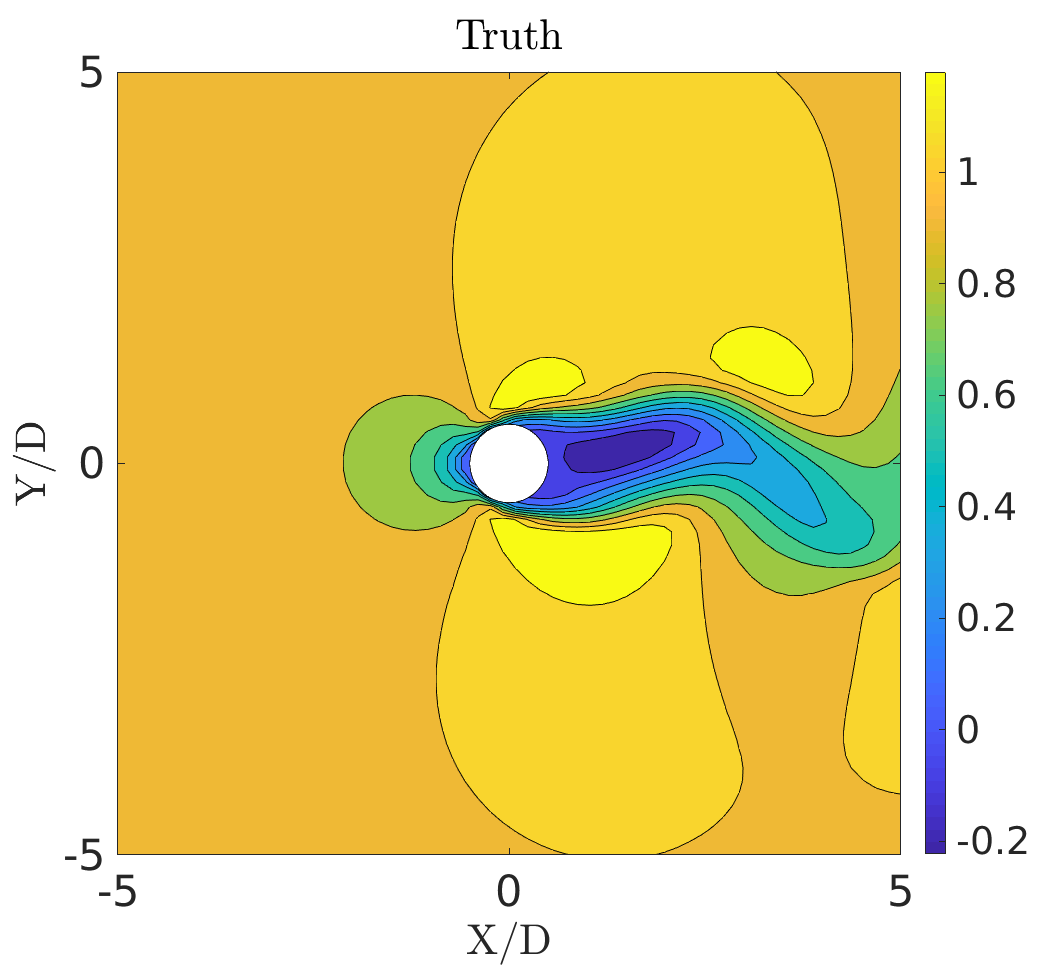}
\hspace{0.01\textwidth}
\includegraphics[width = 0.32\textwidth]{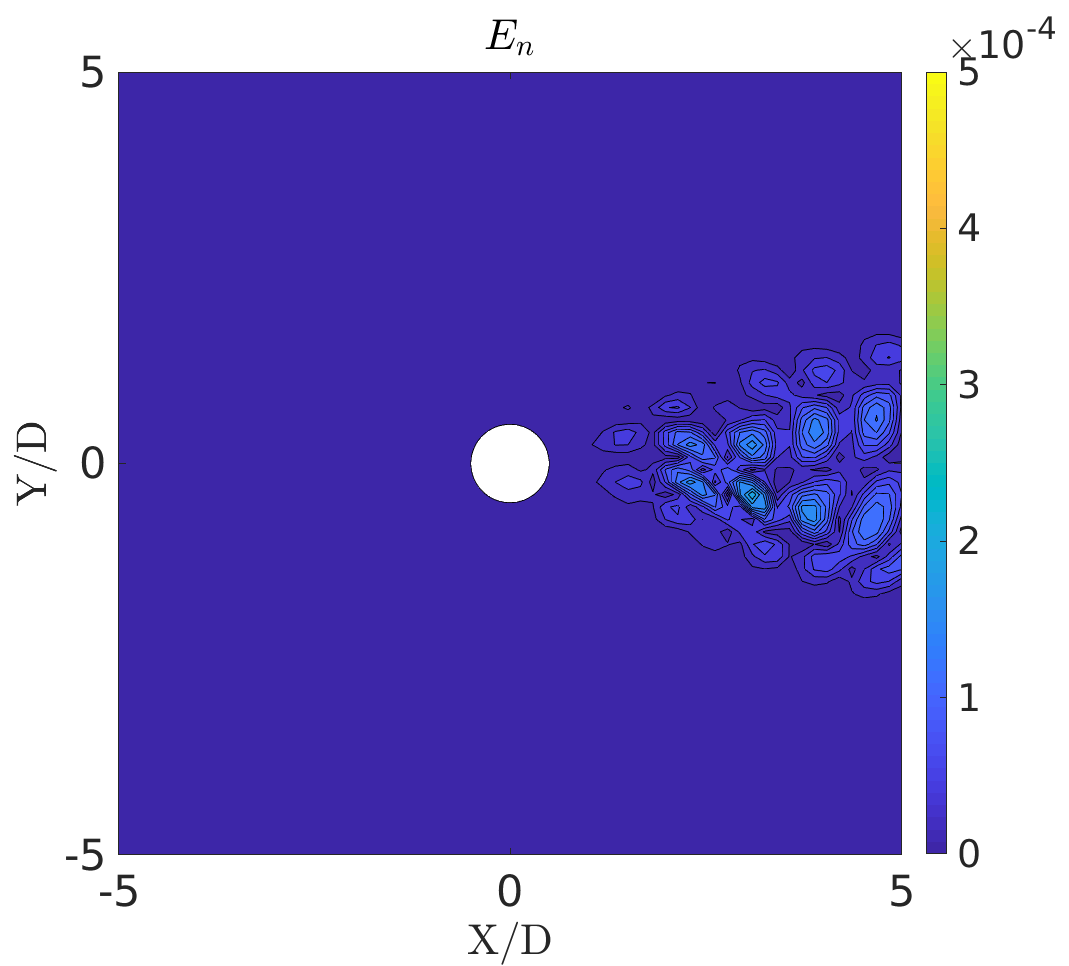}}
\\
\vspace{0.05\textwidth}
\subfloat[]{
\includegraphics[width = 0.32\textwidth]{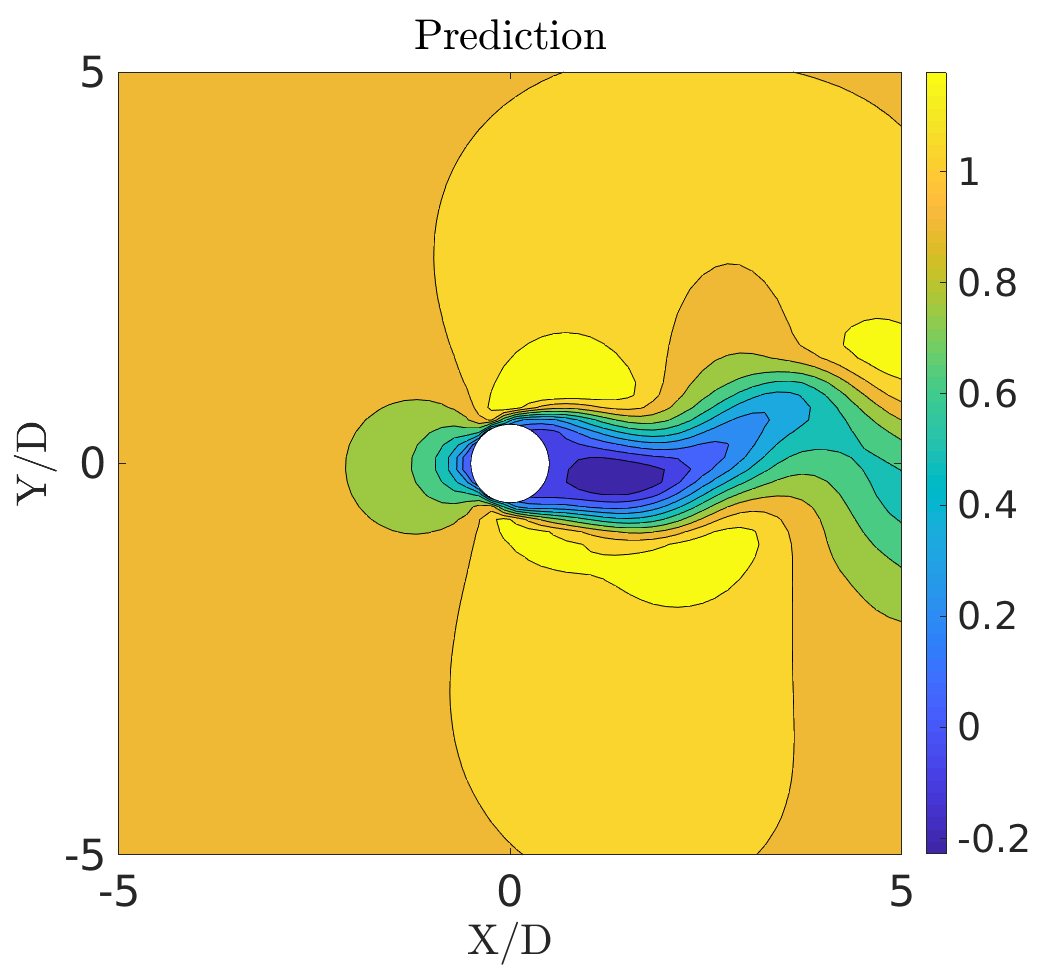}
\hspace{0.01\textwidth}
\includegraphics[width = 0.32\textwidth]{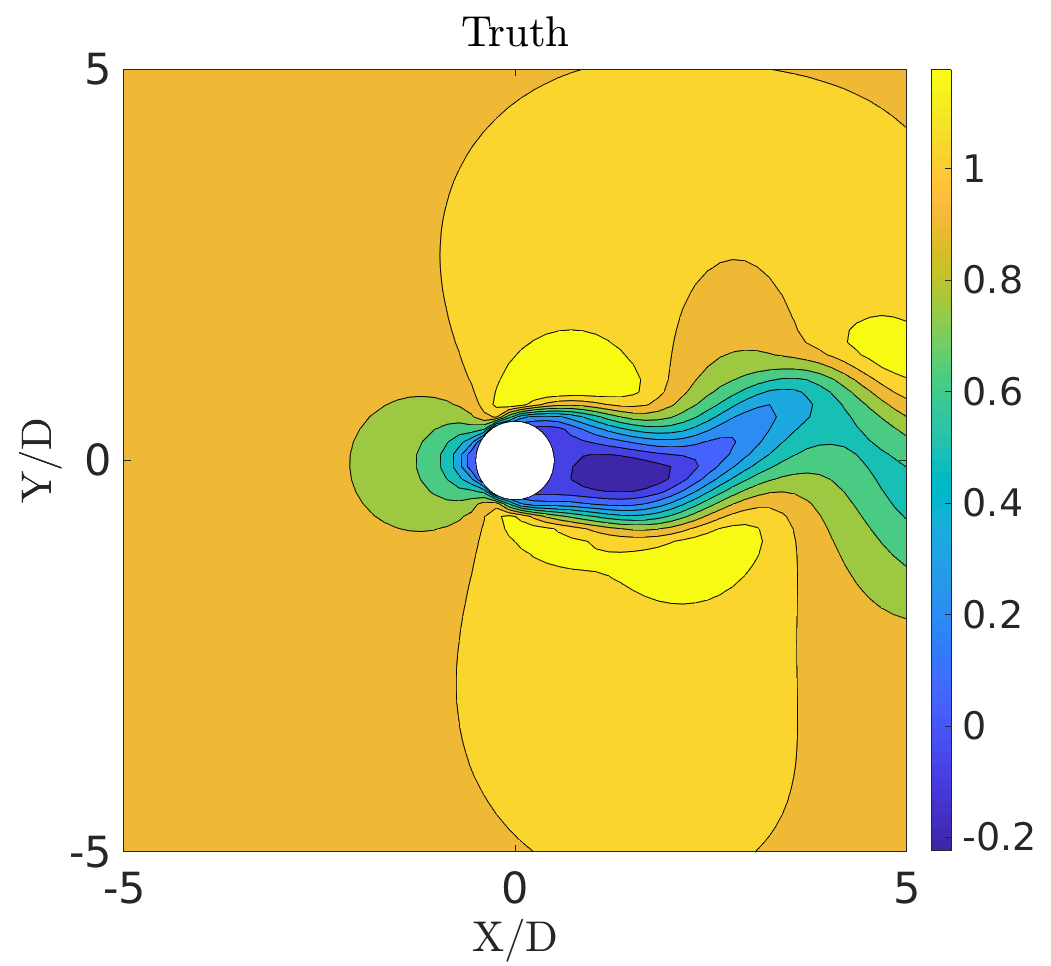}
\hspace{0.01\textwidth}
\includegraphics[width = 0.32\textwidth]{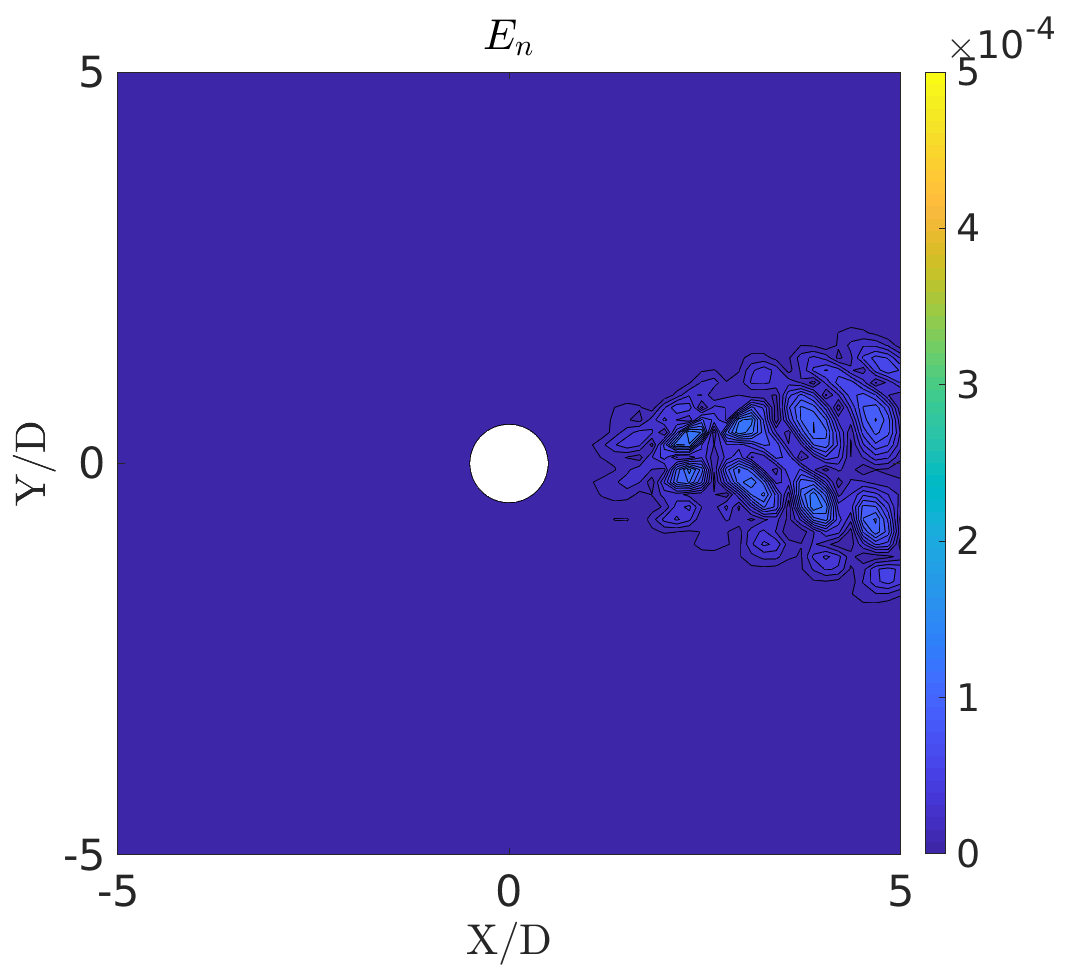}}
\caption{The flow past a cylinder: Comparison of predicted and true fields (POD-RNN model) along with normalized reconstruction error $E_{n}$ at $tU_{\infty}/D =$(a)  925, (b) 1000, (c) 1075 for x-velocity field ($U$)}
\label{fig_flow_comp_u_podrnn_stat}
\end{figure*}

% Field and force prediction
\textit{Field prediction}: The predicted modal coefficients at any time-step,         $\hat{\textbf{A}}_{n} \in \mathbb{R}^{k}$, can simply be reconstructed back to the high dimensional state $\hat{\textbf{S}}_{n}$ using the mean field $\bar{\textbf{S}} \in \mathbb{R}^{m}$ and $k$ spatial POD modes $\boldsymbol{\Phi} \in \mathbb{R}^{m\times k}$ as $\hat{\textbf{S}}_{n} \approx \bar{\textbf{S}} + \boldsymbol{\Phi} \hat{\textbf{A}}_{n}$. Fig. \ref{fig_flow_comp_p_podrnn_stat} and \ref{fig_flow_comp_u_podrnn_stat} depict the comparison of predicted and true values for $P$ and $U$ fields respectively at time-steps $3700$ ($925\;tU_{\infty}/D$), $4000$ ($1000\;tU_{\infty}/D$) and $4300$ ($1075\;tU_{\infty}/D$). The normalized reconstruction error $E_{n}$ is constructed by taking the absolute value of differences between the true $\textbf{S}_{n}$ and predicted $\hat{\textbf{S}}_{n}$ fields at any time-step $n$ and for all points $k$ in the space $(x,y)$ and is given by 

\begin{equation}
    E_{n} = \frac{|\textbf{S}_{n}-\hat{\textbf{S}}_{n}|}{\|\textbf{S}_{n}\|_{2,k}}.
\end{equation}
\\
\textit{Force coefficient prediction}:
The high-dimensional field prediction on all the nodal points allows to carry the reduced numerical integration directly over the fluid-solid boundary on the cylinder. The Cauchy Stress tensor $\bm{\sigma^{f}}$ is constructed with pressure field data and Eq. (\ref{force_eqns}) in section \ref{HDM} is used to integrate it over the fluid-solid boundary to get $\textbf{C}_{\mathrm{D},\text{p}}$ and $\textbf{C}_{\mathrm{L},\text{p}}$.  Fig.~\ref{pred-forces-pod-rnn-stat} depicts the predicted and actual pressure coefficient from time-steps $3501-3900$. Note that the reference length scale is the diameter of cylinder ($D$) and the reference velocity is the inlet velocity ($U_{\infty}$). $t$ denotes the actual simulation time carried at a time-step $0.25s$.

% force 

%\newpage
%\onecolumn
% pressure

% velo 

%\newpage
%\twocolumn
%%%%%%%%%%%%%%%%%%%%%%%%%%%%
\subsection{CRAN model}\label{cran_stat}
\begin{figure}
\centering
\subfloat[]
{\includegraphics[width = 0.238\textwidth]{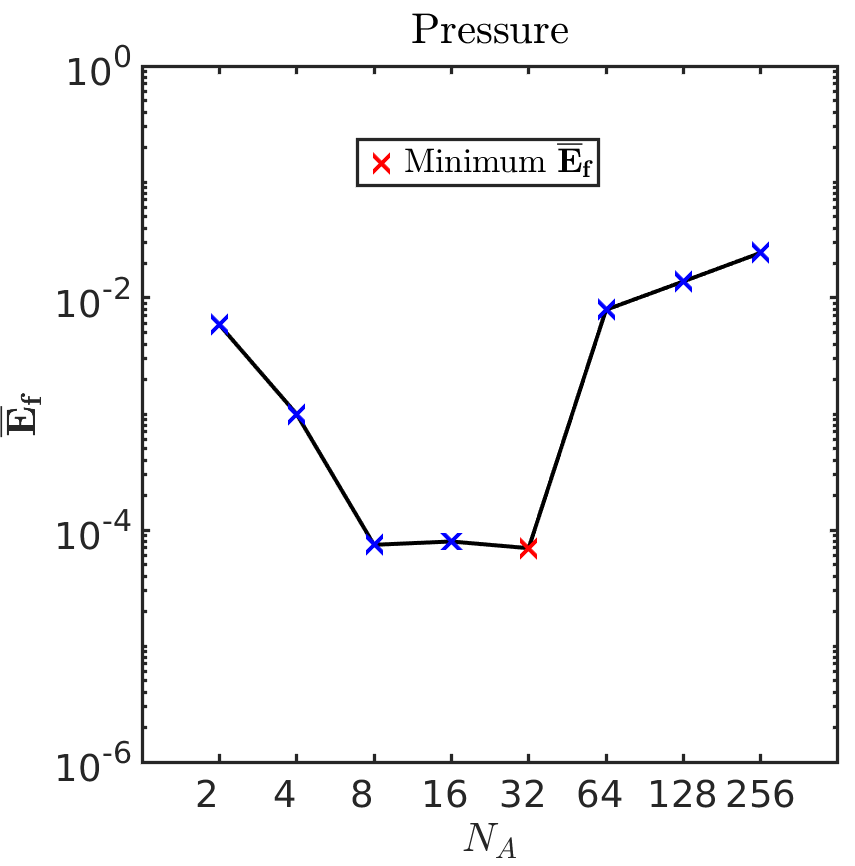}}
\subfloat[]
{\includegraphics[width = 0.238\textwidth]{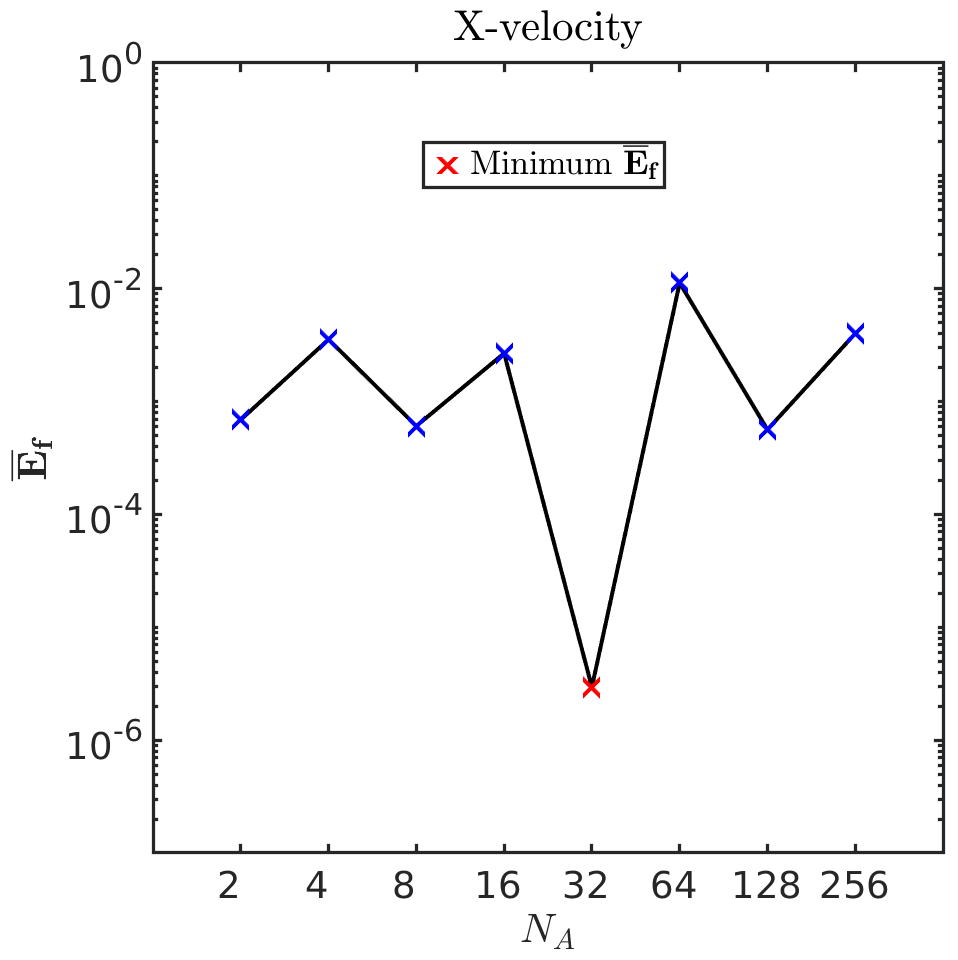}}
\caption{The flow past a cylinder: Mean normalized squared error ($\bar{\mbox{E}}_{\mbox{f}}$) for predicted time-steps with respect to size of low-dimensional space $N_{A}$ for (a) pressure field $P$ and (b) x-velocity field $U$}
\label{cran_stat_ef}
\end{figure}

%%%%%%%%%%%%%%%%%%%%%%%%%%%%
The end-to-end nonlinear model order reduction tool based on CRAN is now applied on the same problem of flow past a cylinder. For uniformity, same set of training (time-steps $501-3500$) and testing (time-steps $3501-4500$) data is selected for this application. The process is outlined as follows:      

\begin{enumerate}
% Step 1 : reference grid 
\item Consider any high-dimensional unstructured data snapshots from a flow solver $\boldsymbol{\mathcal{S}} = \left\lbrace\textbf{S}_{1}\;\textbf{S}_{2}\dots\;\textbf{S}_{N}\;\right\rbrace \in \mathbb{R}^{m\times N}$  (here, $m=26114$ and $N=4000$). To get a spatial uniformity in the unstructured data, SciPy's $"griddata"$ function \cite{SciPy} is used to map the $m$ dimensional unstructured data on a 2-d reference grid of size $N_{x} \times N_{y}$ (here, $N_{x}=N_{y}=64$). Note that the reference grid for this case is similar to Fig. \ref{figunstrct-strct}, but of size $10D \times 10D$ and the single cylinder positioned at the center. The snapshot data-set hence obtained is $\boldsymbol{\mathcal{S}} = \left\lbrace\textbf{s}_{1}\;\textbf{s}_{2}\dots\;\textbf{s}_{N}\;\right\rbrace \in \mathbb{R}^{N_{x}\times N_{y}\times N}$. The $N$, 2-d snapshots are divided into training $n_{tr}=3000$ and testing $n_{ts}=1000$. 

% Step 2 : data set
\item The $n_{tr}$ training data is broken into $N_{s}$ batches, each of finite time-step size $N_{t}$. Note that $n_{tr} = N_{s} N_{t}$. The procedure described in section \ref{unsupervised-t-s} is followed to generate the feature scaled data for training, 
% Data Set 
\begin{equation}
    \mathcal{S} = \{\textbf{S}^{'1}_{s},\dots,\textbf{S}^{'N_{s}}_{s\\}\}\in [0,1]^{N_{x}\times N_{y}\times N_{t}\times N_{s}},
\end{equation}

where each training sample $\textbf{S}^{'i}_{s} = [\textbf{s}^{'1}_{s;i},\dots,\textbf{s}^{'N_{t}}_{s;i}]$. In this analysis, $N_{x},\;N_{y} = 64$ and $N_{t} = 20$, $N_{s} = 150$. Note that, $\textbf{S}^{'i}_{s}$ can be pressure or velocity data. 

% Step 3 : training 
\item Train a range of convolutional recurrent autoencoder networks (CRANs) using the data set described above with the low-dimensional evolver LSTM state $\textbf{A}$ of different sizes (refer Fig.~\ref{figcnnrnn} and \ref{figcnnrnn1}). Since $\textbf{A}$ is a latent abstract feature space, we experiment predictions with different size of low-dimensional state $N_{A} = 2,\;4,\;8,\;16,\;32,\;64,\;128,\;256$ for both pressure $P$ and x-velocity $U$. All the models are trained on a single Intel E5-2690v3 (2.60GHz, 12 cores) CPU node for $N_{train} = 500,000$ iterations. We took sufficiently long iterations to account for a non-convex optimisation, although, we did observe the objective function (Eq. (\ref{loss})) to reach a nearly steady value of approximately $10^{-6}$ in less than \\
$100,000$ iterations. It should be noted that we are using a closed recurrent neural network in the CRAN model (see Fig.~\ref{closedloop}) i.e., the prediction from previous step is used as an input for the next prediction. The main advantage of using such a model is that, there is no requirement of the true data to be fed into the network in a time delayed fashion. One can predict for a finite time horizon with a single true sample (infer data) which is used for initialization of the network.

% Step 4 : prediction and E_f...
\item The online prediction is carried for $n_{ts}=1000$ time-steps (from $3501-4500$) using just one time-step ($3500$) with the trained CRAN models. Fig. \ref{cran_stat_ef} (a) and (b) depict the mean normalized squared error ($\bar{\mbox{E}}_{\mbox{f}}$) for 1000 predicted time-steps for $P$ and $U$ respectively.
\begin{equation}
    \overline{\mbox{E}}_{\mbox{f}} = \frac{  \sum_{n}\frac{\|\textbf{s}_{n}-\hat{\textbf{s}}_{n}\|^{2}_{2,k}}{\|\textbf{s}_{n}\|^{2}_{2,k}+\epsilon}}{n_{ts}},
   \label{mnse}
\end{equation}
where $n$ denotes a time instance from $3501$ to $4500$ time-steps and $k$ denotes a point $(x,y)$ in space. $\|\dots\|_{2,k}$ denotes the $L_2$ norm over spatial dimension.
$\textbf{s}_{n}$ and $\hat{\textbf{s}}_{n}$ are respectively the true and predicted fields. The tuning of the hyper-parameter $N_{A}$ reveals that, there exist an optimal value of $N_{A}$ for all input data. For the pressure data $P$, as shown in Fig.~\ref{cran_stat_ef} (a), there exists approximately two order magnitude difference between $N_{A} = 8,16,32$ and $N_{A} = 2,64,128$. Similarly, we observe around three order magnitude difference for $N_{A}=32$ for x-velocity $U$ compared to the rest. Hence, the CRAN tunes for the best performance at $N_{A}=32$, for both $P$ and $U$.      

\end{enumerate}

\begin{figure*}
\centering
\subfloat[]
{\includegraphics[width = 0.32\textwidth]{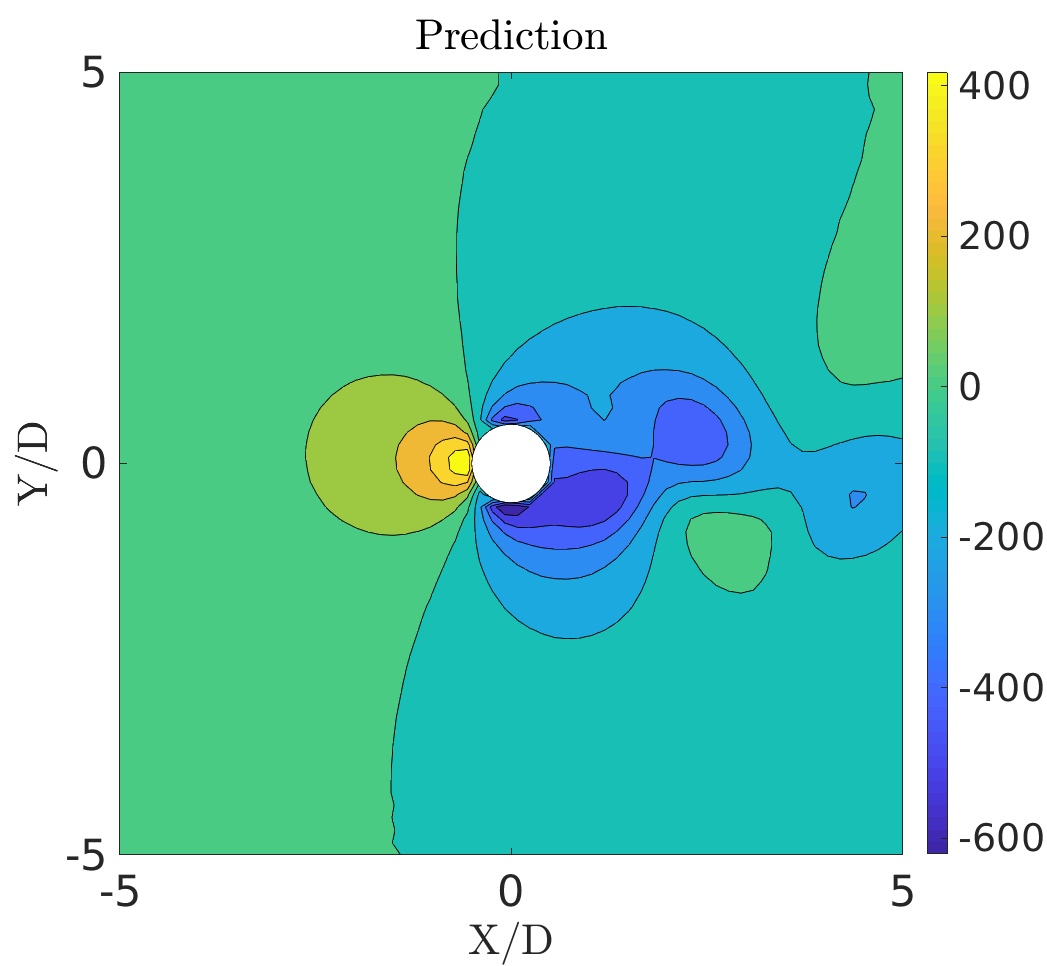}
\hspace{0.01\textwidth}
\includegraphics[width = 0.32\textwidth]{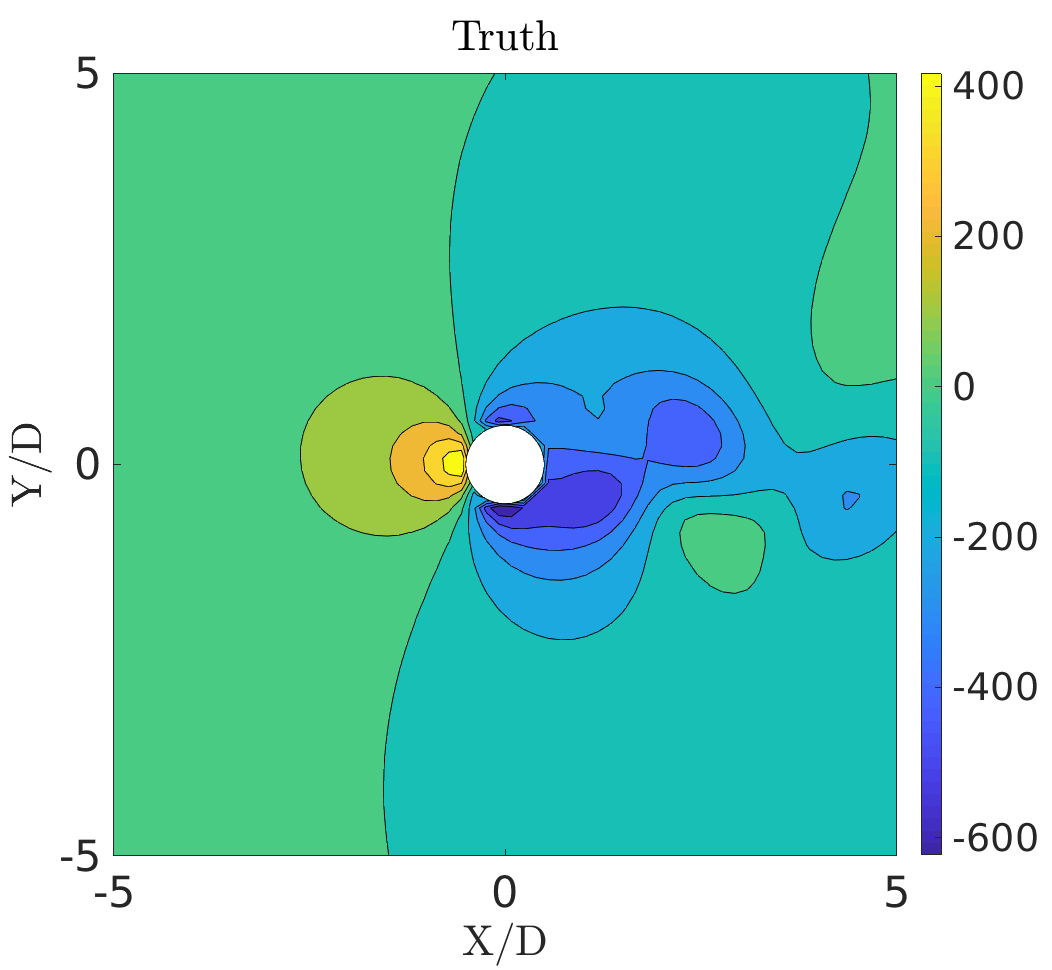}
\hspace{0.01\textwidth}
\includegraphics[width = 0.32\textwidth]{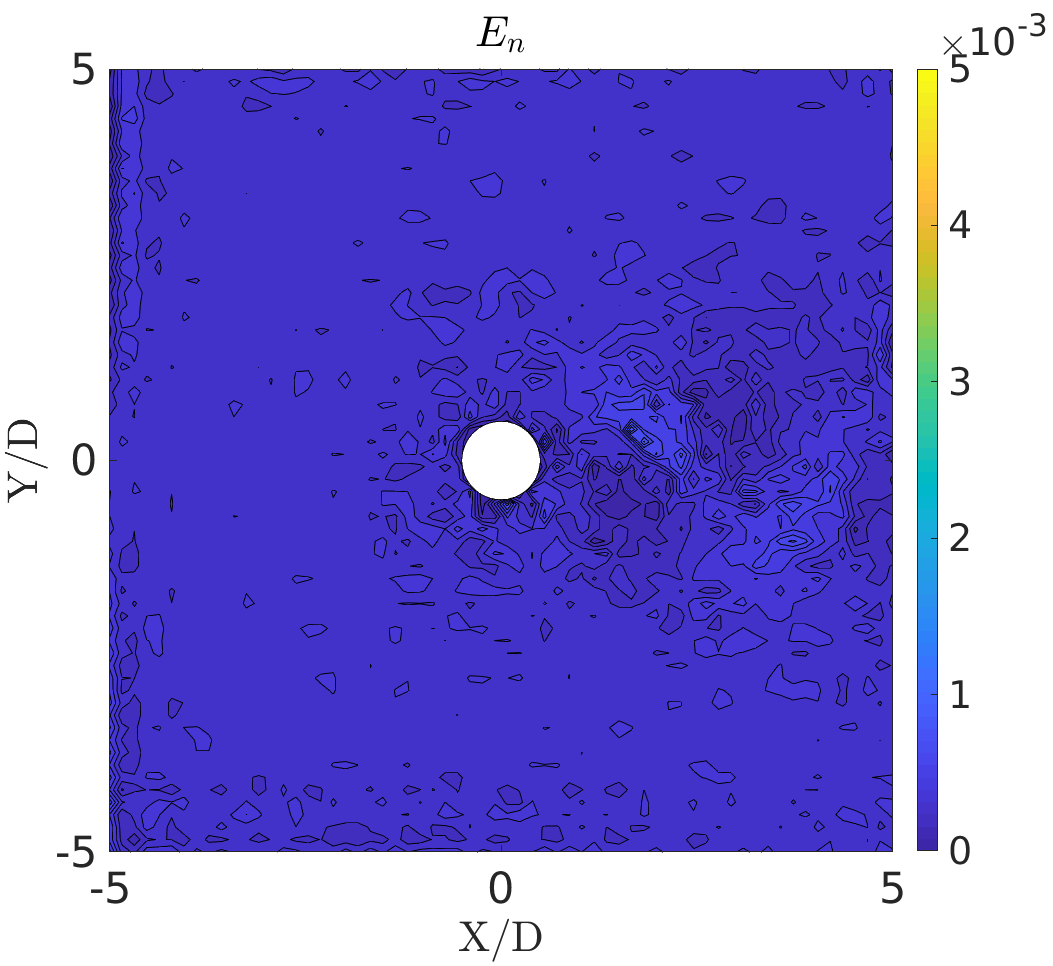}}
\\
\vspace{0.05\textwidth}
\subfloat[]
{\includegraphics[width = 0.32\textwidth]{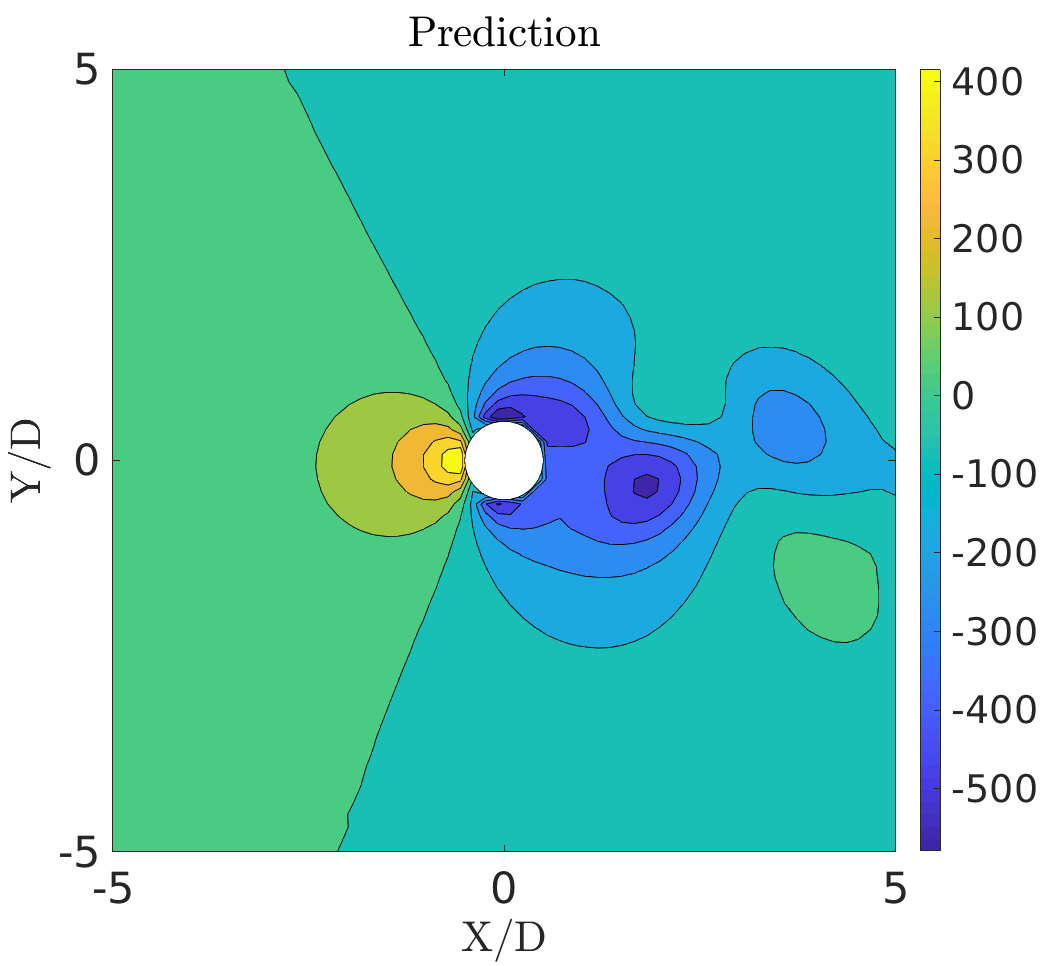}
\hspace{0.01\textwidth}
\includegraphics[width = 0.32\textwidth]{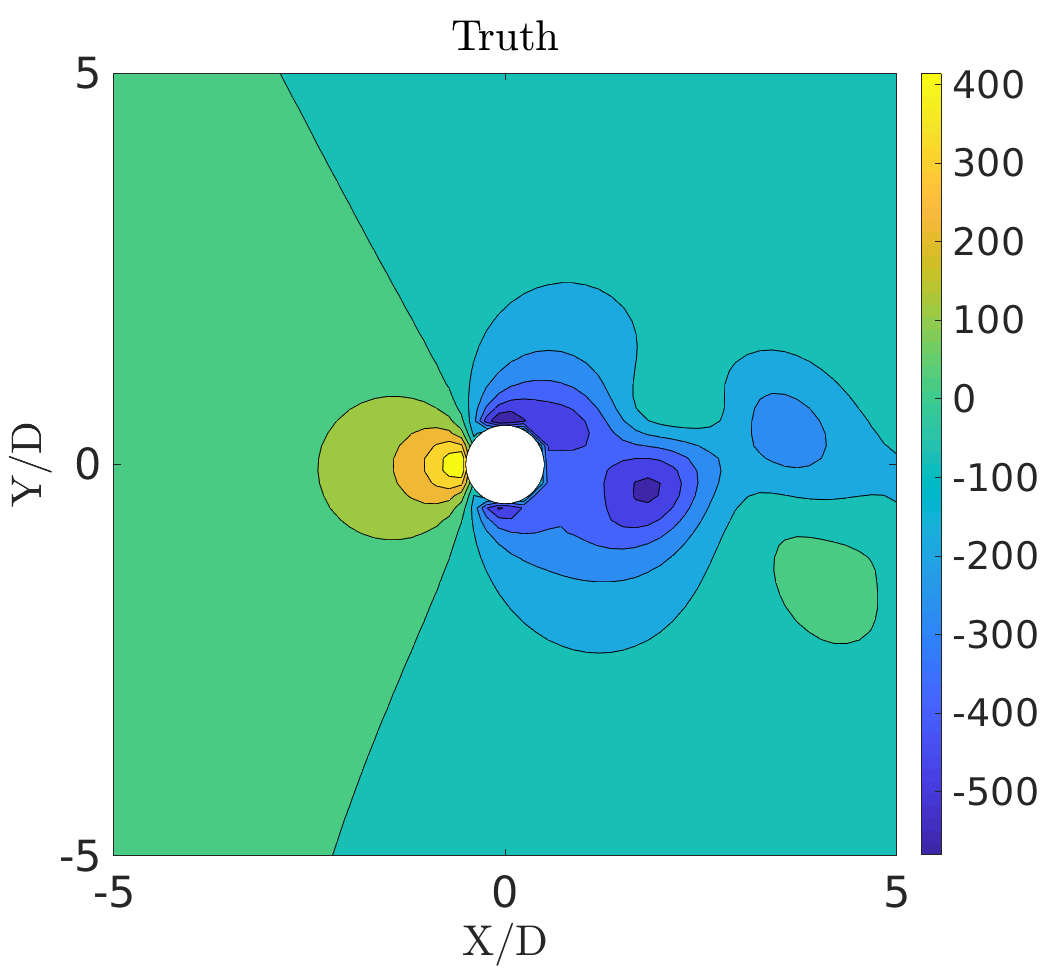}
\hspace{0.01\textwidth}
\includegraphics[width = 0.32\textwidth]{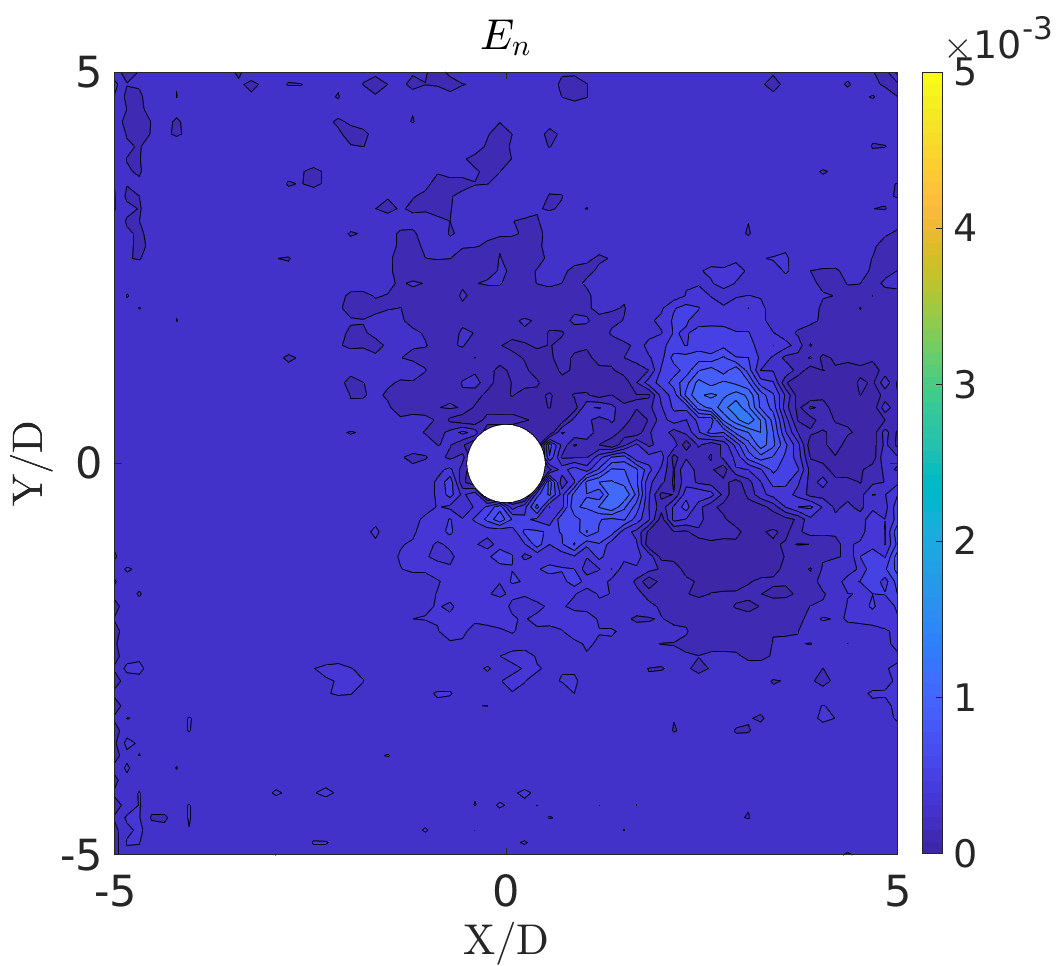}}
\\
\vspace{0.05\textwidth}
\subfloat[]
{\includegraphics[width = 0.32\textwidth]{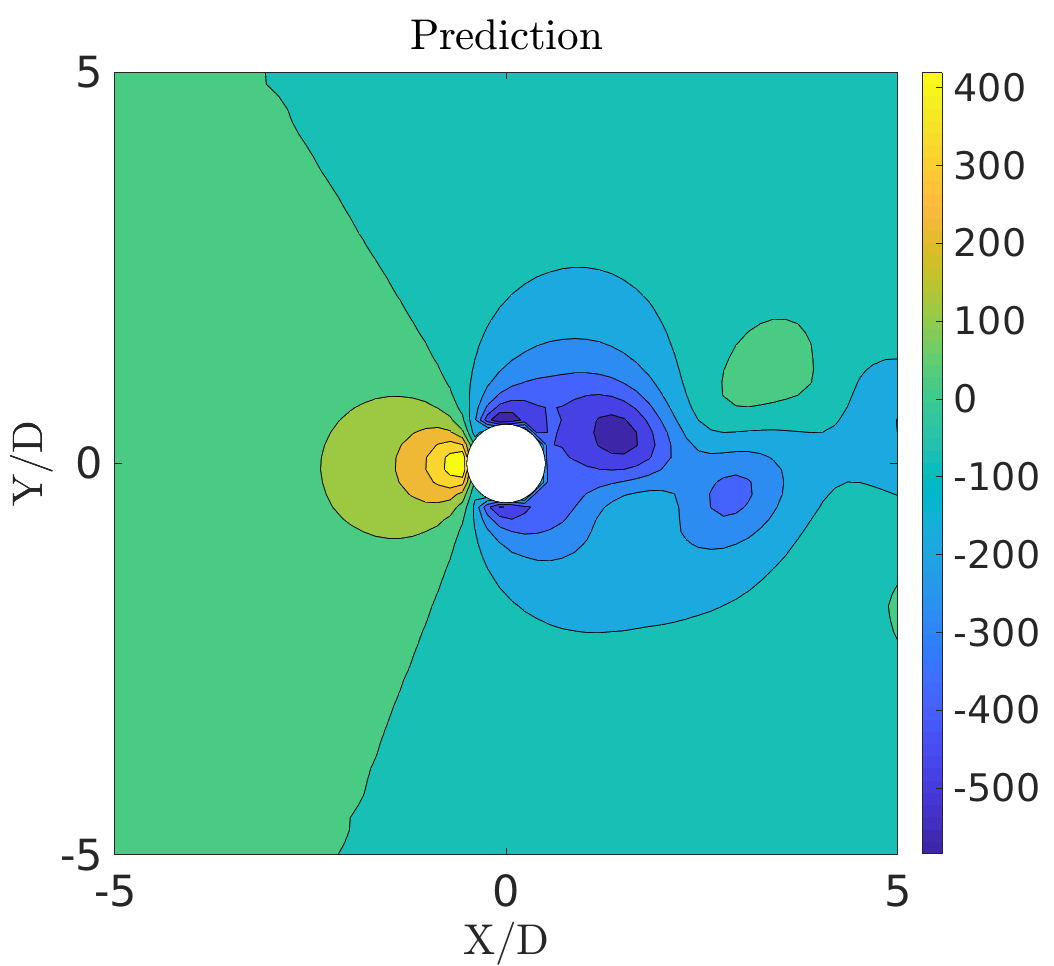}
\hspace{0.01\textwidth}
\includegraphics[width = 0.32\textwidth]{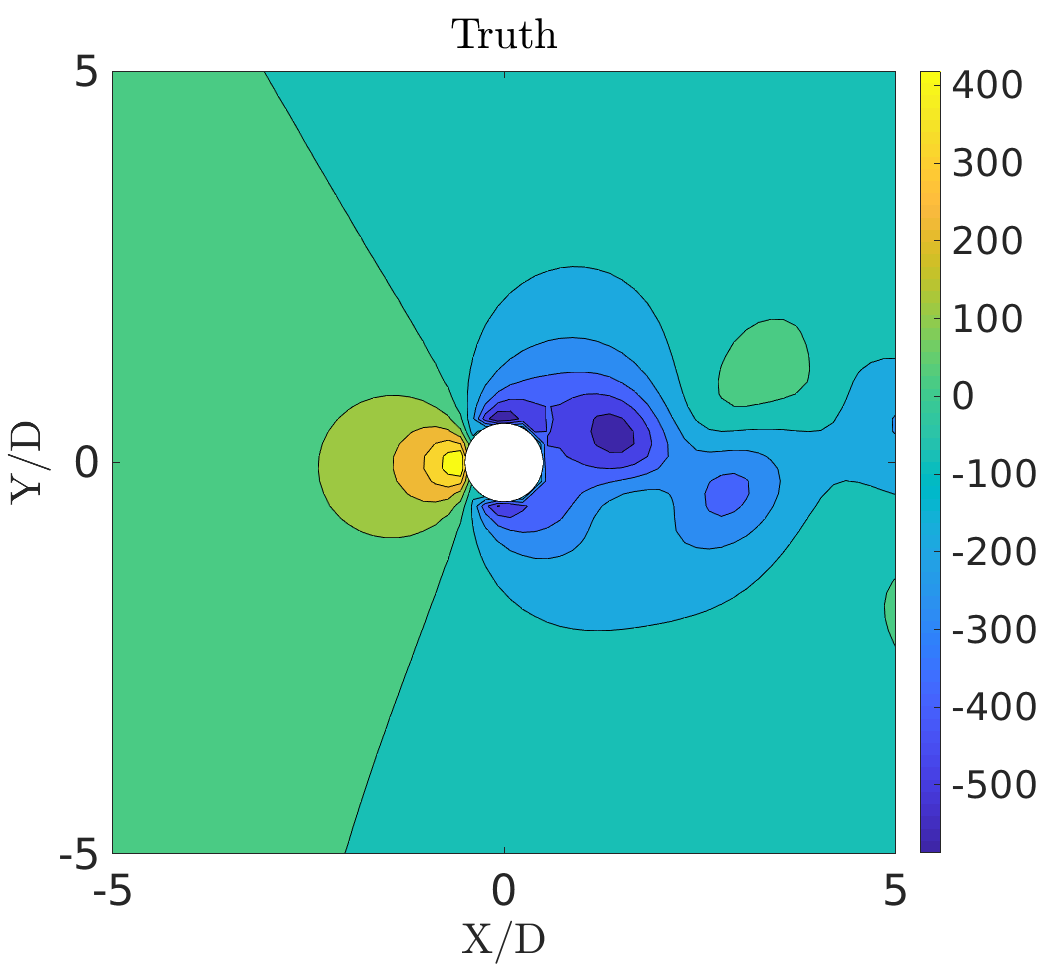}
\hspace{0.01\textwidth}
\includegraphics[width = 0.32\textwidth]{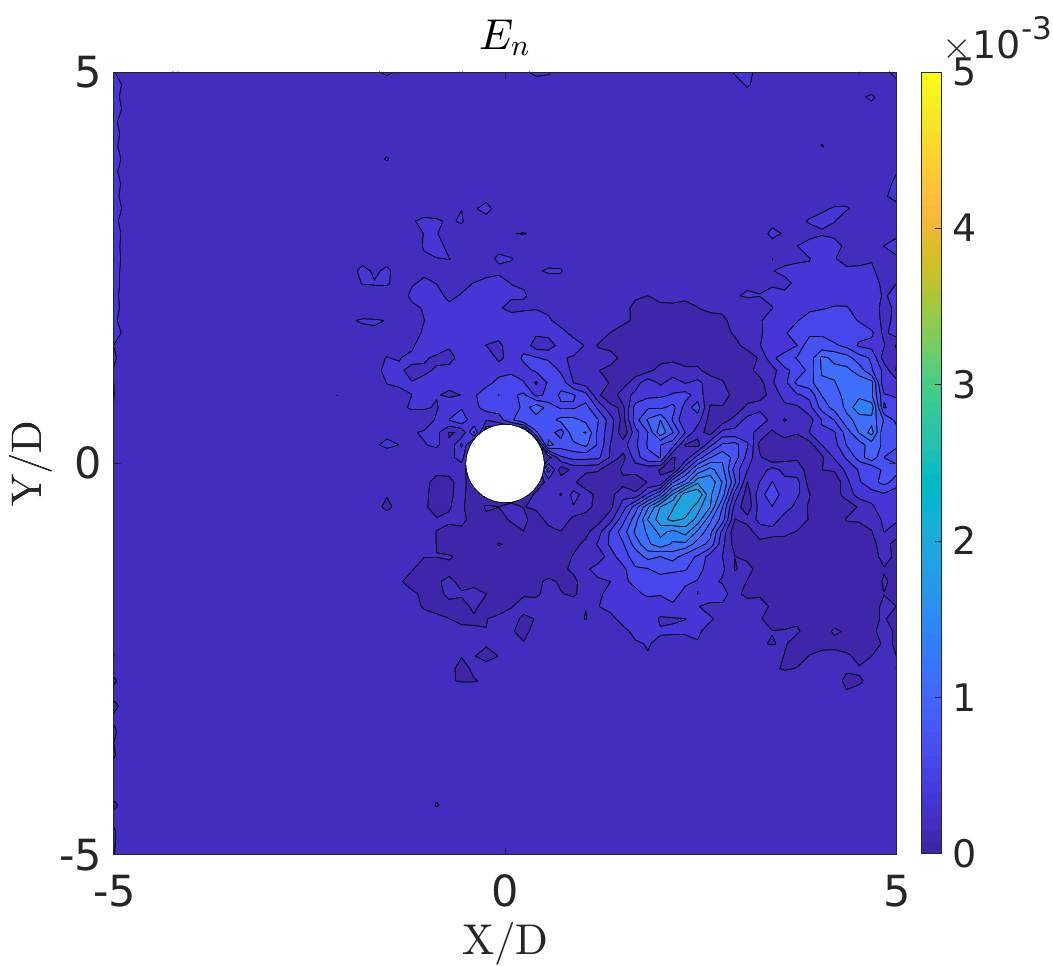}}

\caption{The flow past a cylinder: Comparison of predicted and true fields (CRAN model) along with normalized reconstruction error $E_{n}$ at $\;tU_{\infty}/D =$ (a)  925, (b) 1000, (c)  1075 for pressure field ($P$) }
\label{cran_stat_pred_p}
\end{figure*}

\begin{figure*}
\centering
\subfloat[]
{\includegraphics[width = 0.32\textwidth]{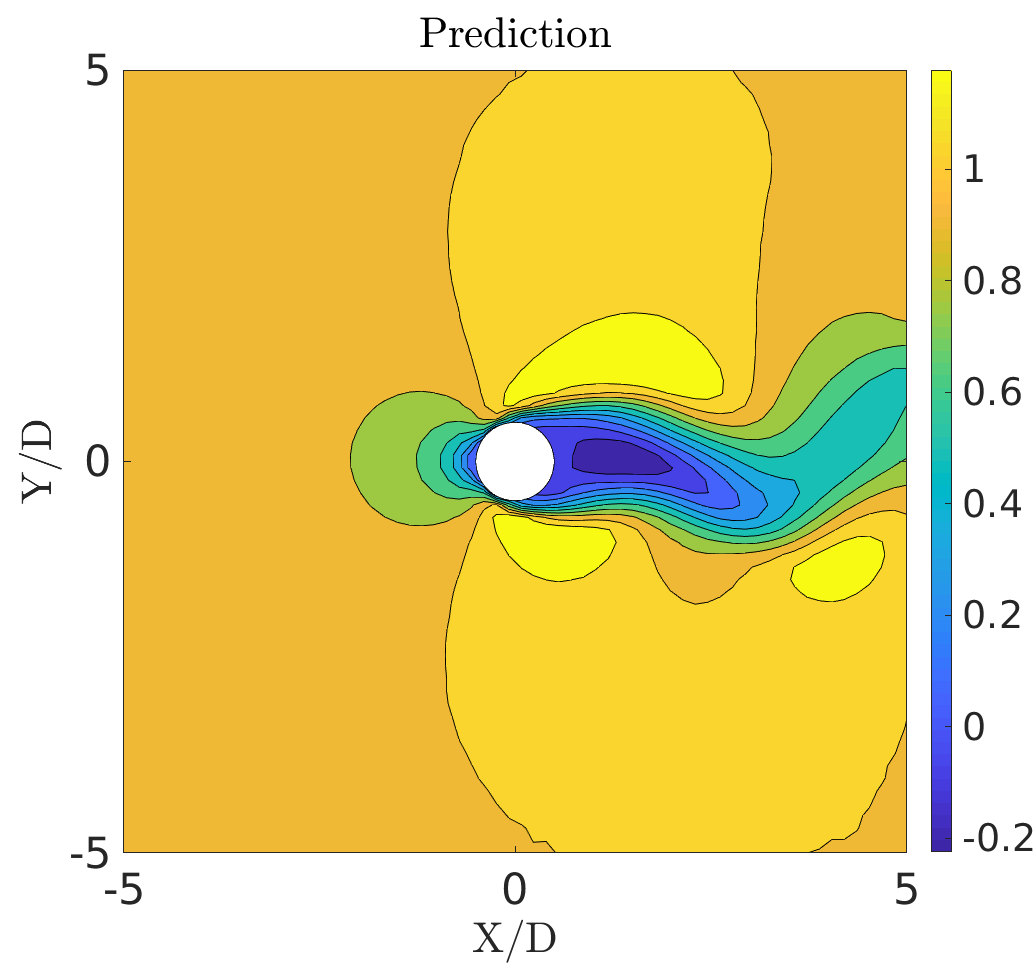}
\hspace{0.01\textwidth}
\includegraphics[width = 0.32\textwidth]{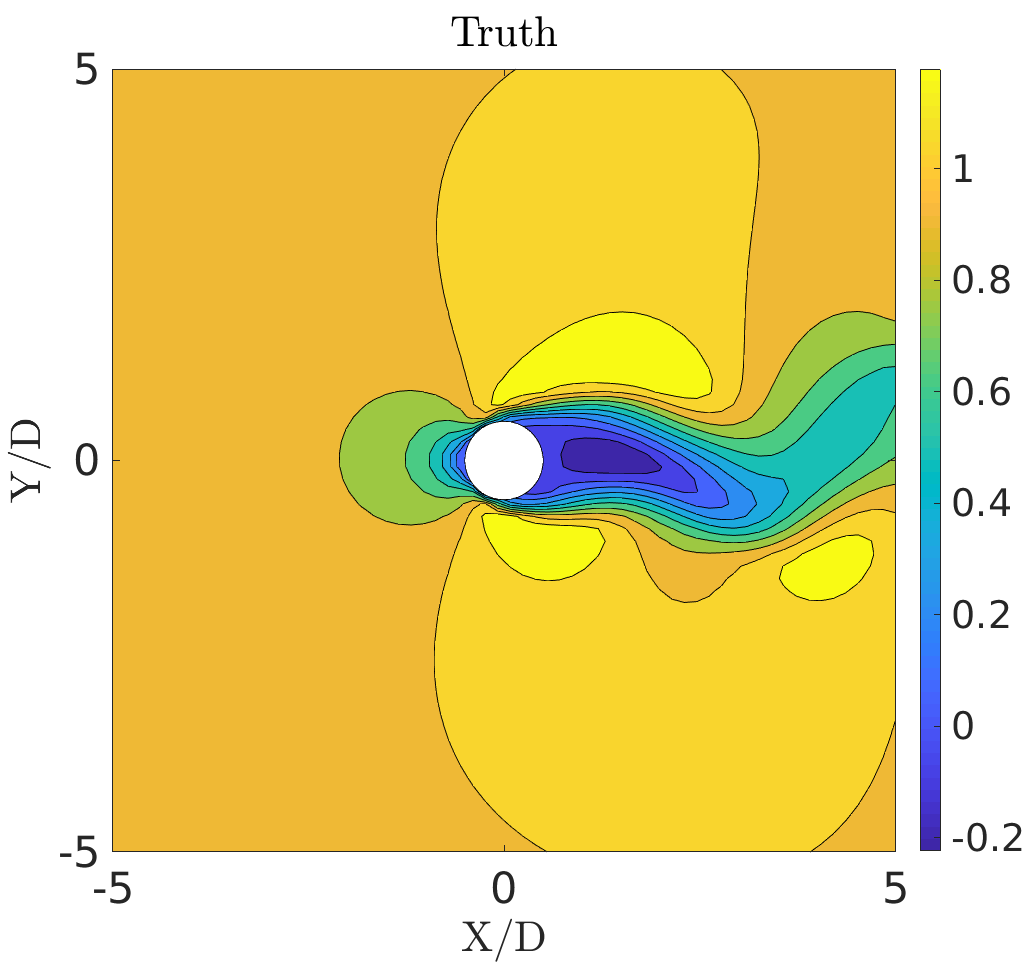}
\hspace{0.01\textwidth}
\includegraphics[width = 0.32\textwidth]{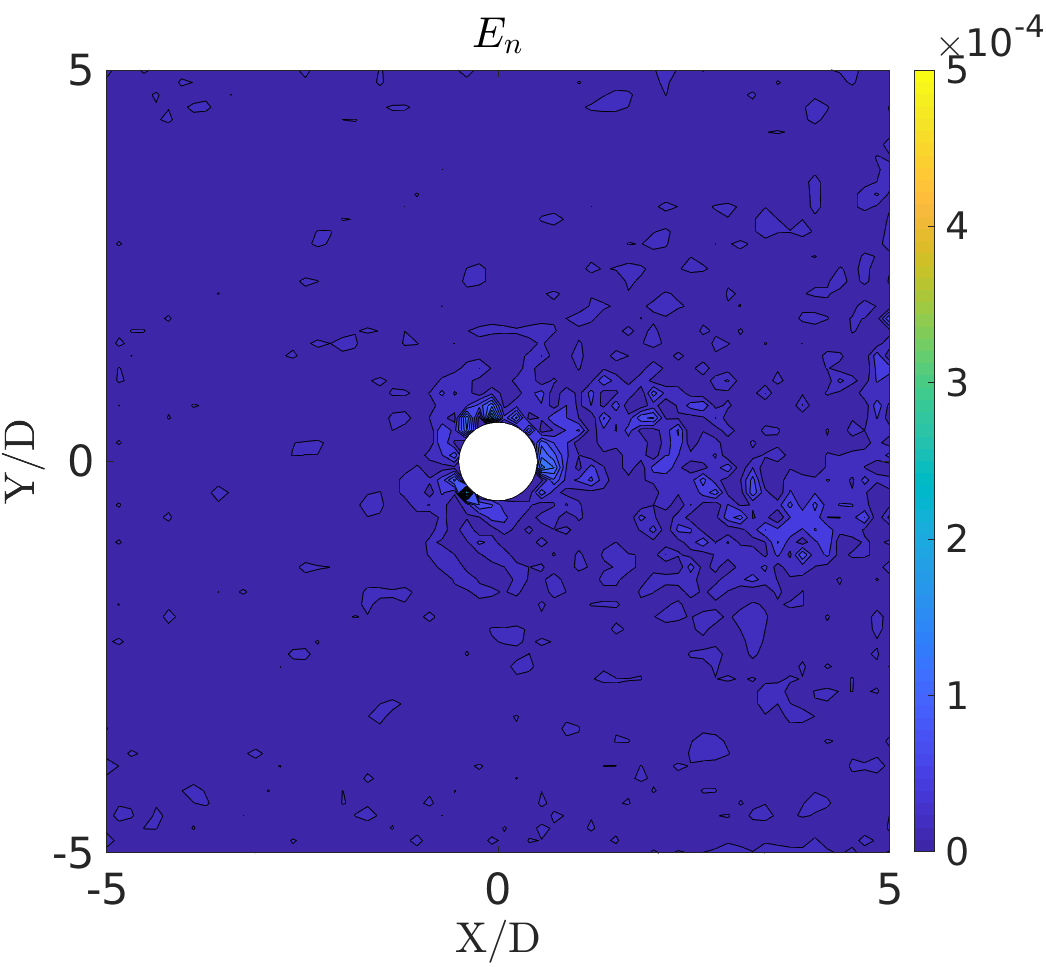}}
\\
\vspace{0.05\textwidth}
\subfloat[]
{\includegraphics[width = 0.32\textwidth]{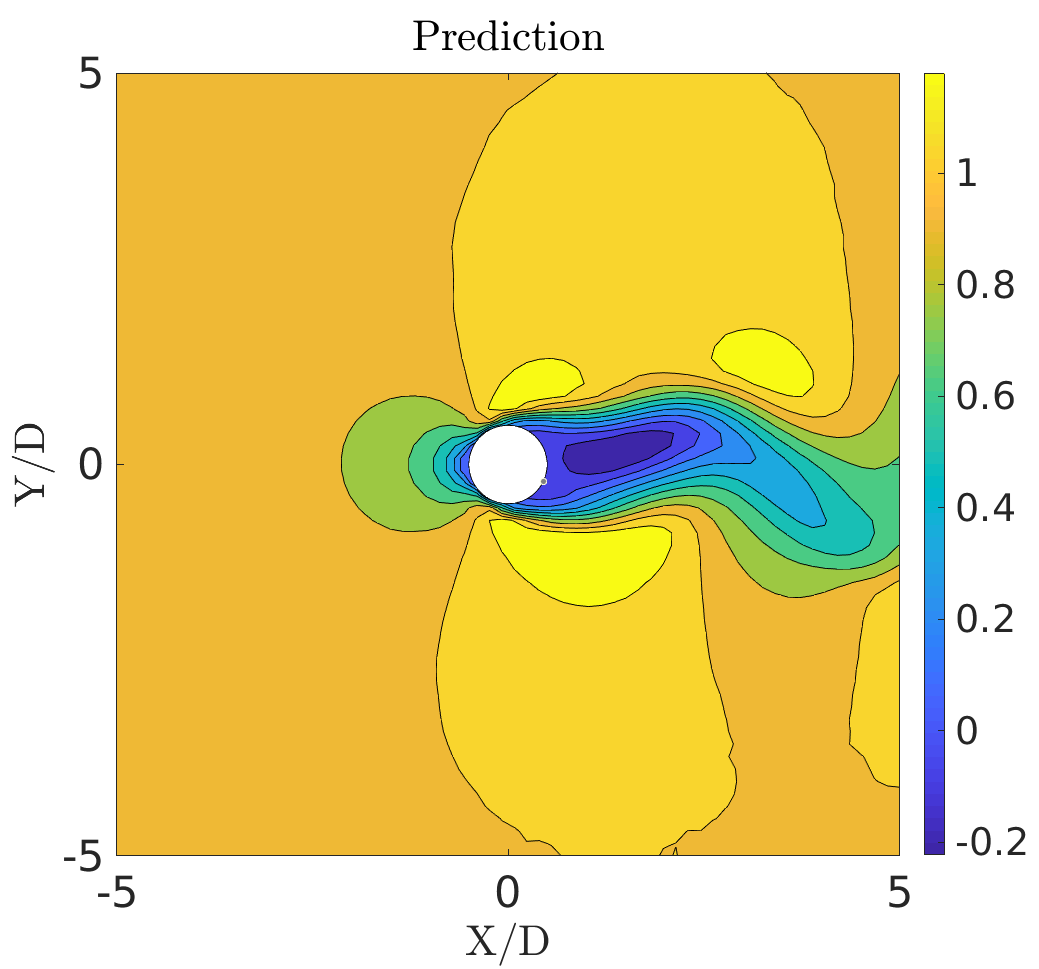}
\hspace{0.01\textwidth}
\includegraphics[width = 0.32\textwidth]{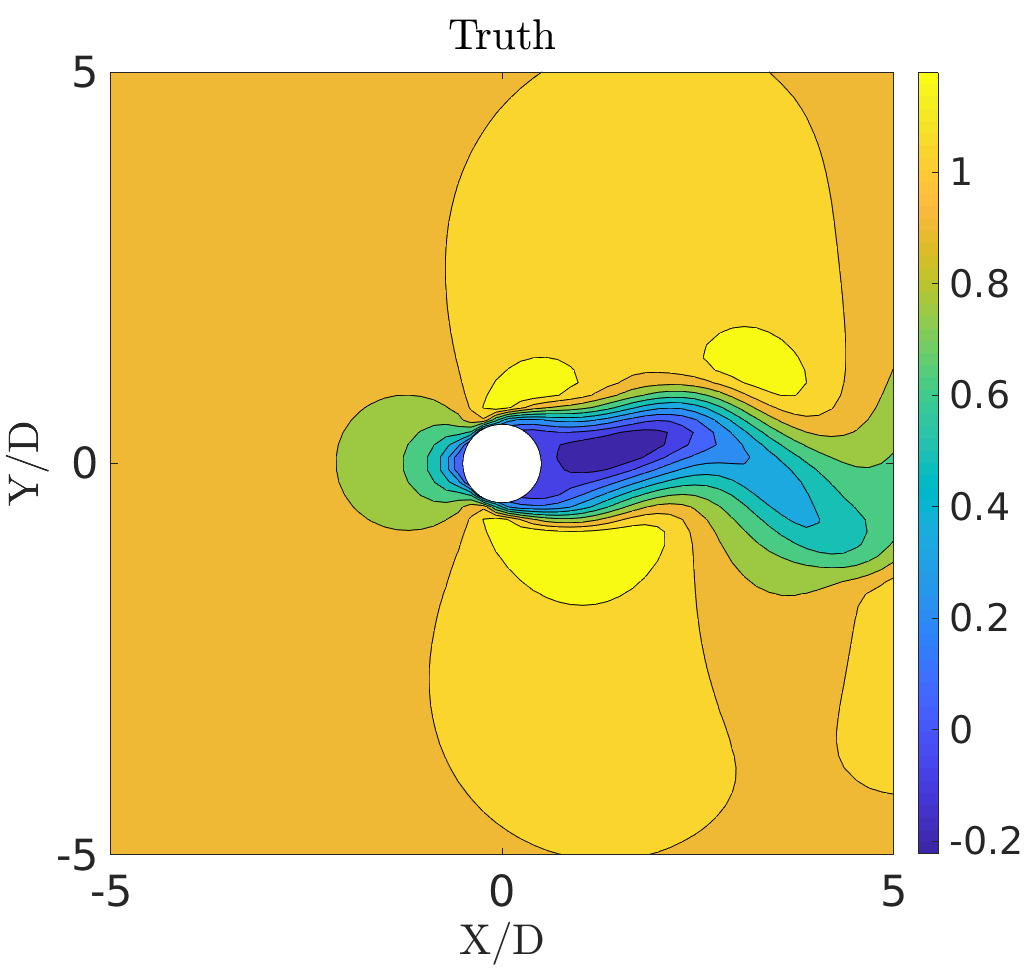}
\hspace{0.01\textwidth}
\includegraphics[width = 0.32\textwidth]{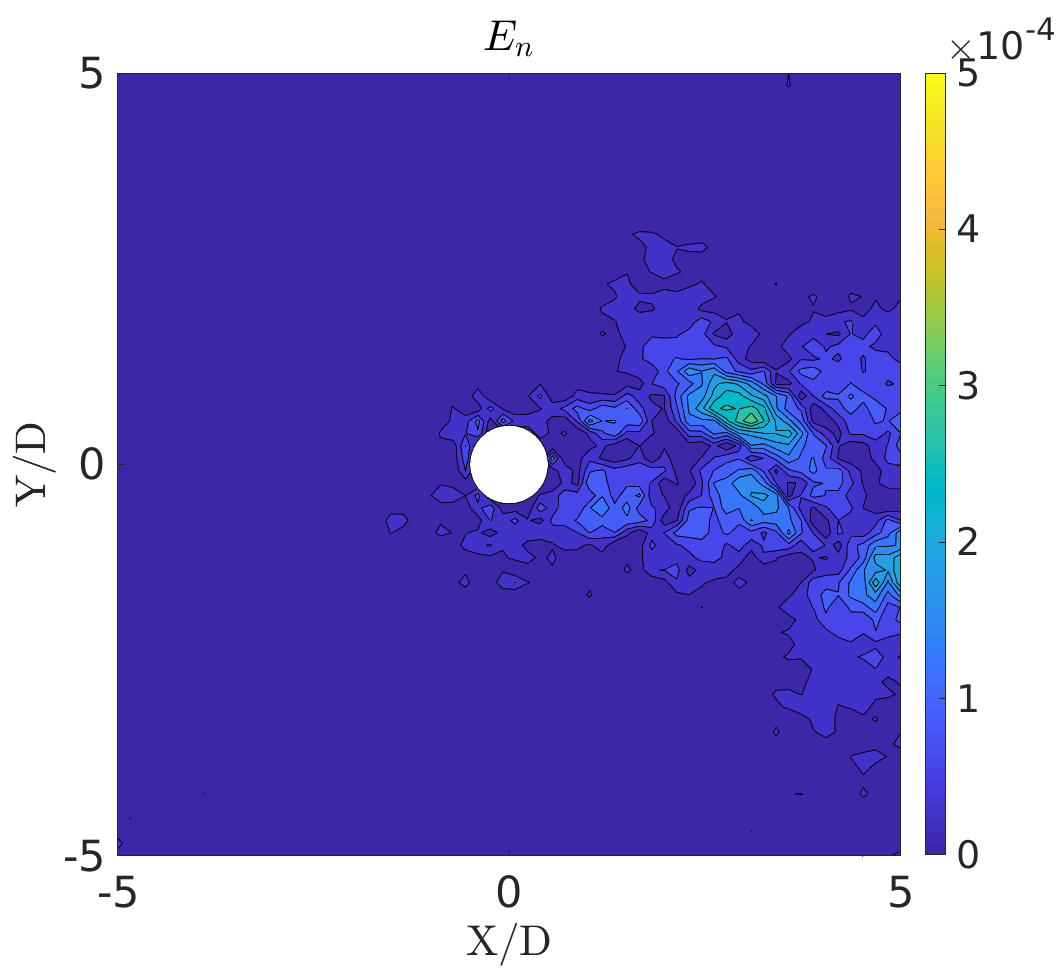}}
\\
\vspace{0.05\textwidth}
\subfloat[]
{\includegraphics[width = 0.32\textwidth]{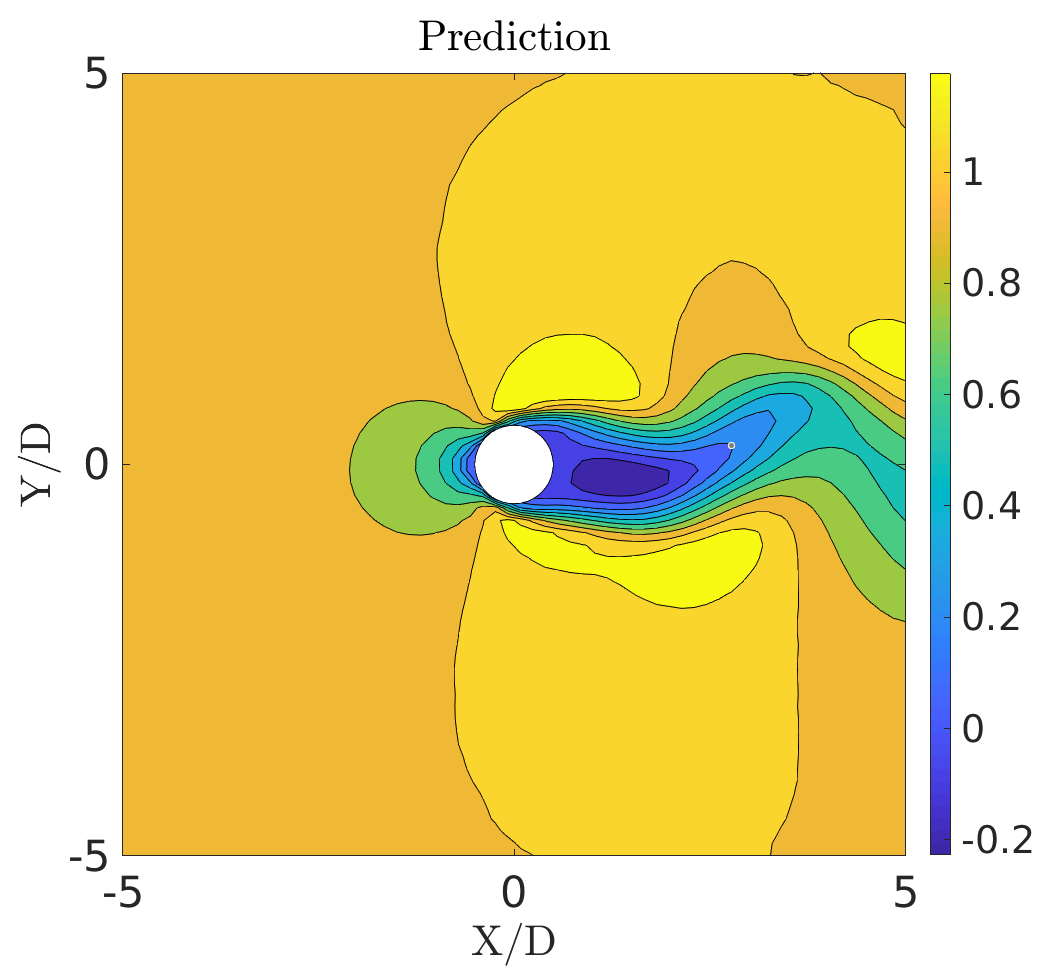}
\hspace{0.01\textwidth}
\includegraphics[width = 0.32\textwidth]{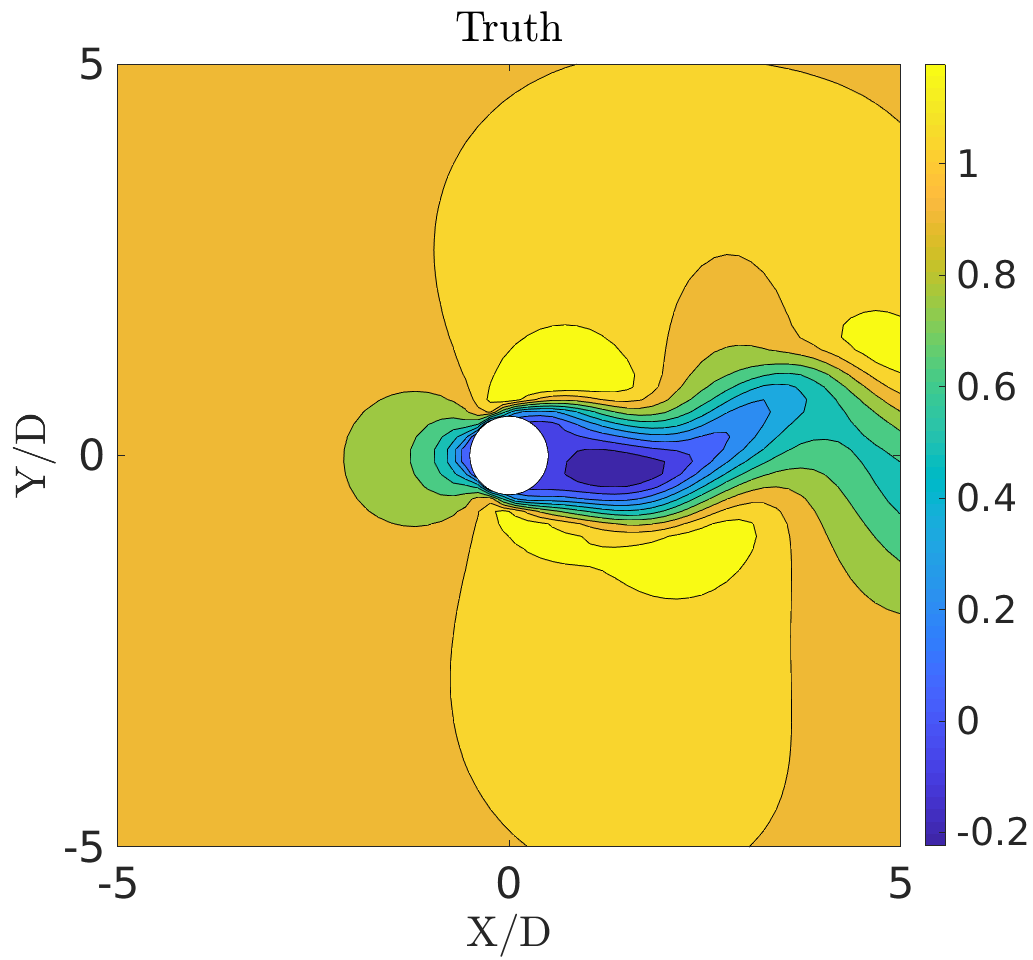}
\hspace{0.01\textwidth}
\includegraphics[width = 0.32\textwidth]{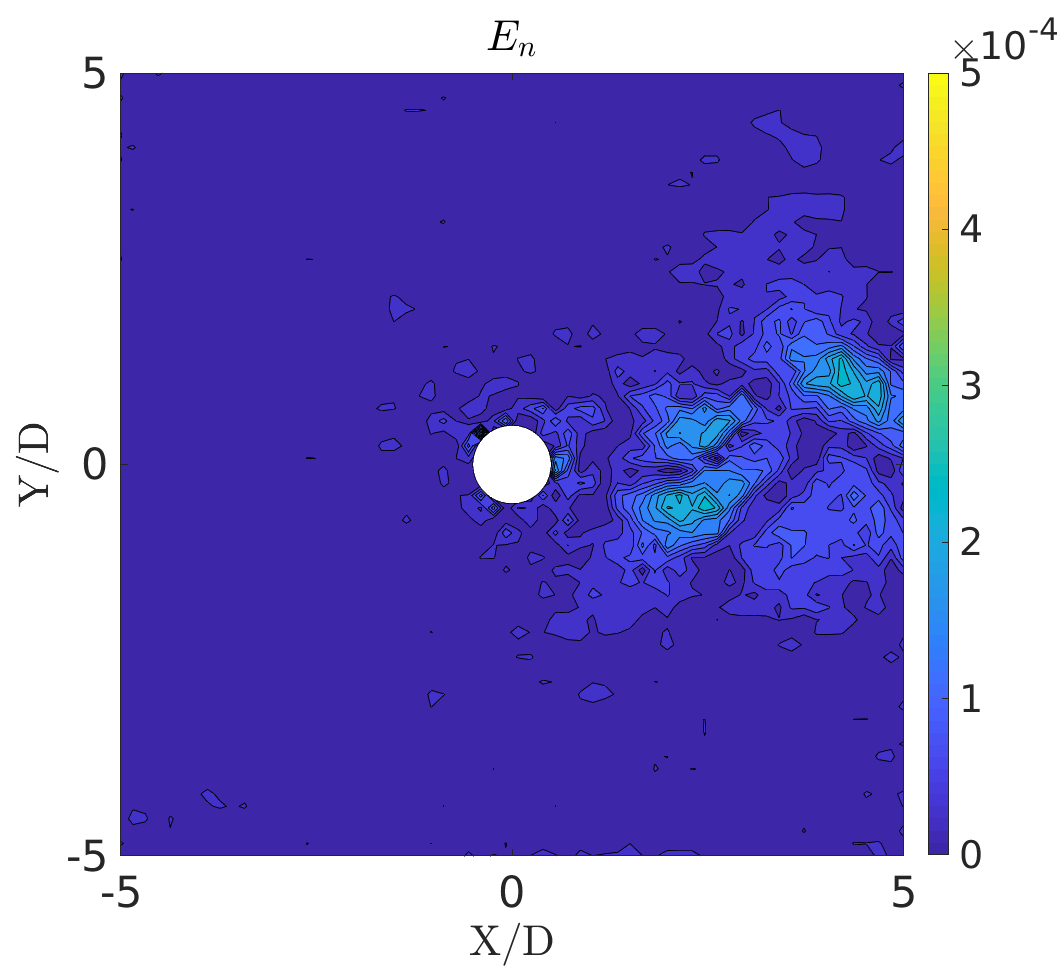}}

\caption{The flow past a cylinder: Comparison of predicted and true fields (CRAN model) along with normalized reconstruction error $E_{n}$ at $\;tU_{\infty}/D =$ (a)  925, (b)  1000, (c)  1075 for x-velocity field ($U$) }
\label{cran_stat_pred_u}
\end{figure*}

%Field Prediction
% Pressure
\textit{Field prediction}: Here, $N_{A}=32$ trained CRAN model is employed to demonstrate the field predictions for both pressure and x-velocity using just one time-step 3500. Fig.~\ref{cran_stat_pred_p} and \ref{cran_stat_pred_u} depict the comparison
of predicted and true values for P and U fields respectively at time-steps $3700$ ($925\;tU_{\infty}/D$), $4000$ ($1000\;tU_{\infty}/D$) and $4300$ ($1075\;tU_{\infty}/D$). The normalized reconstruction error $E_{n}$ is, similarly, constructed by taking the absolute value of differences between the true $\textbf{s}_{n}$ and predicted field $\hat{\textbf{s}}_{n}$ and normalizing it with $L_{2}$ norm of the true data and is given by

\begin{equation}
    E_{n} = \frac{|\textbf{s}_{n}-\hat{\textbf{s}}_{n}|}{\|\textbf{s}_{n}\|_{2,k}}.
\end{equation}

\begin{figure}[H]
\centering
\subfloat[]
{\includegraphics[width = 0.235\textwidth]{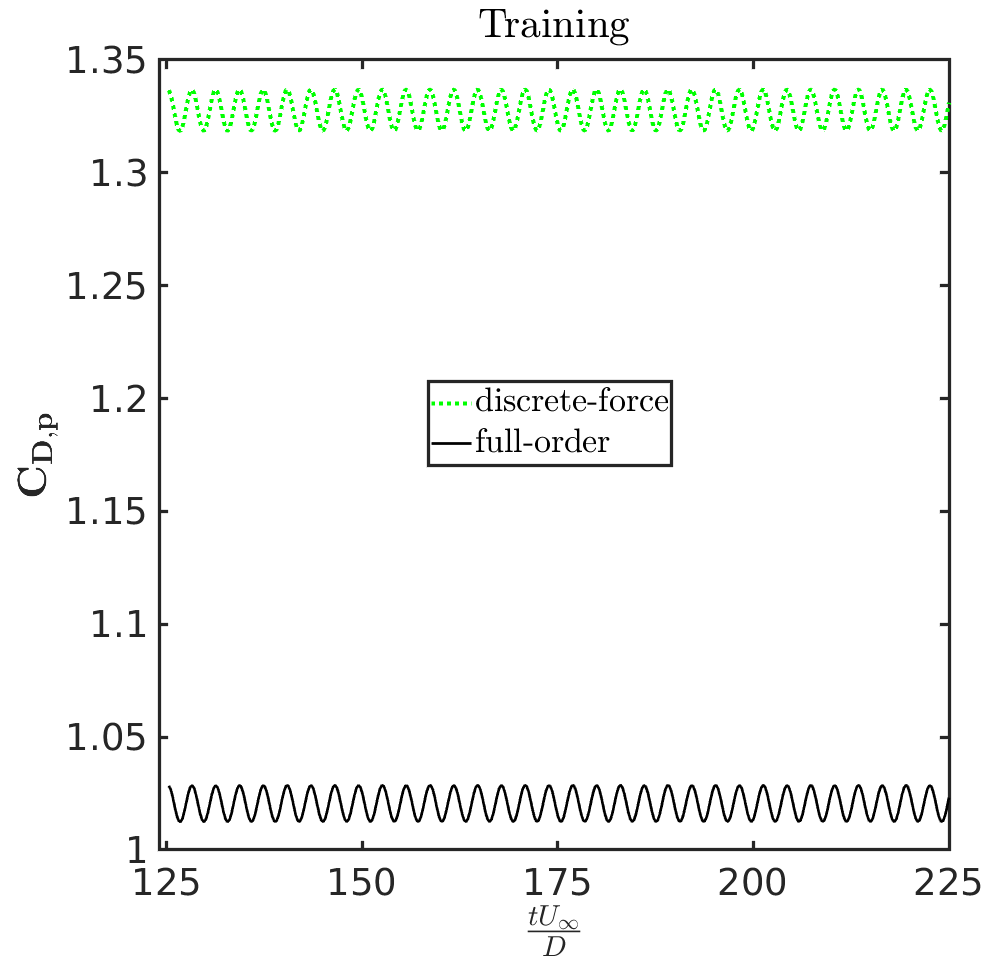}}
\subfloat[]
{\includegraphics[width = 0.235\textwidth]{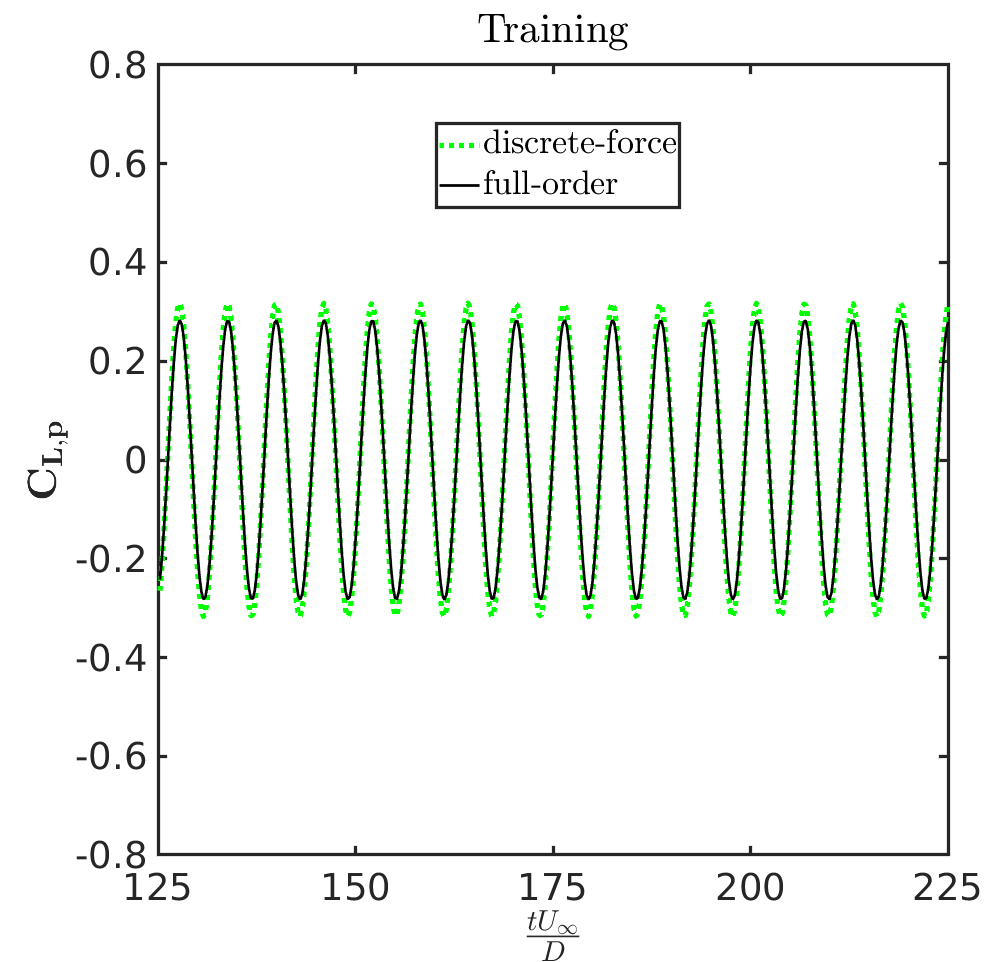}} \\
{\includegraphics[width = 0.243\textwidth]{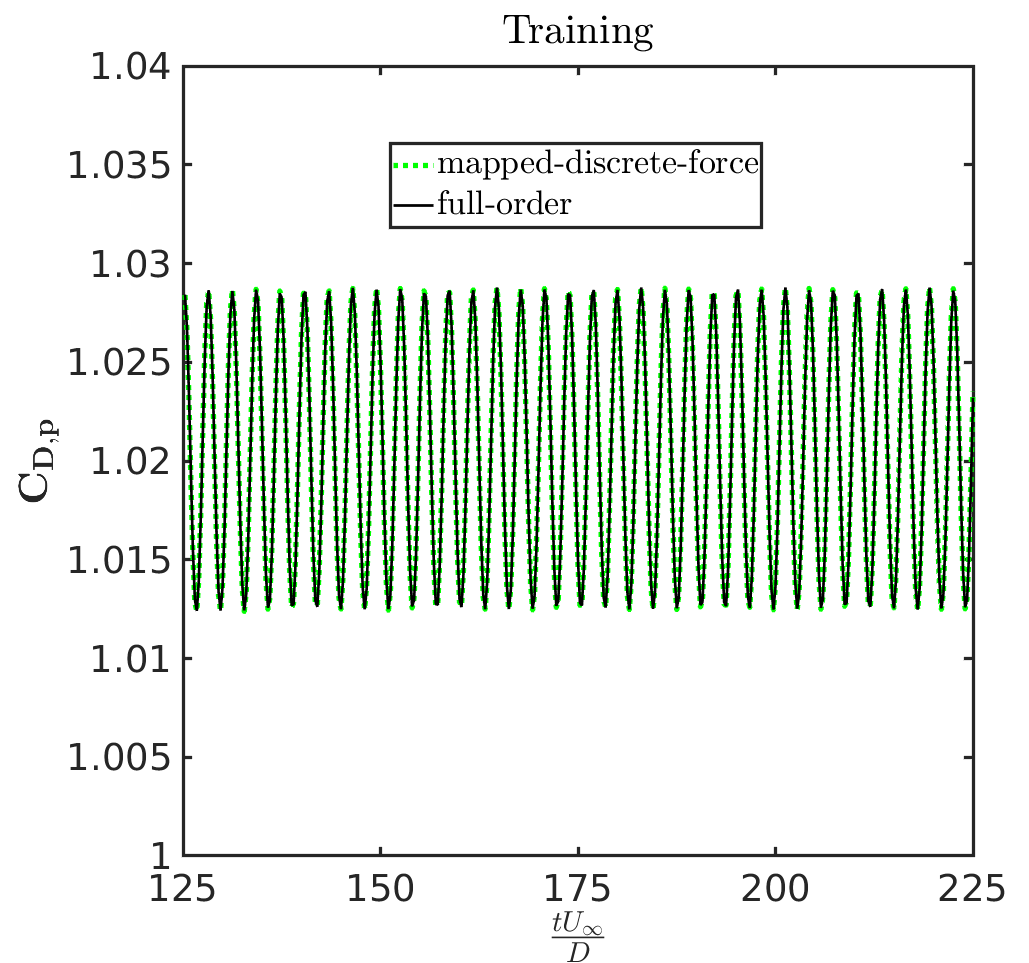}}
\subfloat[]
{\includegraphics[width = 0.233\textwidth]{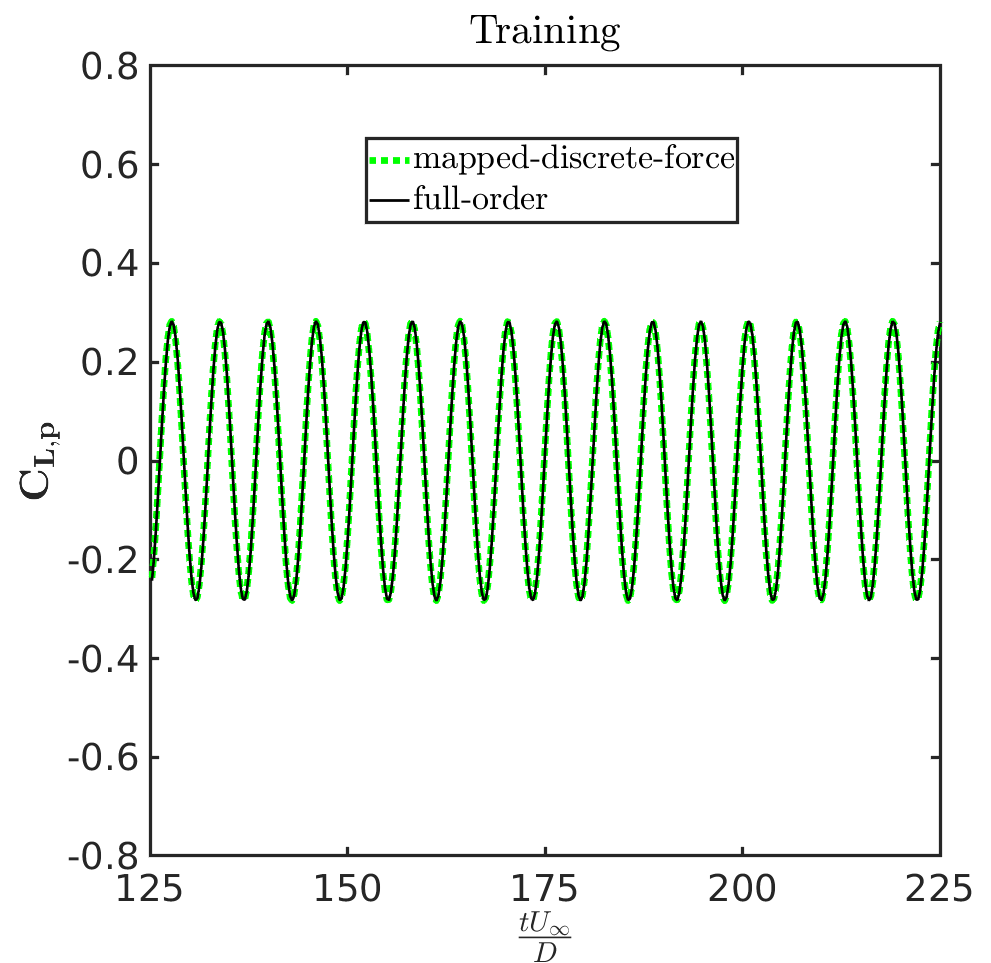}}
\caption{The flow past a cylinder: Comparison of discrete force and mapped discrete force with respect to full-order force coefficients on the training data due to pressure. (a),(b) represent the discrete drag and lift coefficients respectively. (c),(d) represent the reconstructed (mapped) discrete drag and lift coefficients respectively}
\label{discrete-rb-pres-stat}
\end{figure}

\begin{figure}[H]
\centering
\subfloat[]
{\includegraphics[width = 0.243\textwidth]{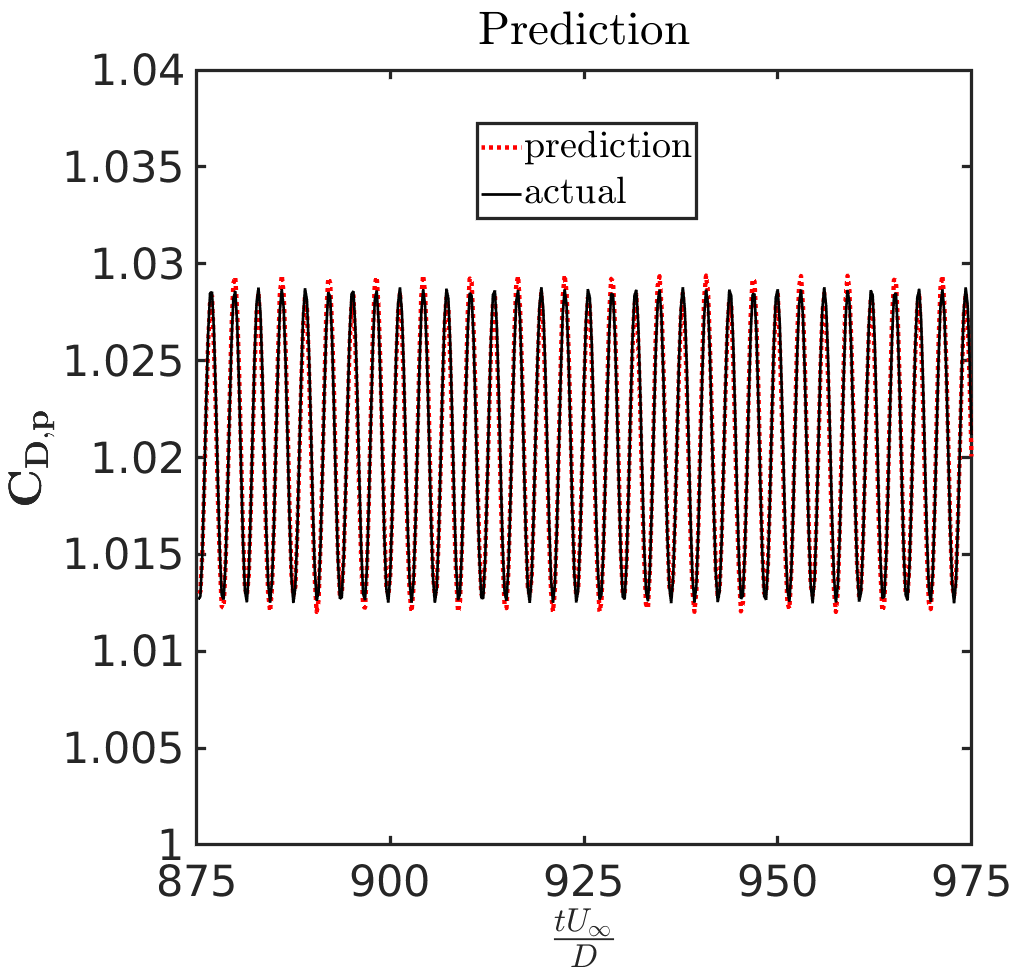}}
\subfloat[]
{\includegraphics[width = 0.233\textwidth]{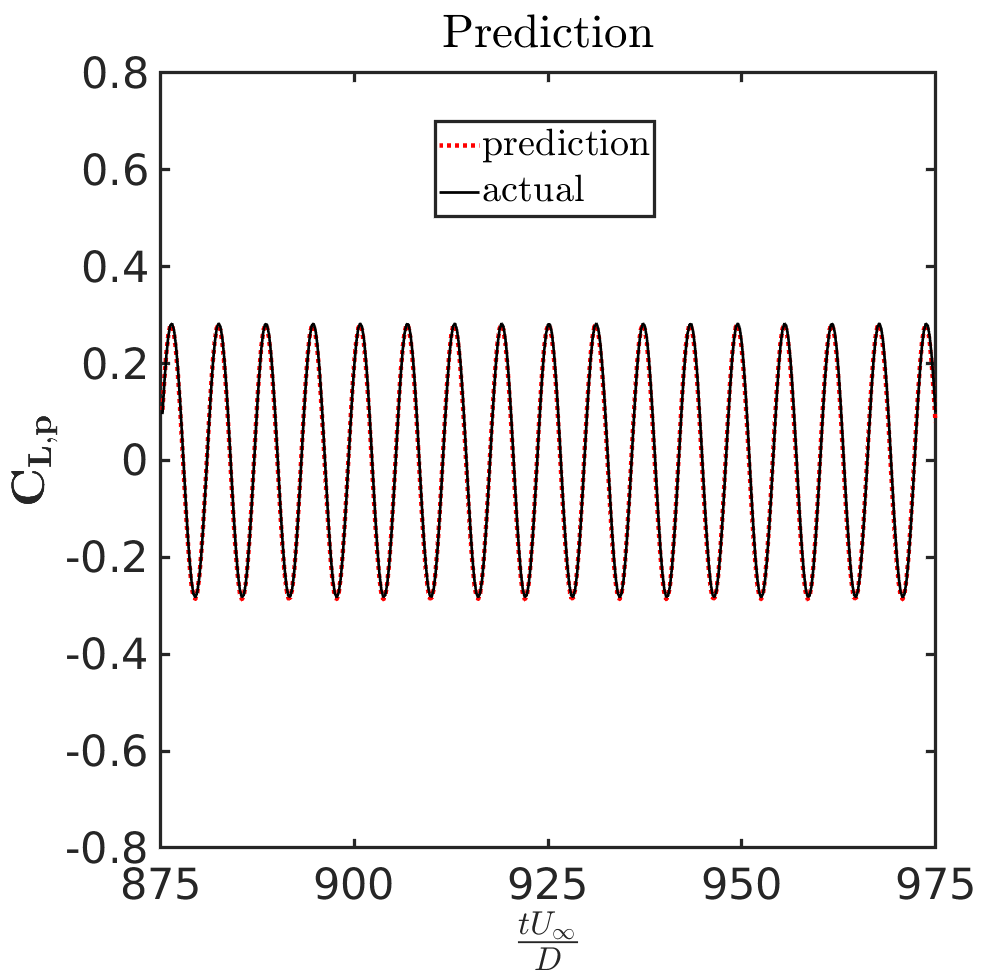}}
\caption{The flow past a cylinder: Predicted and actual (CRAN model) (a) drag and (b) lift force coefficients due to the pressure field ($P$) on the cylinder (shown from $875$ to $975\;tU_{\infty}/D$)}
\label{predicted-discrete-rb-pres-stat}
\end{figure}

% Force 
\textit{Force coefficient prediction}: 
Now, we demonstrate the methodology highlighted in section \ref{Section:ForceCalc} to predict the pressure force coefficient on the cylinder boundary. Here, $\textbf{C}_{\mathrm{D},\mathrm{p}}$ and $\textbf{C}_{\mathrm{L},\mathrm{p}}$  denote the drag and lift force experienced by the stationary cylinder in a pressure field, normalized by $0.5\rho D U_{\infty}^{2}$. Using Eqs. (\ref{presForce}) and (\ref{totalDiscForce}), we first calculate the discrete drag and lift force from the pressure training data (3000 snapshots from time $125\;tU_{\infty}/D$ to $875\;tU_{\infty}/D$). Fig. \ref{discrete-rb-pres-stat} (a) and (b) depict the drag and lift discrete force coefficients (calculated on low-resolution reference grid) with respect to the full-order (calculated on high-resolution unstructured grid) respectively. For visualisation convenience, the plot from $125\;tU_{\infty}/D$ till $225\;tU_{\infty}/D$ is shown. Clearly, there is a mean shift and/or higher amplitudes observed in the discrete force compared to full-order. This is accounted for the fact of interpolation error (coarsening effect) that is introduced on the reference grid. However, the discrete force tends to capture the force propagation trend with mean shift and/or amplitude difference. This peculiarity, itself, is observed in the training data and needs to be corrected. This is done by the reconstruction mapping $\psi$, which is calculated from discrete and full-order forces in the training data. Figs. \ref{discrete-rb-pres-stat} (c) and (d) represent the reconstructed discrete force and the full-order coefficients on the same training data from $125\;tU_{\infty}/D$ to $225\;tU_{\infty}/D$ using the corrector mapping $\psi$. 
Thus, getting the correct force prediction reduces to the task of getting the discrete force from predicted fields. The $\psi$ calculated from the training data is then used to reconstruct the predicted discrete force to full-order (Eq. (\ref{psitotalDiscForce})). The results are presented in Fig. \ref{predicted-discrete-rb-pres-stat}.

%\newpage
%\onecolumn

%\newpage
%\twocolumn

%%%%%%%%%%%%%%%%%%%%%%%%%%%
\subsection{Discussion}\label{disc_stat}
%%%%%%%%%%%%%%%%%%%%%%%%%%%

% point 1
 Table \ref{tab_comp_pc_stat} lists out the differences in the attributes of both the models for flow past a cylinder. We have employed a closed-loop recurrent network in both the models to evolve the low-dimensional subspace. This also includes the fact that only one time-step is sufficient to predict long range of future time-states (here, 1000 for example) for both the hybrid models. This is expected since the problem achieves a periodic behavior. Thus, the compounding effect of the errors are negligible in time. POD-RNN model provides a selective control on the number of modes (features) in the low-dimensional space based on energy distribution and a simple algebraic reconstruction using time-advanced coefficients, modes and mean field. This by-passes the costly encoder and decoder network parameter space which is utilised in the CRAN model to achieve a similar process. In that sense, one can argue that the CRAN model does not provide any significant improvements over POD-RNN model for this problem. However, the same inference is expected since the problem at hand is one of the most simplified fluid flow problems and any standard prediction tool should be able to perform at equal capabilities. 

\begin{table}[H]
    \centering
    \begin{tabular}{|c|c|c|}
    \hline
          & POD-RNN & CRAN \\ \hline
         Encoder/decoder space &  POD  & $\boldsymbol{\theta}\approx$ 3 x $10^{5}$\\
         Evolver features $\textbf{A}$ & $k=5$ & $N_{A}=32$ \\
         Recurrent network type & closed-loop & closed-loop \\ 
         Training time (CPU) & 2 hours & 16 hours \\
         Prediction ($\textbf{I}$, $\textbf{P}$) & $\textbf{I}:1$, $\textbf{P}:1000$ & $\textbf{I}:1$, $\textbf{P}:1000$ \\
         \hline
    \end{tabular}
    $\boldsymbol{\theta}$: trainable parameters \\
    $\textbf{I}$: input time-steps,  $\textbf{P}$: predicted time-steps
    \caption{The flow past a cylinder: Comparison of POD-RNN with convolutional recurrent autoencoder network (CRAN)}
    \label{tab_comp_pc_stat}
\end{table}

\begin{figure}[H]
\centering
\subfloat[]
{\includegraphics[width = 0.238\textwidth]{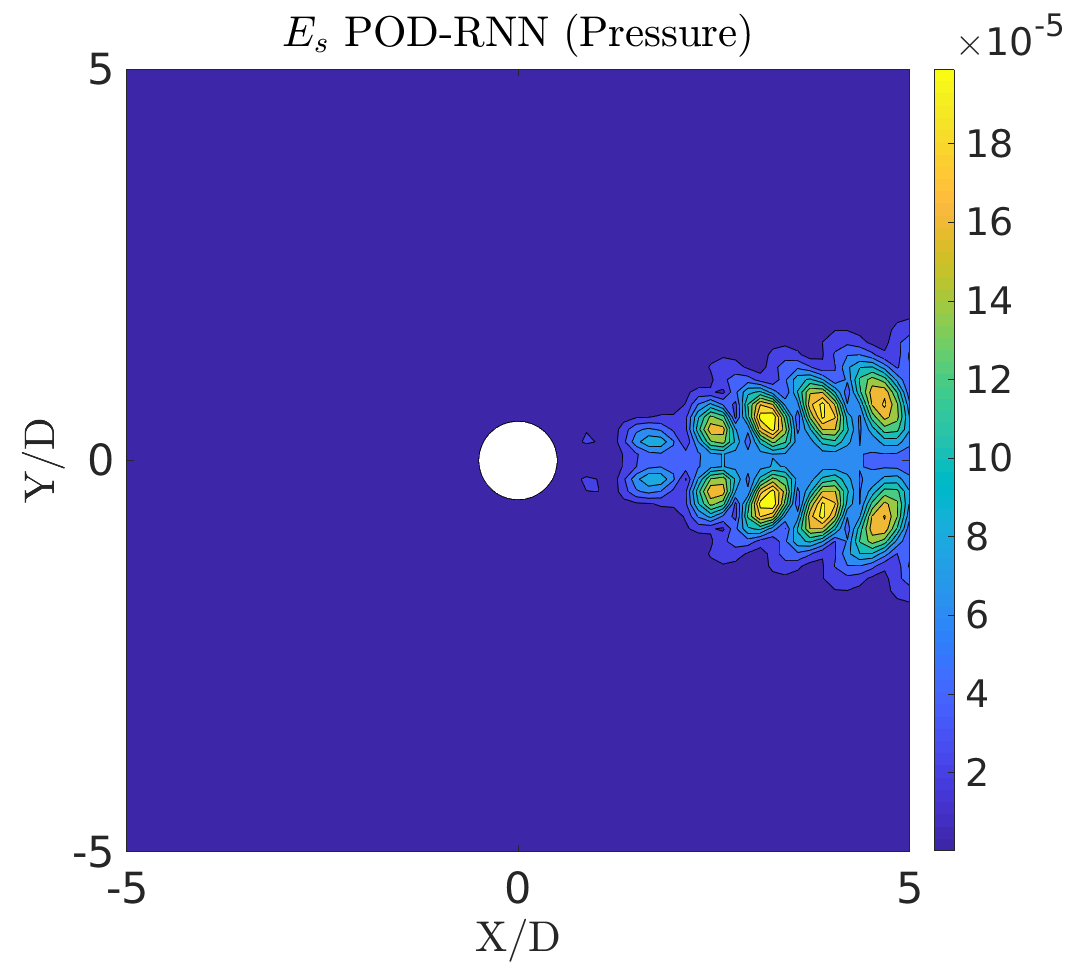}}
\subfloat[]
{\includegraphics[width = 0.238\textwidth]{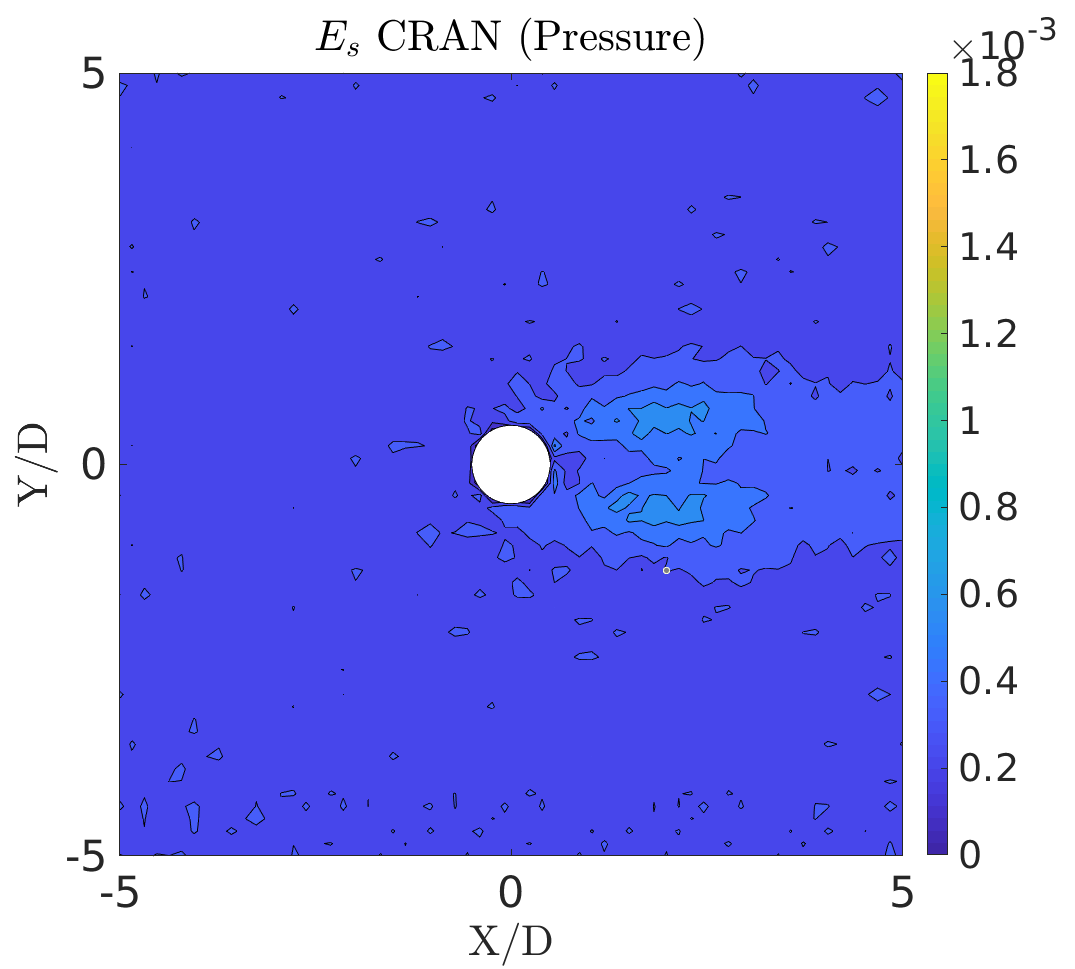}}\\
\subfloat[]
{\includegraphics[width = 0.238\textwidth]{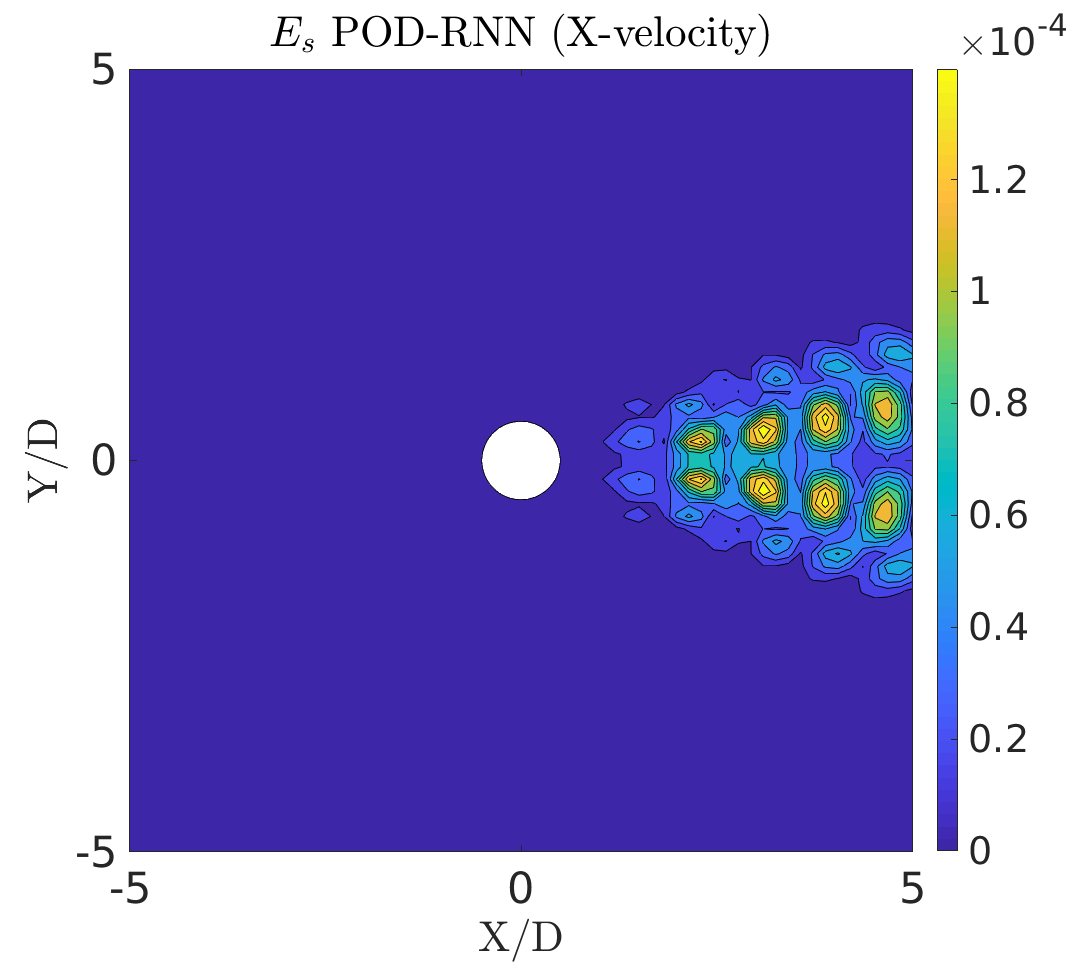}}
\subfloat[]
{\includegraphics[width = 0.238\textwidth]{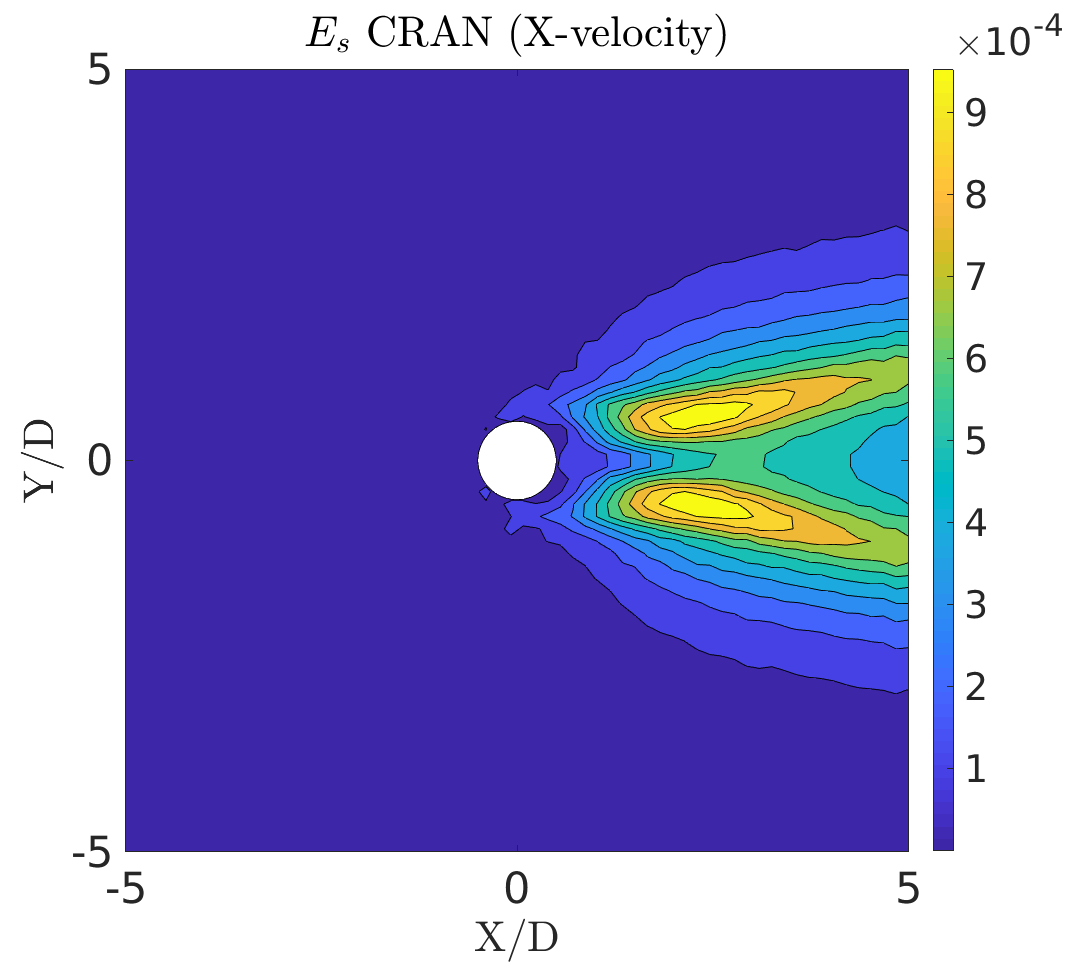}}

\caption{The flow past a cylinder: Comparison between the temporal mean error $E_{s}$ for POD-RNN and CRAN model. (a),(b) for pressure and (c),(d) for x-velocity}
\label{comp_es_stat}
\end{figure}

% point 2
The temporal mean of normalized reconstruction error over the predicted time-steps is given by 
\begin{equation}
    E_{s} = \sum_{n} \frac{E_{n}}{n_{ts}},  
\end{equation}
where $E_{n}$ is the spatial reconstruction error for a predicted time-step $n$ and $n_{ts}$ refer to the number of test time-steps. Fig.~\ref{comp_es_stat} depicts this temporal mean error $E_{s}$ for the predicted fields using POD-RNN and CRAN models. For both the models, majority of the errors are concentrated in the non-linear wake region of the cylinder. For POD-RNN, $E_{s}$ is in the order of $10^{-4}$ for both the field variables. In contrast, the order of errors in CRAN are a bit higher compared to POD-RNN case, with $E_{s} \approx 10^{-3}$ for pressure and x-velocity. \\ \\

% story continues....
 It is worth mentioning that in flow past side-by-side cylinders a closed-loop recurrent architecture can be inefficient to get very long time-series prediction due to compounding effects of error, since the problem in hand is bi-stable. In that case, we have employed an encoder-decoder type network in POD-RNN and ground data as a demonstrator in CRAN to achieve longer time-series of prediction. This application is systematically explored in the next section.  

%\newpage

%%%%%%%%%%%%%%%%%%%%%%%%%%%%%%%%%%%%%%%%%%%%%%%%%%%%%%%%%%%%%%%%%%%%%%%%%%%%%%%%%%%%%%%%%%%%%%
\section{Application \texorpdfstring{\MakeUppercase{\romannumeral 2}}: Flow past side-by-side cylinders} \label{APP2SBS}
%%%%%%%%%%%%%%%%%%%%%%%%%%%%%%%%%%%%%%%%%%%%%%%5%5%%%%%%%%%%%%%%%%%%%%%%%%%%%%%%%%%%%%%%%%%%%%
% intro 
The canonical problem of flow past side-by-side cylinders can act as a representative test case for many multi-body systems. The dynamics associated with this problem is still a popular topic of research among the fluid mechanics community. Hence, as a second application, we apply both POD-RNN and CRAN models for such a bi-stable problem to test the capabilities of the prediction tools. This includes prediction of flow field and integrated pressure force coefficient on the cylinders.

\begin{figure}[H]
\centering
\subfloat[]{\includegraphics[width =0.4\textwidth]{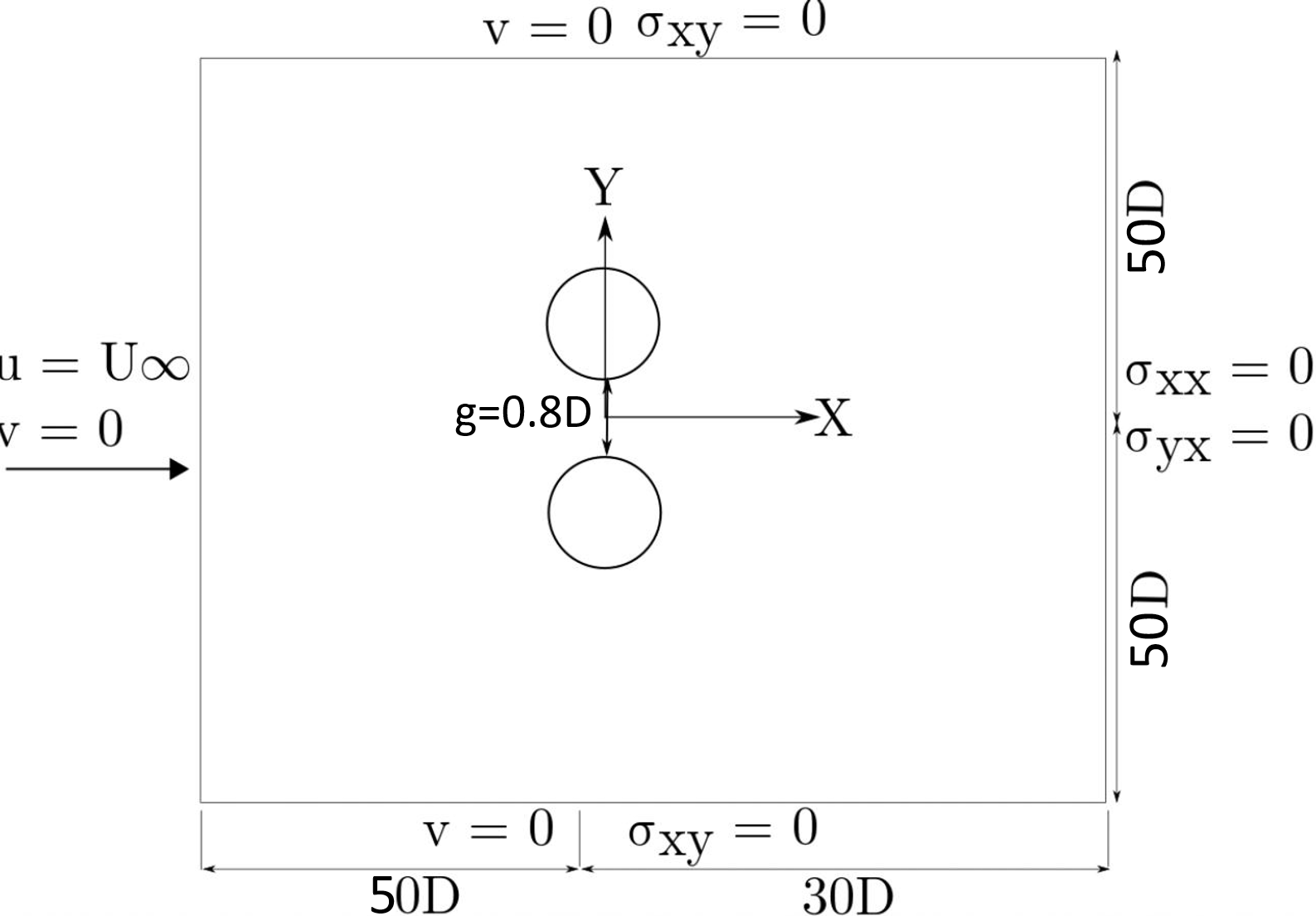}}\\
\subfloat[]{\includegraphics[width = 0.24\textwidth]{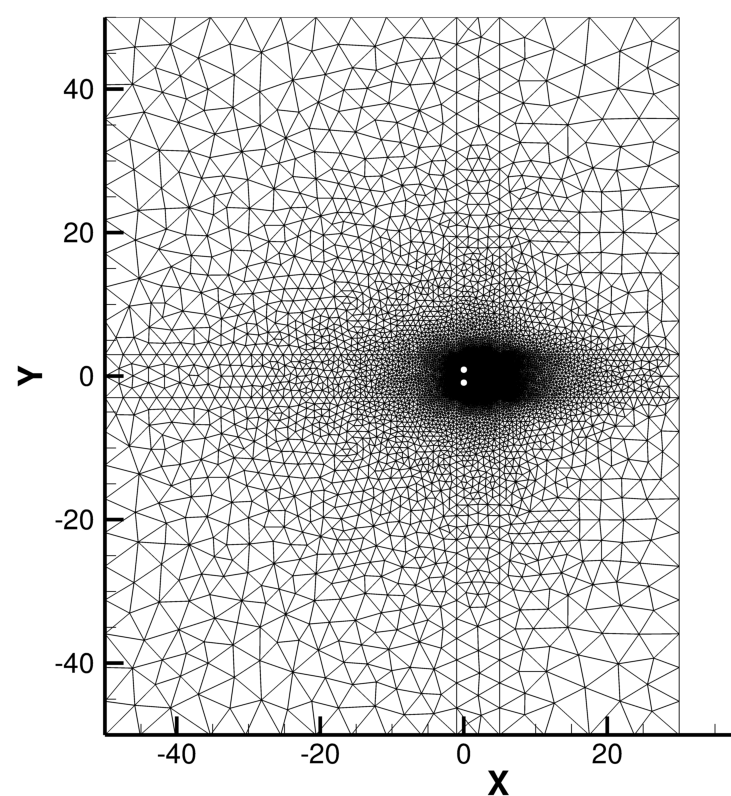}}
\subfloat[]{\includegraphics[width = 0.243\textwidth]{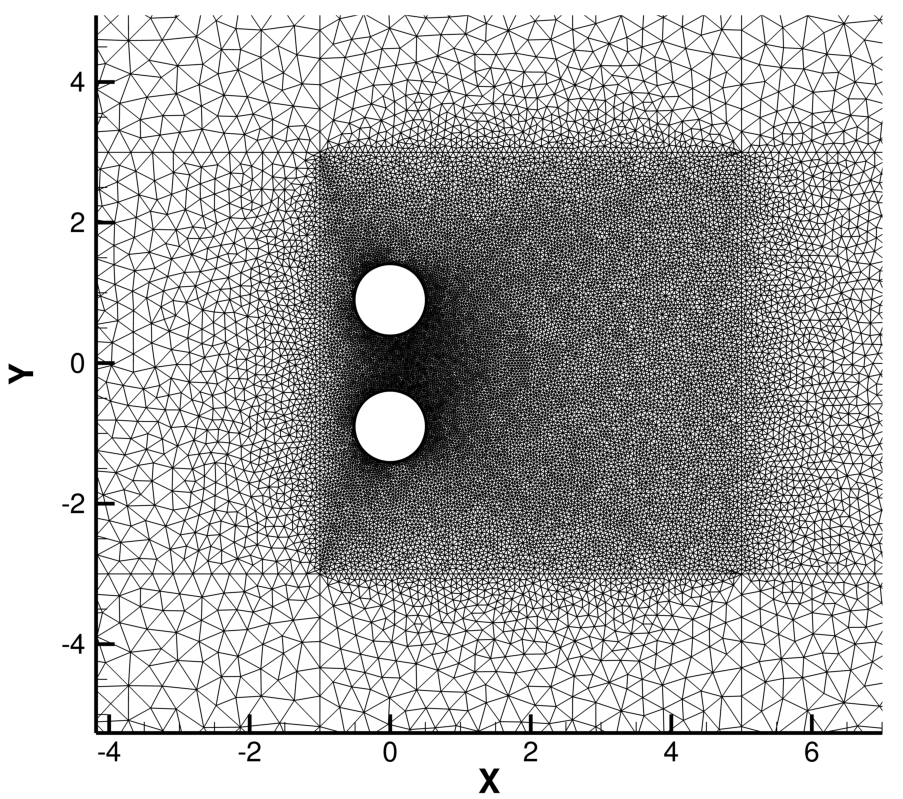}}
\caption{The flow past side-by-side cylinders: (a) Schematic of the problem set-up, (b) full-domain computational mesh view and (c) close-up computational mesh view}
\label{schematic_sbs_fo}
\end{figure}

\textit{Problem objective}: As in the case of a single cylinder, the objective here is to also sufficiently learn the phenomenon of vortex shedding for flow past side-by-side cylinders on a given time-set (training) and extrapolate the behavior based on learned parameters (prediction). However, flow past side-by-side cylinders is known to be a more complex (chaotic) problem with a bi-stable solution, flip-flopping regime and gap flow \cite{liu2016interaction}. The idea is to study the performance of this problem for both types of neural architectures. 
\\
\textit{Full-order data}: The schematic of the problem set-up is given in Fig.~\ref{schematic_sbs_fo} (a), where all the information about the domain boundaries and the boundary conditions are clearly outlined. The final mesh which is obtained after following standard rules of mesh convergence is presented in Fig.~\ref{schematic_sbs_fo} (b) and (c) which contains a total of $49762$ triangular elements with $25034$ nodes. The numerical simulation is carried out via finite element Navier-Stokes solver to generate the train and test data. The Reynolds number of the problem is set at $Re = 100$. The full-order simulation is carried out for a total of $950\;tU_{\infty}/D$ with a time-step of $0.25\;tU_{\infty}/D$. A total of $3800$ snapshots of the simulation are collected at every $0.25\;tU_{\infty}/D$ for the pressure field $P$ and the x-velocity $U$. Of those 3800 snapshots, 3200 (from 301 to 3500 steps) are used for training and 300 (from 3501 to 3800 steps) are kept as testing. The total train and test snapshots $N=3500$.  

Likewise, we systematically assess the complete procedure for prediction of the flow field and pressure force coefficients using POD-RNN (section \ref{podrnnsbs}) and CRAN (section \ref{cransbs}) for flow past side-by-side cylinders. A summary highlighting the major differences in both the models is described for a comparative study in section \ref{discussion_sbs}. 

%%%%%%%%%%%%%%%%%%%%%%%%%%%%
\subsection{POD-RNN model}\label{podrnnsbs}
%%%%%%%%%%%%%%%%%%%%%%%%%%%%
The POD-RNN model is applied on the flow past side-by-side cylinders as follows: 
\begin{enumerate}

% Step 1
\item The POD algorithm (section \ref{pod-rnn-architecture}) is applied on all the train and test time snapshots for field $\boldsymbol{\mathcal{S}} = \left\lbrace\textbf{S}_{1}\;\textbf{S}_{2}\dots\;\textbf{S}_{N}\;\right\rbrace \in \mathbb{R}^{m\times N}$  (here, $m=25034$ and $N=3500$) to get the reduced order dynamics on the data-set. Get the mean field $\bar{\textbf{S}} \in \mathbb{R}^{m}$, the fluctuation matrix $\tilde{\textbf{S}} \in \mathbb{R}^{m\times N}$ and the $N$ spatially invariant POD modes $\boldsymbol{\Phi} \in \mathbb{R}^{m\times N}$.  

% Step 2
\item Get the eigen values $\Lambda_{N\times N}$ (energy of POD modes) of the covariance matrix $\tilde{\textbf{S}}^{T}\tilde{\textbf{S}} \in \mathbb{R}^{N\times N}$. The POD energy spectrum for both pressure and x-velocity fields is plotted in Fig.~\ref{sbspodspectrum}. From the spectrum, it is observed that around $95\%$ of energy is concentrated in first $21$ modes for pressure and first $19$ for x-velocity. 
\begin{figure}[H]
\centering
\subfloat[]{\includegraphics[width = 0.245\textwidth]{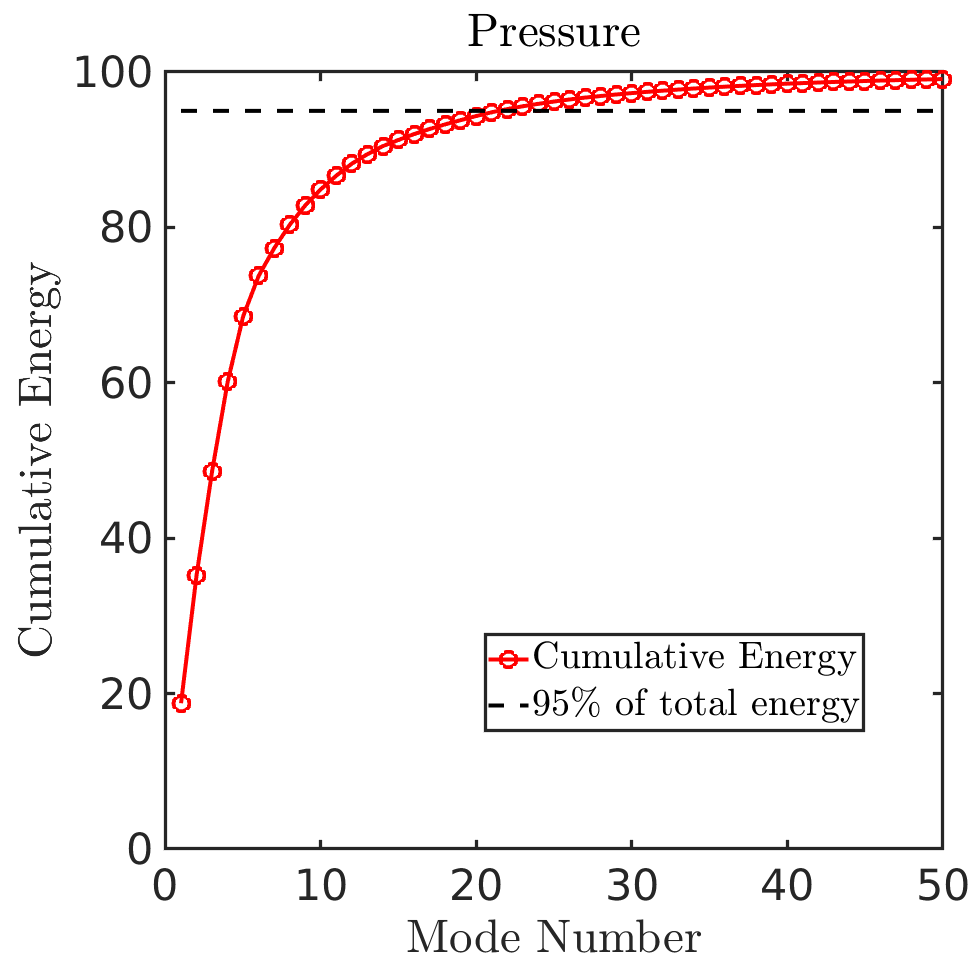}}
\subfloat[]{\includegraphics[width = 0.245\textwidth]{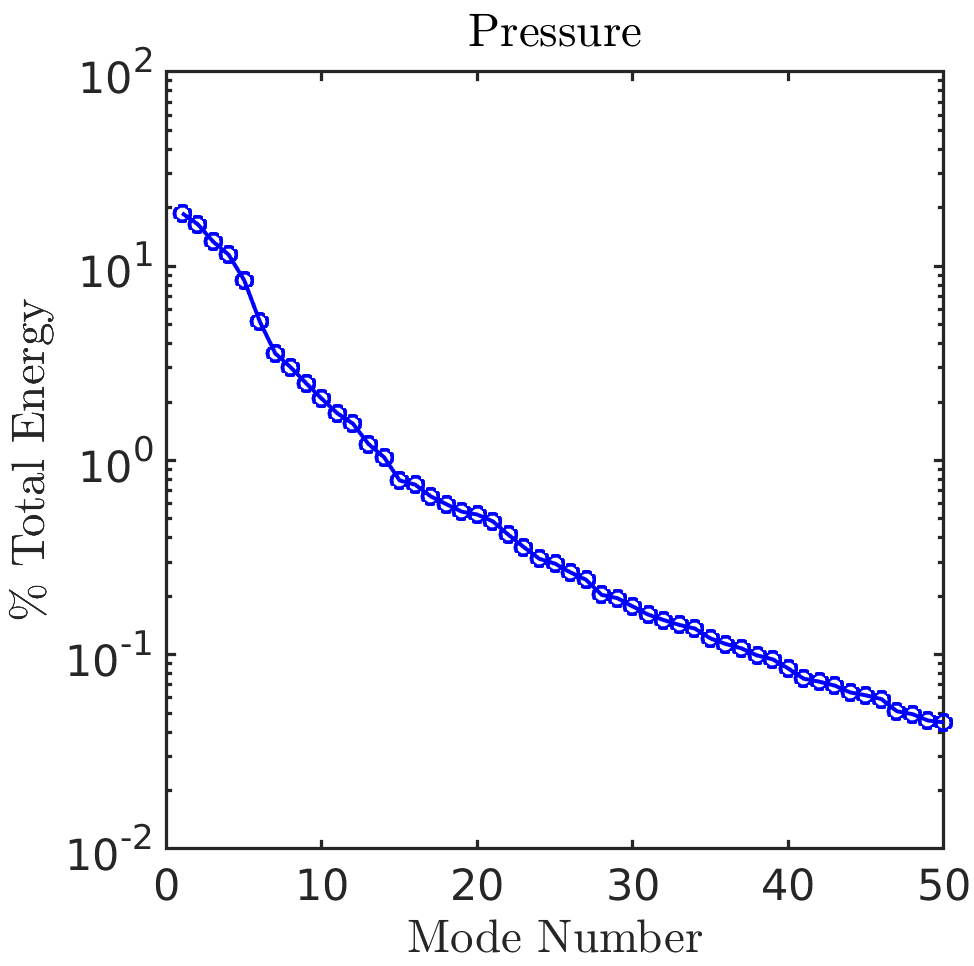}} \\
\subfloat[]{\includegraphics[width = 0.245\textwidth]{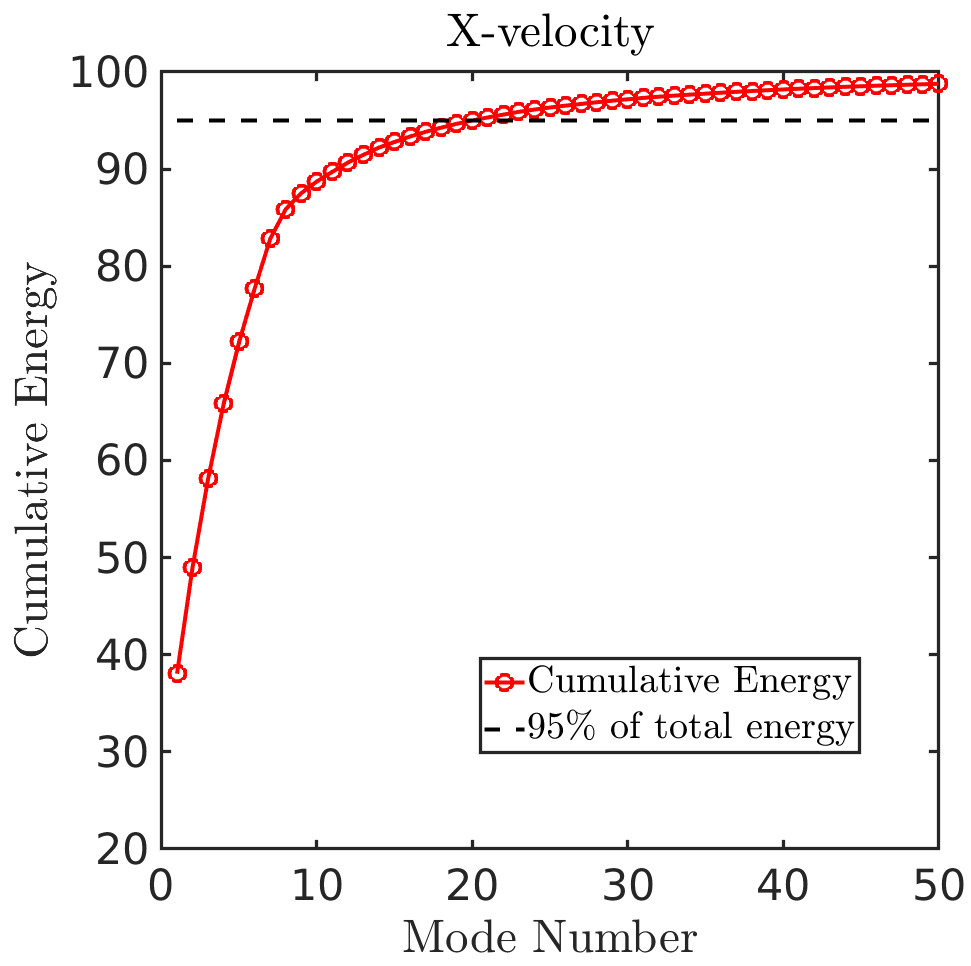}}
\subfloat[]{\includegraphics[width = 0.245\textwidth]{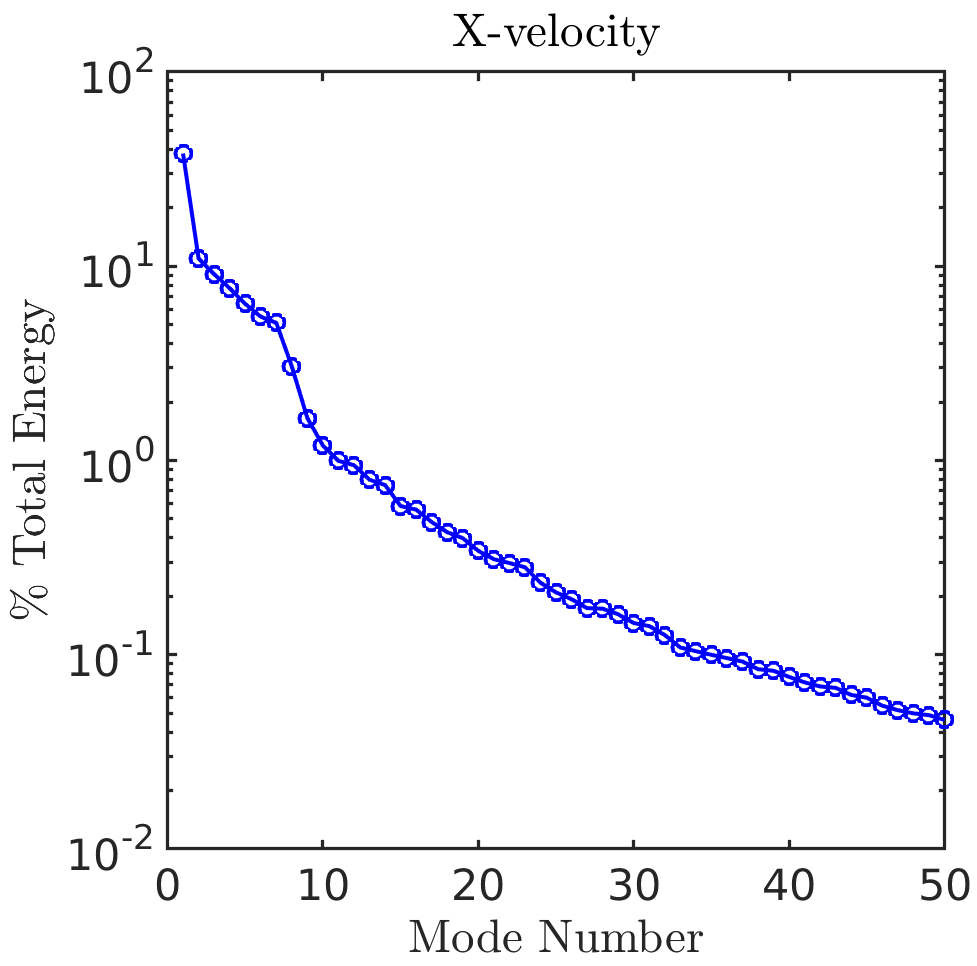}}
\caption{The flow past side-by-side cylinders: Cumulative and percentage of modal energies. (a)-(b) for pressure field $P$, (c)-(d) for x-velocity field $U$}
\label{sbspodspectrum}
\end{figure}

% Step 3
\item Construct the reduced order system dynamics by selecting first few energetic POD modes in the analysis (here, $k=25$). These 25 modes account for nearly $96\%$ of energy for both pressure and $x$-velocity. The $N$ POD modes are now simply approximated with these $k$ modes ($k \ll N$), reducing the POD system to $\boldsymbol{\Phi} \in \mathbb{R}^{m\times k}$. For instance, the first four spatial POD modes for pressure and x-velocity fields along with $\%$ total energy are depicted in Fig.~\ref{sbspodmodes}. 
\begin{figure}
\centering
\subfloat[]{\includegraphics[width = 0.245\textwidth]{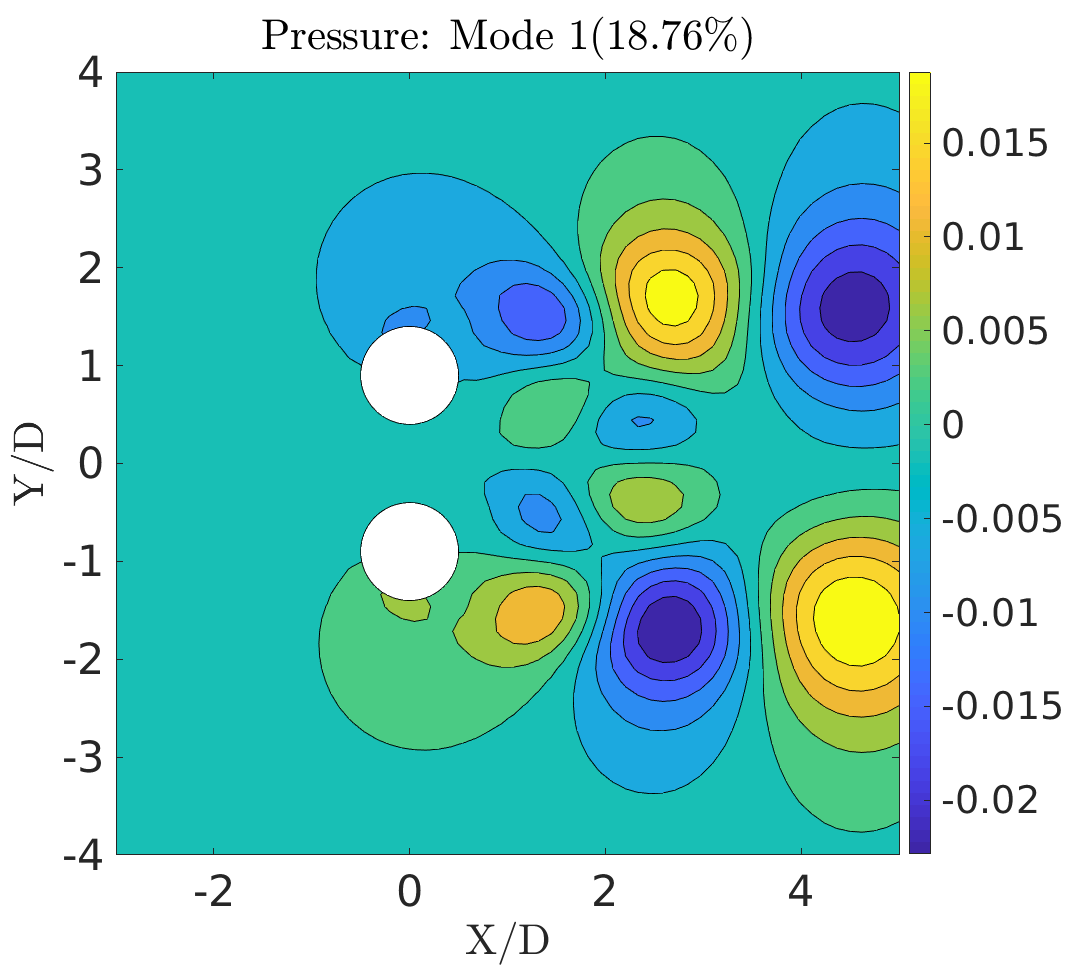}}
\subfloat[]{\includegraphics[width = 0.245\textwidth]{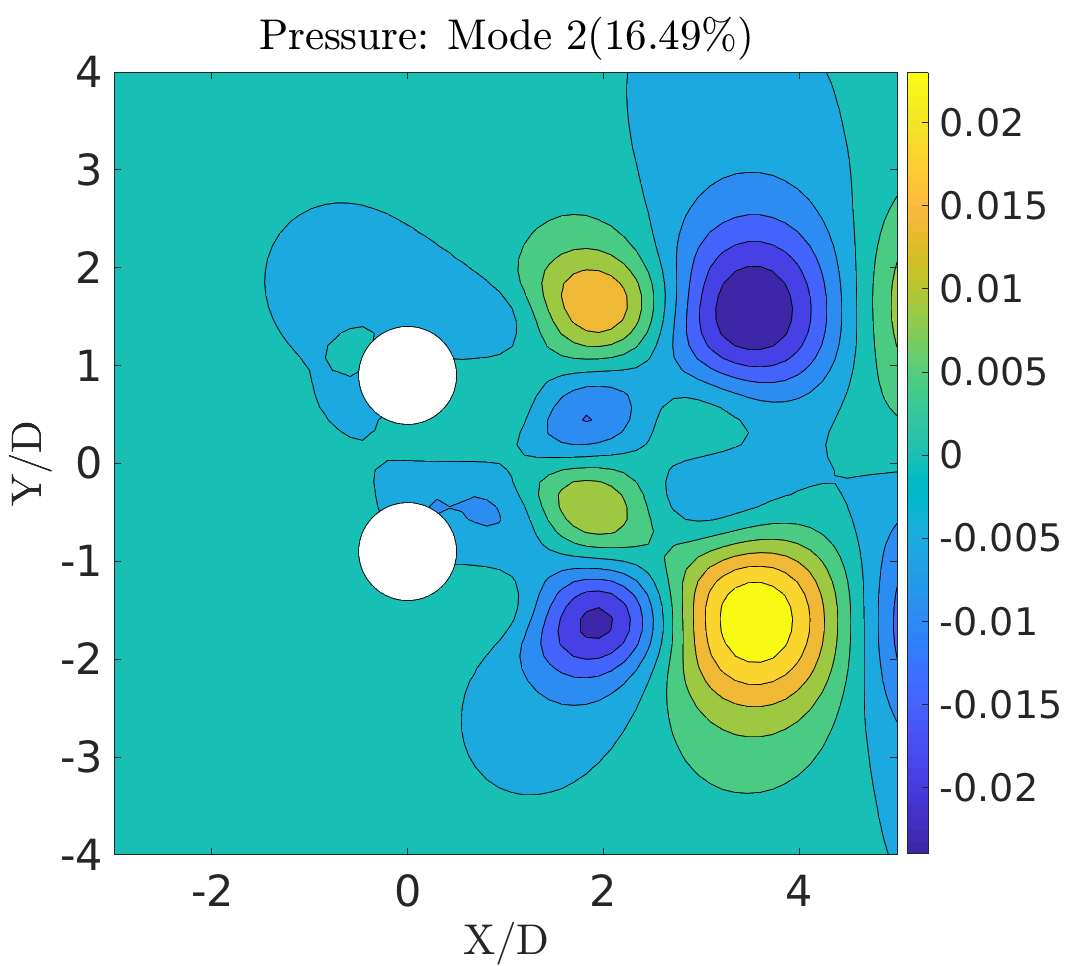}}\\
\subfloat[]{\includegraphics[width = 0.245\textwidth]{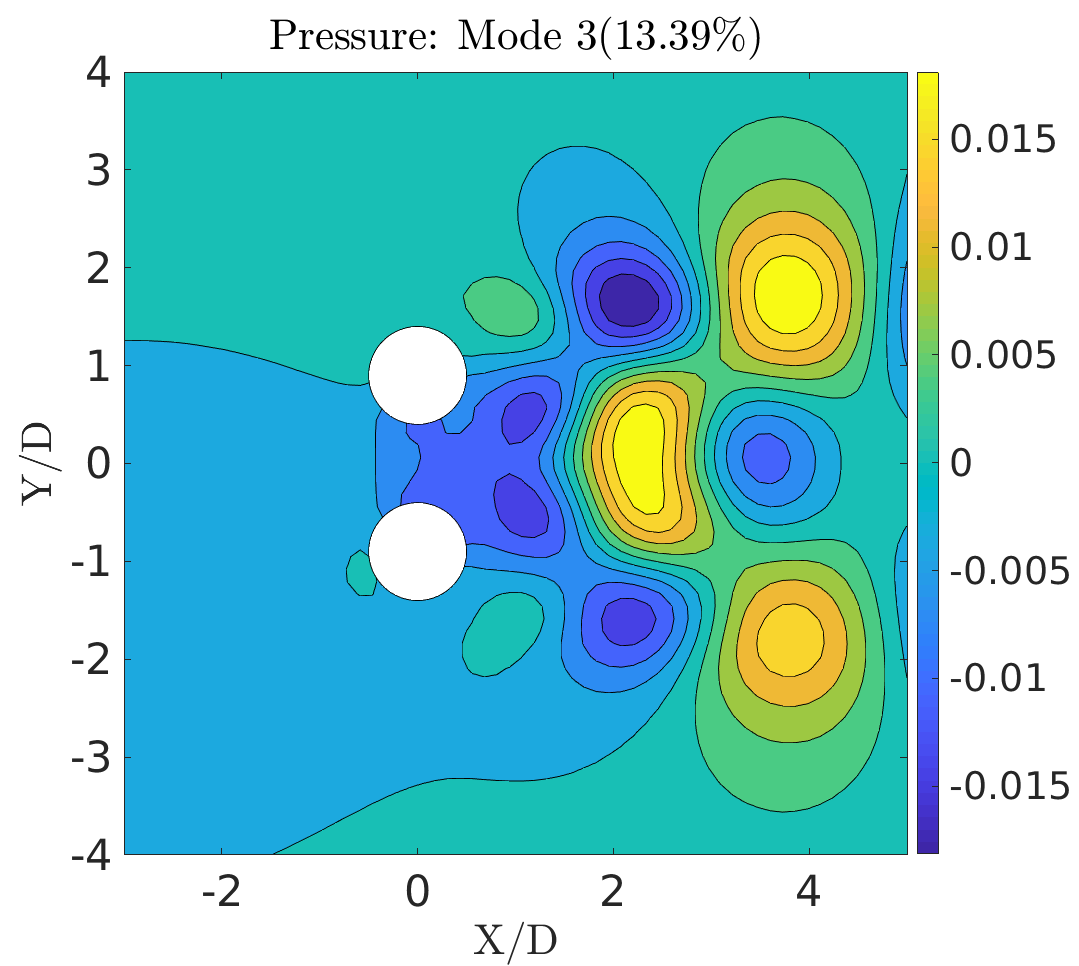}}
\subfloat[]{\includegraphics[width = 0.245\textwidth]{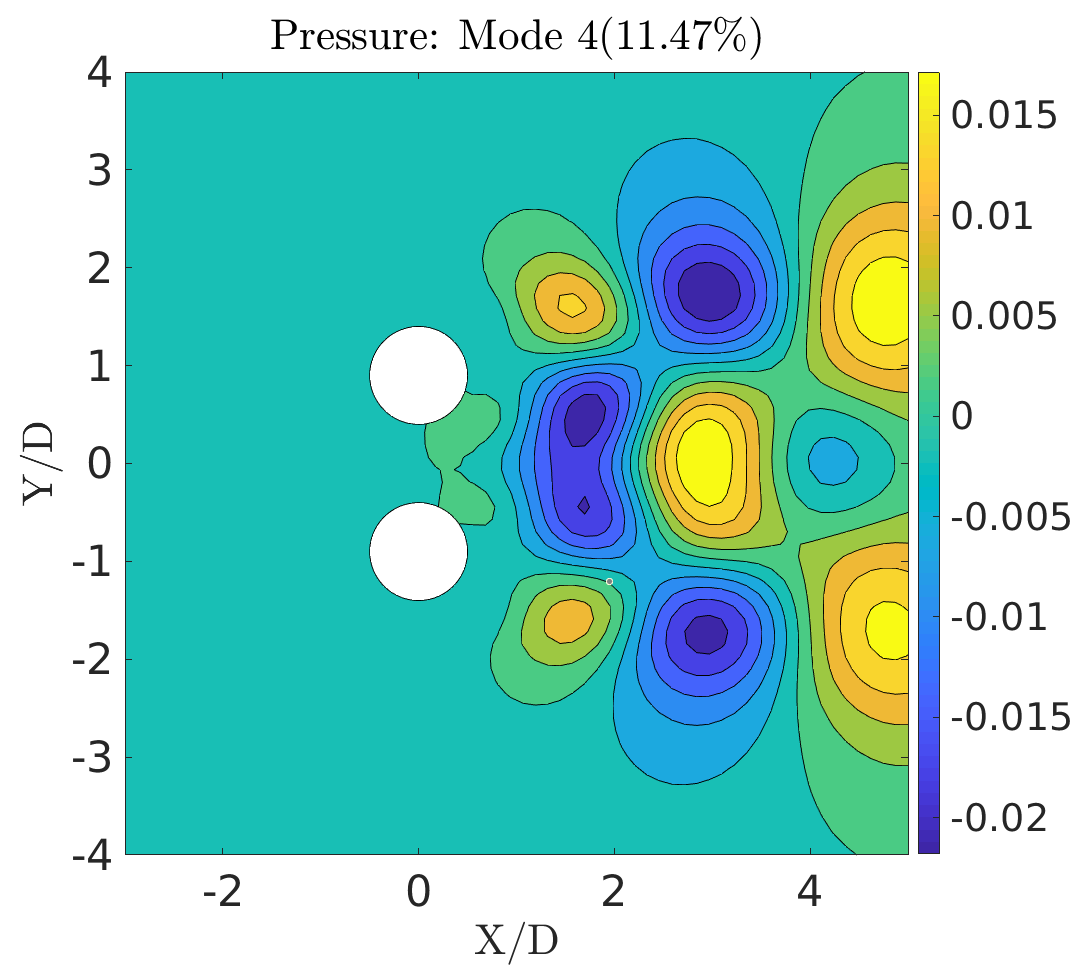}}\\
\subfloat[]{\includegraphics[width = 0.245\textwidth]{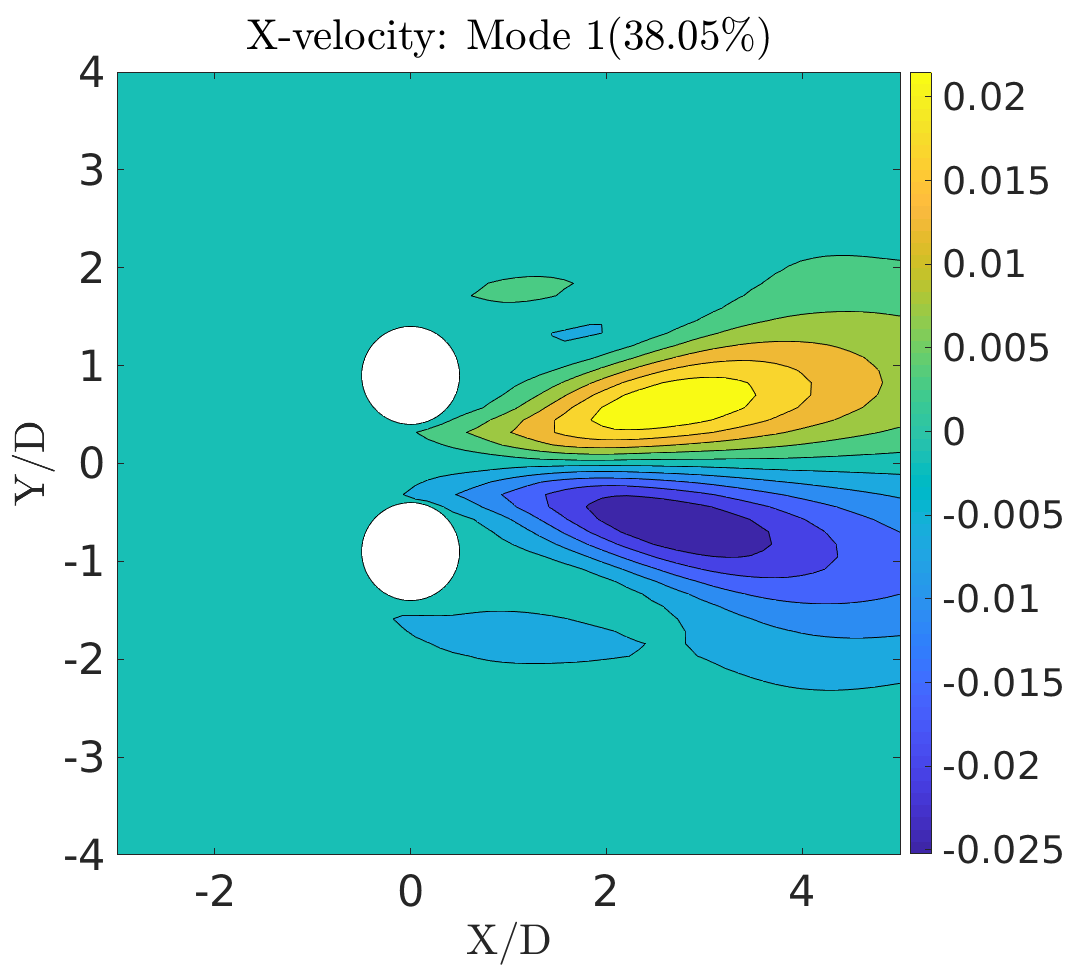}}
\subfloat[]{\includegraphics[width = 0.245\textwidth]{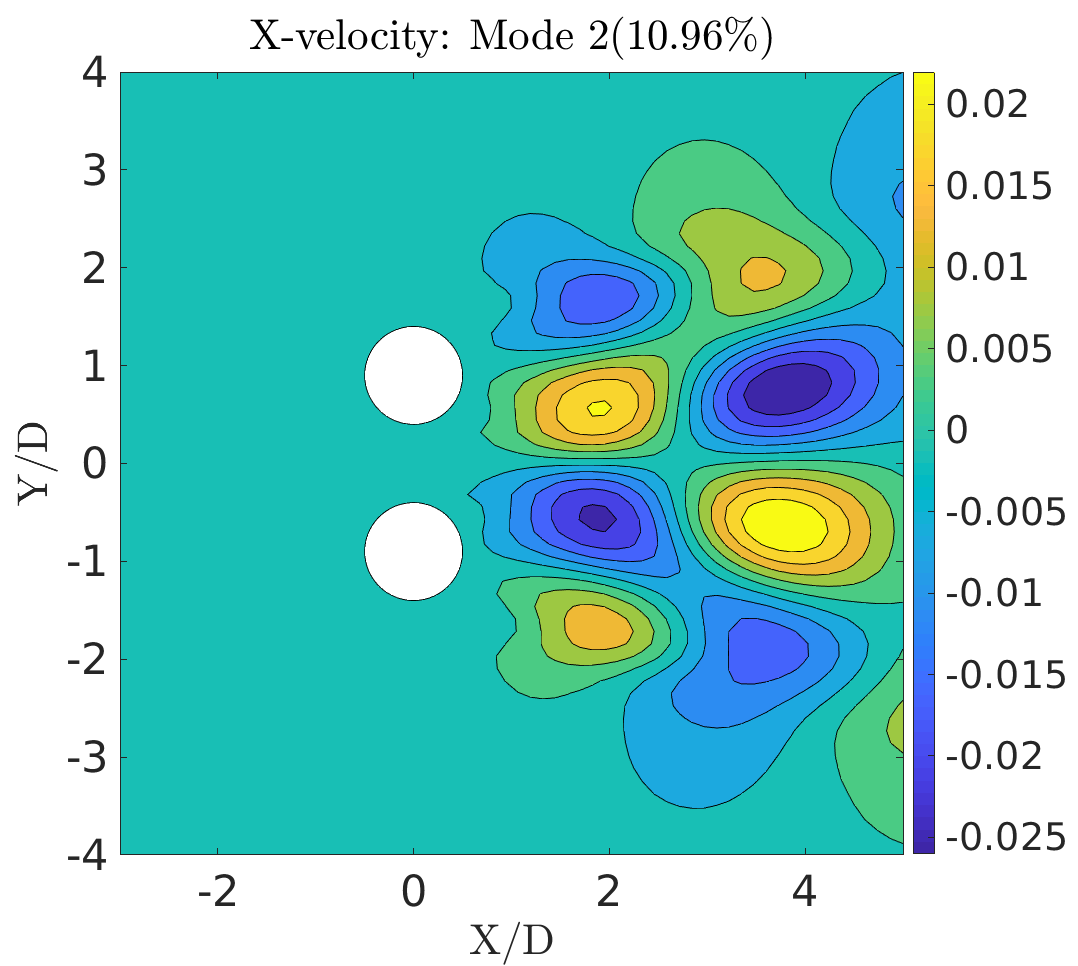}}\\
\subfloat[]{\includegraphics[width = 0.245\textwidth]{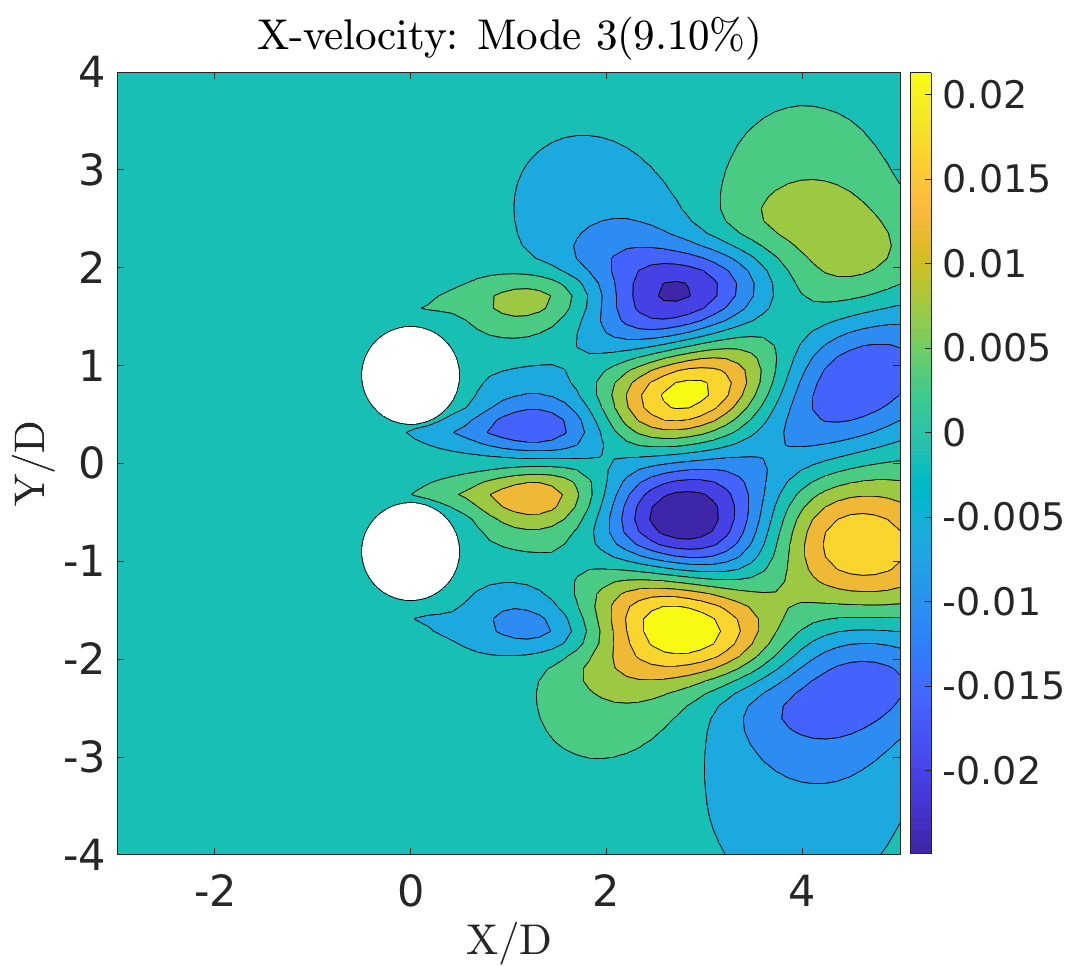}}
\subfloat[]{\includegraphics[width = 0.245\textwidth]{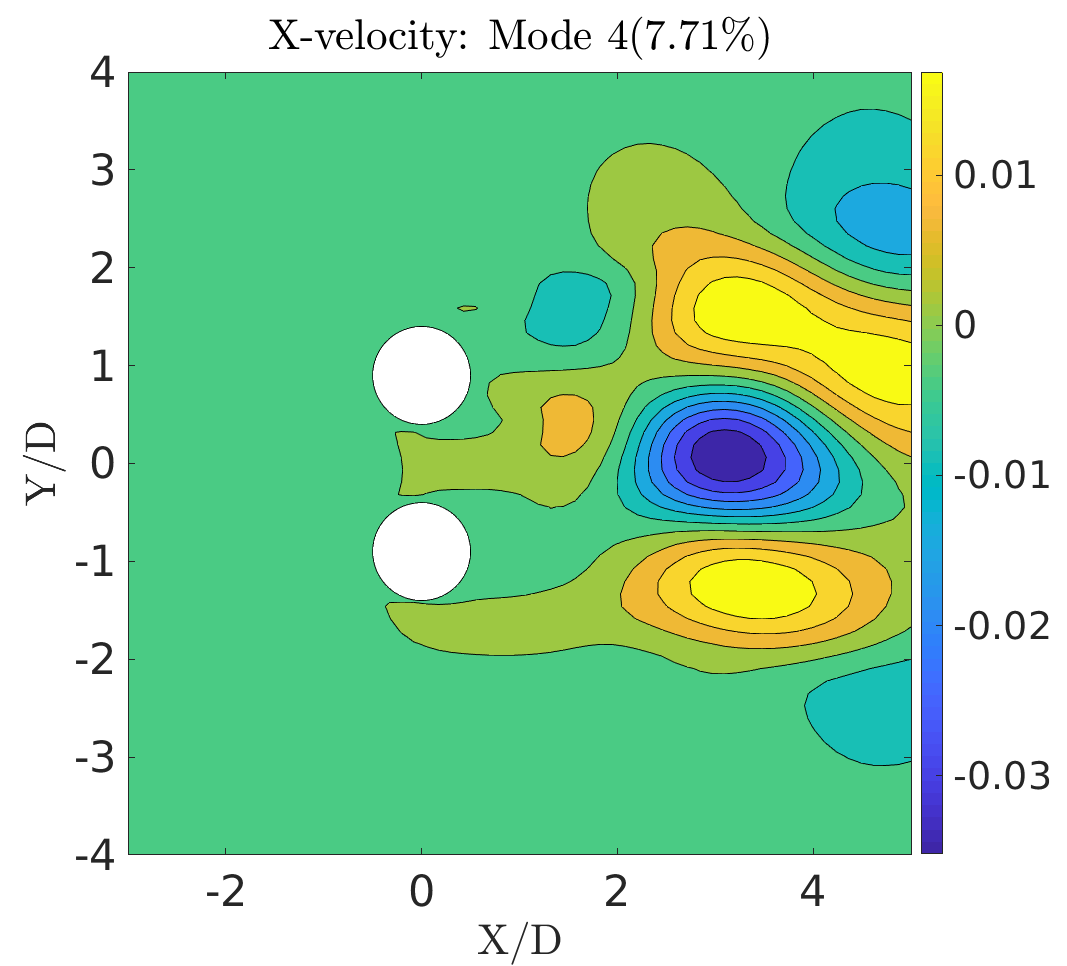}}
\caption{The flow past side-by-side cylinders: First four most energetic time-invariant spatial modes obtained from POD along with $\% $ of total energy. (a)-(b)-(c)-(d) for pressure field $P$ and (e)-(f)-(g)-(h) for x-velocity field $U$}
\label{sbspodmodes}
\end{figure}

% Step 4
\item The temporal variations of these $k$ modes are obtained by $\mathbf{A} = \boldsymbol{\Phi}^{T} \tilde{\textbf{S}}$. Here, $\mathbf{A} \in \mathbb{R}^{k\times N}$, $\boldsymbol{\Phi} \in \mathbb{R}^{m\times k}$ and $\tilde{\textbf{S}} \in \mathbb{R}^{m\times N}$. These temporal modes are divided into training ($n_{tr}$ steps) and testing part ($n_{ts}=N-n_{tr}$ steps). Here, $n_{tr}=3200$ and $n_{ts}=300$. The temporal coefficients from $301-3500$ time-steps are considered for training and $3501-3800$ are kept for testing. For illustration, the time history of first eight POD modes for pressure and x-velocity fields from $301-1300$ ($75-325\;tU_{\infty}/D$) time-steps is depicted in Fig.~\ref{podsbsth}. 
\begin{figure}
\centering
\subfloat[]{\includegraphics[width =0.5\textwidth]{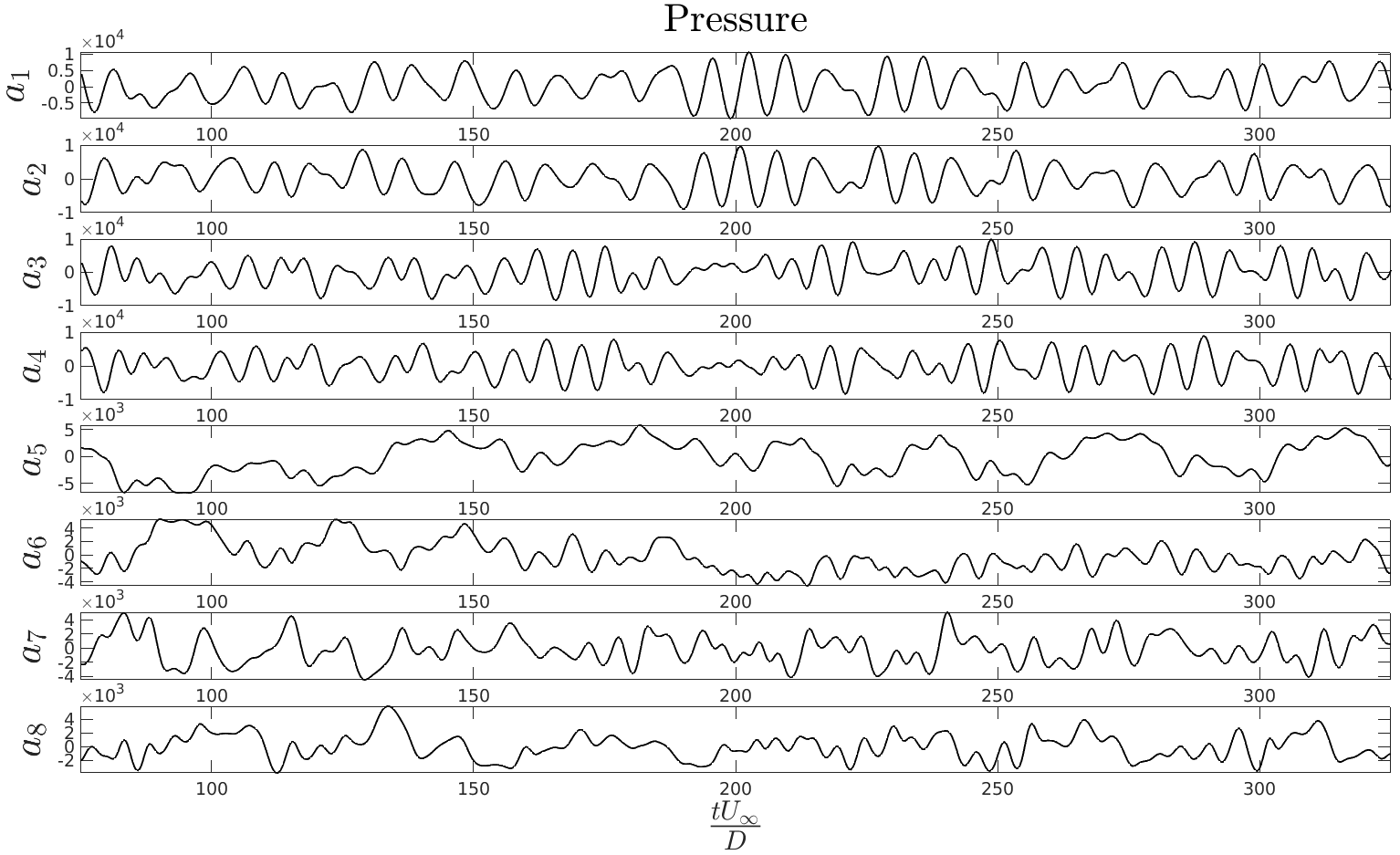}}\\
\subfloat[]{\includegraphics[width =0.5\textwidth]{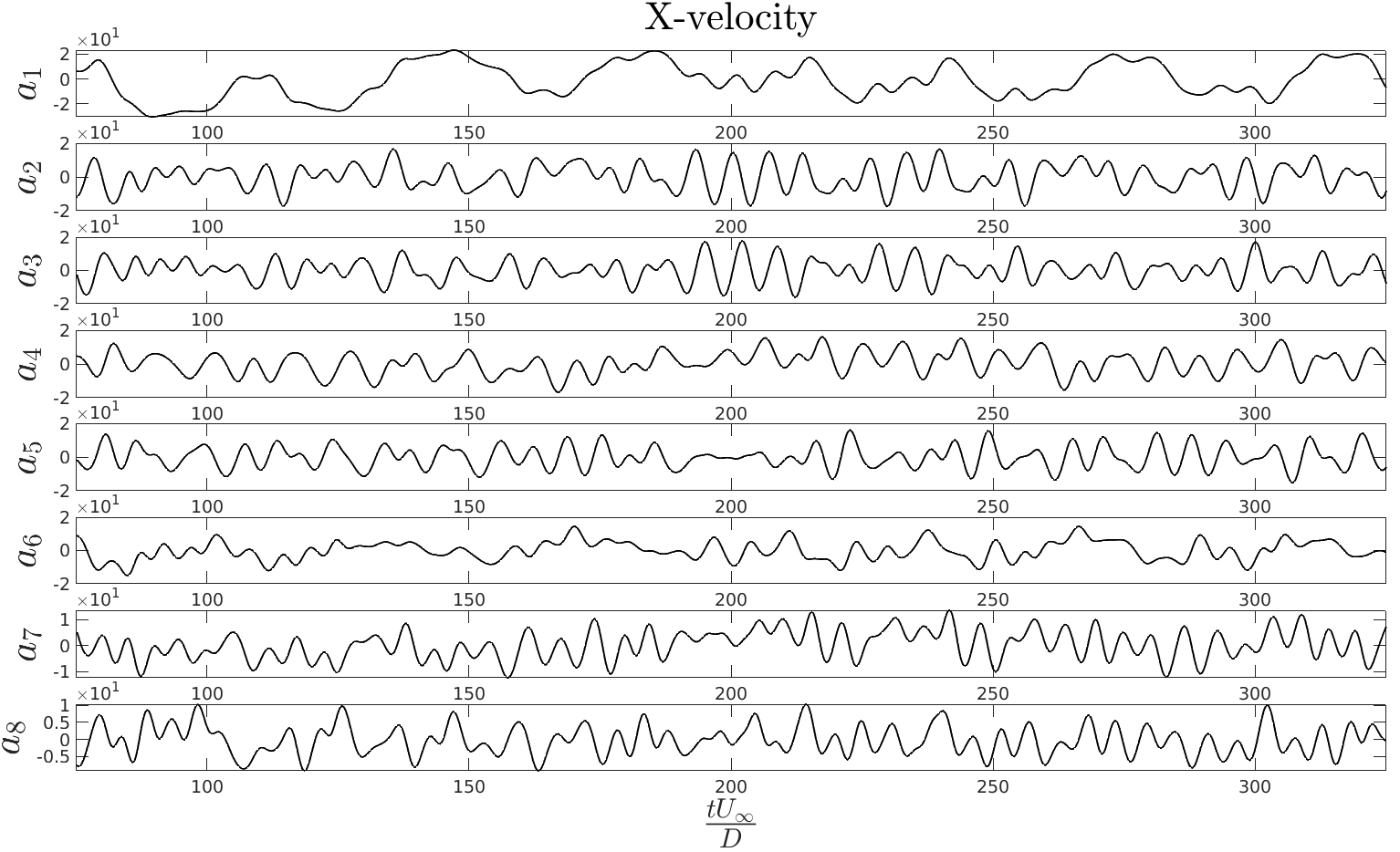}}
\caption{The flow past side-by-side cylinders: Time history of modal coefficients from $75$ to $325\;tU_{\infty}/D$): (a) pressure field $P$ and (b) x-velocity field $U$}
\label{podsbsth}
\end{figure}

% Step 5
\item The temporal behavior of these modal coefficients (for both pressure and x-velocity) is learned via an encoder-decoder type recurrent neural network (see Fig.~\ref{encdnc}) as it found to work better compared to a closed-loop type. Note that before proceeding with training, the modal coefficients are normalized between -1 to 1 to prevent vanishing/exploding gradients. The primary feature of this recurrent architecture is to encode a finite sequence of time-steps (of length, say, $n_{i}$) and decode it to predict the next set of time-steps (of length $n_{o}$). For example, time-steps 1 to 25 can be encoded to predict 26 until 30 steps and so on. The training data consists of finite input-output time sequences of $k$ modes generated in random batches of size $n_{b}$ in every epoch (here, $n_{b}=50$, $n_{i}=n_{o}=25$, $k=25$). The parametric details of the network for $P$ and $U$ is outlined in the Table \ref{podrnn_sbs_na}. The test time-steps of 300 are organized into six equal sets of input-output pairs. The size of the input ($n_{i}$) and output ($n_{o}$) time-window is 25 here.  
\begin{table}[H]
\centering
\begin{tabular}{|c|c|}
\hline
Parameter & Value \\ \hline
Input features  & $k=25$ \\
Output features & $k=25$  \\
Hidden dimension & 512 \\
RNN cell & LSTM \\
Time-steps per batch  & $n_{i}=n_{o}=25$  \\
Number of batches & $n_{b}=50$ (random) \\
Initial learning rate & 0.0005\\
L-2 regularization factor & 0.003 \\
Iterations & 15000 epochs \\
Optimizer & ADAM \\ \hline
\end{tabular}
\caption{The flow past side-by-side cylinders: Network and parameter details for the encoder-decoder type recurrent network}
\label{podrnn_sbs_na}
\end{table}

% Step 6
\item The result of the temporal modal predictions from the encoder-decoder type recurrent network is plotted in Fig.~\ref{podsbsthpredict} for (a) pressure and (b) x-velocity for the test time-steps from $3501$ till $3800$ ($875-950\;tU_{\infty}/D$). For visualisation convenience, only first $8$ modal coefficients are shown, although the network predicts the temporal trend in all $k=25$ input modes within reasonable accuracy. The network input, prediction and true values are depicted by green, red and black lines respectively in Fig.~\ref{podsbsthpredict}. In spite of the drastic differences between the POD modes, one trained encoder-decoder network is suffice to capture such a chaotic behavior in all the modes for finite time-steps. The normalised root mean square error 
$RMSE = \sqrt{\frac{\operatornamewithlimits{\sum}_{n=1}^{i=T}(\textbf{A}_{n}-\hat{\textbf{A}}_{n})^2}{T}}$ is plotted in Fig. \ref{podsbsrmse} for all the predicted $25$ modes. Here, $\textbf{A}_{n}$ and $\hat{{\textbf{A}}}_{n}$ are the true and predicted modal coefficients respectively. $T$ is the total length of the output time-sequences (here, $T=6 n_{o}=150$). 
\end{enumerate}

\begin{figure}[H]
\centering
\subfloat[]{\includegraphics[width = 0.245\textwidth]{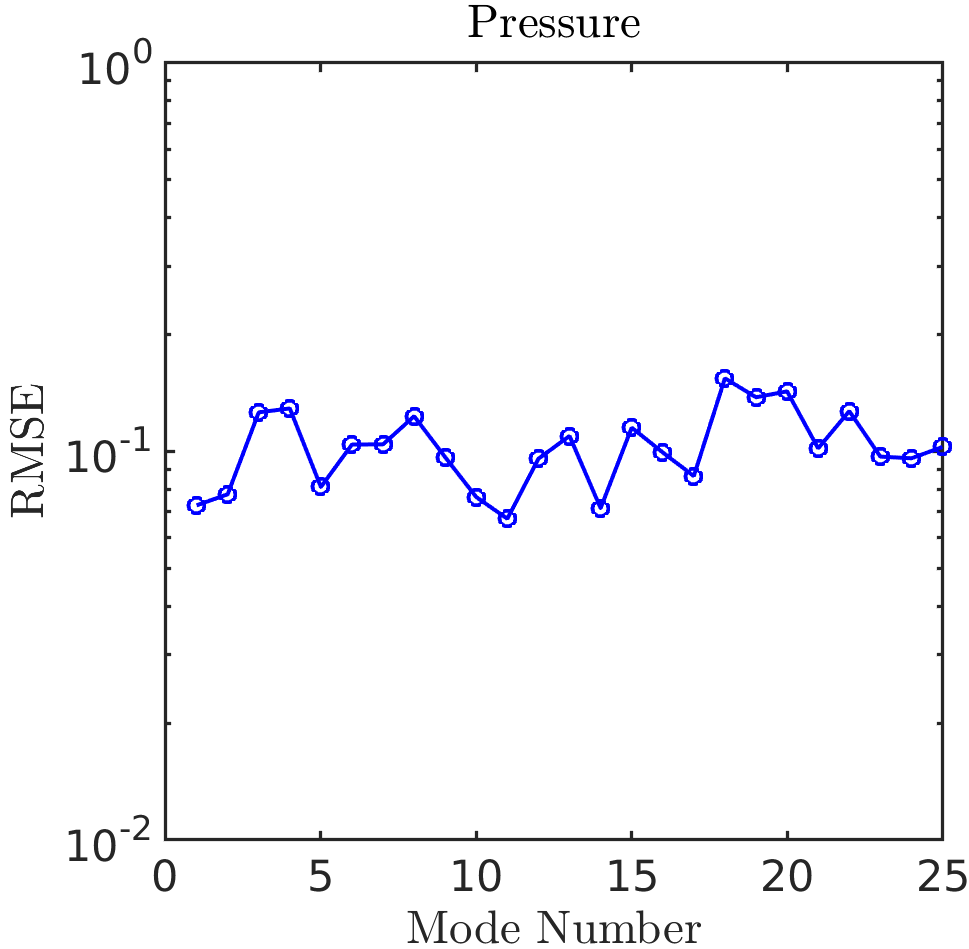}}
\subfloat[]{\includegraphics[width = 0.245\textwidth]{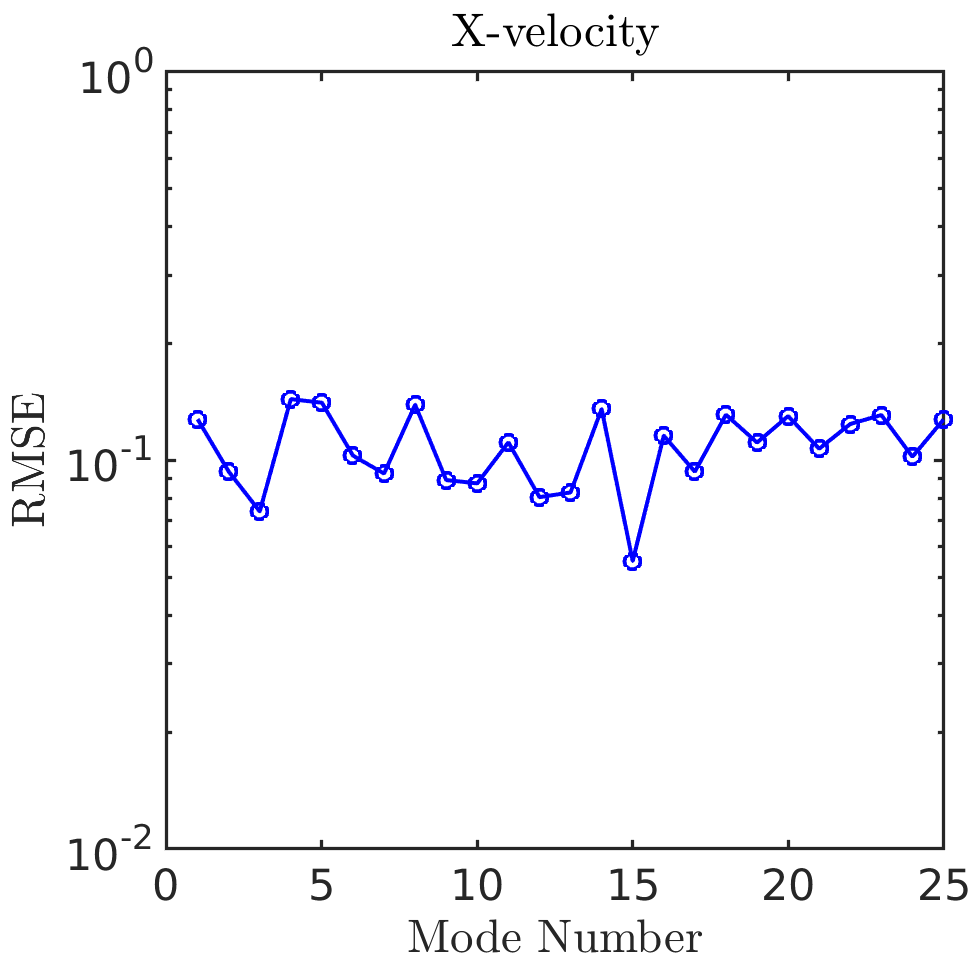}}
\caption{The flow past side-by-side cylinders: RMSE for the predicted against true (normalised) modal coefficients for (a) pressure (b) x-velocity}
\label{podsbsrmse}
\end{figure}

\begin{figure}[H]
\centering
\subfloat[]{\includegraphics[width = 0.48\textwidth]{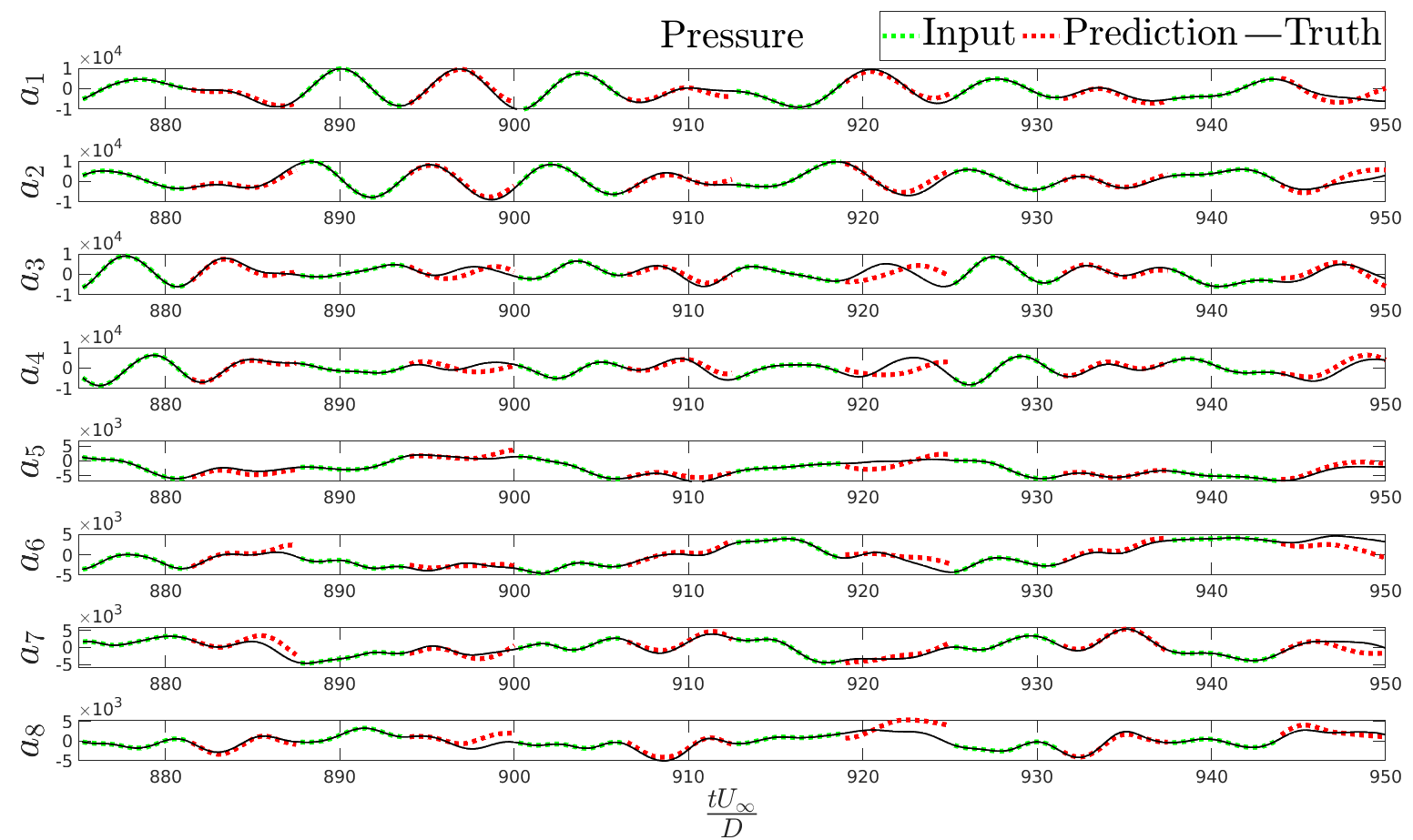}}\\
\subfloat[]{\includegraphics[width = 0.48\textwidth]{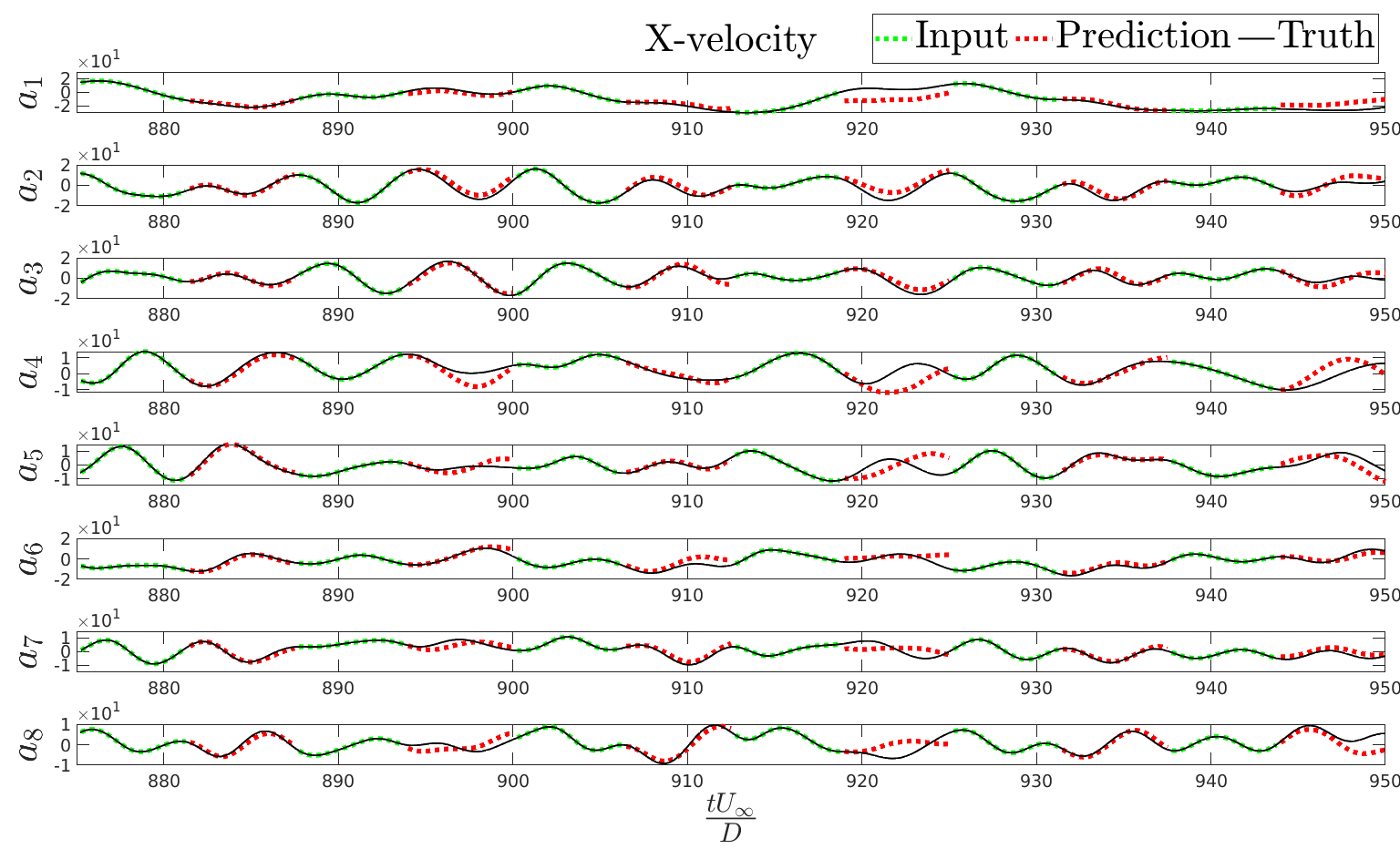}}
\caption{The flow past side-by-side cylinders: Temporal evolution (prediction) of modal coefficients (from $875$ till $950\;tU_{\infty}/D$) for (a) pressure field $P$ and (b) x-velocity field $U$}
\label{podsbsthpredict}
\end{figure}

%\newpage 

\begin{figure*}
\centering
\subfloat[]
{\includegraphics[width = 0.32\textwidth]{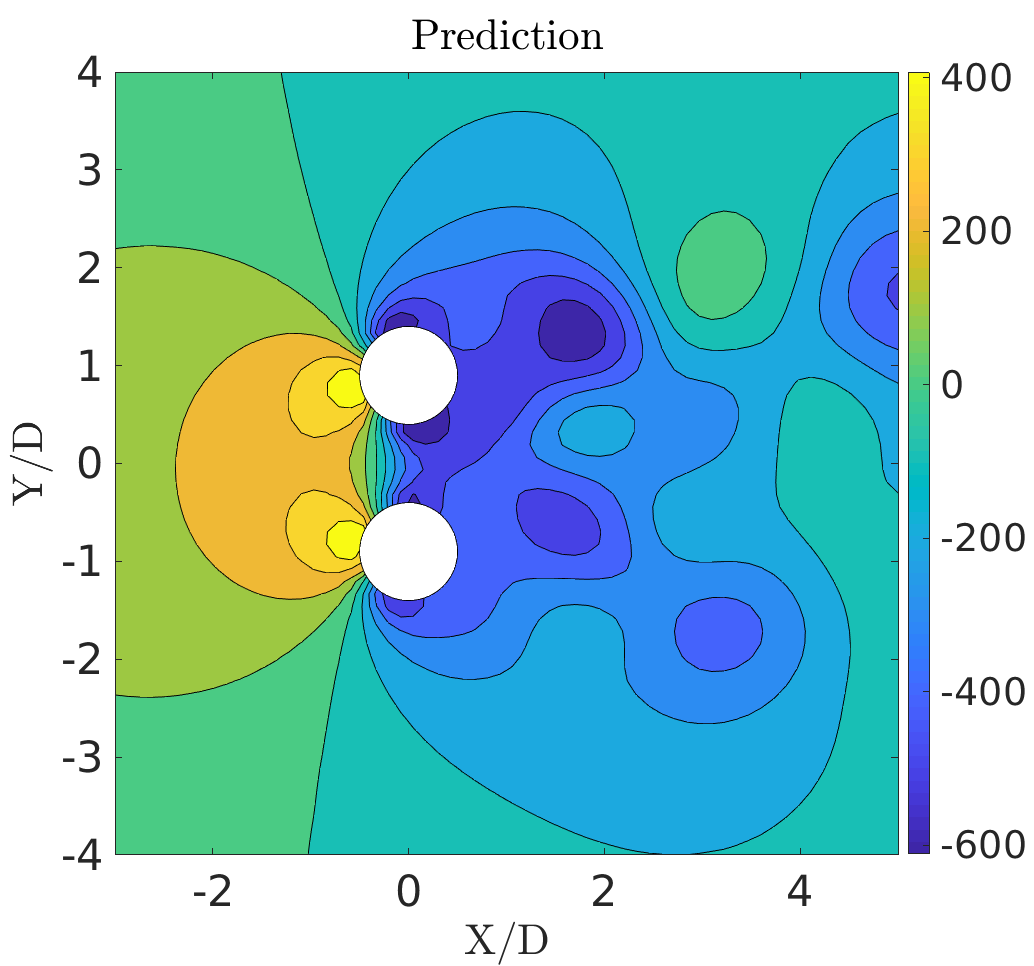}
\hspace{0.02\textwidth}
\includegraphics[width = 0.32\textwidth]{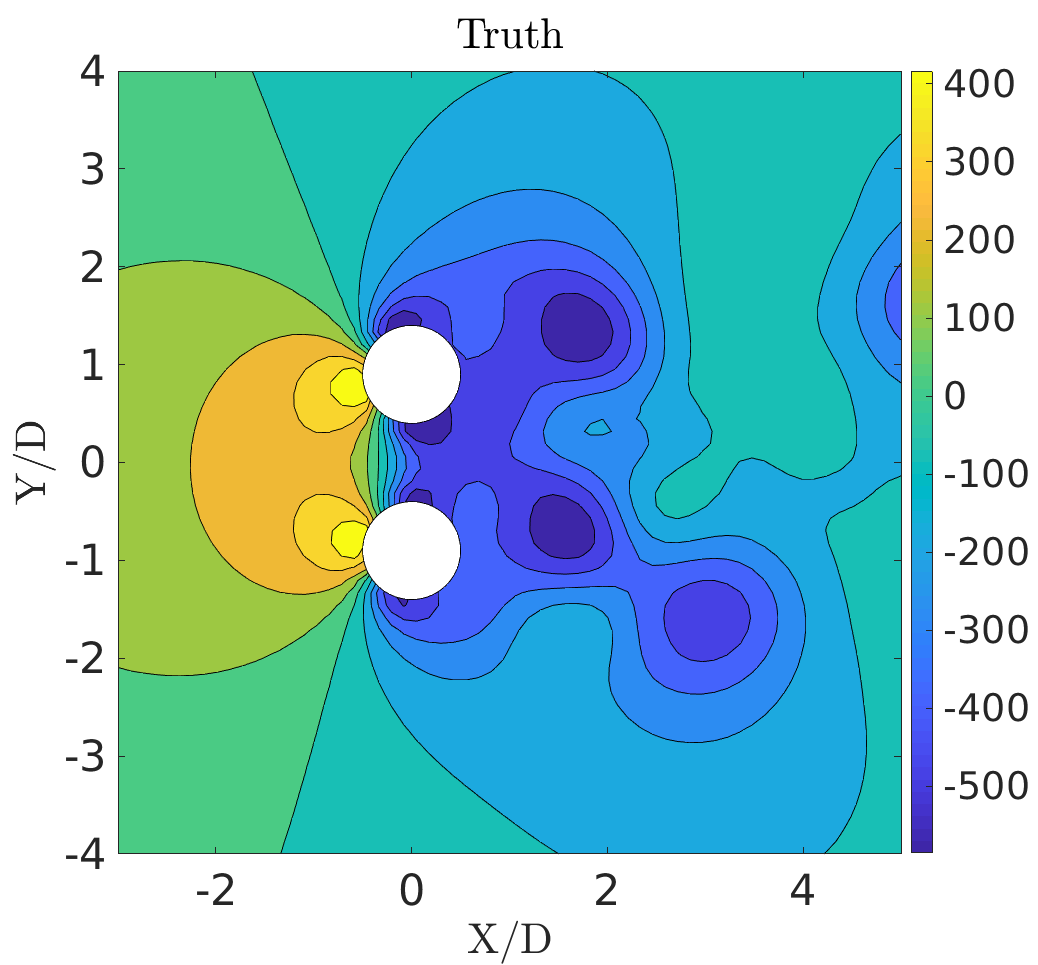}
\hspace{0.02\textwidth}
\includegraphics[width = 0.32\textwidth]{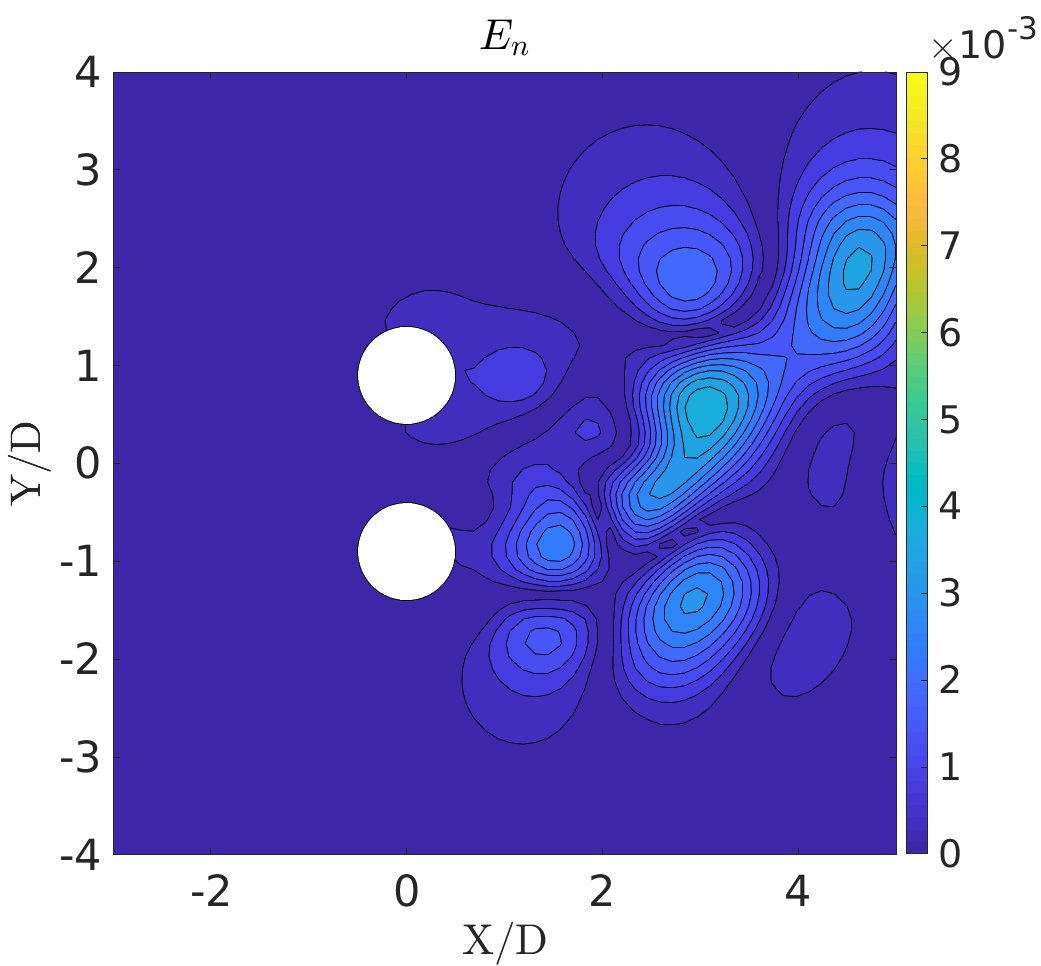}}
\\
\vspace{0.025\textwidth}
\subfloat[]
{\includegraphics[width = 0.32\textwidth]{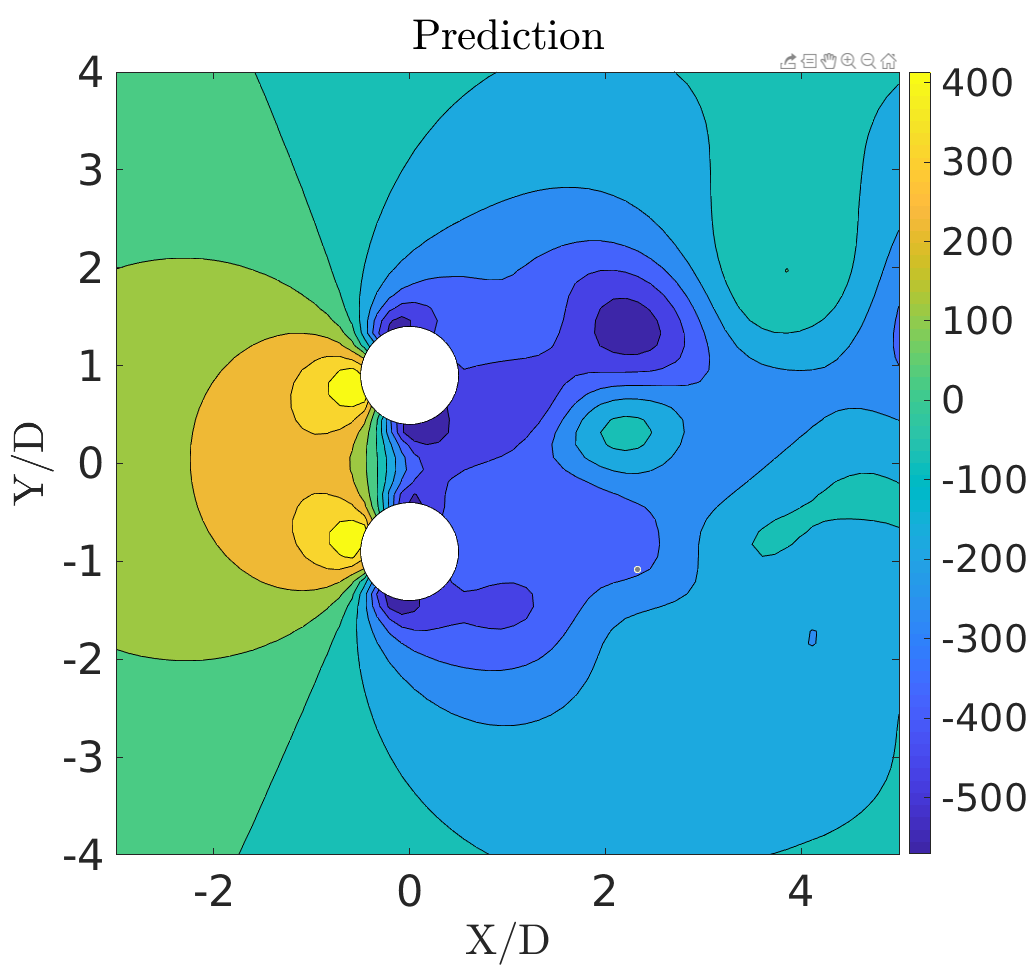}
\hspace{0.02\textwidth}
\includegraphics[width = 0.32\textwidth]{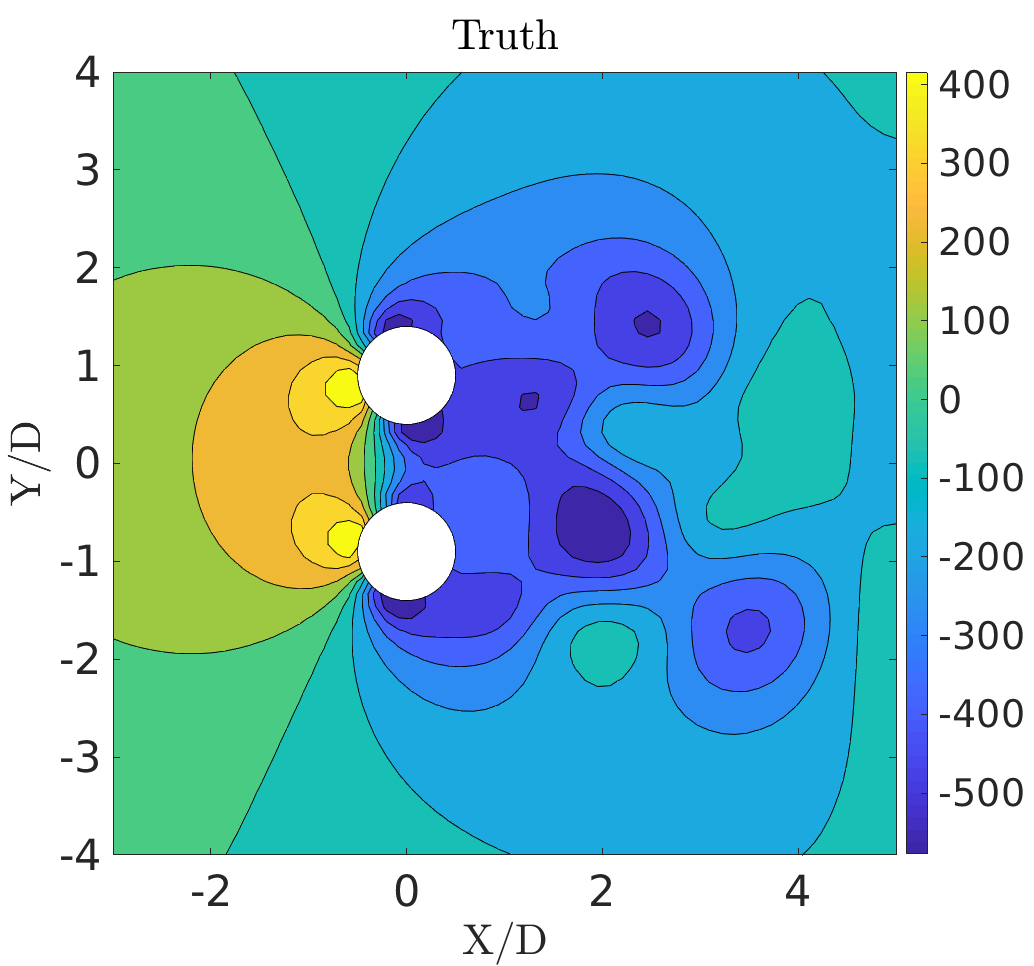}
\hspace{0.02\textwidth}
\includegraphics[width = 0.32\textwidth]{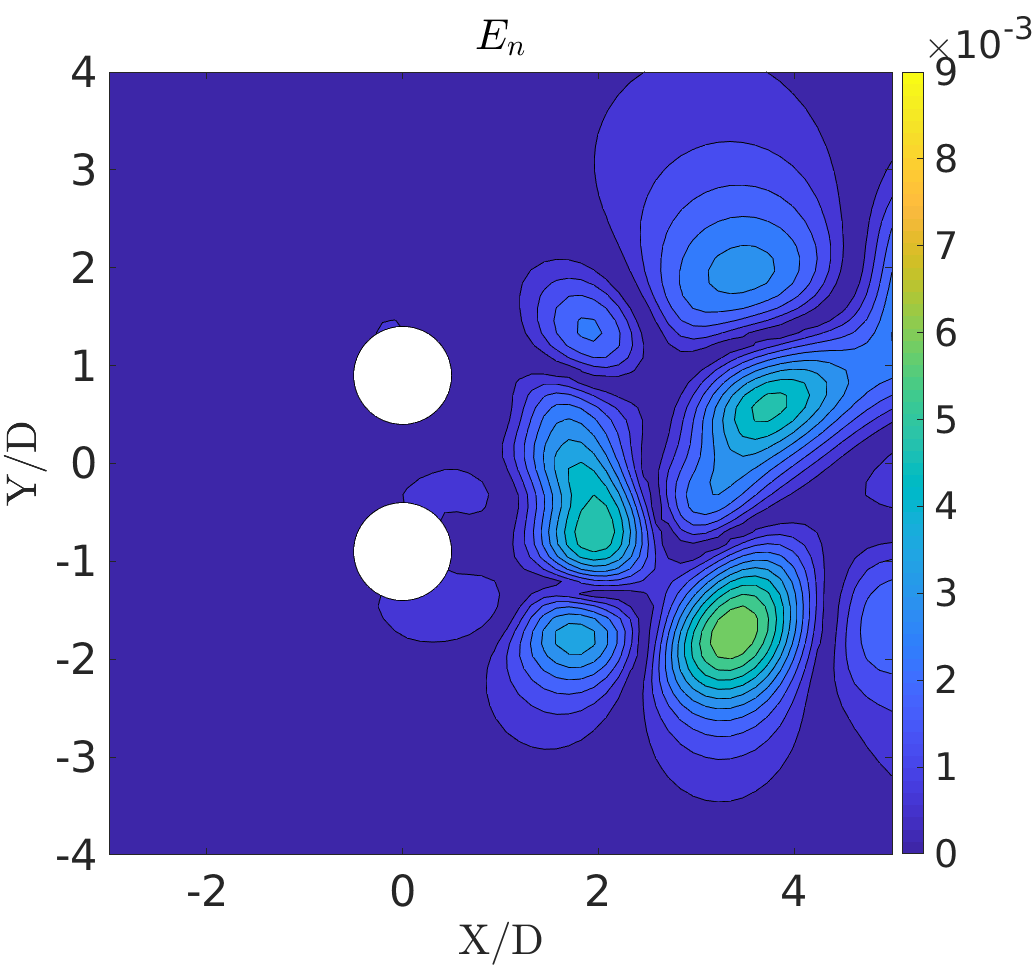}}
\\
\vspace{0.025\textwidth}
\subfloat[]
{\includegraphics[width = 0.32\textwidth]{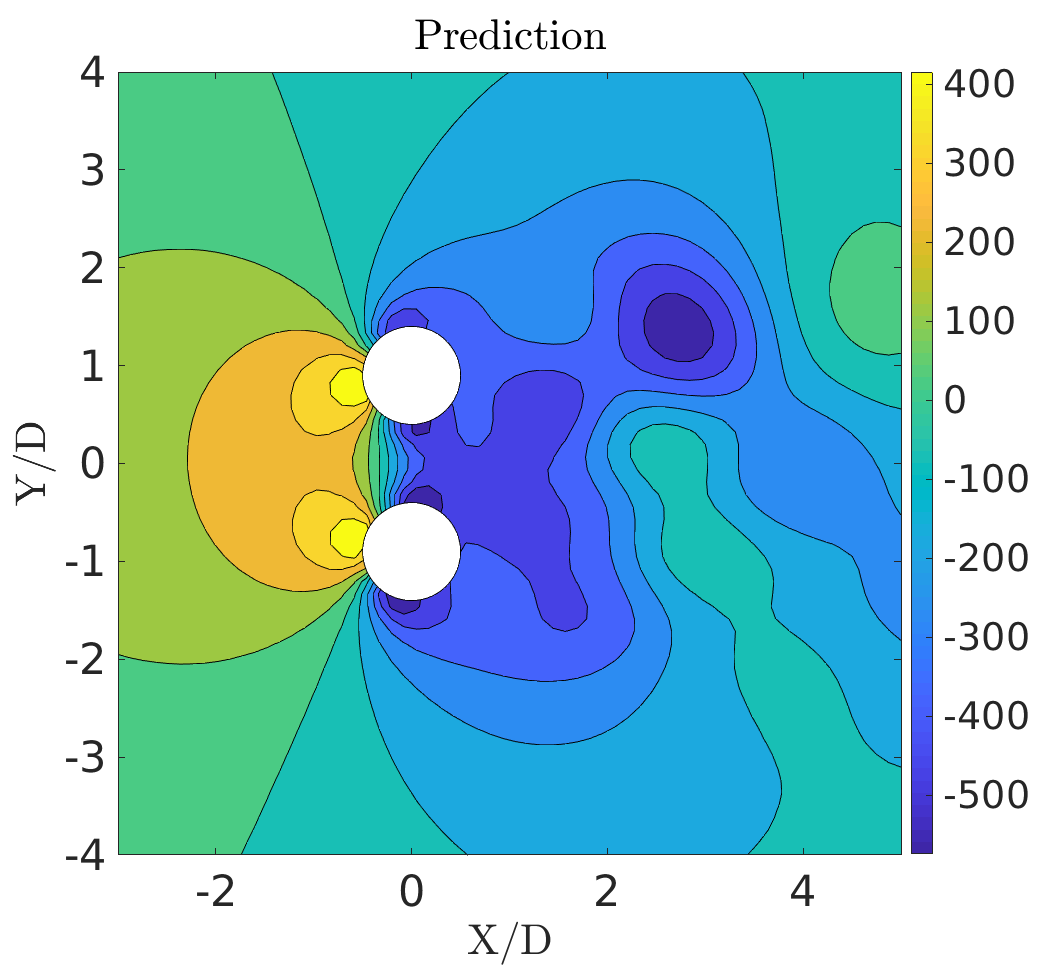}
\hspace{0.02\textwidth}
\includegraphics[width = 0.32\textwidth]{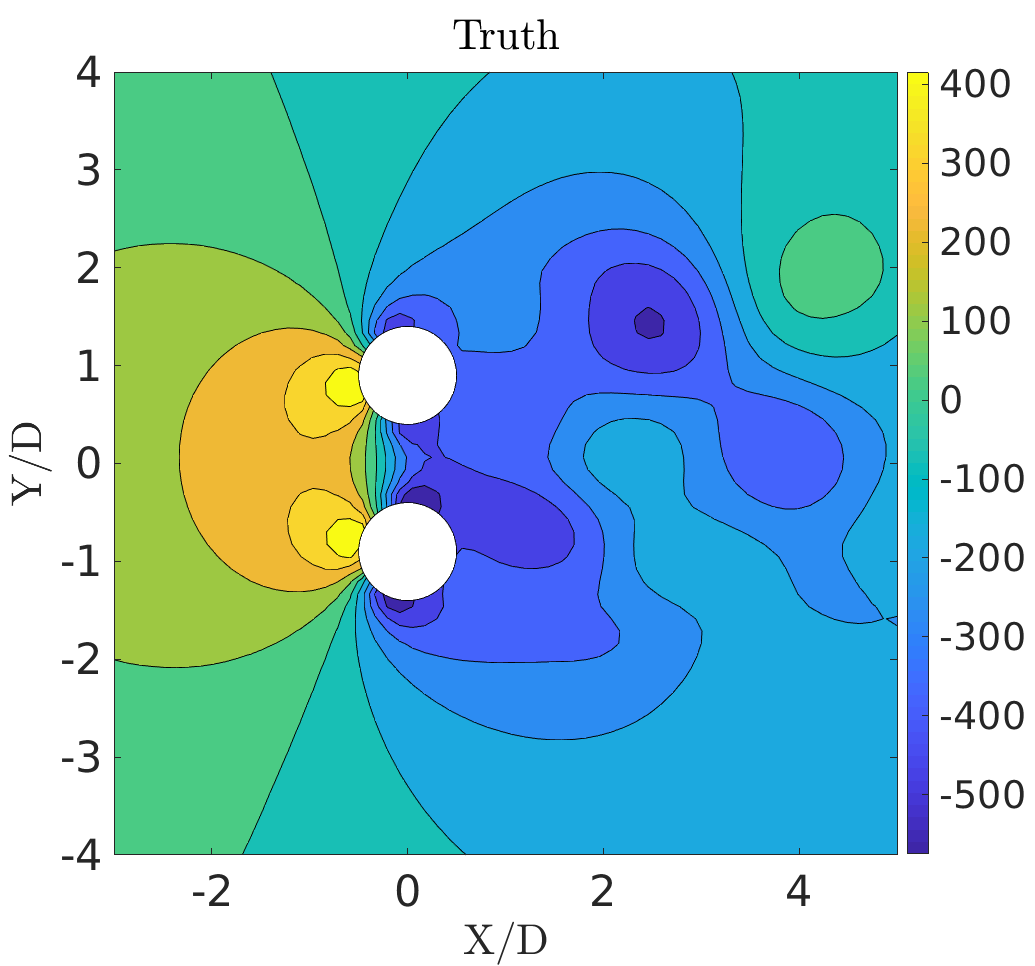}
\hspace{0.02\textwidth}
\includegraphics[width = 0.32\textwidth]{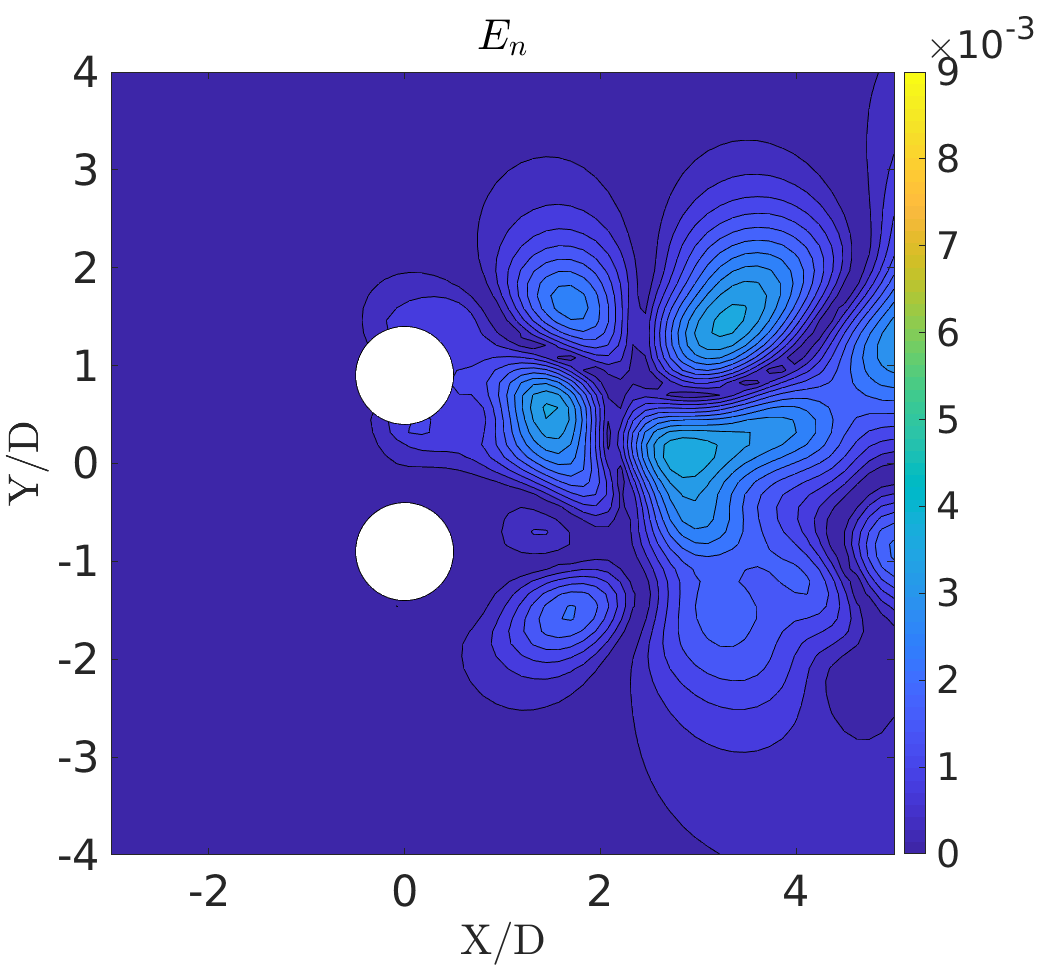}}
\caption{The flow past side-by-side cylinders: Comparison of predicted and true fields (POD-RNN model) along with normalized reconstruction error $E_{n}$ at (a) $tU_{\infty}/D = 898$, (b) $tU_{\infty}/D = 923$, (c) $tU_{\infty}/D = 948$ for pressure field ($P$)}
\label{podsbs_flow_comp_p}
\end{figure*}

\begin{figure*}
\centering
\subfloat[]
{\includegraphics[width = 0.32\textwidth]{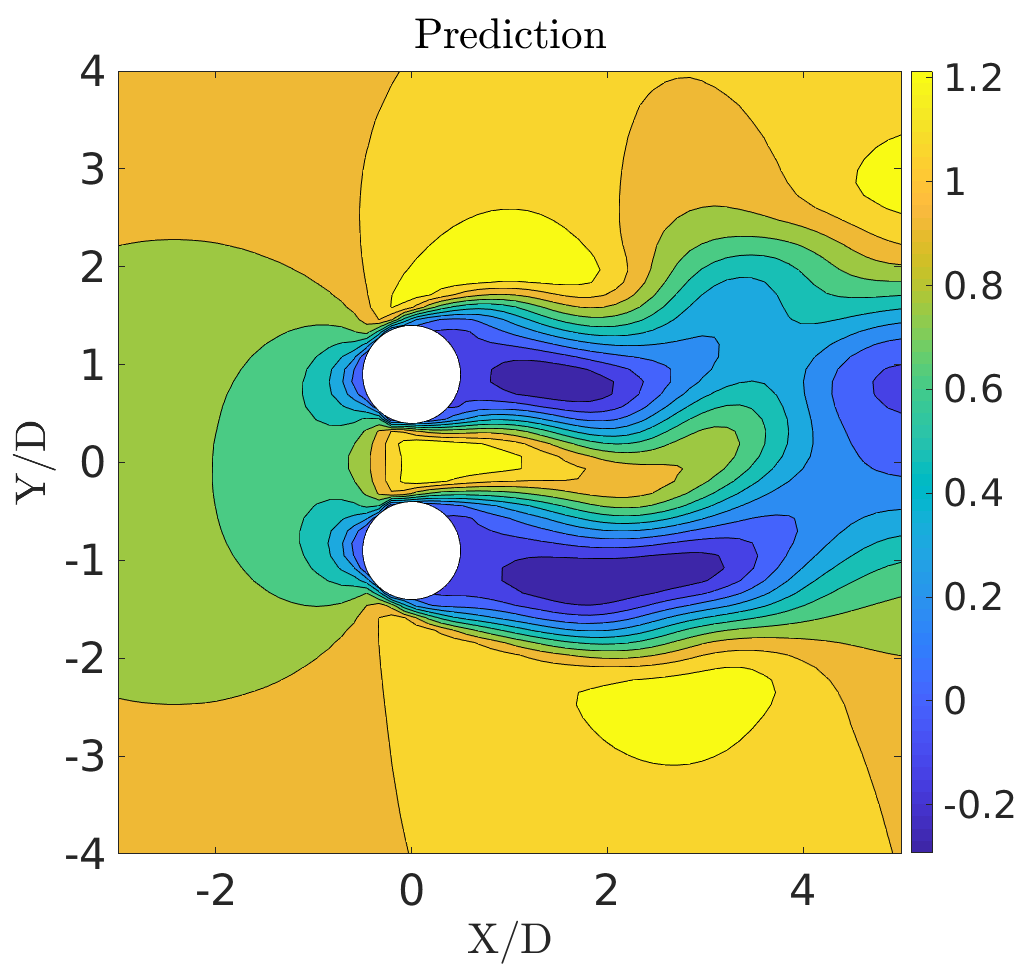}
\hspace{0.02\textwidth}
\includegraphics[width = 0.32\textwidth]{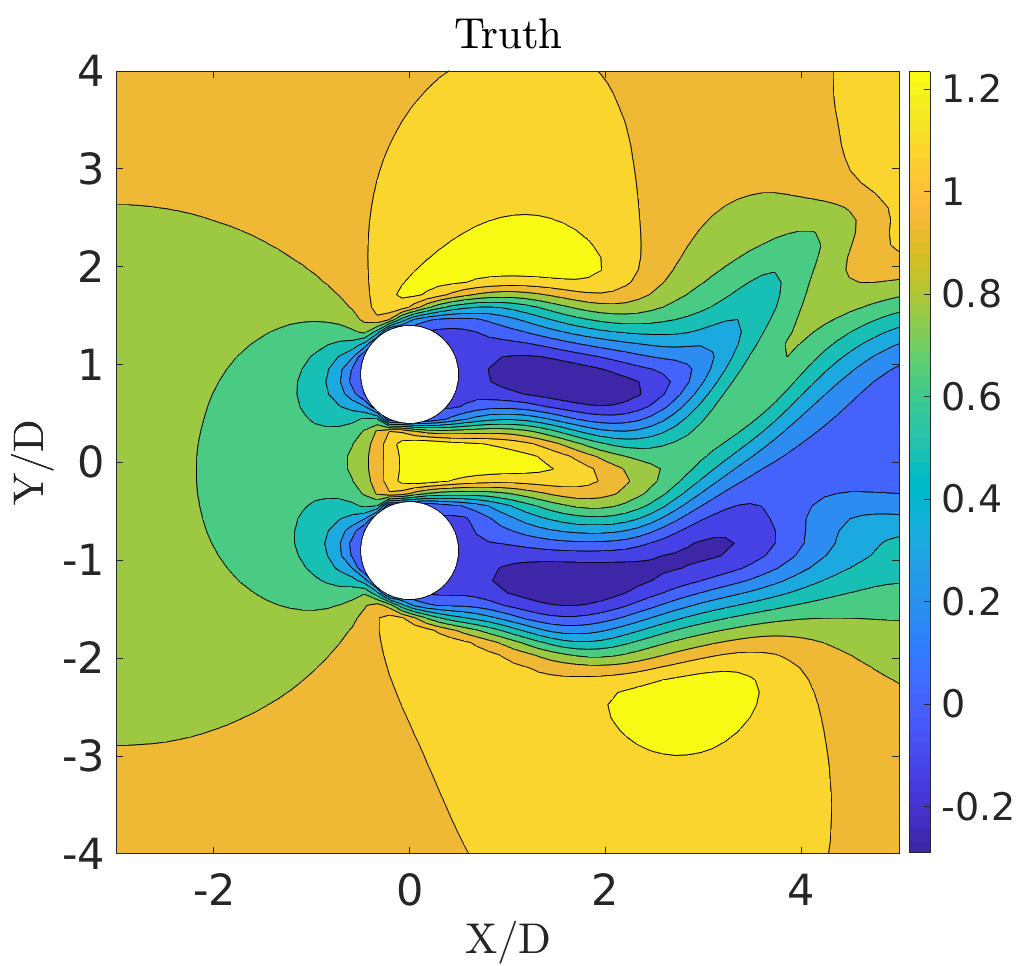}
\hspace{0.02\textwidth}
\includegraphics[width = 0.32\textwidth]{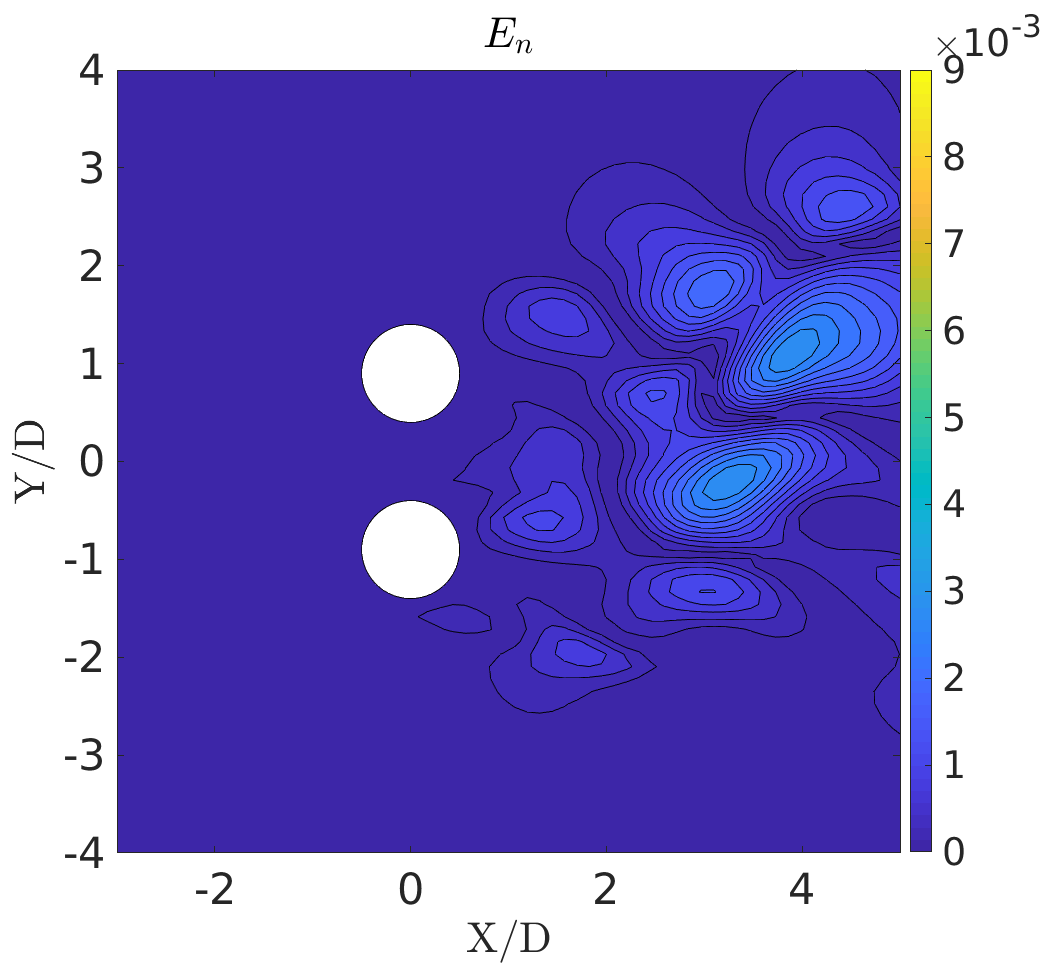}}
\\
\vspace{0.02\textwidth}
\subfloat[]
{\includegraphics[width = 0.32\textwidth]{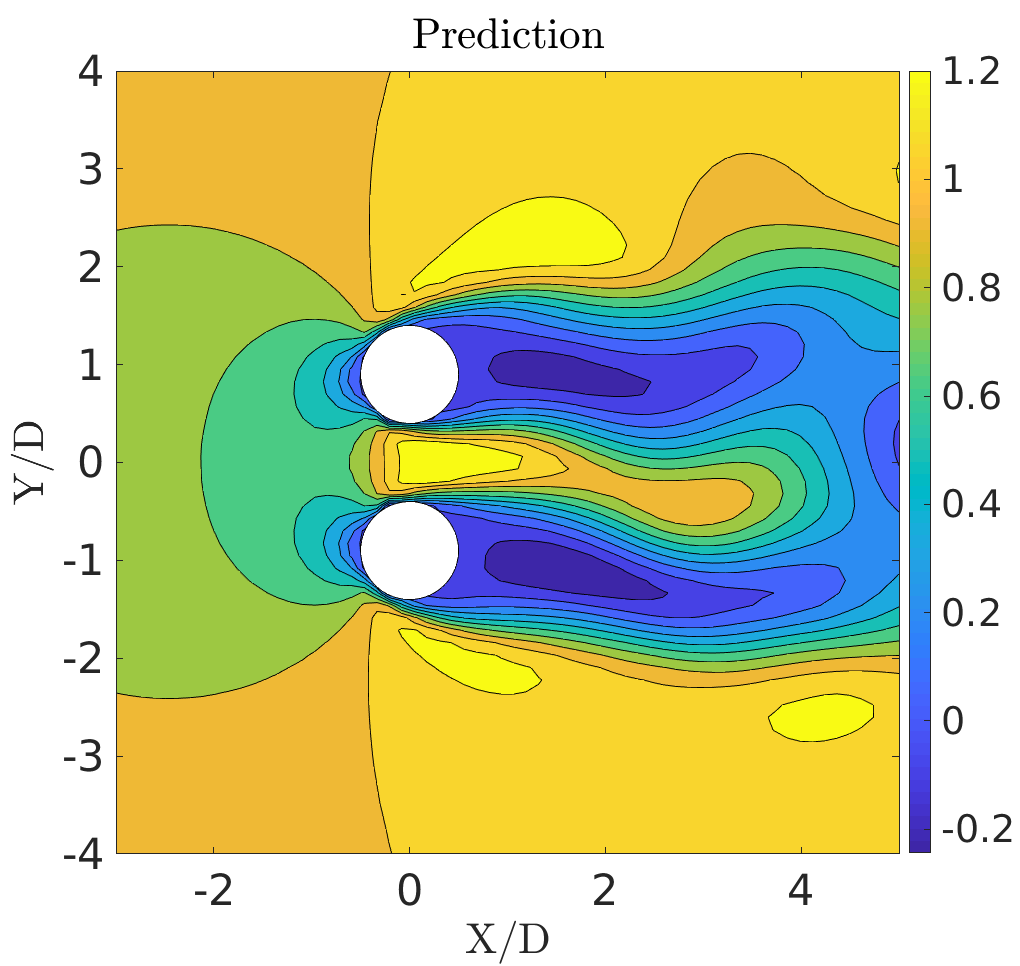}
\hspace{0.02\textwidth}
\includegraphics[width = 0.32\textwidth]{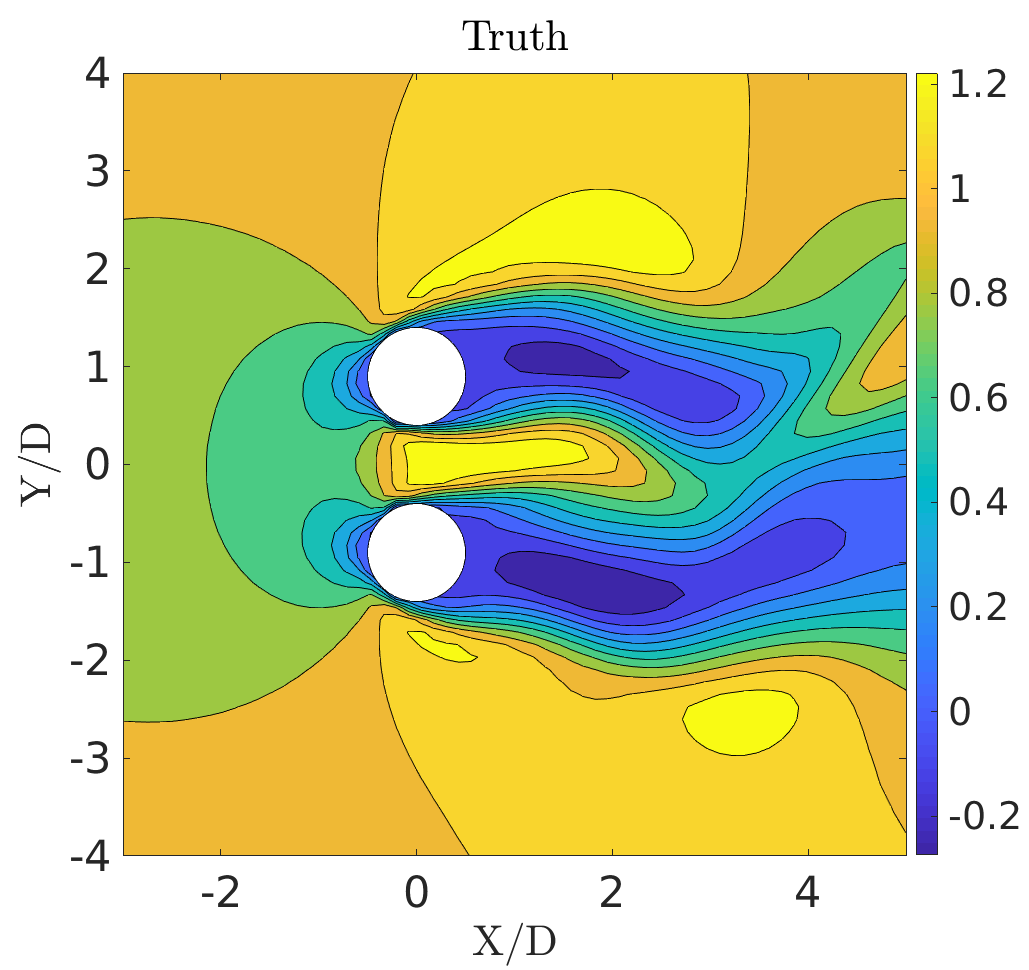}
\hspace{0.02\textwidth}
\includegraphics[width = 0.32\textwidth]{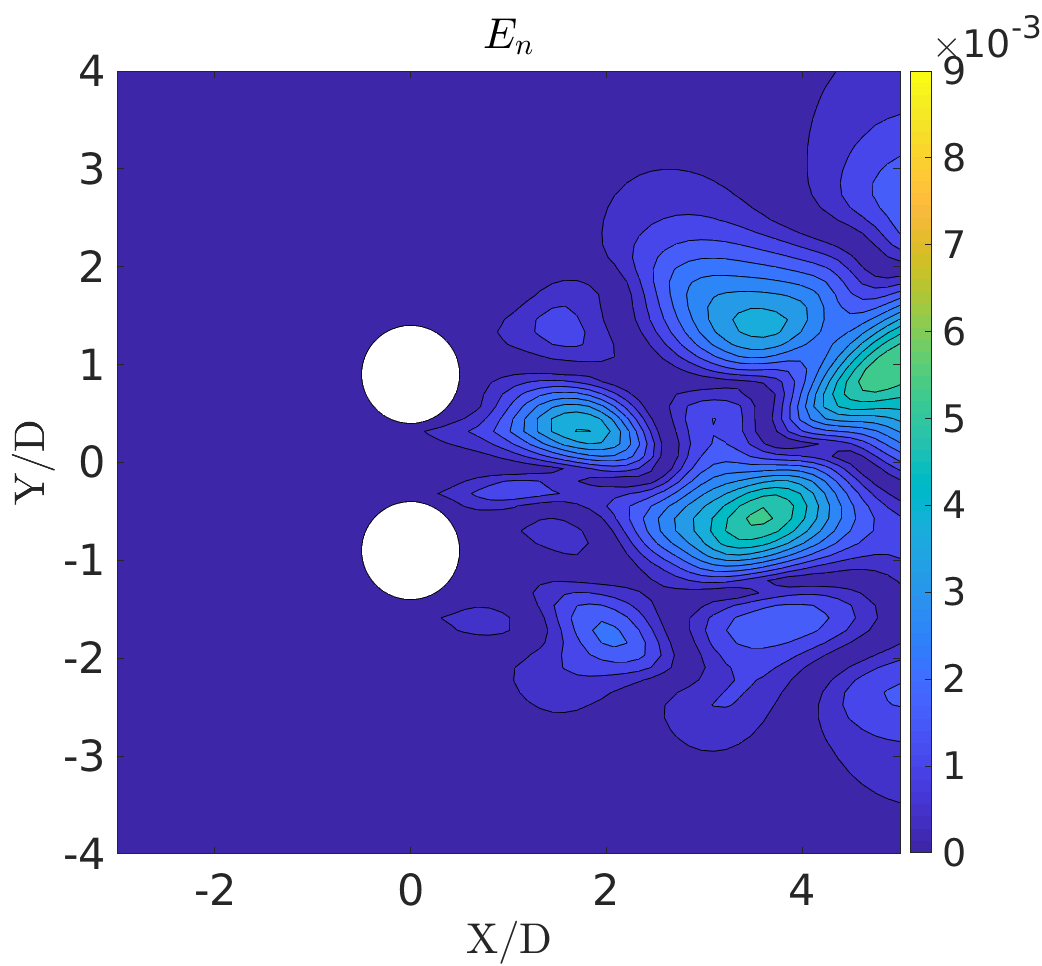}}
\\
\vspace{0.025\textwidth}
\subfloat[]
{\includegraphics[width = 0.32\textwidth]{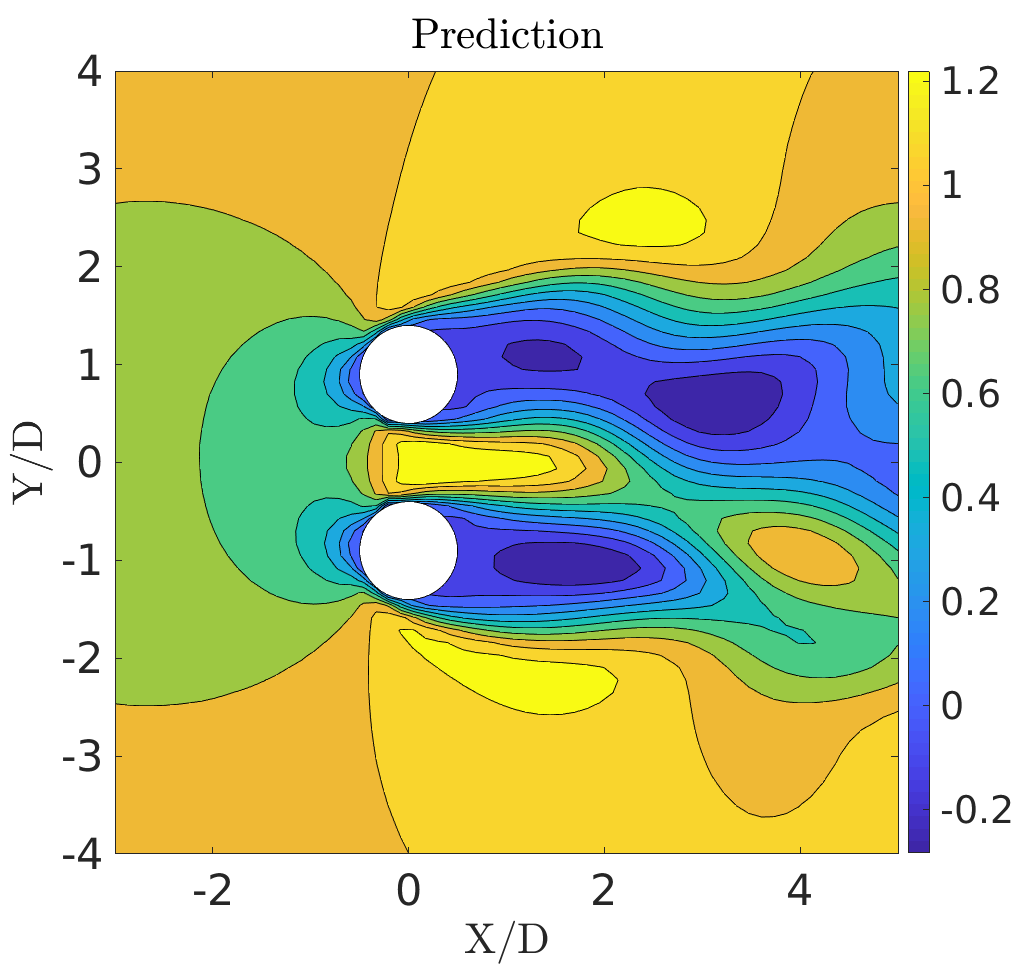}
\hspace{0.02\textwidth}
\includegraphics[width = 0.32\textwidth]{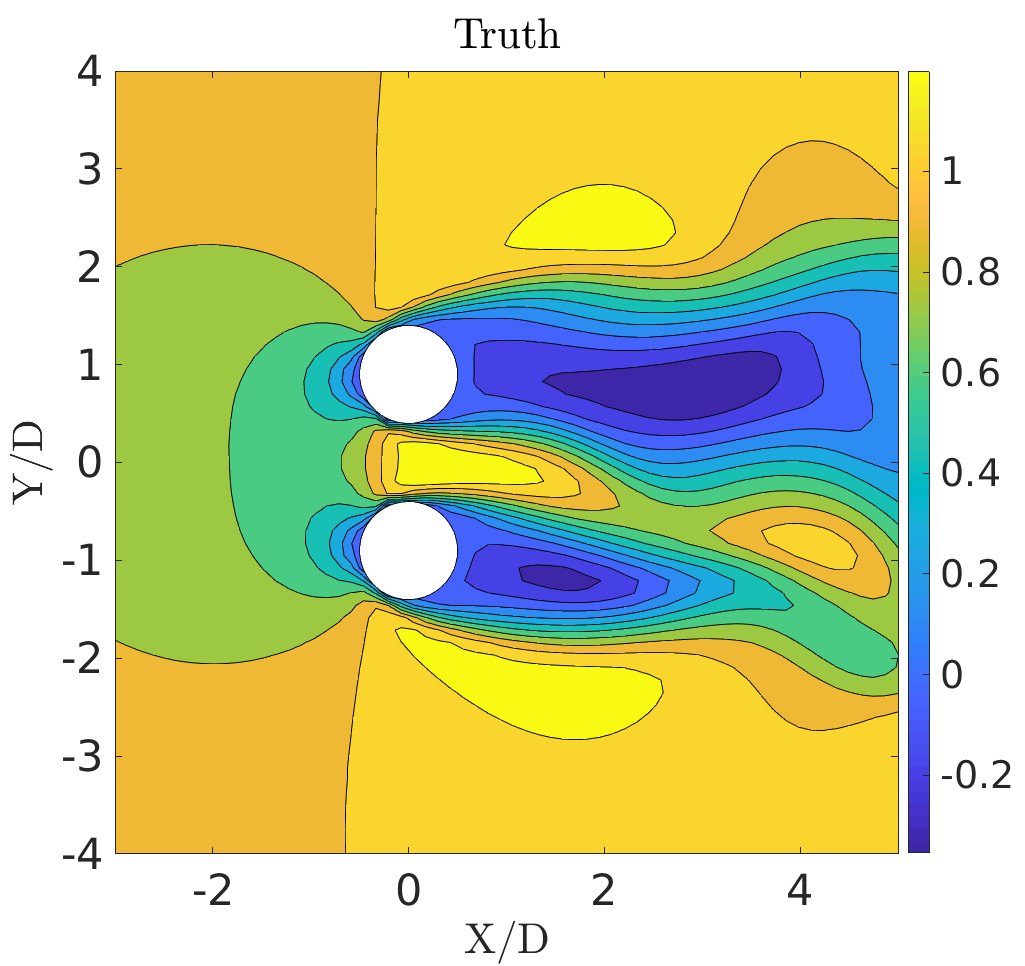}
\hspace{0.02\textwidth}
\includegraphics[width = 0.32\textwidth]{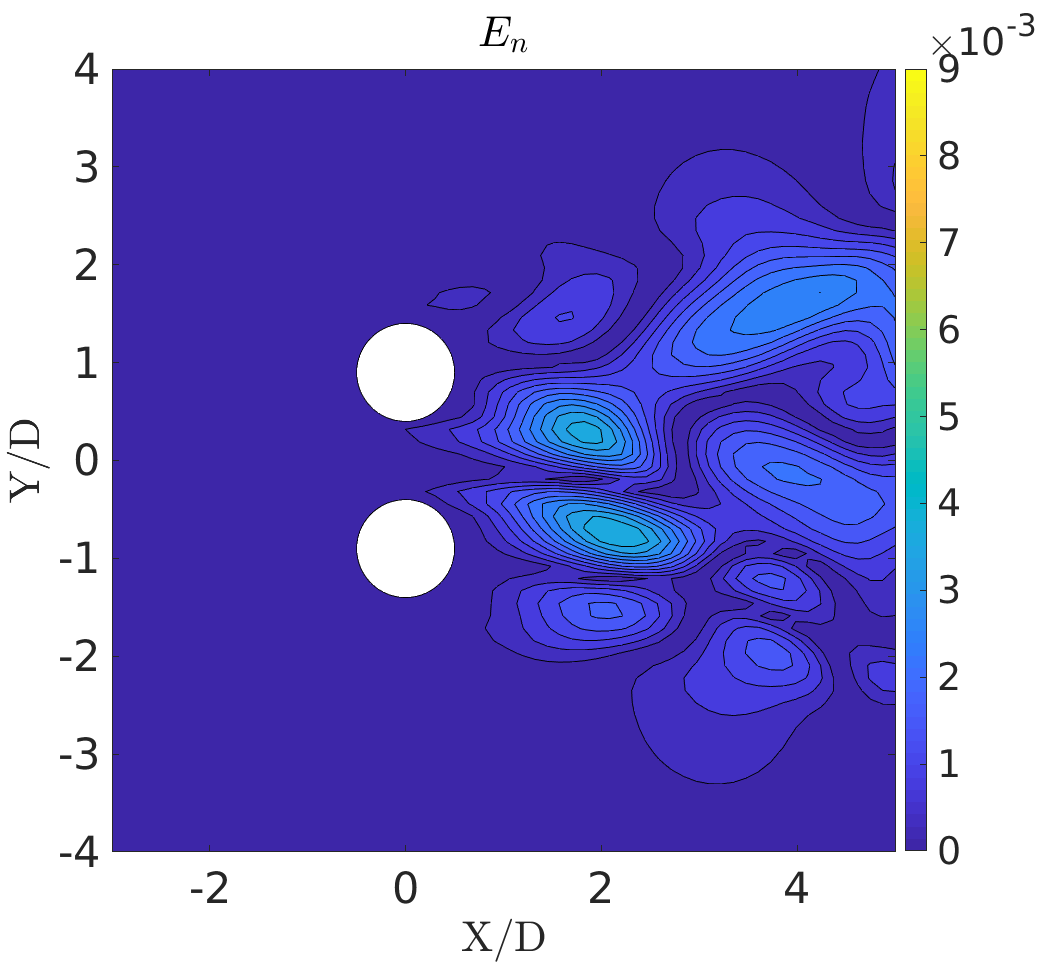}}
\caption{The flow past side-by-side cylinders: Comparison of predicted and true fields (POD-RNN model) along with normalized reconstruction error $E_{n}$ at (a) $tU_{\infty}/D = 898$, (b) $tU_{\infty}/D = 923$, (c) $tU_{\infty}/D = 948$ for x-velocity field ($U$)}
\label{podsbs_flow_comp_u}
\end{figure*}

\begin{figure*}
\centering
\subfloat[]
{\includegraphics[width = 0.32\textwidth]{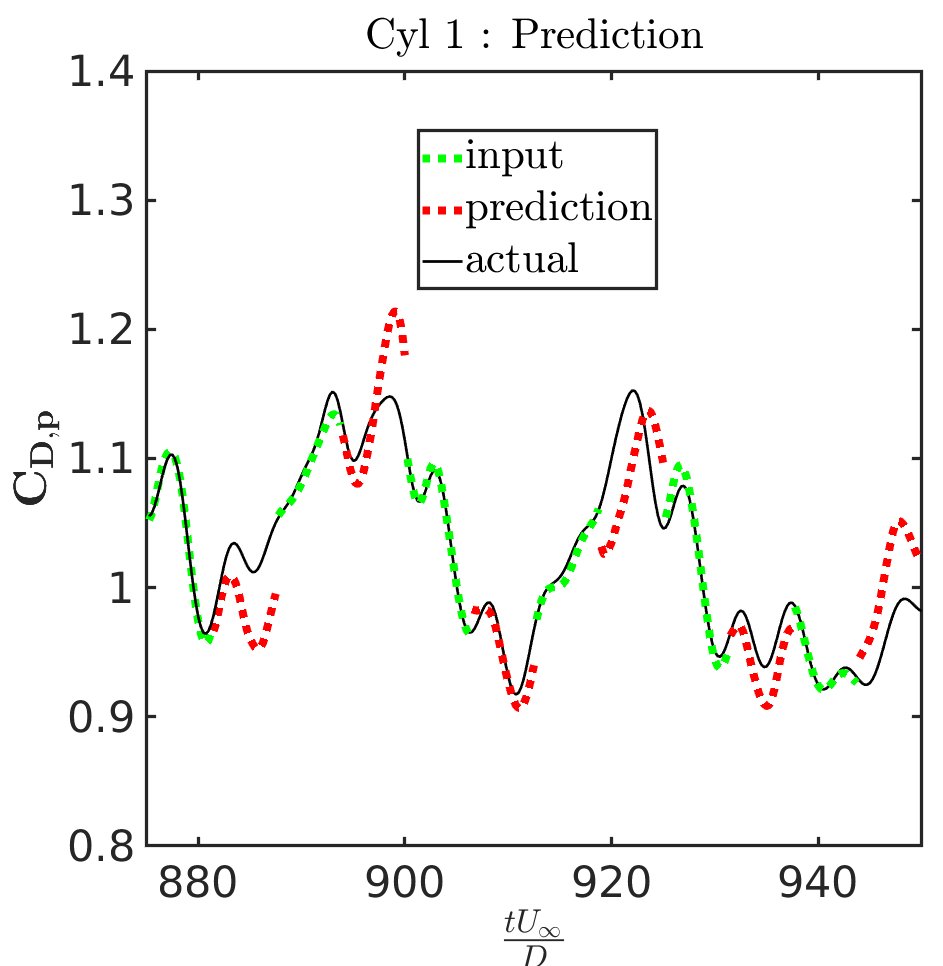}}
\subfloat[]
{\includegraphics[width = 0.32\textwidth]{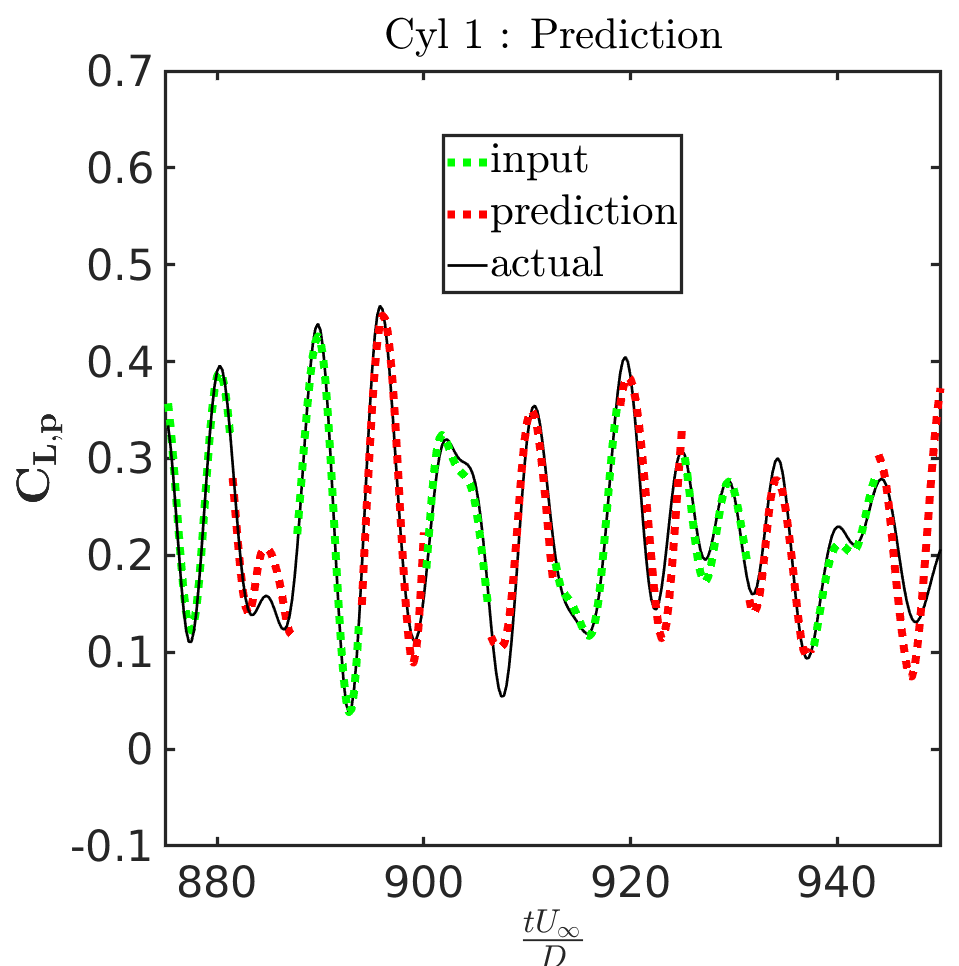}} \\
\subfloat[]
{\includegraphics[width = 0.32\textwidth]{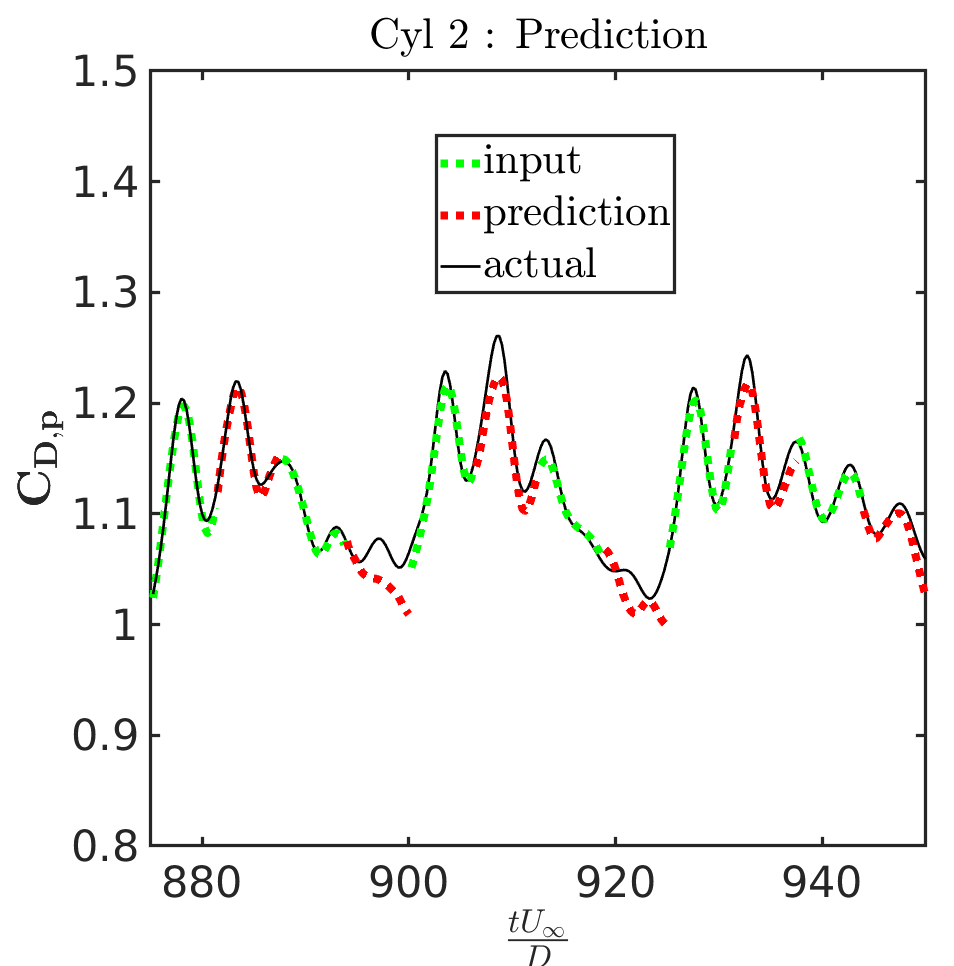}}
\subfloat[]
{\includegraphics[width = 0.32\textwidth]{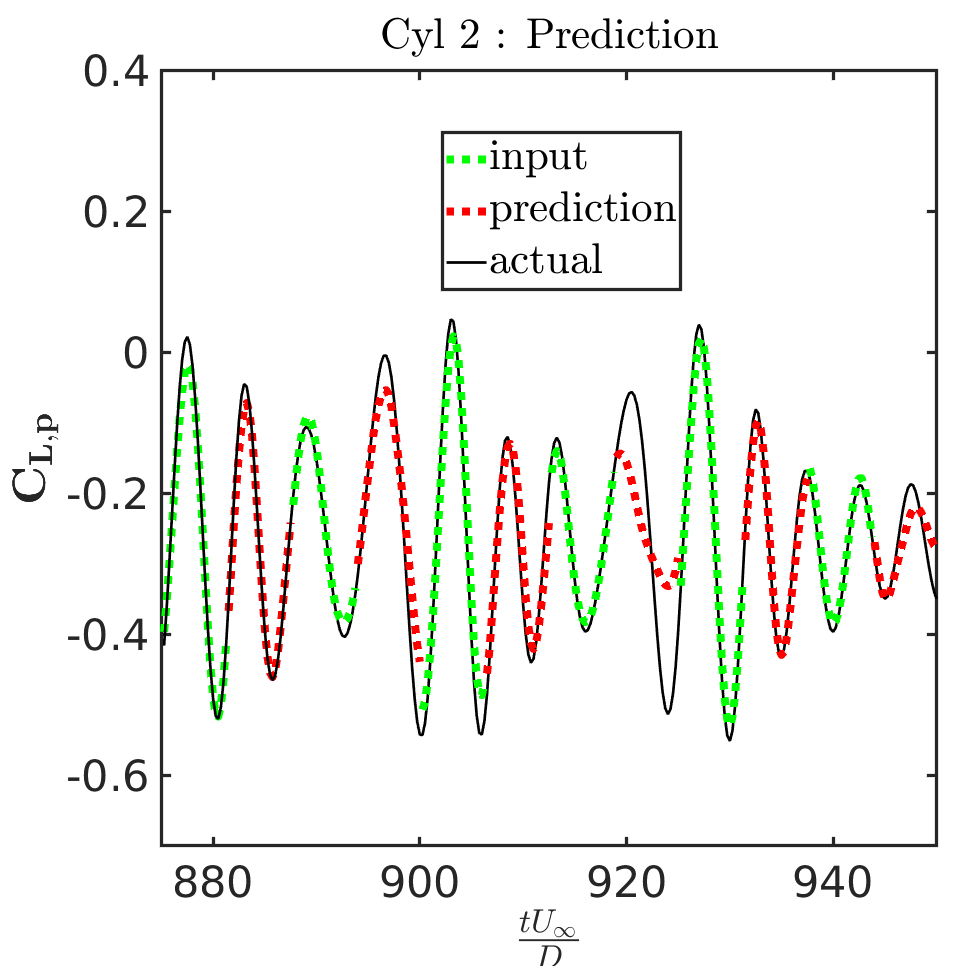}}
\caption{The flow past side-by-side cylinders: Predicted and actual (POD-RNN model) pressure force coefficients. (a) Drag (Cylinder 1), (b) lift (Cylinder 1), (c) drag (Cylinder 2), (d) lift (Cylinder 2)}
\label{sbspodforces}
\end{figure*}

% Field and force prediction
\textit{Field prediction}: The predicted modal coefficient $\hat{\textbf{A}}_{n} \in \mathbb{R}^{k}$ can simply be reconstructed back to the high dimensional state $\hat{\textbf{S}}_{n}  \in \mathbb{R}^{m}$ using the mean field $\bar{\textbf{S}} \in \mathbb{R}^{m}$ and $k$ spatial POD modes $\boldsymbol{\Phi} \in \mathbb{R}^{m\times k}$ as $\hat{\textbf{S}}_{n} \approx \bar{\textbf{S}} + \boldsymbol{\Phi} \hat{\textbf{A}}_{n}$. Figs. \ref{podsbs_flow_comp_p} and \ref{podsbs_flow_comp_u} depict the comparison of predicted and true values for $P$ and $U$, respectively at time-steps $3592$ ($898\;tU_{\infty}/D$), $3692$ ($923\;tU_{\infty}/D$) and $3792$ ($948\;tU_{\infty}/D$). The normalized reconstruction error $E_{n}$ is constructed by taking the absolute value of differences between the true $\textbf{S}_{n}$ and predicted $\hat{\textbf{S}}_{n}$ field at any time-step $n$ and normalizing it with $L_{2}$ norm of the true data and is given by
\begin{equation}
    E_{n} = \frac{|\textbf{S}_{n}-\hat{\textbf{S}}_{n}|}{\|\textbf{S}_{n}\|_{2,k}}.
\end{equation}
\\
\textit{Force coefficient prediction}:
The high-dimensional field prediction on all the nodal points allows to carry the reduced numerical integration directly over the fluid-solid boundary on the cylinders. The Cauchy Stress tensor $\bm{\sigma^{f}}$ is constructed with pressure field data and Eq. (\ref{force_eqns}) in section \ref{HDM} is used to integrate it over the fluid-solid boundary to get $\textbf{C}_{\mathrm{D},\text{p}}$ and $\textbf{C}_{\mathrm{L},\text{p}}$. Fig.~\ref{sbspodforces} depicts the predicted and actual pressure coefficient from time-steps $3501-3800$ for the upper (Cylinder 1) and lower (Cylinder 2) cylinders. Note that the reference length scale is the diameter of cylinder ($D=1$) and the reference velocity is the inlet velocity ($U_{\infty}$=1). $t$ denotes the actual simulation time carried at a time-step $0.25s$.     

%\newpage
%\onecolumn
% pressure

% velo 

%\newpage

% force 

%\clearpage
%\newpage
%\twocolumn
%%%%%%%%%%%%%%%%%%%%%%%%%%%%
\subsection{CRAN model}\label{cransbs}
%%%%%%%%%%%%%%%%%%%%%%%%%%%%

The end-to-end nonlinear model order reduction tool based on CRAN is now applied on the same problem of flow past side-by-side cylinders. Same set of training (time-steps $301-3500$) and testing (time-steps $3501-3800$) data is selected for this application. For the sake of completeness the overall process is outlined as follows:      

\begin{enumerate}
% Step 1 : reference grid 
\item Consider high-dimensional unstructured data from a flow solver $\boldsymbol{\mathcal{S}} = \left\lbrace\textbf{S}_{1}\;\textbf{S}_{2}\dots\;\textbf{S}_{N}\;\right\rbrace \in \mathbb{R}^{m\times N}$  (here, $m=25034$ and $N=3500$). To achieve spatial uniformity in the unstructured data, SciPy's $"griddata"$ function \cite{SciPy} is used to map the $m$ dimensional unstructured data on a 2-d reference grid of size $N_{x} \times N_{y}$ (here, $N_{x}=N_{y}=64$). The details of the reference grid and the fluid-solid boundary is represented in Fig.~\ref{figunstrct-strct}. The snapshot data-set hence obtained is $\boldsymbol{\mathcal{S}} = \left\lbrace\textbf{s}_{1}\;\textbf{s}_{2}\dots\;\textbf{s}_{N}\;\right\rbrace \in \mathbb{R}^{N_{x}\times N_{y}\times N}$. These $N$, 2-d snapshots are divided into training $n_{tr}=3200$ and testing $n_{ts}=N-n_{tr}=300$. 

% Step 2 : data set
\item The $n_{tr}$ training data is broken into $N_{s}$ batches, each of finite time-step size $N_{t}$. Note that $n_{tr} = N_{s} N_{t}$. The procedure described in section \ref{unsupervised-t-s} is followed to generate the feature scaled data for training, 
% Data Set 
\begin{equation}
    \mathcal{S} = \{\textbf{S}^{'1}_{s},\dots,\textbf{S}^{'N_{s}}_{s\\}\}\in [0,1]^{N_{x}\times N_{y}\times N_{t}\times N_{s}},
\end{equation}

where each training sample $\textbf{S}^{'i}_{s} = [\textbf{s}^{'1}_{s;i},\dots,\textbf{s}^{'N_{t}}_{s;i}]$ with $N_{x},\;N_{y} = 64$ and $N_{t} = 25$, $N_{s} = 128$. $\textbf{S}^{'i}_{s}$ can be pressure or velocity data.

% Step 3 : training 
\item Convolutional recurrent autoencoder networks (CRANs) are trainied using the data set described above and the low-dimensional evolver LSTM state $\textbf{A}$ of different sizes. Predictions are experimented with different sizes $N_{A} = 2,\;4,\;8,\;16,\;32,\;64,\;128,\;\\
256$ for both pressure $P$ and x-velocity $U$. All the models are trained on a single \\
Intel E5-2690v3 (2.60GHz, 12 cores) CPU node for $N_{train} = 500,000$ epochs. The objective function (Eq. (\ref{loss})) was observed to reach a nearly steady value of approximately $10^{-5}$ at the end of iterations. It should be noted that we are employing a closed-loop recurrent neural network in the CRAN model for this problem (see Fig.~\ref{closedloop}) i.e., the prediction from previous step is used as an input for the next prediction. 

% Step 4 : prediction and E_f...
\item The closed-loop recurrent net uses prediction from previous time-step to generate a new prediction. Due to the bistable nature of the problem, the errors are compounded at every time-step, thus, causing prediction trajectory to deviate after a few finite time-steps ($\approx N_{t}=25$ time-steps). In order to address this issue, ground truth data is fed after every $N_{t}$ predicted time-steps as a demonstrator for the net to follow the actual trajectory and avoid divergence. We follow the $1-to-N_{t}$ rule for prediction. This means that any one ground truth snapshot predicts $N_{t}$ time-steps from a trained model. Thus, in testing, we input $\approx n_{ts} / N_{t}$ ground truth data to predict $n_{ts}$ time-steps. For instance, time-step $3500$ predicts sequence $3501-3525$ steps, time-step $3525$ predicts $3526-3550$ and so on.  

Fig. \ref{sbs-cran-ef} (a) and (b) depict the normalised mean squared error ($\overline{\mbox{E}}_{\mbox{f}}$) for the first set of $N_{t}$ predicted time-steps for $P$ and $U$ respectively with respect to the size of low-dimensional feature $N_{A}$.
\begin{equation}
   \overline{\mbox{E}}_{\mbox{f}} = \frac{  \sum_{n}\frac{\|\textbf{s}_{n}-\hat{\textbf{s}}_{n}\|^{2}_{2,k}}{\|\textbf{s}_{n}\|^{2}_{2,k}+\epsilon}}{N_{t}},
   \label{mnse}
\end{equation}
where $n$ denotes a time instance from $3501$ to $3525$ time-steps and $k$ denotes a point $(x,y)$ in space. $\|\dots\|_{2,k}$ denotes the $L_2$ norm over spatial dimension.
$\textbf{s}_{n}$ and $\hat{\textbf{s}}_{n}$ are respectively the true and predicted fields. The tuning of the hyper-parameter $N_{A}$ reveals the existence of an optimal value of $N_{A}$ for the field data. Referring to Fig.~\ref{sbs-cran-ef}, the variation of $\overline{\mbox{E}}_{\mbox{f}}$ vs $N_{A}$ follows a nearly convex shaped optimisation curve with the minimum at $N_{A} = 32$ for pressure and $N_{A}=16$ for x-velocity. Based on this discovery, we utilise $N_{A}=32$ and $N_{A}=16$ trained CRAN models for pressure and x-velocity respectively to predict all the $n_{ts}=300$ test time-steps. 
\begin{figure}
\centering
\subfloat[]
{\includegraphics[width = 0.238\textwidth]{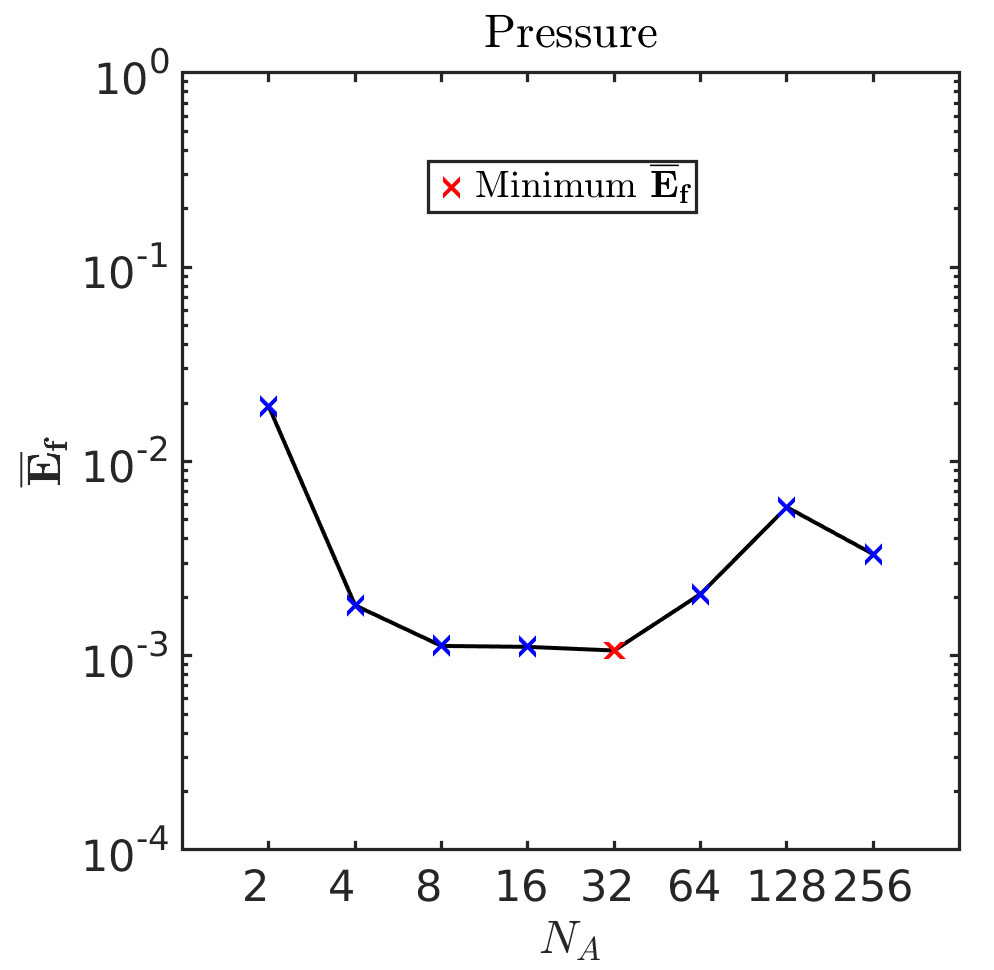}}
\subfloat[]
{\includegraphics[width = 0.238\textwidth]{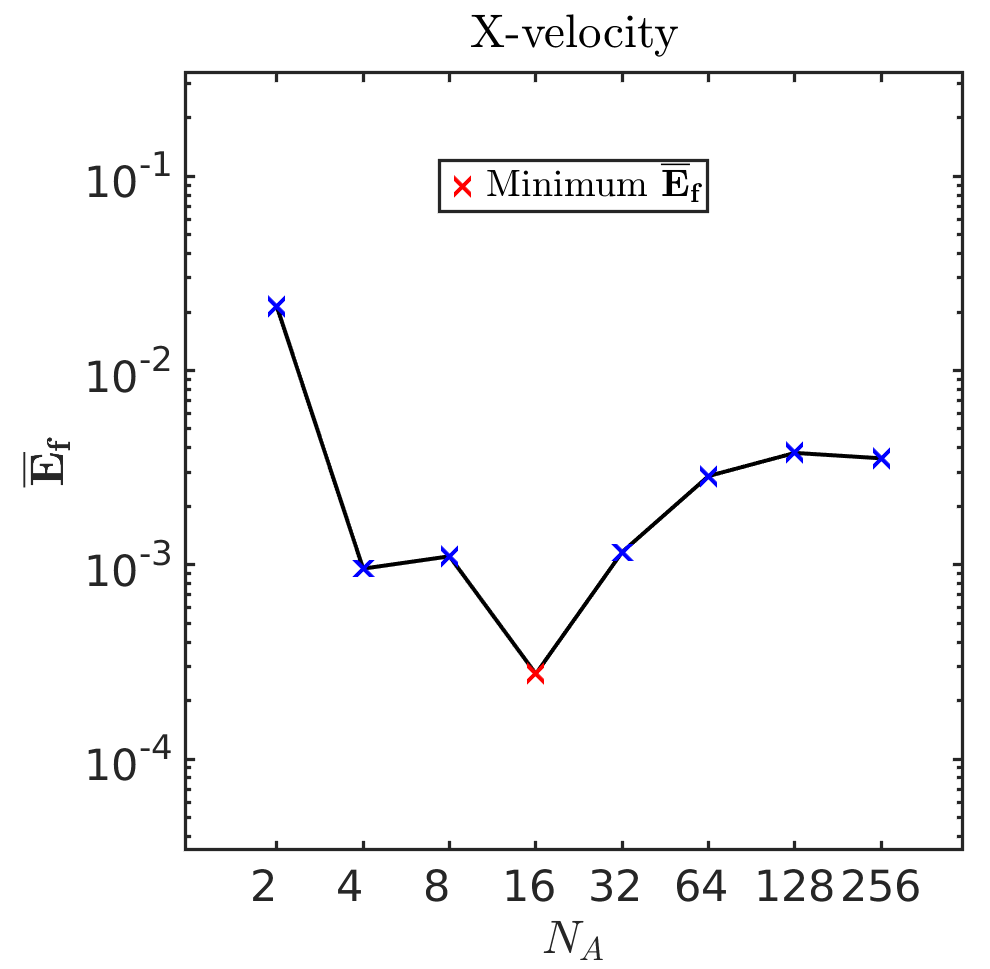}}
\caption{The flow past side-by-side cylinders: Mean normalized squared error ($\overline{\mbox{E}}_{\mbox{f}}$) ) for first
$N_{t}$ predicted time-steps with respect to size of low-dimensional space $N_{A}$ for (a) pressure field P and (b) x-velocity field}
\label{sbs-cran-ef}
\end{figure}
\end{enumerate}

\begin{figure*}
\centering
\subfloat[]{
\includegraphics[width = 0.32\textwidth]{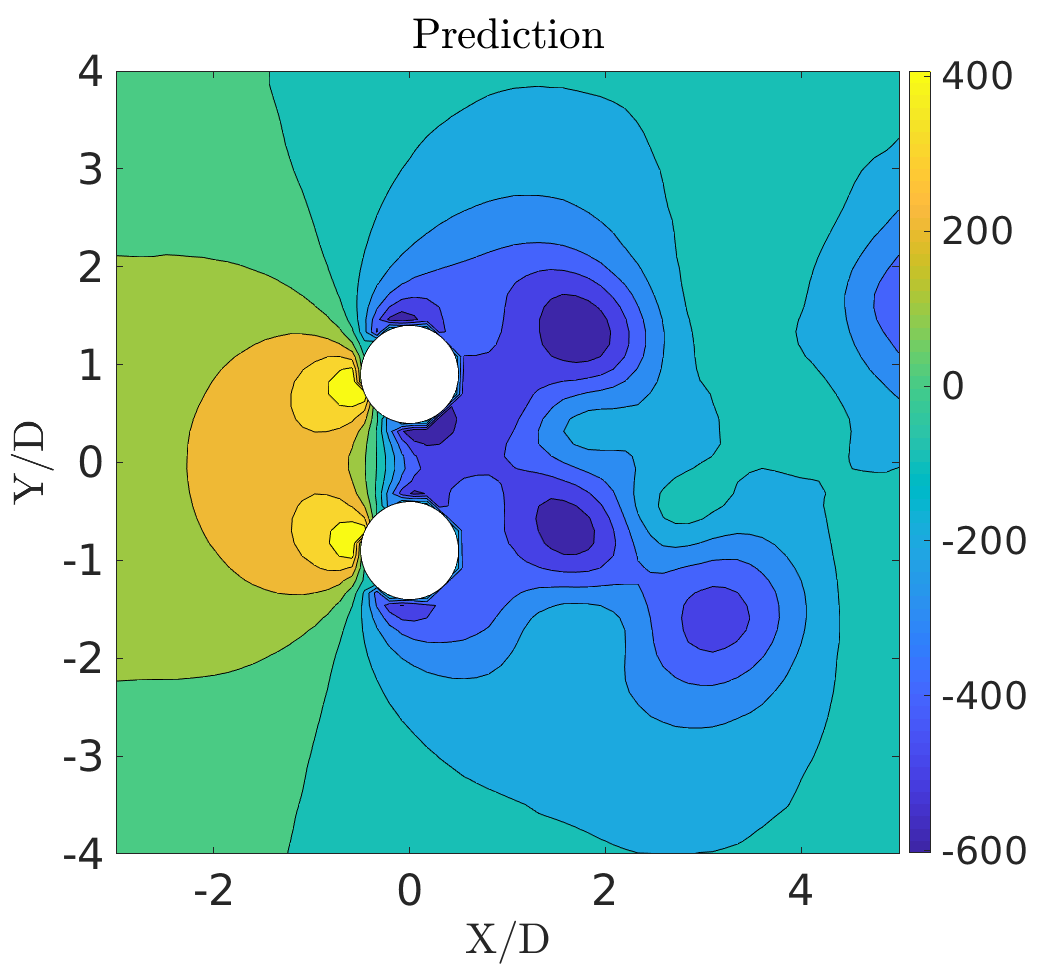}
\hspace{0.02\textwidth}
\includegraphics[width = 0.32\textwidth]{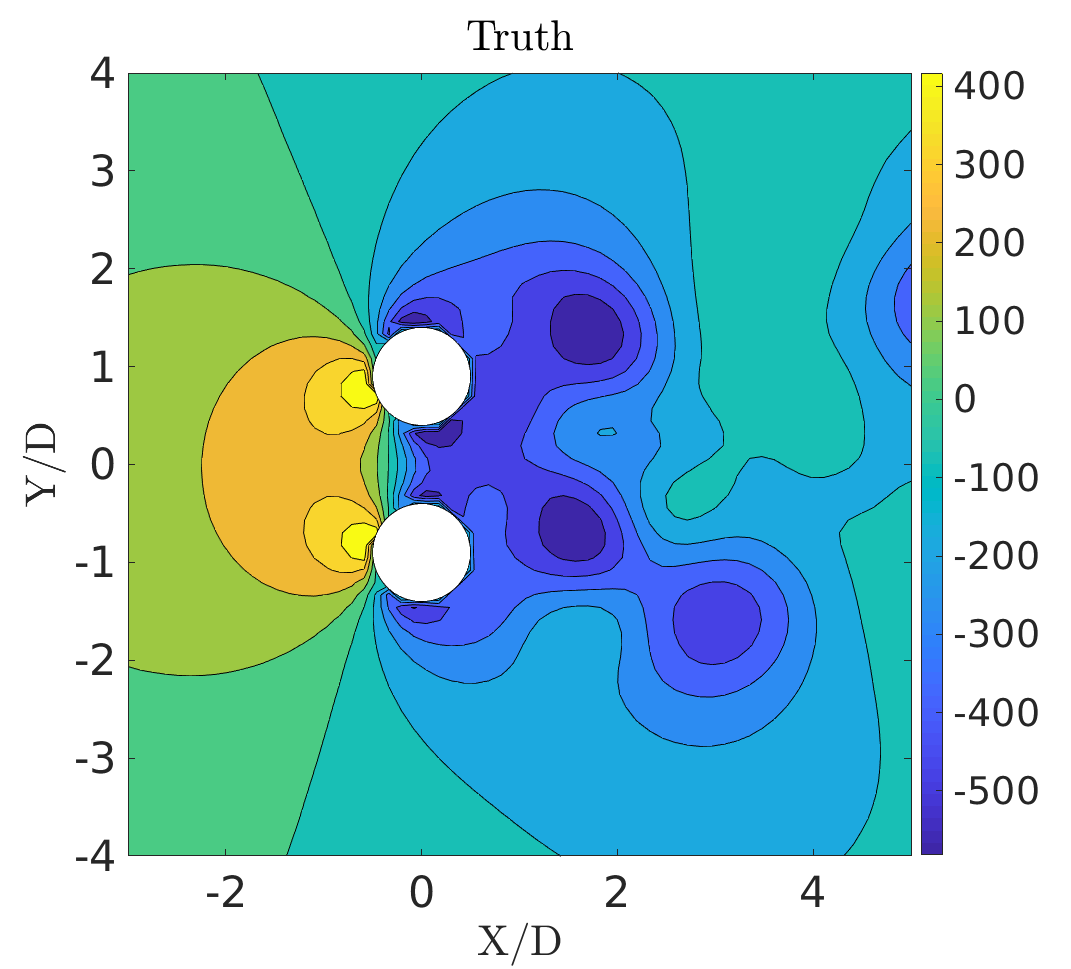}
\hspace{0.02\textwidth}
\includegraphics[width = 0.32\textwidth]{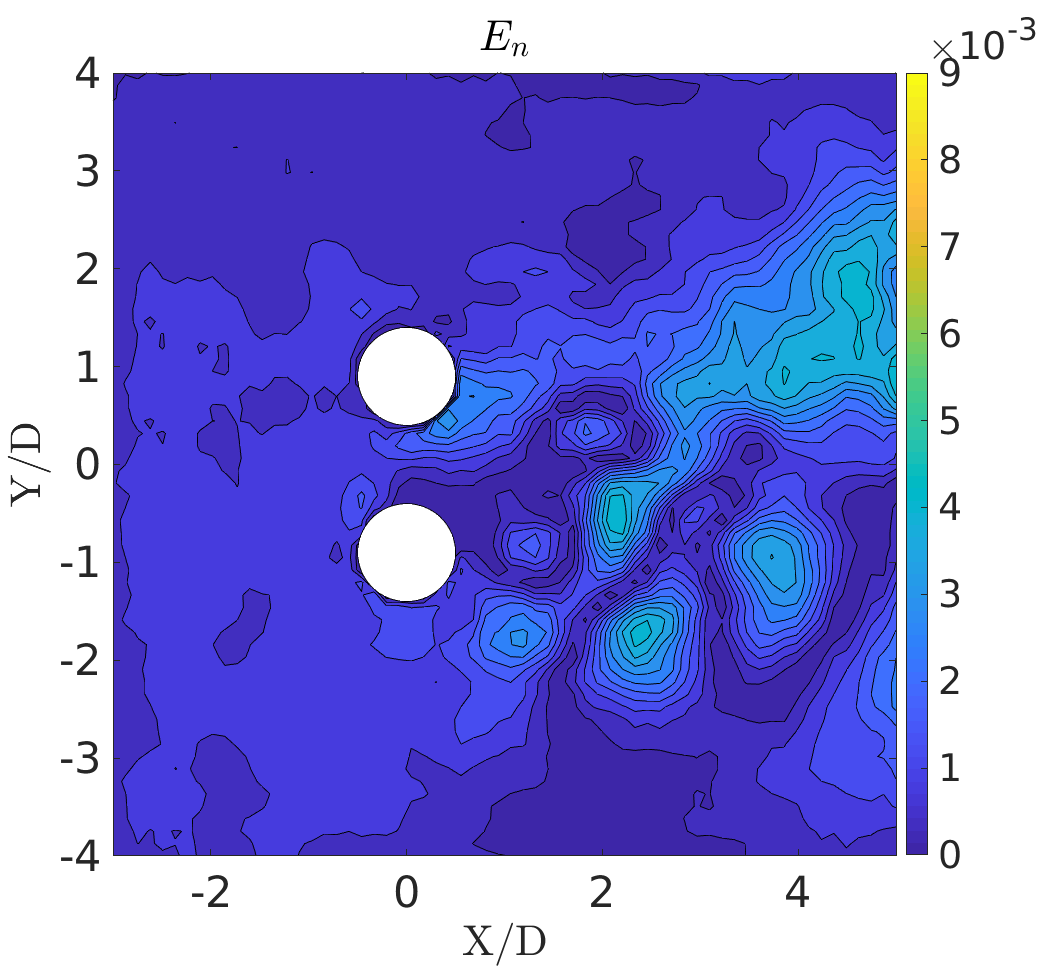}}
\\
\vspace{0.025\textwidth}
\subfloat[]{
\includegraphics[width = 0.32\textwidth]{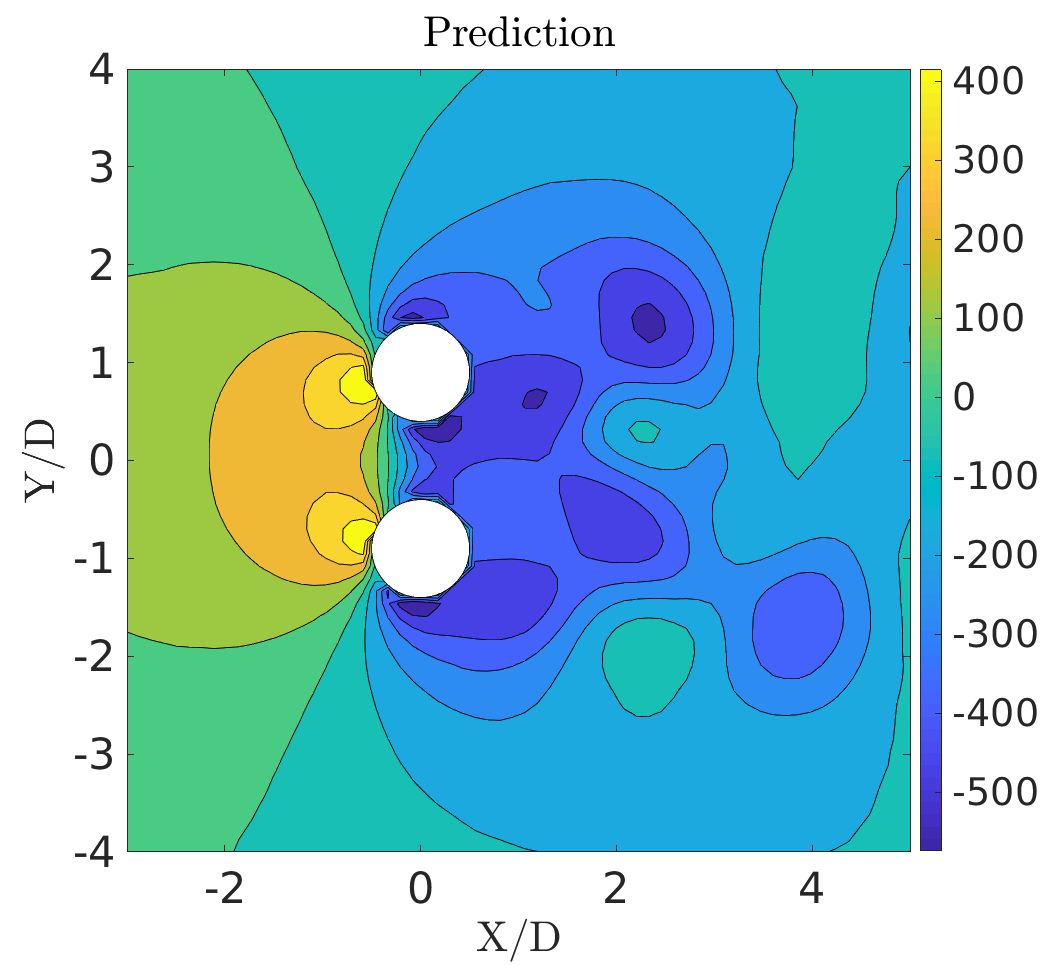}
\hspace{0.02\textwidth}
\includegraphics[width = 0.32\textwidth]{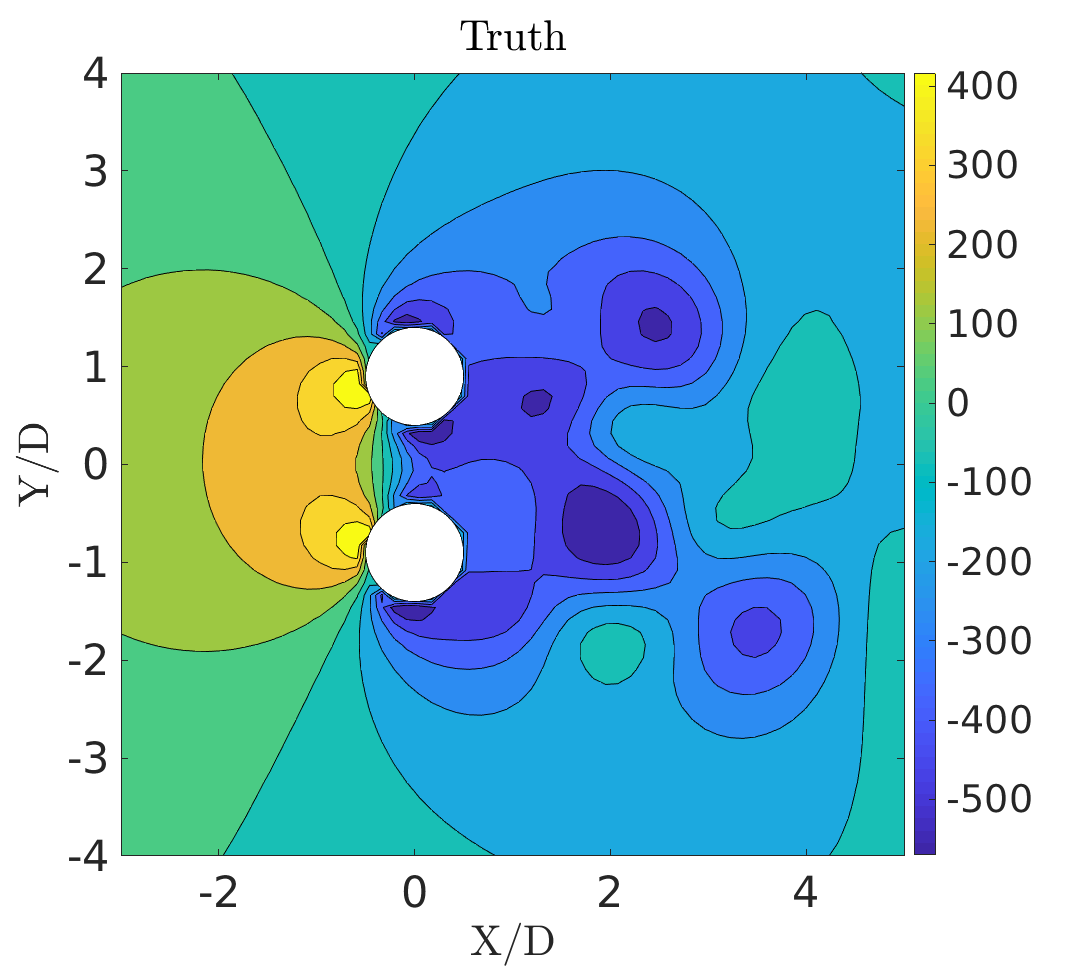}
\hspace{0.02\textwidth}
\includegraphics[width = 0.32\textwidth]{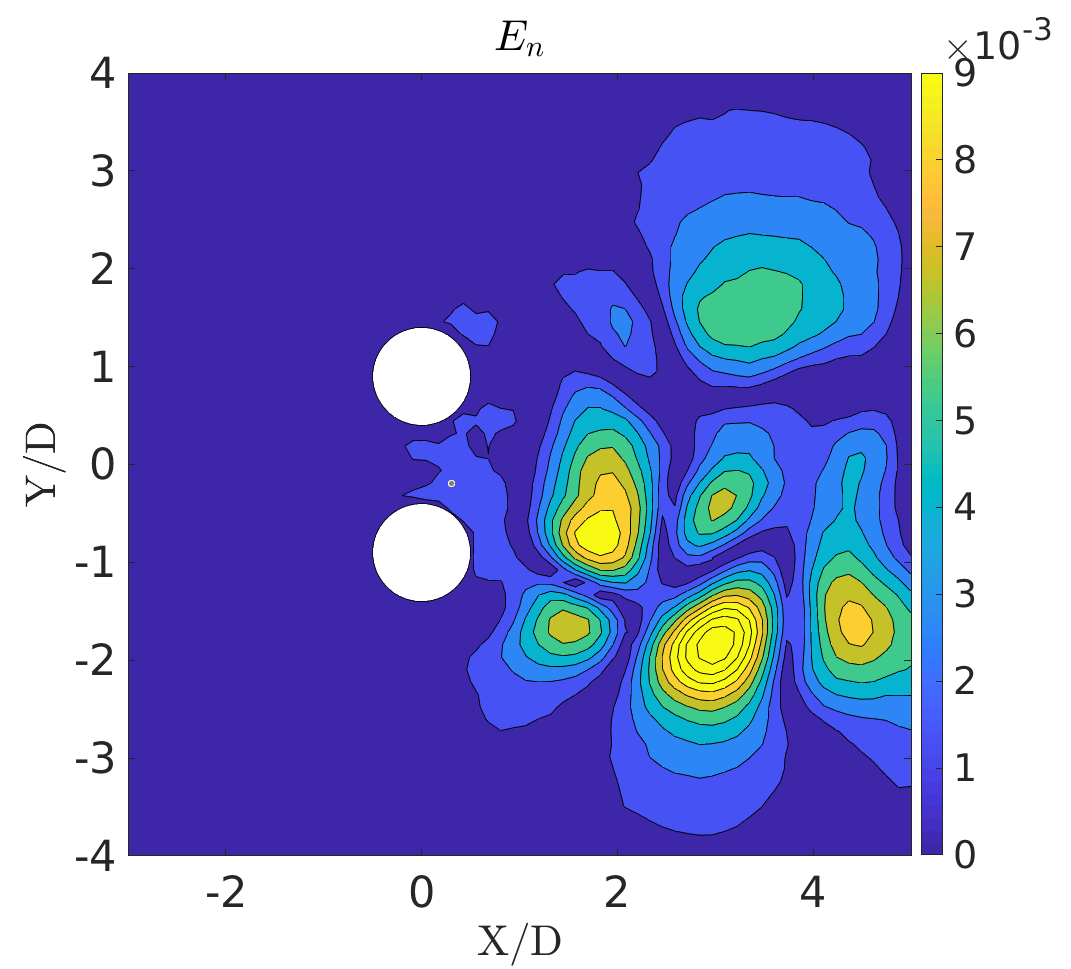}}
\\
\vspace{0.025\textwidth}
\subfloat[]{
\includegraphics[width = 0.32\textwidth]{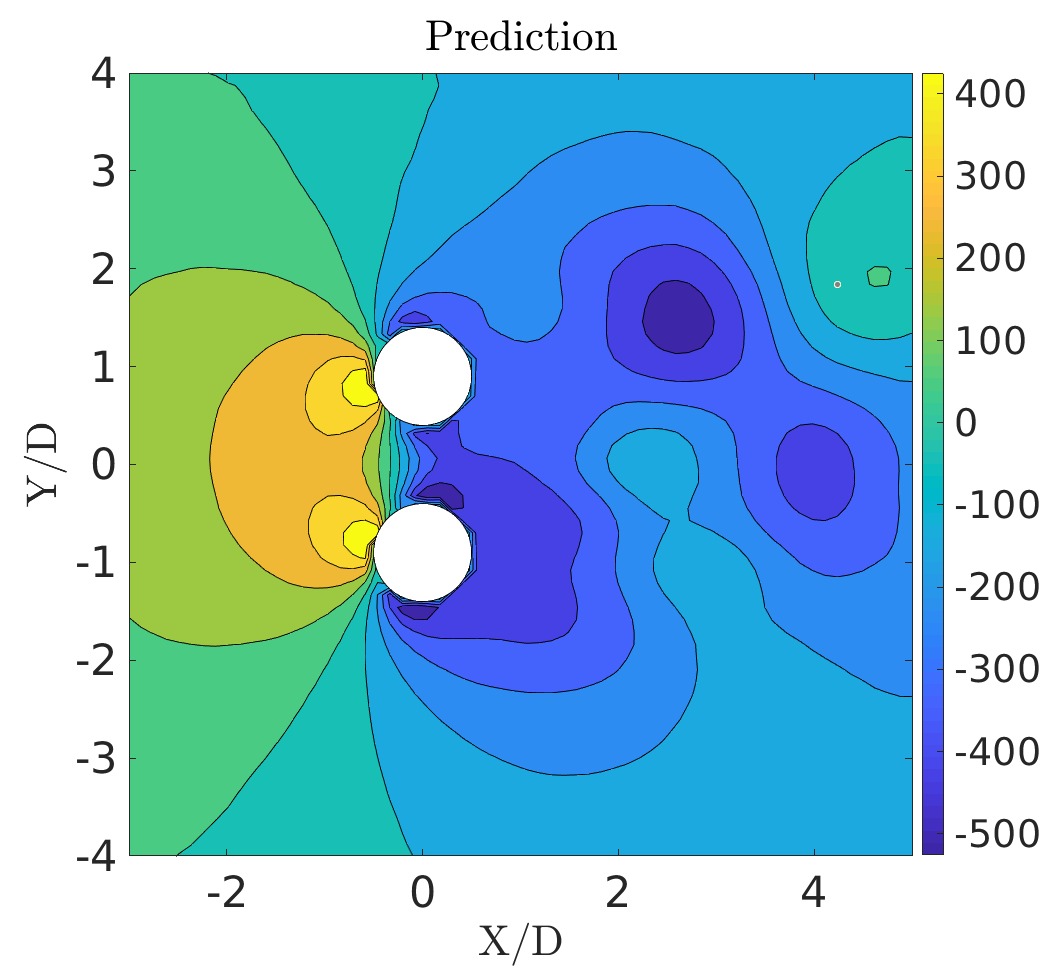}
\hspace{0.02\textwidth}
\includegraphics[width = 0.32\textwidth]{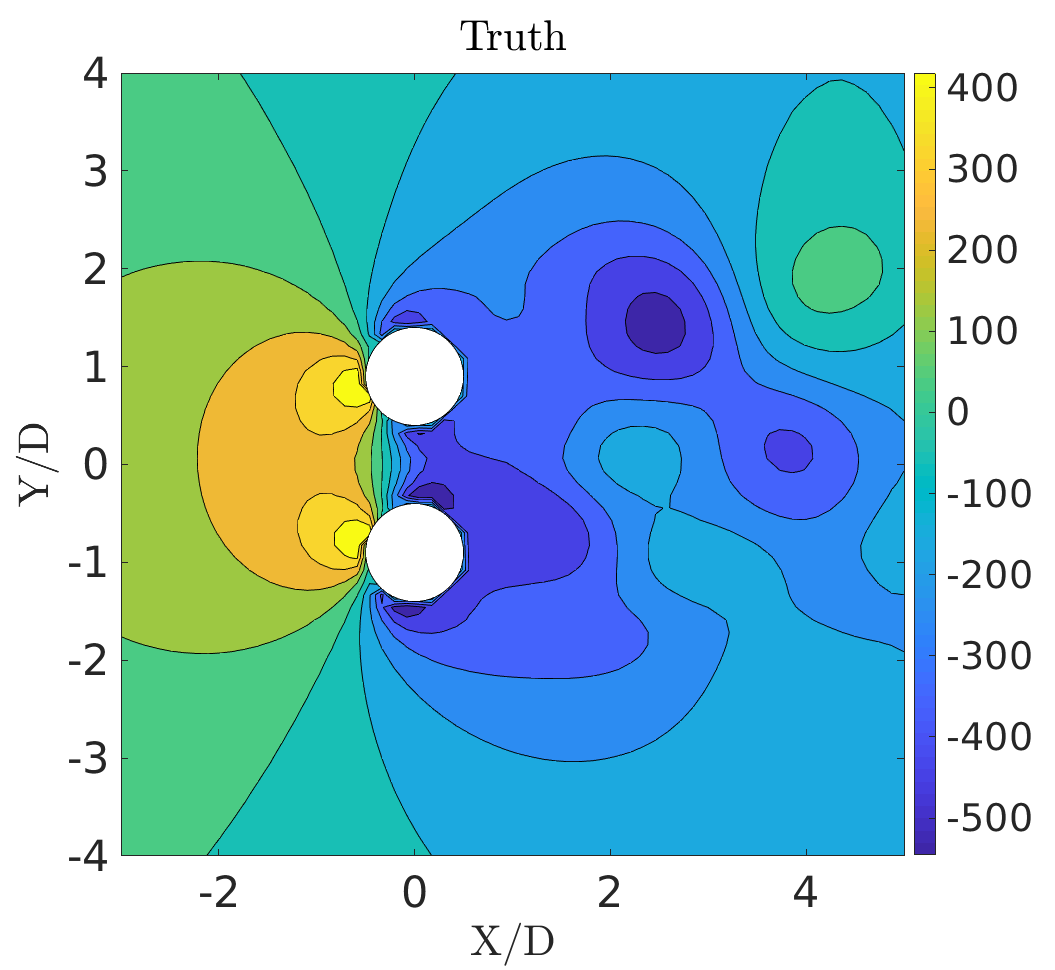}
\hspace{0.02\textwidth}
\includegraphics[width = 0.32\textwidth]{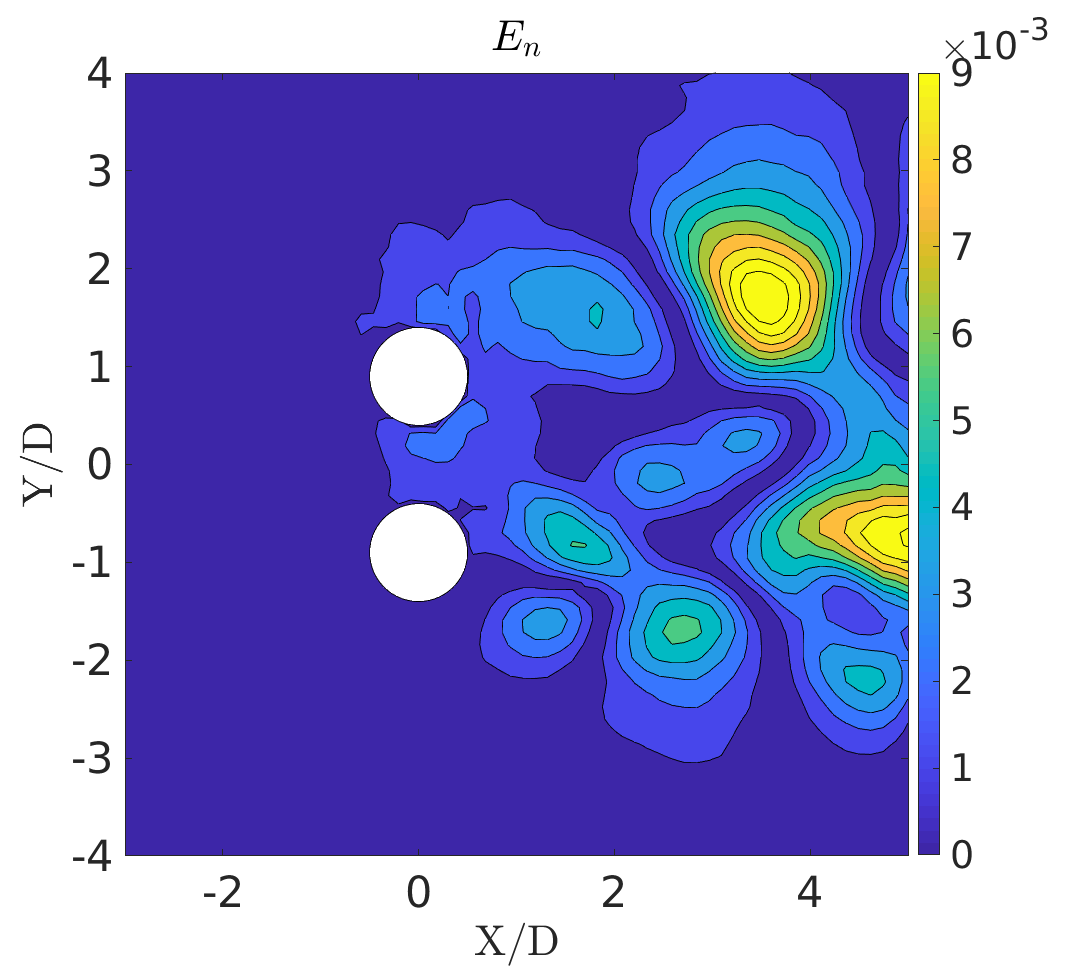}}

\caption{The flow past side-by-side cylinders: Comparison of predicted and true fields (CRAN model) along with normalized reconstruction error $E_{n}$ at (a) $tU_{\infty}/D = 898$, (b) $tU_{\infty}/D = 923$, (c) $tU_{\infty}/D= 948$ for pressure field ($P$)}
\label{cran-sbs-pres-pred}
\end{figure*}

\begin{figure*}
\centering
\subfloat[]{
\includegraphics[width = 0.32\textwidth]{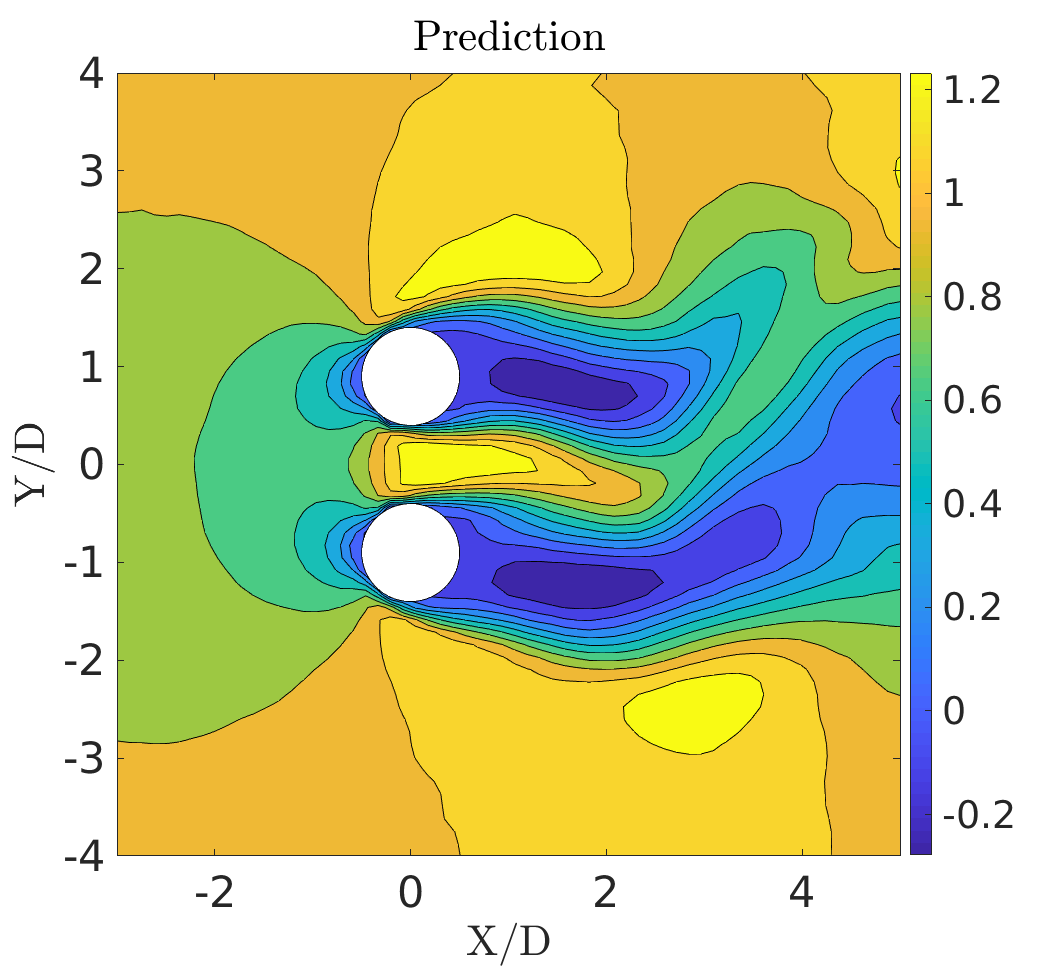}
\hspace{0.02\textwidth}
\includegraphics[width = 0.32\textwidth]{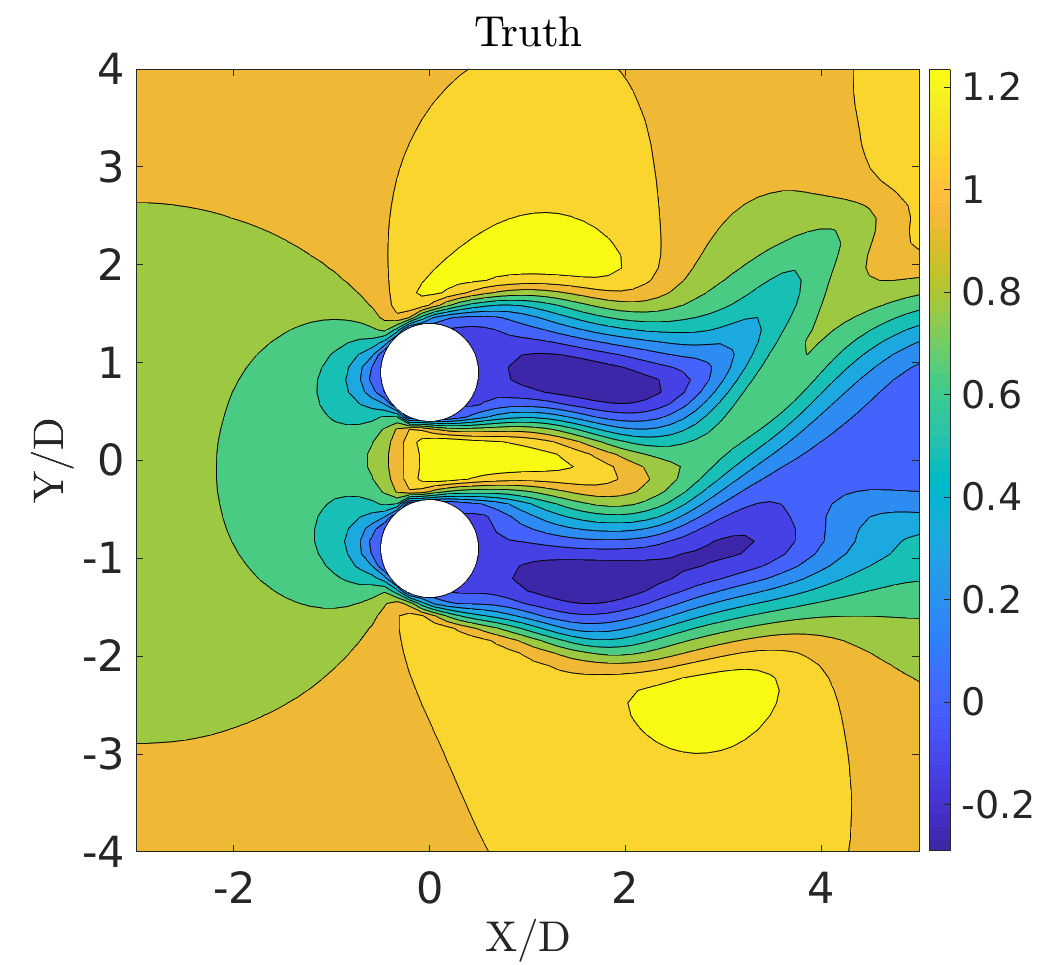}
\hspace{0.02\textwidth}
\includegraphics[width = 0.32\textwidth]{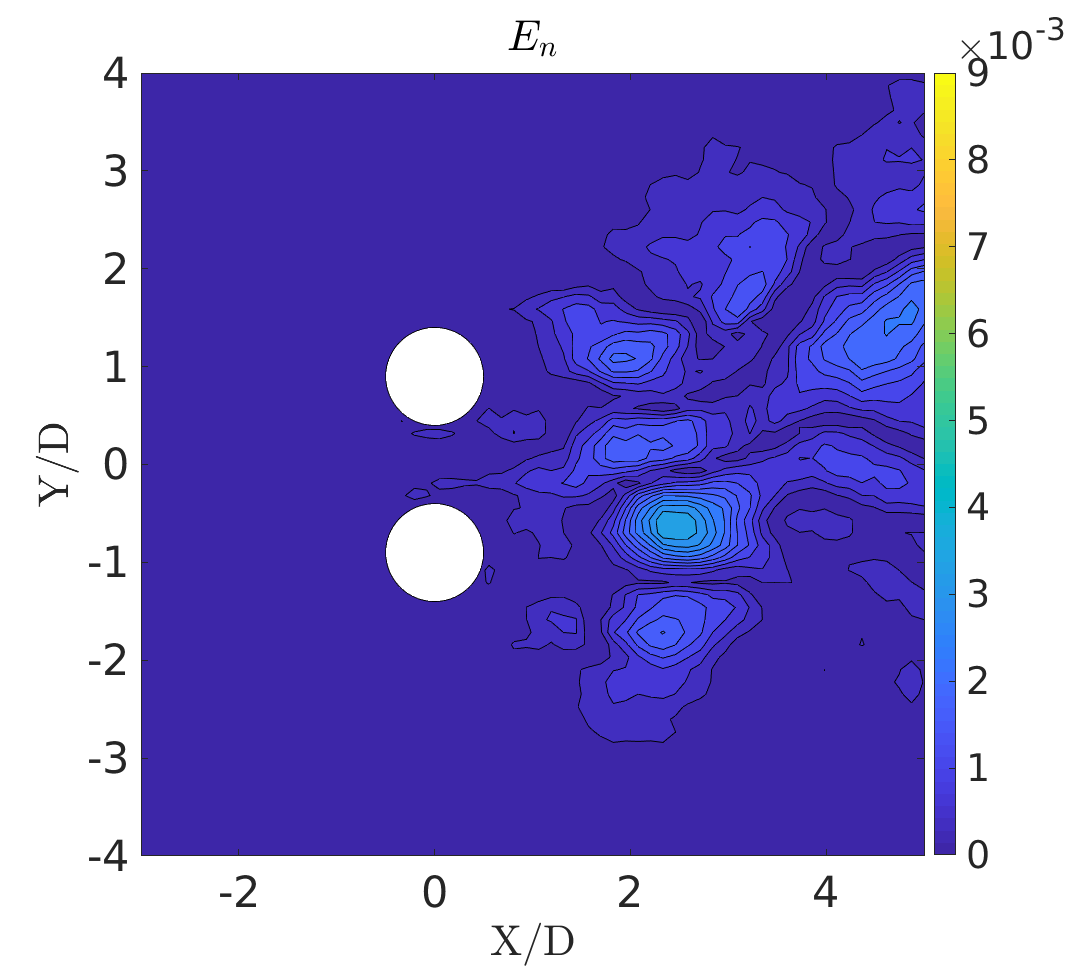}}
\\
\vspace{0.025\textwidth}
\subfloat[]{
\includegraphics[width = 0.32\textwidth]{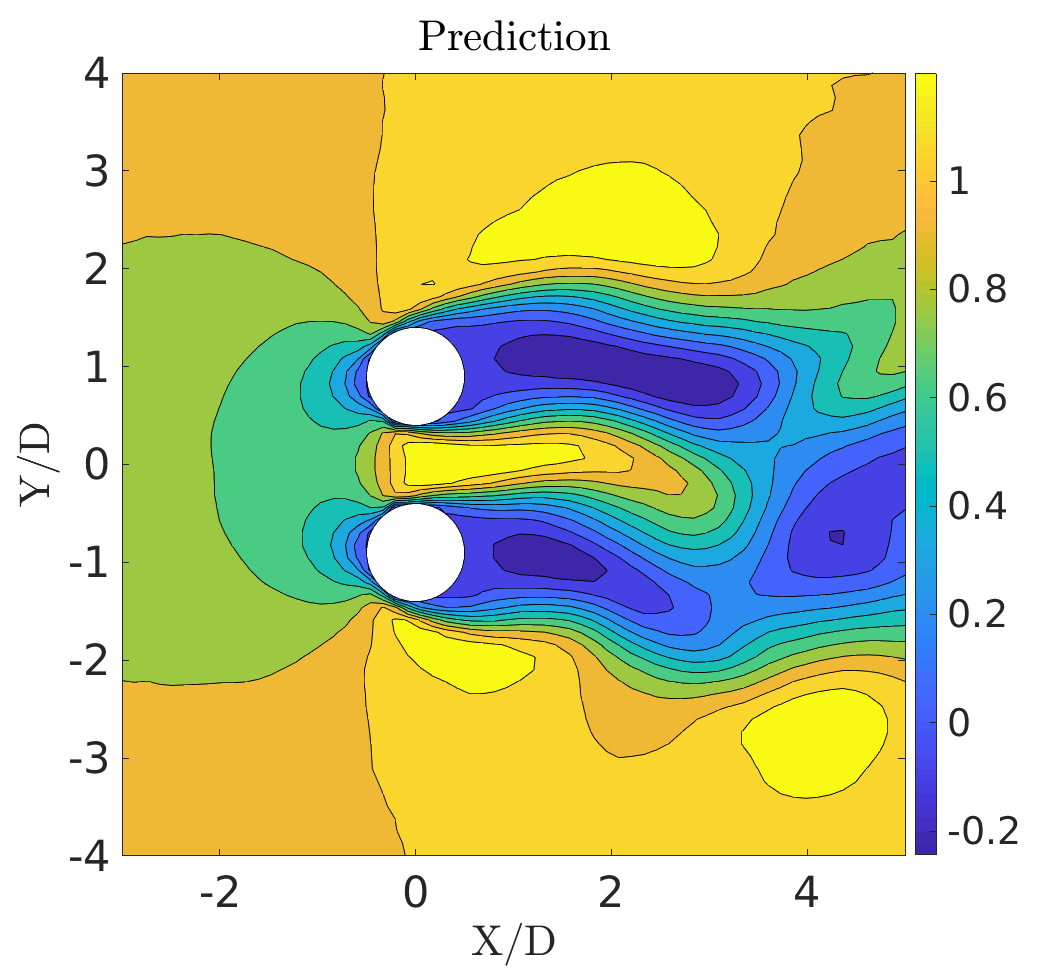}
\hspace{0.02\textwidth}
\includegraphics[width = 0.32\textwidth]{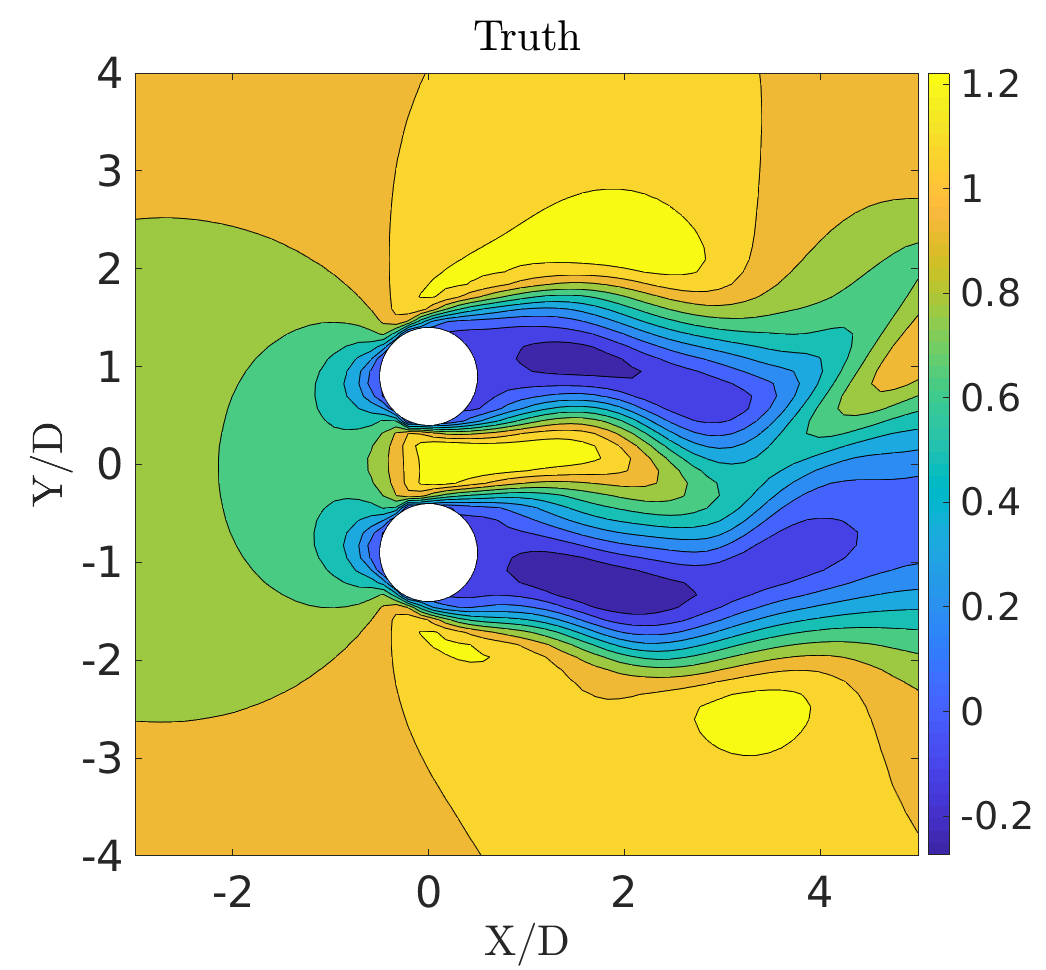}
\hspace{0.02\textwidth}
\includegraphics[width = 0.32\textwidth]{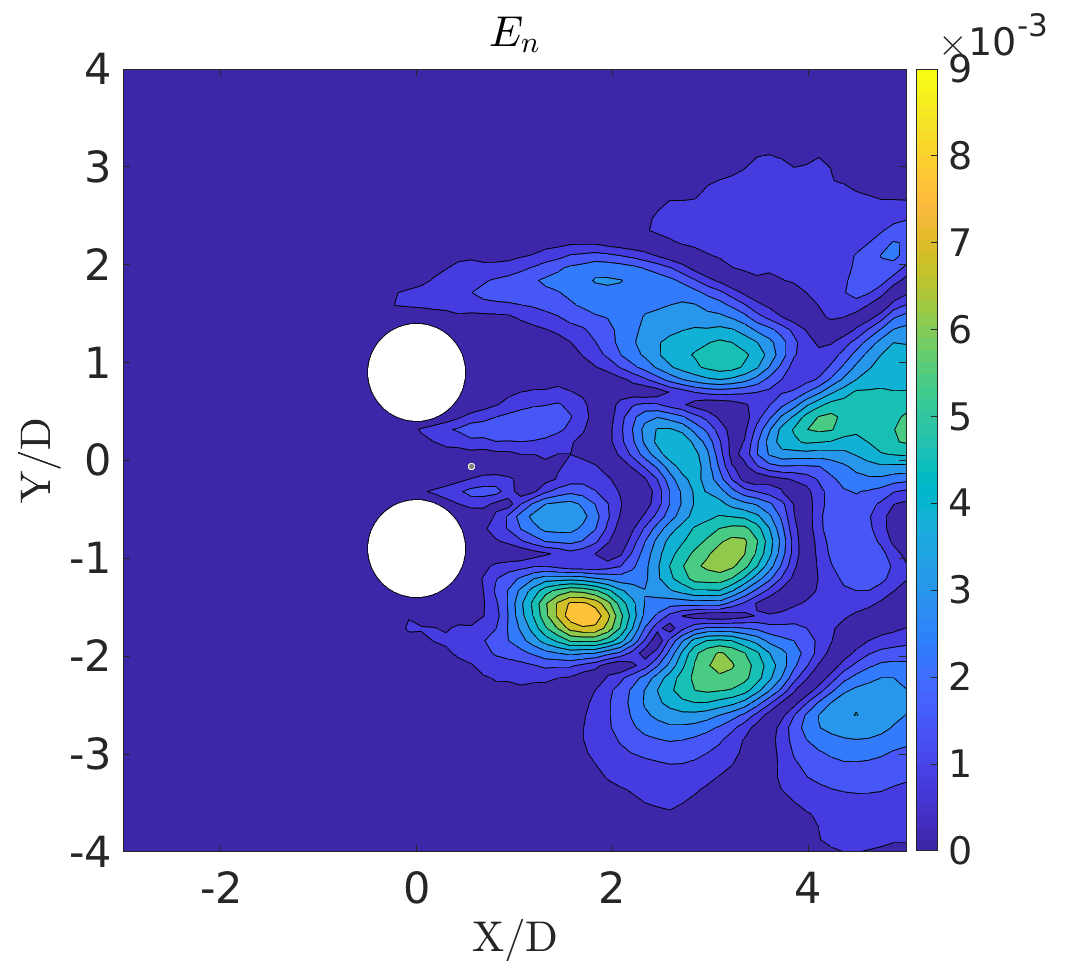}}
\\
\vspace{0.025\textwidth}
\subfloat[]{
\includegraphics[width = 0.32\textwidth]{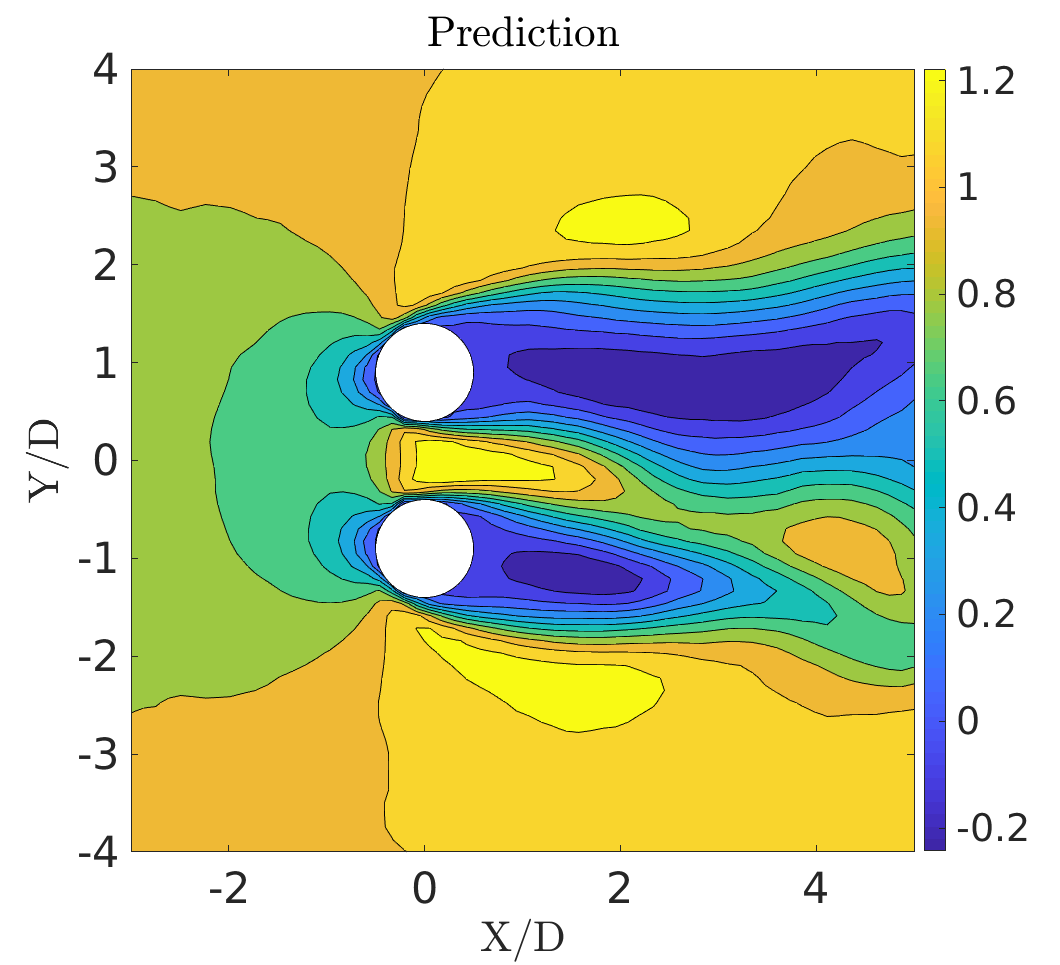}
\hspace{0.02\textwidth}
\includegraphics[width = 0.32\textwidth]{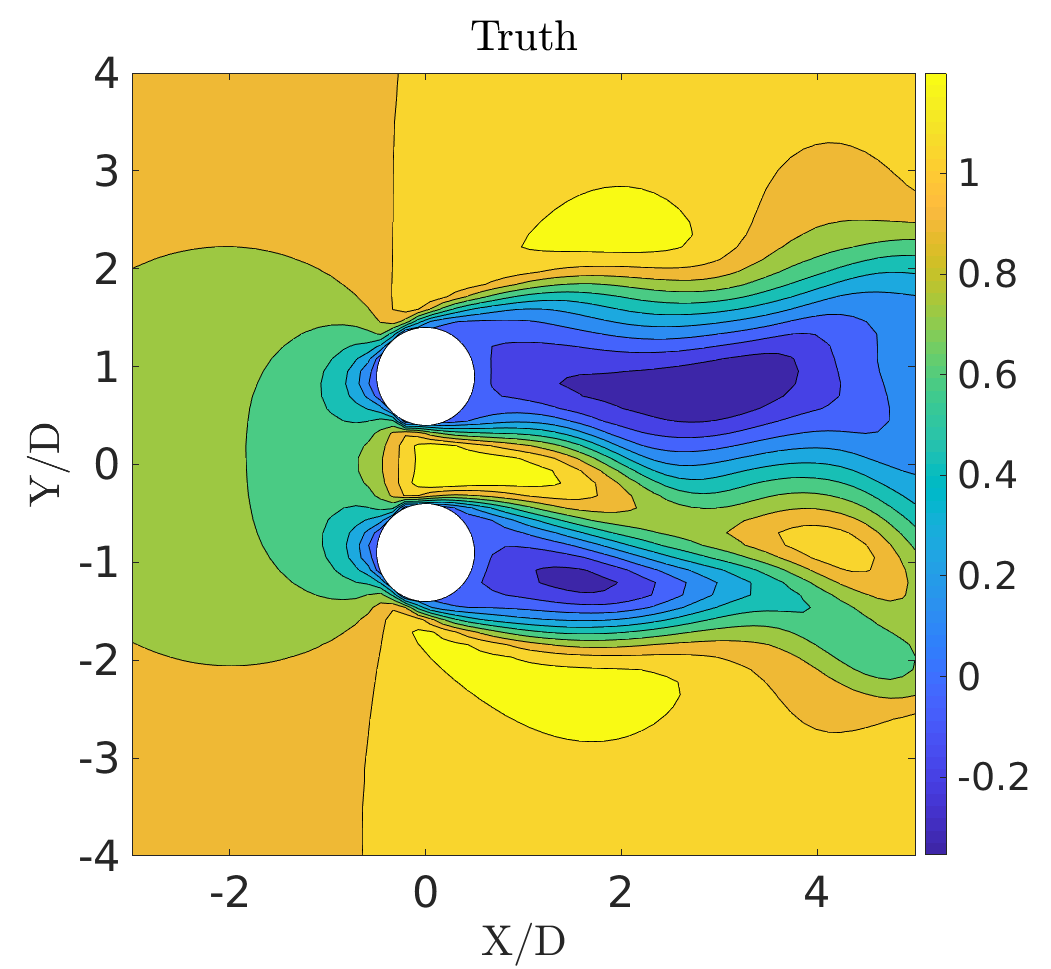}
\hspace{0.02\textwidth}
\includegraphics[width = 0.32\textwidth]{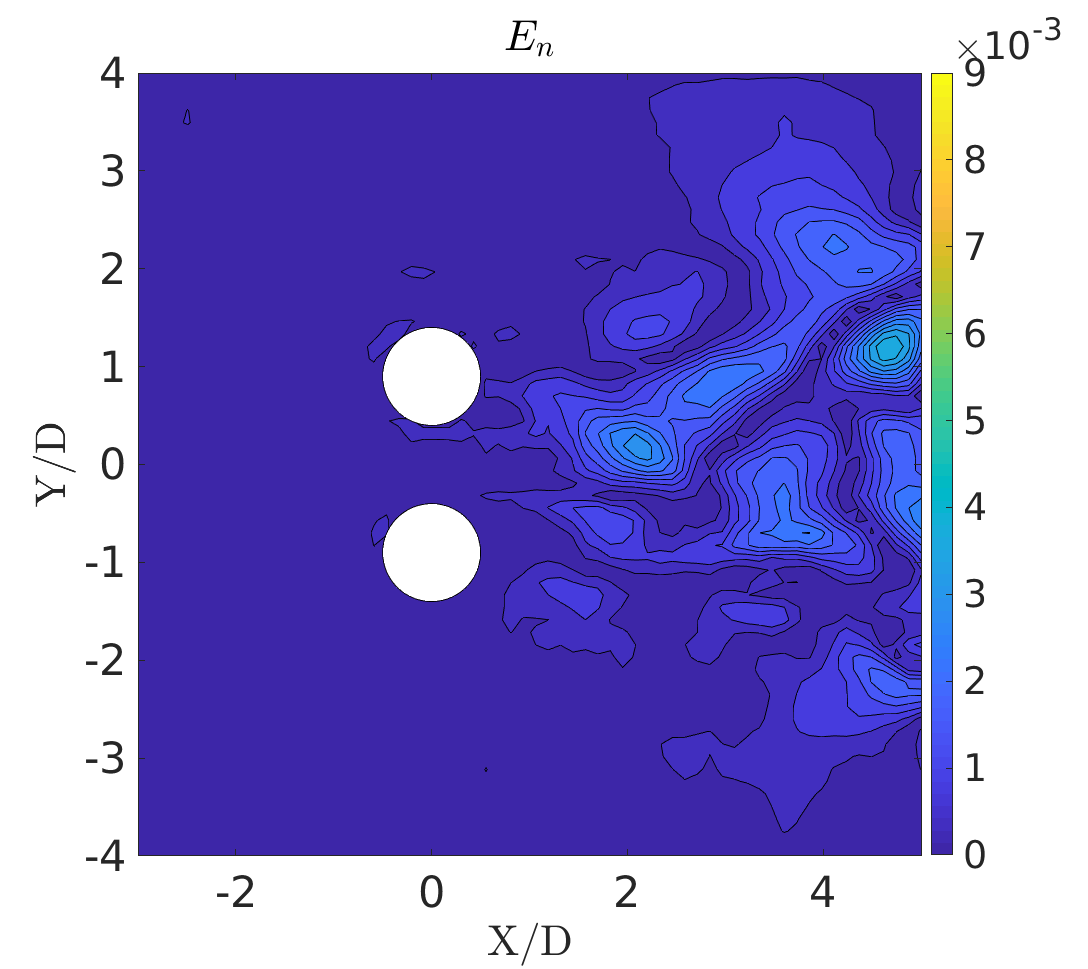}}

\caption{The flow past side-by-side cylinders: Comparison of predicted and true fields (CRAN model) along with normalized reconstruction error $E_{n}$ at (a) $tU_{\infty}/D = 898$, (b) $tU_{\infty}/D = 923$, (c) $tU_{\infty}/D = 948$ for x-velocity field ($U$)}
\label{cran-sbs-velx-pred}
\end{figure*}

%Field Prediction
\textit{Field prediction}: Fig.~\ref{cran-sbs-pres-pred} and \ref{cran-sbs-velx-pred} depict the comparison of predicted and true values for the pressure and x-velocity field, respectively, at time-steps $3592$ ($898\;tU_{\infty}/D$), $3692$ ($923\;tU_{\infty}/D$) and $3792$ ($948\;tU_{\infty}/D$). The normalized reconstruction error $E_{n}$ is constructed by taking the absolute value of differences between and true and predicted values and normalizing it with $L_{2}$ norm of the truth data and is given by

\begin{equation}
    E_{n} = \frac{|\textbf{s}_{n}-\hat{\textbf{s}}_{n}|}{\|\textbf{s}_{n}\|_{2,k}}.
\end{equation}

\begin{figure}
\centering
\subfloat[]
{\includegraphics[width = 0.235\textwidth]{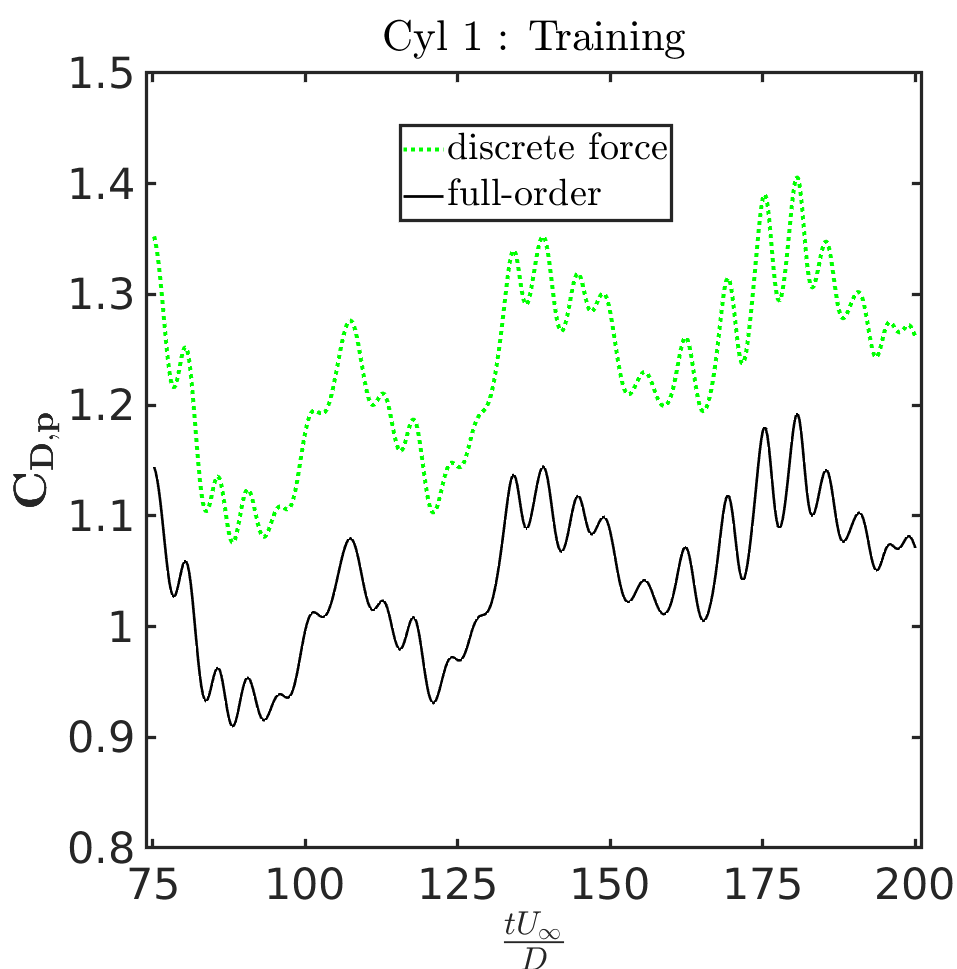}}
\subfloat[]
{\includegraphics[width = 0.231\textwidth]{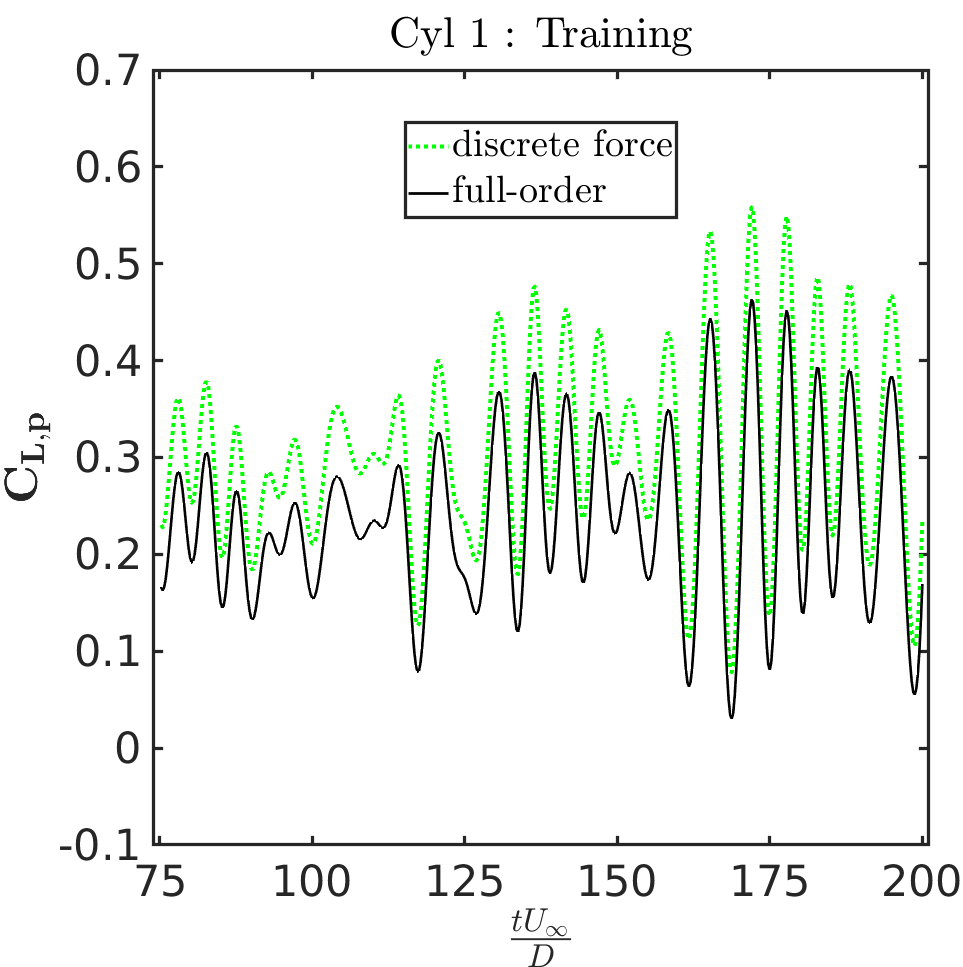}} \\
\subfloat[]
{\includegraphics[width = 0.235\textwidth]{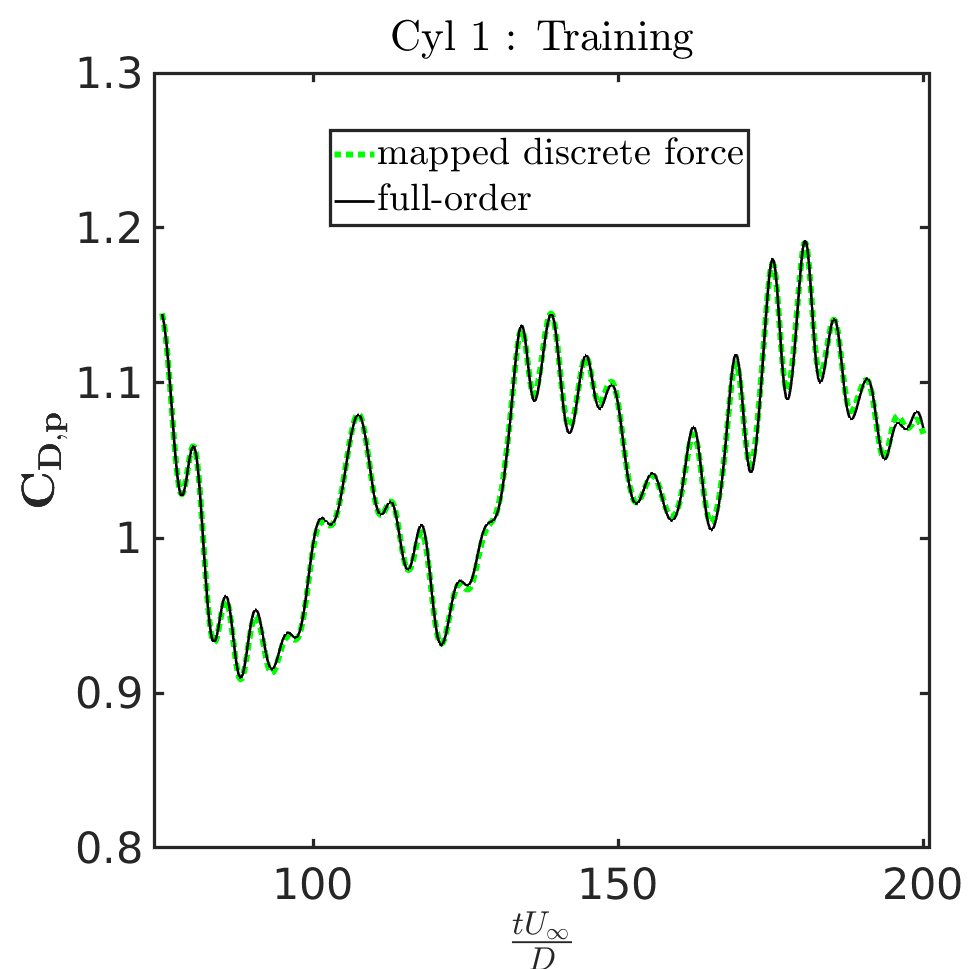}}
\subfloat[]
{\includegraphics[width = 0.235\textwidth]{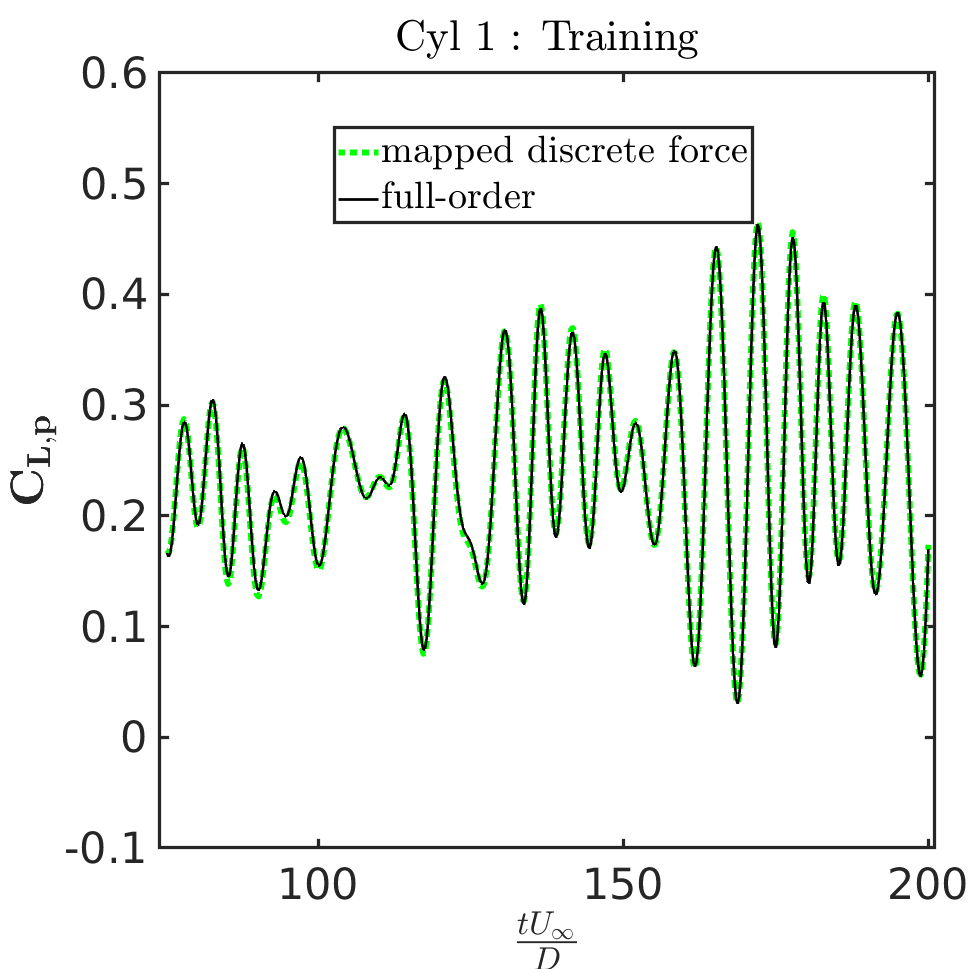}}\\
\subfloat[]
{\includegraphics[width = 0.235\textwidth]{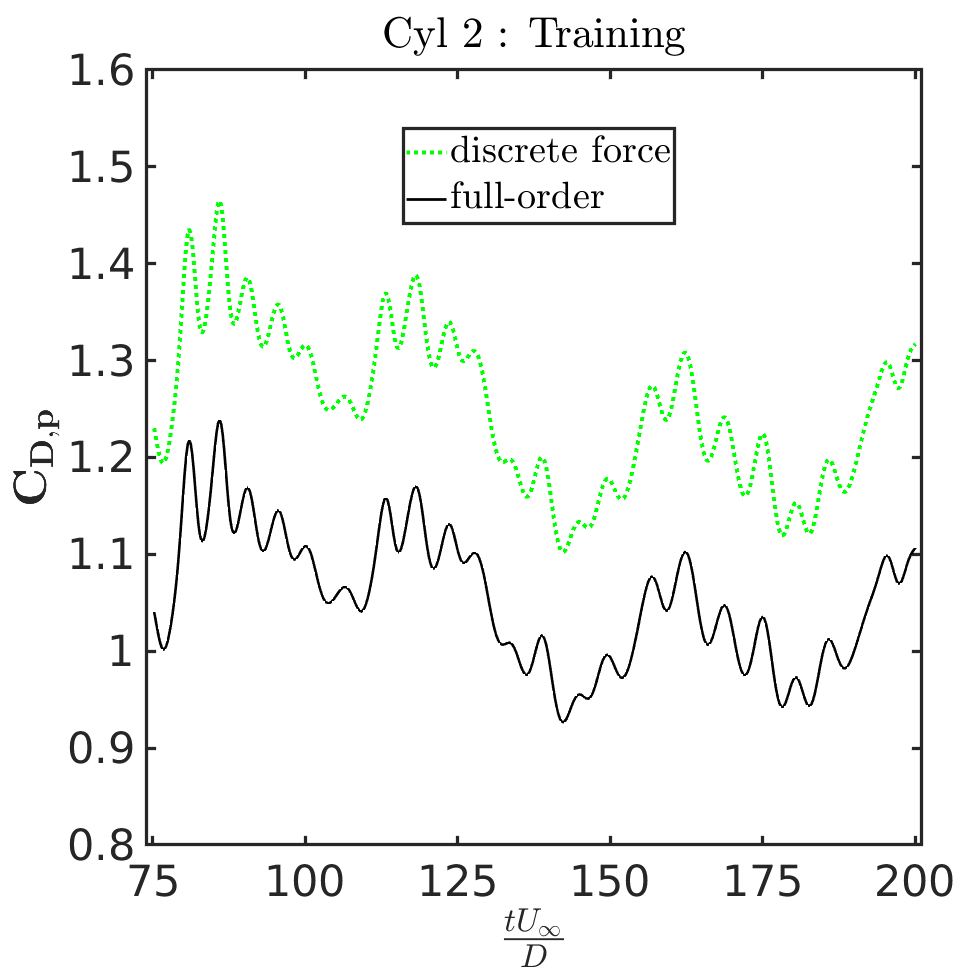}}
\subfloat[]
{\includegraphics[width = 0.235\textwidth]{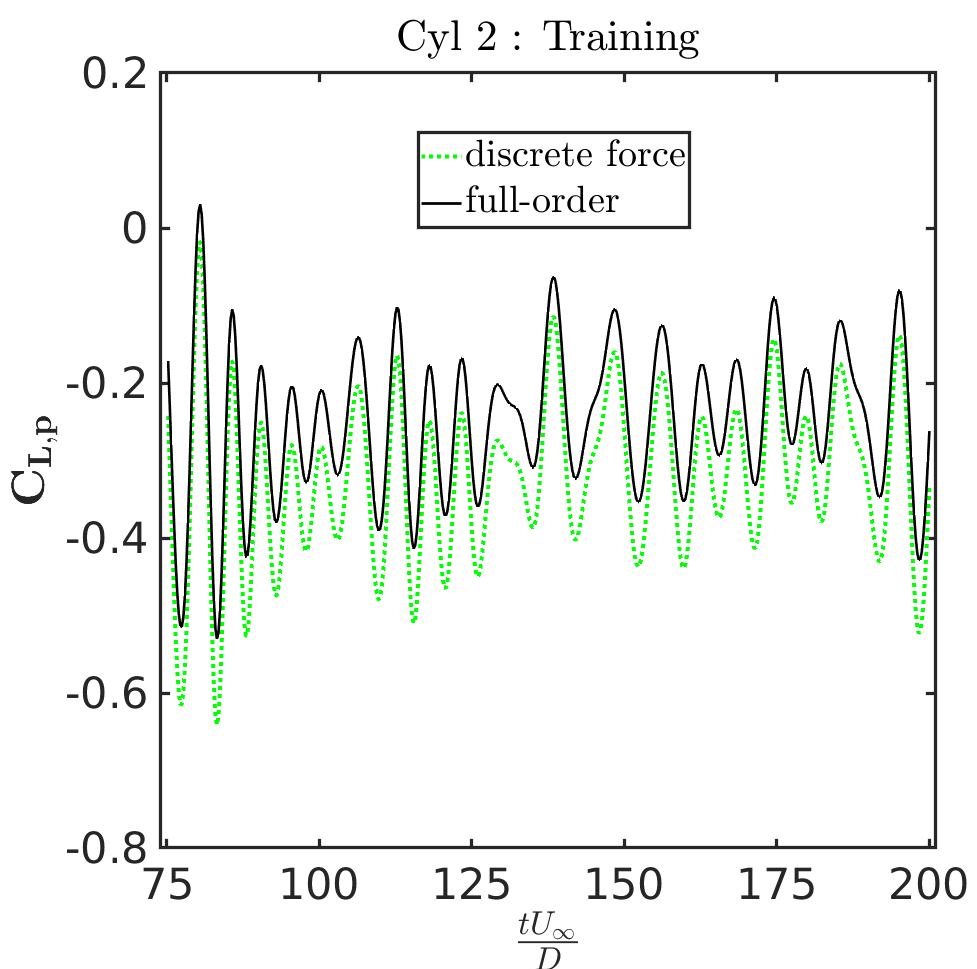}} \\
\subfloat[]
{\includegraphics[width = 0.235\textwidth]{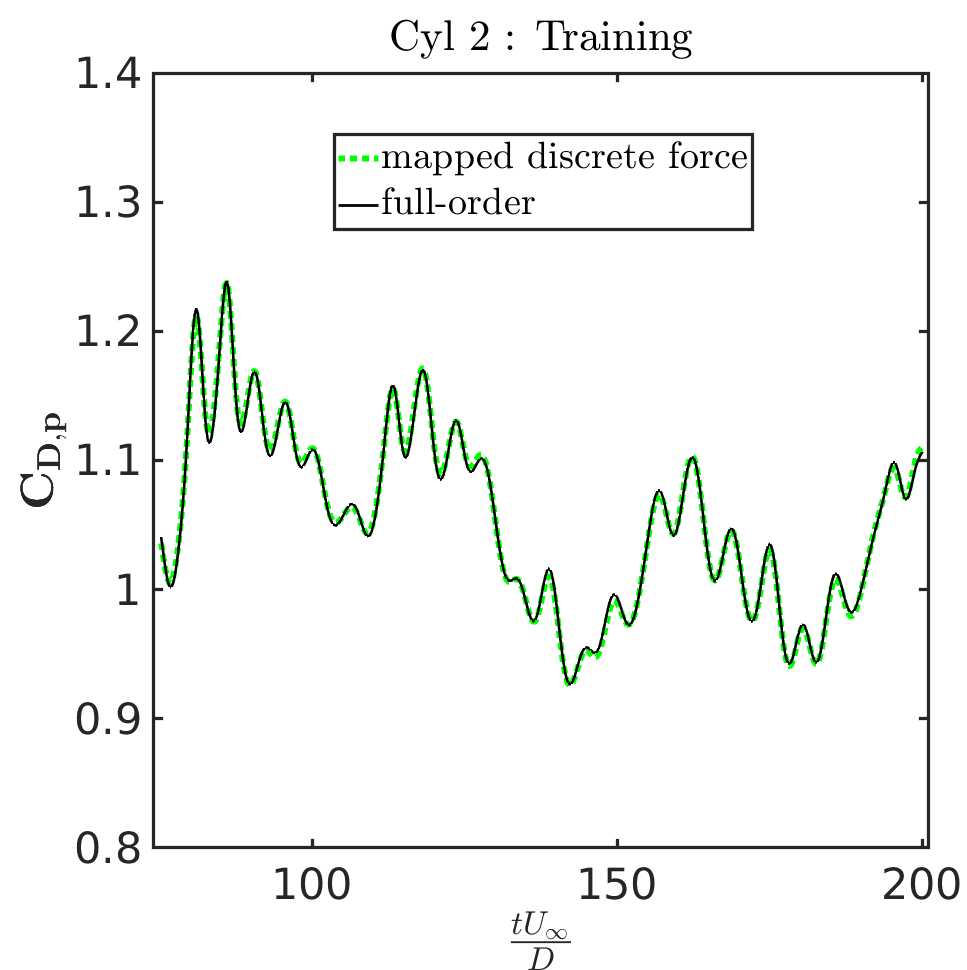}}
\subfloat[]
{\includegraphics[width = 0.235\textwidth]{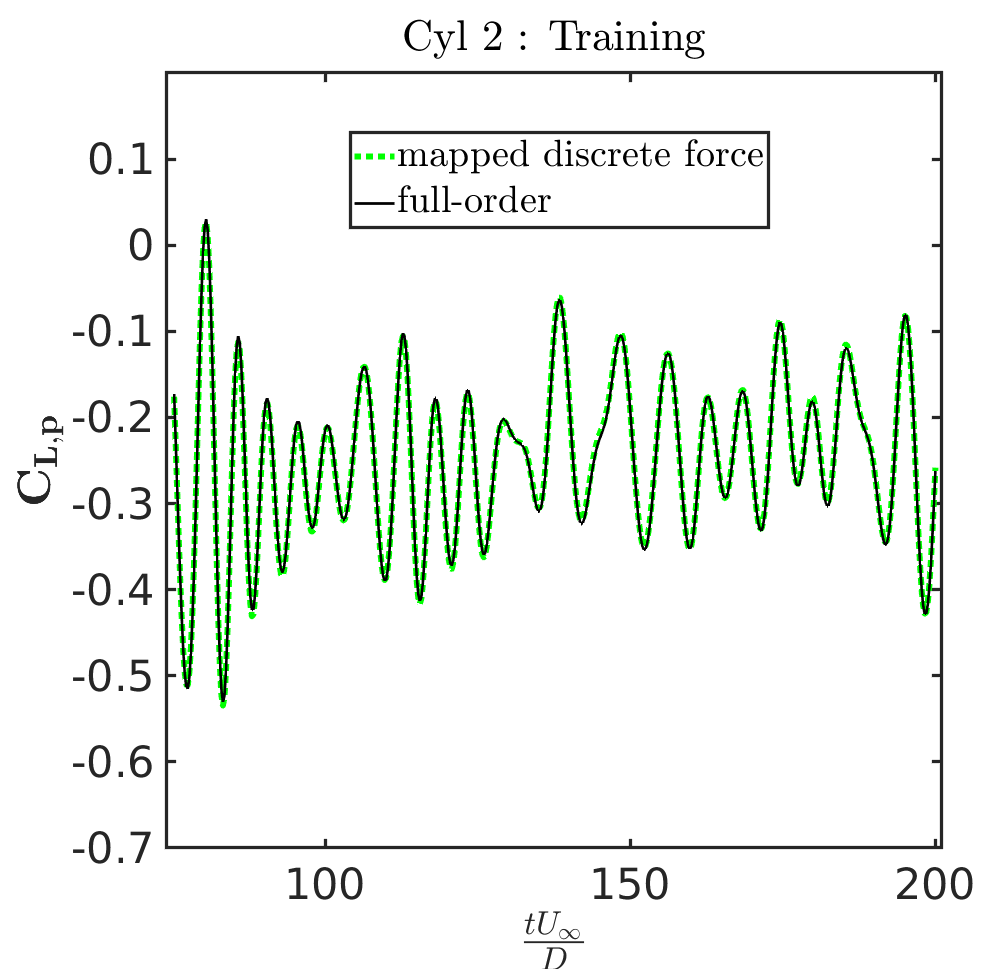}}
\caption{The flow past side-by-side cylinders: Comparison of discrete force, mapped discrete force and full-order force coefficients on the training data due to pressure for Cylinder 1 ((a)-(b)-(c)-(d)) and Cylinder 2 ((e)-(f)-(g)-(h))}
\label{discrete-sbs-cyls-pres}
\end{figure}

\begin{figure*}
\centering
\subfloat[]
{\includegraphics[width = 0.32\textwidth]{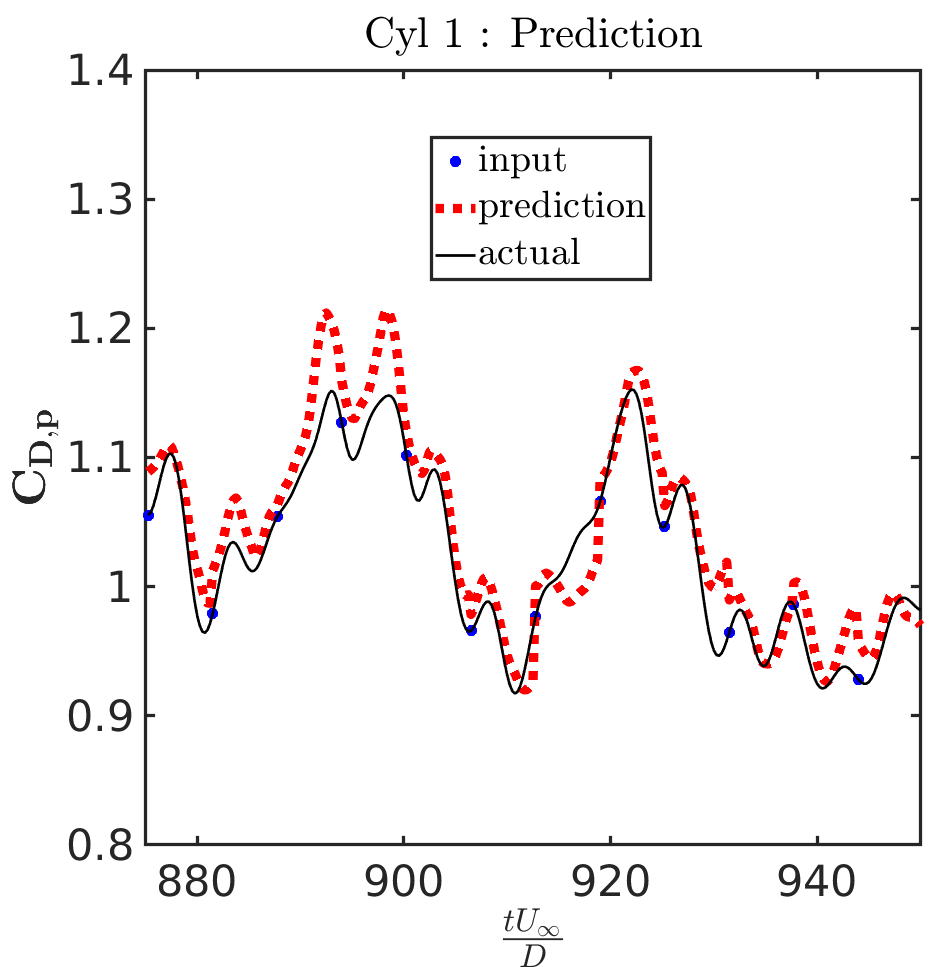}}
\subfloat[]
{\includegraphics[width = 0.32\textwidth]{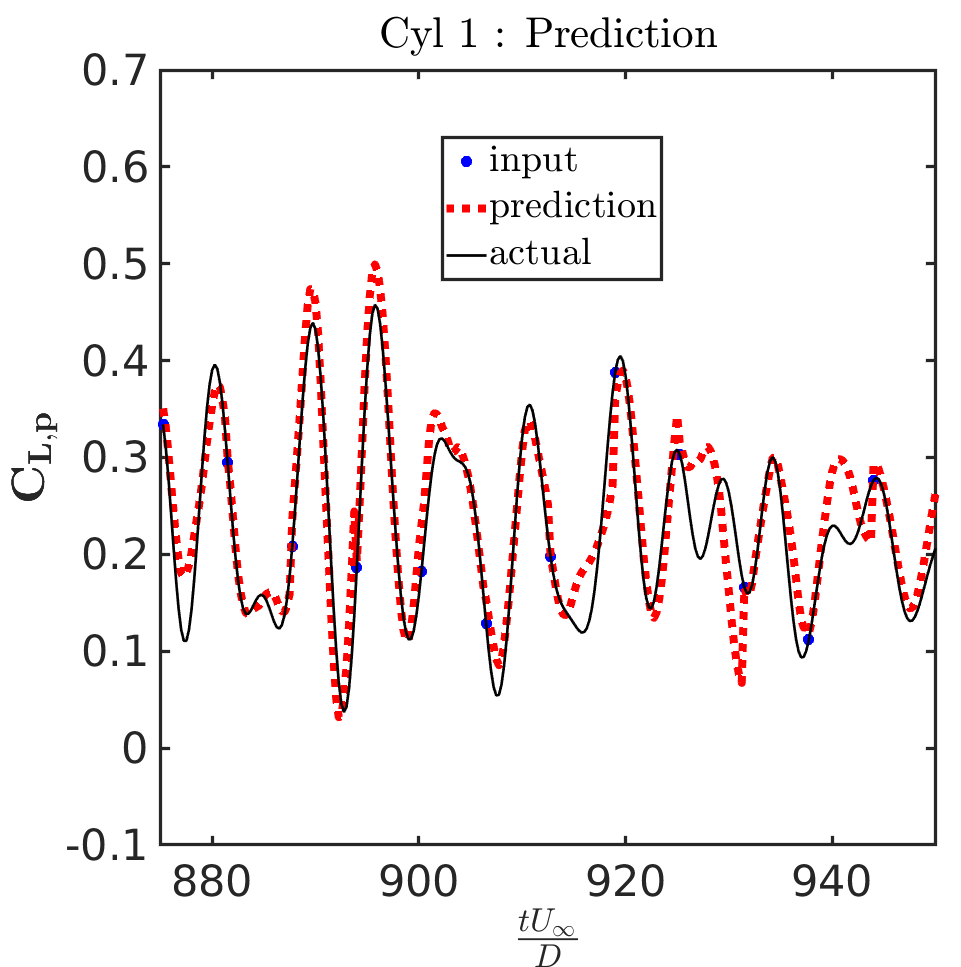}} \\
\subfloat[]
{\includegraphics[width = 0.32\textwidth]{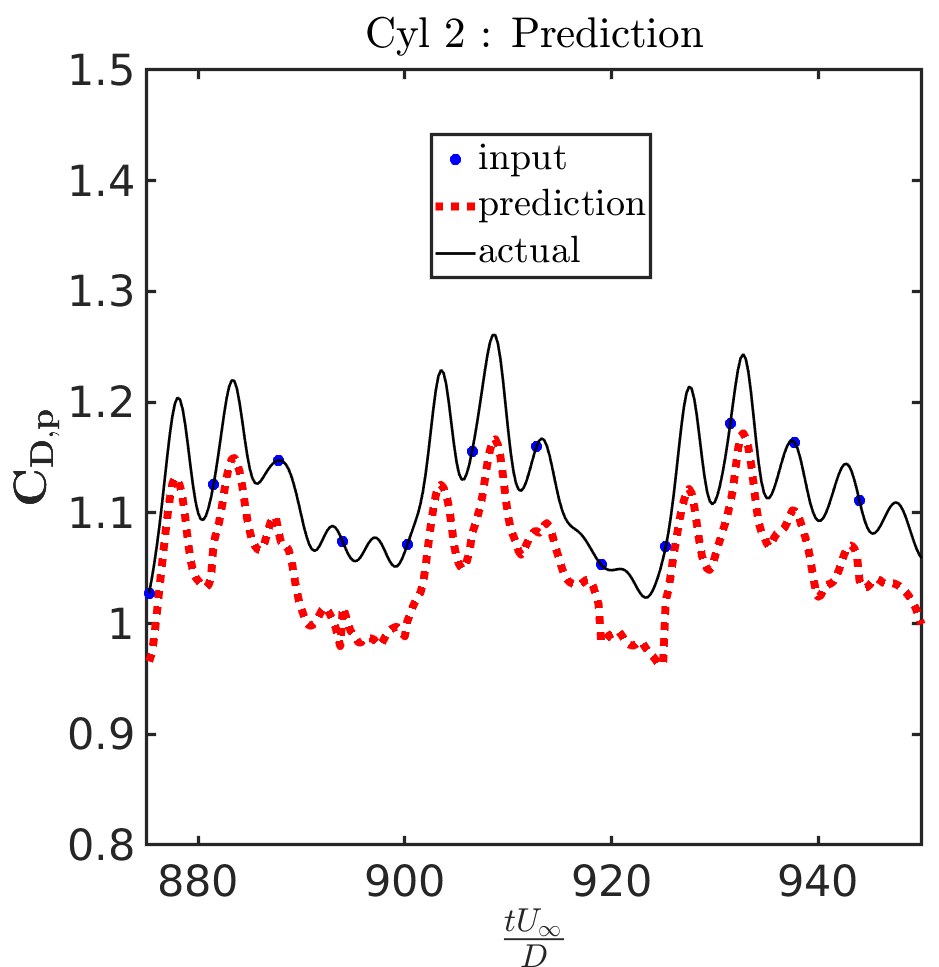}}
\subfloat[]
{\includegraphics[width = 0.32\textwidth]{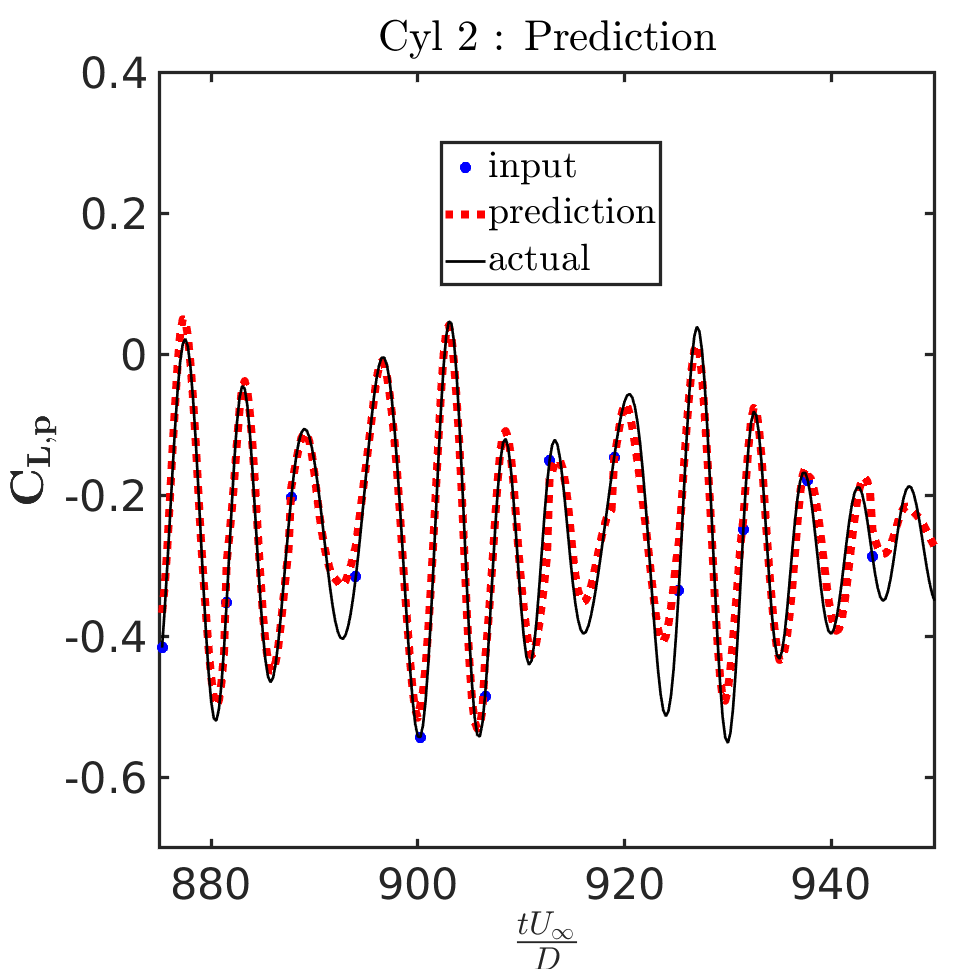}}
\caption{The flow past side-by-side cylinders: Predicted and actual (CRAN model) pressure force coefficients for all test time-steps: (a) Drag (Cylinder 1), (b) lift (Cylinder 1), (c) drag (Cylinder 2), (d) lift (Cylinder 2). Note that at a time, one input time-step (blue dot) is used to predict the next sequence of $N_{t}=25$ steps, until a new input is fed. This helps in reducing the compounding effect of the errors while still achieving predictions in closed-loop recurrent fashion}
\label{cran-pred-sbs-cyls-force}
\end{figure*}

% Force 
\textit{Force coefficient prediction}: 
We use the methodology highlighted in section \ref{Section:ForceCalc} to predict the pressure force coefficients on the boundary of the cylinders. Here, $\textbf{C}_{\mathrm{D},\mathrm{p}}$ and $\textbf{C}_{\mathrm{L},\mathrm{p}}$ denote the drag and lift force coefficient experienced by the stationary cylinder in a pressure field. Using Eqs. (\ref{presForce}) and (\ref{totalDiscForce}), we first calculate the discrete drag and lift force (coefficient) from the pressure training data (3200 snapshots from time 75s to 875s). Fig. \ref{discrete-sbs-cyls-pres} (a) and (b) depict the discrete lift and drag force coefficients on the Cylinder 1 and the effect of reconstruction $\psi$ on these data is shown in Fig. \ref{discrete-sbs-cyls-pres} (c) and (d). Similarly, the discrete and mapped-discrete force coefficients on Cylinder 2 are shown in Fig. \ref{discrete-sbs-cyls-pres} (e),(f),(g),(h). These plots are shown for $75\;tU_{\infty}/D$ till $200\;tU_{\infty}/D$ for visualisation point of view. The reconstruction accuracy is nearly $99.5\%$. Recovering the correct force prediction boils down to the task of extracting the discrete force from accurate predicted fields. The $\psi$ calculated from the training data recovers the missing force data in prediction (Eq. (\ref{psitotalDiscForce})). The predicted force coefficients are shown in Fig.~\ref{cran-pred-sbs-cyls-force} for both the cylinders for all the test time-steps (3501 till 3800 time-steps).       

% cyl 1 and 2 force training data

%%%% fields
%\newpage
%\onecolumn
% pressure

% velocity 

%\newpage

% cyl 1 and 2 force prediction 

%\newpage
%\twocolumn 
%%%%%%%%%%%%%%%%%%%%%%%%%%%
\subsection{Discussion}\label{discussion_sbs}
%%%%%%%%%%%%%%%%%%%%%%%%%%%
% Some remarks 
It is noted that the POD-RNN model with the closed-loop recurrent neural network is inefficient in predicting the flow past side-by-side cylinders. For that purpose, we employ an encoder-decoder type recurrent net instead of a closed-loop type. Whereas, the convolutional recurrent autoencoder network (CRAN) model can predict the flow fields in a closed-loop fashion for longer-time steps with limited ground data as a demonstrator. In the case of the POD-RNN model, the spatial and temporal parts of the problem are dealt with independently which might work for simple problems of flow past a cylinder. However, such linear decomposition is less effective in combating highly nonlinear problems such as side-by-side cylinders. The improvement compared to the POD-RNN model can be attributed to the low-dimensional features obtained by CNN of the CRAN model. The non-linear kernels in CNN are very powerful in identifying dominant local flow features \cite{tharindu2018}. Also, the complete end-to-end architecture of the network, which enables us to integrate the encoding, evolution and decoding in a complete nonlinear fashion, is the primary reason for this model to work for the problem. This is a very promising result and motivates us to take forward this concept of convolutional recurrent autoencoder models and its variants (such as variational autoencoders, registration/interpretable autoencoders \cite{rambod2020,taddei2020}) for unsteady fluid flow problems with strong convection effect and moving boundaries. 

Table \ref{tab_comp_pc_sbs} lists out the differences in the attributes of both the models for flow past side-by-side cylinders for both the fields. It is worth mentioning that CRAN outperforms POD-RNN in terms of longer time-series prediction. In this analysis, POD-RNN requires $\approx n_{ts} / 2 = 150$ test time-steps as input for predicting remaining 150 time-steps, while CRAN utilises $\approx n_{ts} / N_{t} = 12$ data demonstrators to predict all $n_{ts}$ time-steps. Note that $n_{ts}=300$ denotes the test time-steps. 
\begin{table}[H]
    \centering
    \begin{tabular}{|c|c|c|}
    \hline
          & POD-RNN & CRAN \\ \hline
         Encoder/decoder space &  POD  & $\boldsymbol{\theta}\approx$ 3 x $10^{5}$\\
         Evolver features $\textbf{A}$ ($P$,$U$) & $k=25$ & $N_{A}=32,16$ \\
         Recurrent network type & encoder-decoder & closed-loop \\ 
         Training time (CPU) & 1 hour & 16 hours \\
         Prediction ($\textbf{I}$, $\textbf{P}$) & $\textbf{I}:25$, $\textbf{P}:25$ & $\textbf{I}:1$, $\textbf{P}:25$ \\
         \hline
    \end{tabular}
    $\boldsymbol{\theta}$: trainable parameters \\
    $P$: pressure field, $U$: x-velocity field \\ 
    $\textbf{I}$: input time-steps,  $\textbf{P}$: predicted time-steps
    \caption{The flow past side-by-side cylinders: Comparison of POD-RNN with convolutional recurrent autoencoder network (CRAN)}
    \label{tab_comp_pc_sbs}
\end{table}

\begin{figure}
\centering
\subfloat[]
{\includegraphics[width = 0.238\textwidth]{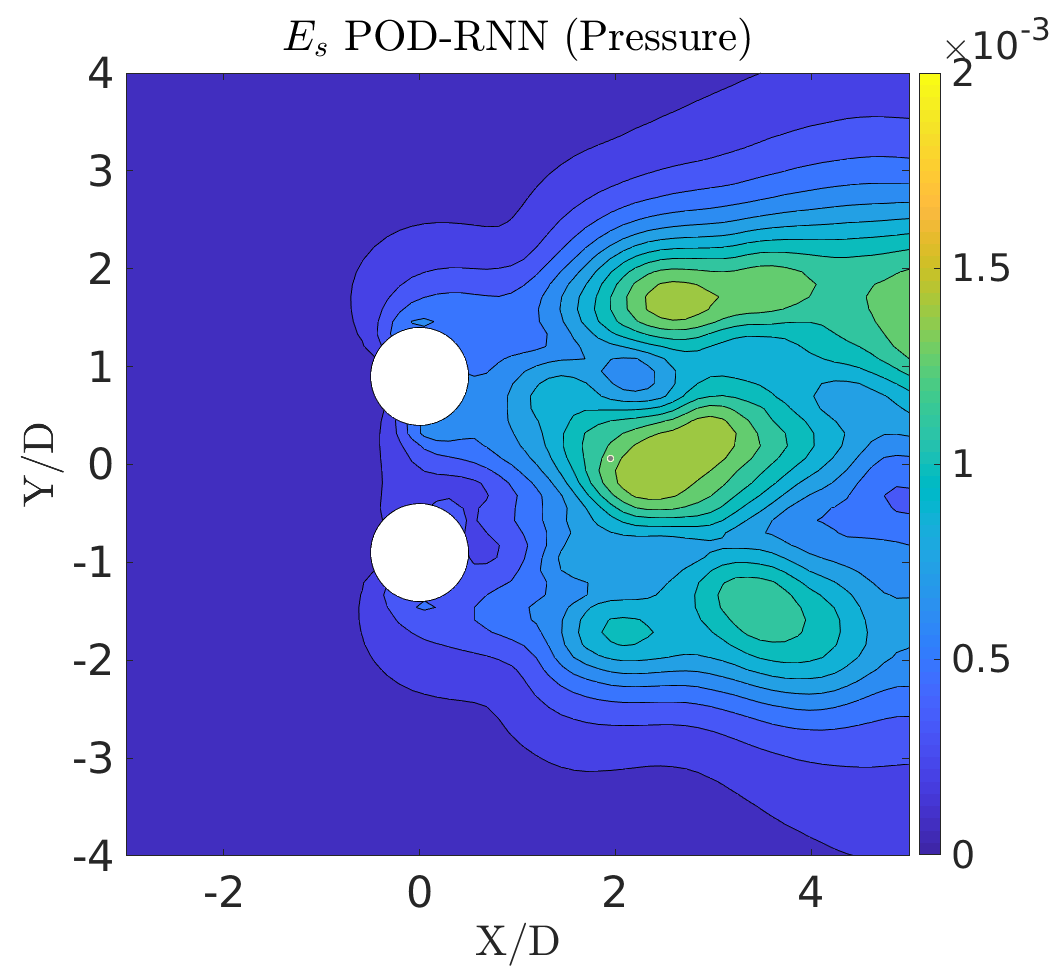}}
\subfloat[]
{\includegraphics[width = 0.238\textwidth]{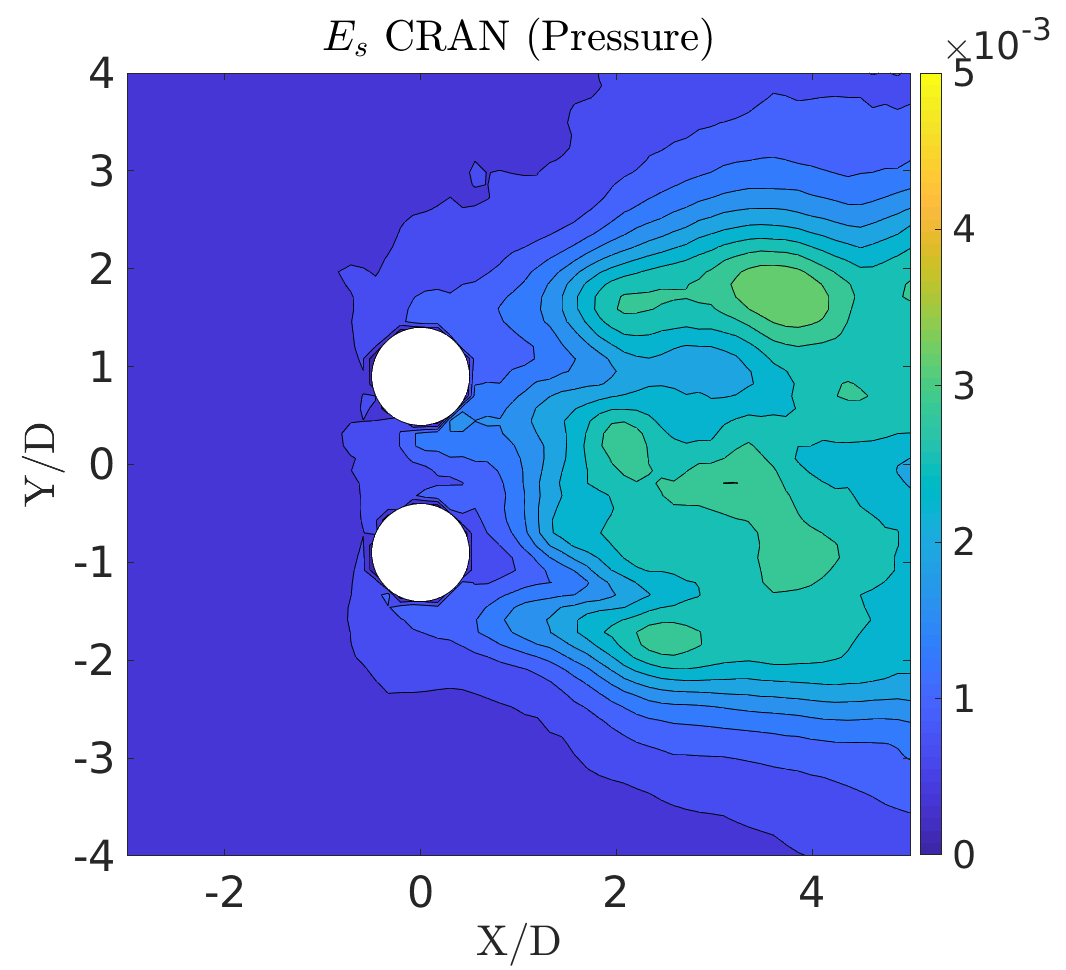}}\\
\subfloat[]
{\includegraphics[width = 0.238\textwidth]{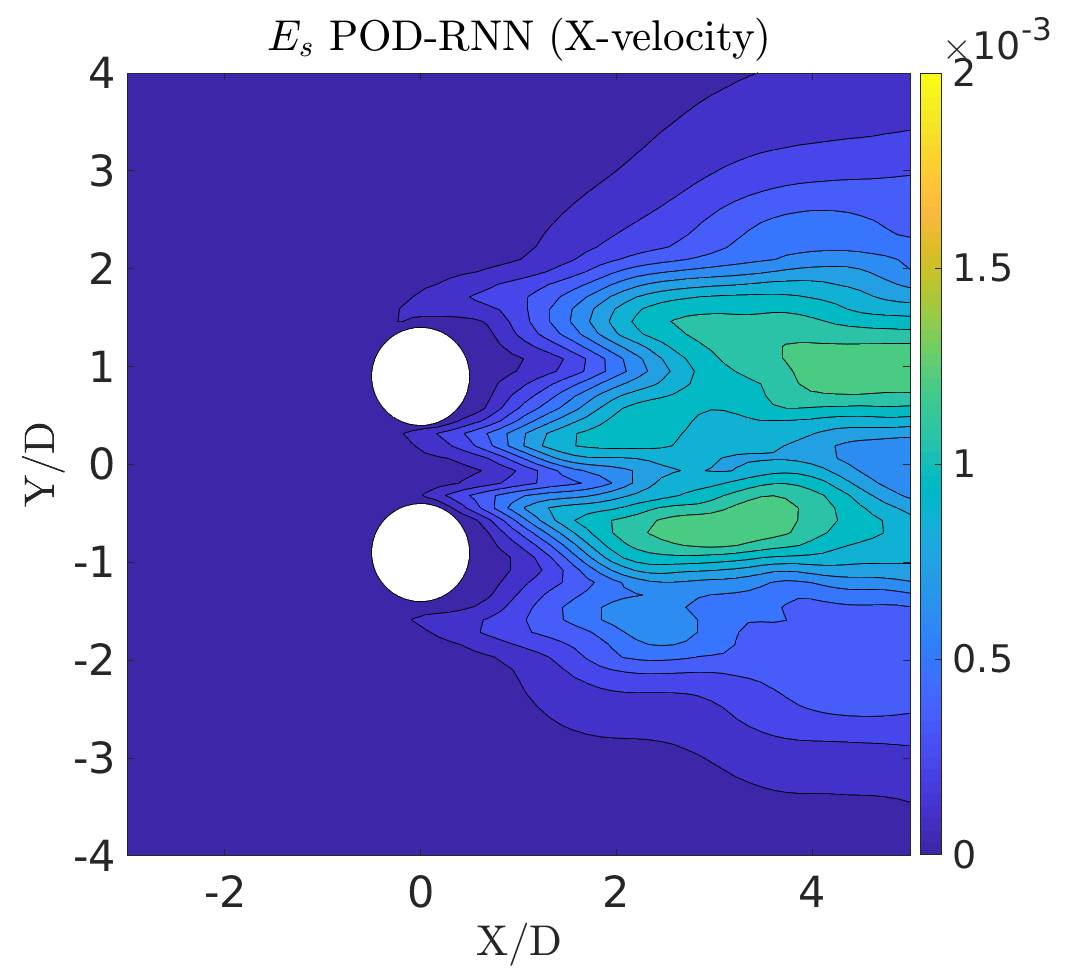}}
\subfloat[]
{\includegraphics[width = 0.238\textwidth]{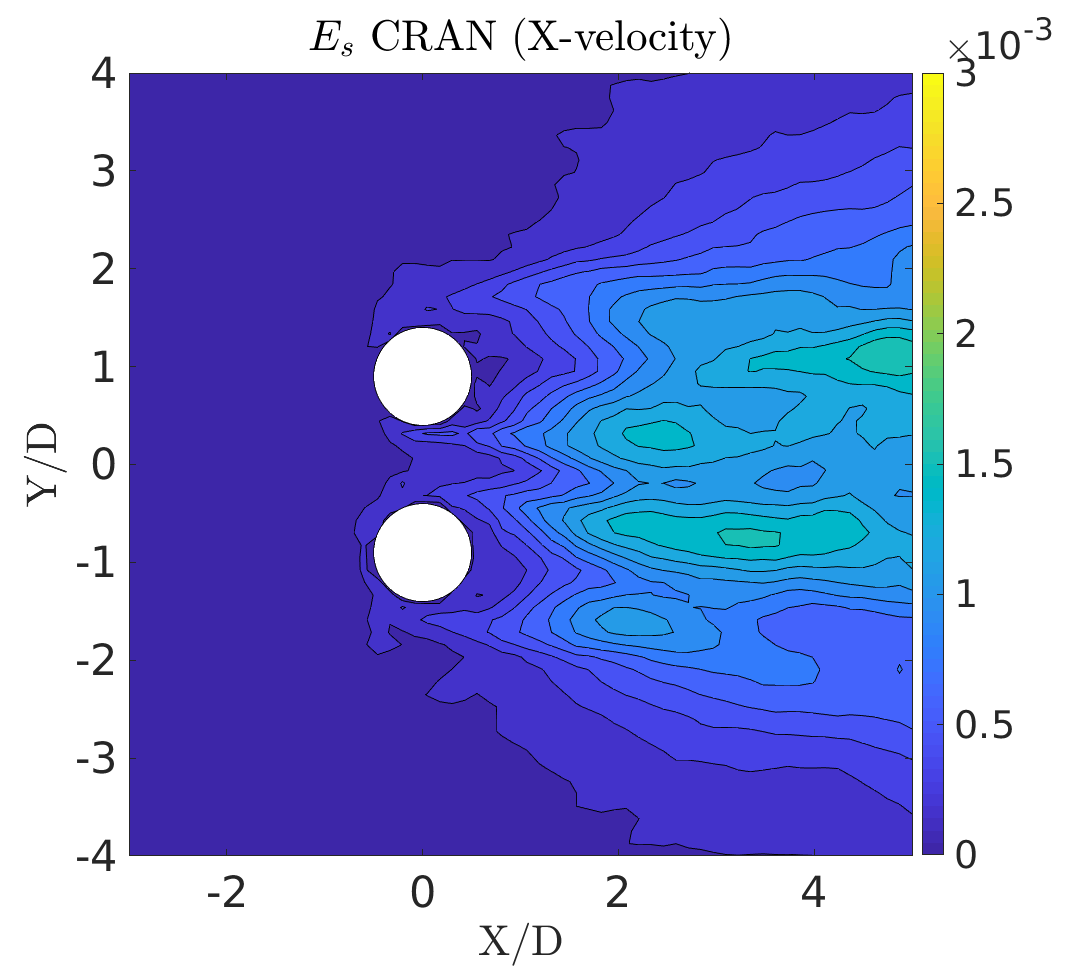}}

\caption{The flow past side-by-side cylinders: Comparison between the temporal mean error $E_{s}$ for POD-RNN and CRAN model. (a),(b) for pressure and (c),(d) for x-velocity.}
\label{comp_es_sbs}
\end{figure}
 
Fig.~\ref{comp_es_sbs} depicts the temporal mean of the reconstruction error $E_{s}$ for both the models over the test-time steps. The temporal mean error $E_{s}$ is quite low in the order of $10^{-3}$ for both the predictions. Similar to the plain cylinder, the majority of the errors are concentrated in the near wake region where there is presence of strong nonlinearity.

%\newpage
%%%%%%%%%%%%%%%%%%%%%%%%%%%%%%%%%%%%%%%%%%%%%%%%%%%%%%%%%%%%%%%%%%%%%%%%%%%%%%%%%%%%%%%%%%%%%%%%
\section{Conclusions}\label{conclusion}
%%%%%%%%%%%%%%%%%%%%%%%%%%%%%%%%%%%%%%%%%%%%%%%%%%%%%%%%%%%%%%%%%%%%%%%%%%%%%%%%%%%%%%%%%%%%%%%%
We have presented two data-driven reduced-order models for the prediction of nonlinear fluid flow past bluff bodies. Both methods share a common data-driven framework. The common principle of both methods is to first obtain a set of low-dimensional features of the high dimensional data coming from the fluid flow problems. These low dimensional features are then evolved in time via recurrent neural networks. In the first method, proper orthogonal decomposition is used to obtain a set of low-dimensional features (POD modes in this case). We have shown a mathematical equivalence of POD with autoencoders and then proceed to present a more general method where the low-dimensional features are obtained by convolutional neural networks. We have selected flow past a plain cylinder and flow past side-by-side cylinder as candidate test cases to evaluate the above methods. POD-RNN seems to be more efficient in terms of computational costs and accuracy of prediction when it comes to the problem of flow past a plain cylinder, CNN-RNN is a bit overkill for this problem. However, it performs extremely well and completely bypasses POD-RNN in terms of accuracy for the complicated problem of flow past side-by-side cylinders. This study shows that convolutional recurrent autoencoders can encompass a broader range of nonlinear fluid flow problems and can be pursued further for more complicated problems. We intend to extend this method to predict nonlinear fluid flow past moving bodies and we target to predict the motions of the bodies interacting with the fluid flow.  

%%%%%%%%%%%%%%%%%%%%%%%%%%%%%%%%%%%%%%%%%%%%%%%%%%%%%%%%%%%%%%%%%%%%%%%%%%%%%%%%%%%%%%%%%%%%%%%%%%%%%
\section*{Acknowledgements}
%%%%%%%%%%%%%%%%%%%%%%%%%%%%%%%%%%%%%%%%%%%%%%%%%%%%%%%%%%%%%%%%%%%%%%%%%%%%%%%%%%%%%%%%%%%%%%%%%%%%%
A part of work has been done at the Keppel-NUS Corporate Laboratory was supported by the National Research Foundation, Keppel Offshore and Marine Technology Centre (KOMtech) and National University of Singapore (NUS). The conclusions put forward to reflect the views of the authors alone and not necessarily those of the institutions within the Corporate Laboratory. 
Second and fourth authors would like to acknowledge the support from the University of British
Columbia (UBC) and the Natural Sciences and Engineering Research Council of Canada (NSERC).
The computational work for this article was performed on resources of the National Supercomputing Centre, Singapore and at Advanced Computing Resources of ARC at UBC.
Some part of the data presented in this work has been taken from our previous publications in \cite{bukka2019data,reddy2019data,bukka2020deep}.

% Authors must disclose all relationships or interests that 
% could have direct or potential influence or impart bias on 
% the work: 
%
\section*{Data Availability}
The data that support the findings of this study are openly available in \url{https://github.com/rachit1307-code/Assessment-of-hybrid-DLROM}
\section*{Conflict of interest}
The authors declare that they have no conflict of interest.

% BibTeX users please use one of
%\bibliographystyle{spbasic}      % basic style, author-year citations
%\bibliographystyle{spmpsci}      % mathematics and physical sciences
%\bibliographystyle{apsrmp4-1.bst}       % APS-like style for physics
\bibliography{template.bib}   % name your BibTeX data base

\end{document}